\documentclass[prd,twocolumn,nofootinbib,preprintnumbers]{revtex4-1}
\usepackage{epsfig,graphicx,color,verbatim,amsmath,amsfonts,amssymb}

\newcommand{\unit}{\leavevmode\hbox{\small1\kern-3.6pt\normalsize1}}

\newcommand{\sigsi}{\sigma^{SI}}
\newcommand{\sigsd}{\sigma^{SD}}

\newcommand{\sigsdp}{\sigma^{SD,\,p}}
\newcommand{\sigsdn}{\sigma^{SD,\,n}}

\newcommand{\sigsinu}{\sigma^{SI,\,N}_0}
\newcommand{\sigsdnu}{\sigma^{SD,\,N}_0}

\newcommand{\mwimp}{m_\chi}

\usepackage{color}
\def\lsim{\mathrel{\rlap{\lower4pt\hbox{\hskip1pt$\sim$}}
    \raise1pt\hbox{$<$}}}                % less than or approx. symbol
\def\gsim{\mathrel{\rlap{\lower4pt\hbox{\hskip1pt$\sim$}}
    \raise1pt\hbox{$>$}}}                % greater than or approx. symbol

\begin{document}

\title{
Nuclear uncertainties in the spin-dependent structure functions for direct dark matter detection
}
\author{D.~G.~Cerde\~no\,$^{1,2}$}
\author{M.~Fornasa\,$^{3}$}
\author{J.-H.~Huh\,$^{1,2,a}$}
\author{M.~Peir\'o\,$^{1,2,b}$}
\affiliation{$^{1}$ Instituto de F\'{\i}sica Te\'{o}rica, UAM/CSIC, Universidad Aut\'{o}noma de Madrid, Cantoblanco, E-28049, Madrid, Spain}
\affiliation{$^{2}$ Departamento de F\'{\i}sica Te\'{o}rica, Universidad Aut\'{o}noma de Madrid, Cantoblanco, E-28049, Madrid, Spain}
\affiliation{$^{3}$ School of Physics and Astronomy, University of Nottingham, University Park, NG7 2RD, United Kingdom}
\preprint{FTUAM-12-101}
\preprint{IFT-UAM/CSIC-12-86}

\begin{abstract}
We study the effect that uncertainties in the nuclear spin-dependent structure functions 
have in the determination of the dark matter (DM) parameters in a 
direct detection experiment. 
We show that different nuclear models that describe the spin-dependent structure function
of specific target nuclei can lead to variations in the reconstructed 
values of the DM mass and scattering cross-section. We propose a 
parametrization of the spin structure functions that allows us to treat these
uncertainties as variations of three parameters, with a central value and 
deviation that depend on the specific nucleus. 
The method is illustrated for germanium and xenon detectors
with an exposure of 300 kg yr, 
assuming a hypothetical detection of DM and studying a series of benchmark points for the DM properties. We 
find that the effect of these uncertainties can be similar in amplitude to that of 
astrophysical uncertainties, especially in those cases where the 
spin-dependent contribution to the elastic scattering cross-section is sizable. 
\end{abstract}

\maketitle

\section{Introduction} 
\label{sec:introduction}

\let\oldthefootnote\thefootnote
\renewcommand{\thefootnote}{\alph{footnote}}
\footnotetext[1]{MultiDark Fellow}
\footnotetext[2]{MultiDark Scholar}
\let\thefootnote\oldthefootnote
\setcounter{footnote}{0}

Direct searches of dark matter (DM) aim to observe this abundant but elusive 
component of the Universe by detecting its recoils off target nuclei of a 
detector (for a recent review, see, e.g. Ref.\,\cite{Cerdeno:2010jj}). 
A large number of experiments have been taking data in the last decades or are 
currently under construction with this objective, leading to a very exciting 
present situation.

In fact, some experiments have claimed potential signals that could be 
compatible with the detection of a weakly-interacting massive particle (WIMP). 
This is the case of the DAMA collaboration \cite{Bernabei:2003za}, which 
observed an annual modulation in the recoil rate on a NaI target that was 
later confirmed by the upgraded DAMA/LIBRA detector \cite{Bernabei:2008yi}. 
Similarly, the CoGeNT collaboration, with a germanium target, reported an 
irreducible excess in their data that could point towards very light WIMPs 
\cite{Aalseth:2010vx} and also observed an annual modulation effect 
\cite{Aalseth:2011wp} although the latter is not easy to reconcile with the 
DAMA/LIBRA result. Finally, the CRESST-II experiment, which uses CaWO$_4$ as a 
target, also reported an excess \cite{Angloher:2011uu} over the expected 
background.
However, these observations are in conflict with the negative results 
obtained in searches by other experimental collaborations. Experiments such 
as CDMS-II \cite{Ahmed:2009zw,Ahmed:2010wy}, XENON10 \cite{Angle:2011th}, 
XENON100 \cite{Aprile:2011hi,Aprile:2012}, EDELWEISS \cite{Armengaud:2011cy}, 
SIMPLE \cite{Felizardo:2011uw}, KIMS \cite{Kim:2012rz}, and a combination of 
CDMS and EDELWEISS data \cite{Ahmed:2011gh} are in strong tension with the 
regions of the parameter space compatible with WIMP signals in DAMA/LIBRA or 
CoGeNT. Moreover, a reanalysis of CDMS data has been performed in order to 
look for annual modulation with negative results \cite{Ahmed:2012vq}.

The elastic scattering cross-section of WIMPs off nuclei can be separated in 
two components, spin-independent (SI) $\sigsinu$, and spin-dependent (SD) 
$\sigsdnu$, which originate from different terms in the Lagrangian describing
the interaction of a DM particle with quarks. The SI term stems from scalar or 
vector couplings and its contribution to the total WIMP-nucleus cross-section 
scales as the nucleon number squared, $A^2$, whereas the SD term originates 
from axial-vector couplings and its total contribution to the cross-section 
off nuclei is only a function of the total nuclear angular momentum and the 
DM spin. Thus, the SI term typically dominates for heavy nuclei.

Constraints are normally expressed in terms of the SI and SD components of the WIMP-nucleon elastic cross-section, $\sigsi$ and $\sigsd$, respectively.
To date, the most stringent constraints on $\sigsi$ are those obtained from 
the XENON100 data \cite{Aprile:2012} that exclude SI cross-sections above 
$\sigsi \approx 2 \times 10^{-8}$ pb for a mass around 50 GeV, as well as 
XENON10 \cite{Angle:2011th} and the low-energy reanalysis of CDMS-II 
\cite{Ahmed:2010wy}, which dominate for light WIMPs.
Regarding the SD contribution, the leading bounds from direct detection 
experiments have been provided by XENON \cite{Angle:2008we} (SD cross-section 
with neutrons, $\sigsdn$) and COUPP \cite{Behnke:2010xt} and PICASSO 
\cite{Archambault:2012pm} (SD cross-section with protons, $\sigsdp$) but 
indirect detection experiments such as SuperKamiokande \cite{Tanaka:2011uf} 
and IceCube \cite{Abbasi:2009uz}, as well as searches for mono-jet \cite{Goodman:2010yf,Goodman:2010ku} and 
mono-photon plus missing energy in Tevatron \cite{Aaltonen:2012jb} and LHC 
\cite{Rajaraman:2011wf,Chatrchyan:2012pa,Fox:2011pm}  lead to even more compelling constraints 
on $\sigsdp$.
Larger and more sophisticated direct detection experiments are currently under
development that will be able to explore the DM parameter space with 
unprecedented sensitivity. This is the case, for example, of the 
SuperCDMS and XENON1T collaborations, which aim at the construction of 1 Ton scale 
detectors based on germanium and xenon, respectively.

In the light of this promising experimental situation, it seems plausible 
that the DM can be discovered in the near future in direct detection 
experiments. In such an event, the study of the signal rate and spectrum 
(differential rate) can be used to determine some of the DM properties, namely 
its mass, $\mwimp$, and elastic scattering cross-section 
\cite{Green:2007rb,Green:2008rd,Drees:2007hr}. 
The precision of this reconstruction is very sensitive to the characteristics 
of the detector and is affected by uncertainties in the parameters describing 
the DM halo, as well as in the nuclear form factors.
Astrophysical uncertainties have been widely discussed in the literature 
\cite{Green:2010gw,Akrami:2010dn,Green:2011bv,Strege:2012kv} and they are 
known to introduce significant errors in the determination of the mass and 
scattering cross-section of DM. Regarding nuclear uncertainties, those in the 
SI form factor have been argued to be relatively small \cite{Chen:2011im}. The effect of 
variations in the SD form factors has not been previously addressed and 
constitutes the objective of this work.

We consider the hypothetical future observation of a DM candidate in a direct 
detection experiment and, 
sampling over the three-dimensional space of
$(\mwimp,\,\sigsi,\,\sigsd)$, we investigate how 
the reconstruction of these quantities is affected by nuclear 
uncertainties in the spin-dependent structure function of the target nucleus. 
In order to do so, we propose a description of structure functions 
based on three parameters, which enlarge the parameter space sampled,
and allow us to incorporate uncertainties in a 
consistent and systematic way.
This provides a general method, applicable to any detector target.
We particularize our analysis for the case of a germanium detector (such as, 
e.g., SuperCDMS), for which we consider the spin-dependent structure functions 
provided by the analysis of various groups 
\cite{Bednyakov:2006ux,Ressell:1993qm,Dimitrov:1994gc}, and for xenon 
detectors (such as, e.g., the future XENON1T), for which we use the structure functions 
derived in Ref.\,\cite{Ressell:1997kx} and \cite{Menendez:2012tm}.

We observe that the effect of nuclear uncertainties in SD structure functions can 
lead to variations in the reconstructed DM mass and SD elastic cross-section, 
the effect being more important in those scenarios in which the SD term in 
the WIMP-nucleus cross-section is the main contribution to the total detection 
rate. 
In such cases uncertainties in the spin-dependent structure functions are similar in amplitude to those induced by astrophysical uncertainties in the DM halo parameters, although the latter also affect the SI component.

The paper is organized as follows. In Sec.\,\ref{sec:DD} we introduce the 
formalism used to compute the recoil event rate, emphasizing the role of SD 
interactions. We concentrate on the case of a germanium detector, introduce 
the models available in the literature that describe the spin-dependent structure function and 
comment on their differences. Sec.\,\ref{sec:Bayesian} describes the 
generation of the simulated data for a set of benchmark models, and the 
implementation of the scanning algorithm to probe the phenomenological 
parameter space. 
In Sec.\,\ref{sec:results} we show the reconstruction of DM parameters for 
each benchmark scenario, using different nuclear models for the SD structure function 
and investigating how this alters the predictions for the DM properties. 
In Sec.\,\ref{sec:parametrization} we present a parametrization of the SD structure function 
that allows us to systematically account for uncertainties
when scanning over our parameter space, and we apply the method to the cases 
of germanium and xenon detectors. Our conclusions are summarised in 
Sec.\,\ref{sec:conclusions}.

\section{Nuclear uncertainties in direct dark matter detection}
\label{sec:DD}

The differential event rate for the elastic scattering of a WIMP with mass 
$\mwimp$ off a nucleus with mass $m_N$ 
is given by
\begin{equation}
	\frac{dR}{dE_R} = \frac{\rho_0}{m_N \,\mwimp}
	\int_{v_{min}}^{v_{esc}} v f(v) \frac{d\sigma}{dE_R}(v,E_R)\, dv,
	\label{eqn:diff_rate}
\end{equation}
where $\rho_0$ is the local WIMP density and $f(v)$ is the WIMP velocity
distribution in the detector frame normalized to unity. The integration over 
the WIMP velocity $v$ is performed from the minimum needed to induce 
a recoil of energy $E_{R}$, $v_{min}=\sqrt{m_N E_R/2\mu_N^2}$, to the 
escape velocity, $v_{esc}$, above which WIMPs are not bound to the Milky 
Way. The WIMP-nucleus elastic scattering cross-section, $d\sigma/dE_R$, is 
expressed as a function of the recoil energy, and 
$\mu_N=m_N \mwimp/(m_N+\mwimp)$ is the reduced mass. The total event rate is 
calculated by integrating Eq.\,(\ref{eqn:diff_rate}) over all the possible 
recoil energies in a window defined by a threshold energy 
$E_T$ and a maximal energy $E_{max}$, both depending on the 
experiment\footnote{In order to take into account the energy 
resolution of the detector, the differential rate is convoluted with a 
Gaussian, whose  standard deviation is a function of the recoil energy, as 
done in Ref.\,\cite{Pato:2010zk}.}.

In general, the WIMP-nucleus cross-section is separated into a SI and a SD
contribution, as follows:
\begin{equation}
	\frac{d\sigma}{dE_R} = \frac{m_N}{2 \mu_N^2 v^2}
	\left(\sigsinu F^2_{SI}(E_R) + 
	\sigsdnu F^2_{SD}(E_R) \right),
	\label{eqn:SI_and_SD}
\end{equation}
where $\sigsinu$ and $\sigsdnu$ are the SI and SD WIMP-nucleus cross-sections 
at zero momentum transfer. $F_{SI}(E_R)$ and $F_{SD}(E_R)$ are the SI and 
SD form factors that account for the coherence loss which leads to a 
suppression of the event rate for heavy WIMPs or heavy nuclei. The 
differential rate, $dR/dE_R$, depends on the recoil energy $E_R$ through the 
form factors and the minimal velocity $v_{min}(E_R)$.

The total number of recoils, as well as their distribution in energy, are 
affected by uncertainties in the nuclear form factors (both SI and SD) and in 
the parameters describing the DM halo (usually referred to as astrophysical 
uncertainties). Determining the impact of these is crucial to understand the 
capability of a DM experiment to reconstruct the properties of the WIMP.

The role of astrophysical uncertainties has been widely addressed in the 
literature. They are known to significantly affect the reconstruction of both 
the mass and scattering cross-section of the DM
\cite{Vergados:2007nc,Kuhlen:2009vh,Lisanti:2010qx,Green:2010gw,Green:2011bv,Fairbairn:2012zs}
Since the subject of our work is to 
study the effect of nuclear uncertainties from the form factors, we do not 
include astrophysical ones. We therefore consider a fixed
model for the the DM halo, namely the Standard Halo Model with a escape 
velocity of $v_{esc}=544$~km s$^{-1}$ , a central velocity $v_0=230$~km s$^{-1}$ 
\cite{Kerr:1986hz,Reid:2009nj,McMillan:2009yr,Bovy:2009dr,Savage:2009mk},
and a local dark matter density $\rho_0=0.4$ GeV cm$^{-3}$ 
\cite{Catena:2009mf,Salucci:2010qr,Pato:2010yq,Iocco:2011jz}.

\subsection{Uncertainties in the SI form factors} 
Regarding SI interactions, the so-called Woods-Saxon form factor is the 
Fourier transform of the nucleon distribution function $\rho_A(x)$,
\begin{equation}
	F_{SI}(q)=\int e^{-iqx} \rho_A(x) d^3x\,,
	\label{eqn:SI_FF}
\end{equation}
where $q=\sqrt{2 m_N E_R}$ is the momentum transfer. The Fermi 
distribution is assumed for the nucleon distribution,
\begin{equation}
	\rho _A(x) \propto \frac{1}{1+\exp[(r-R_A)/a]}\,,
\end{equation}
where $R_A=(1.23A^{1/3}-0.6)$~fm, $A$ is the nucleon number and $a=0.5$~fm the 
surface thickness of the nucleus. 
Although other parametrizations can be found in the literature,
the Wood-Saxon form factor provides a good description of the nuclear 
structure for energies in the range between 1-100~keV, typical of WIMP scatterings.
It has been shown in Ref.\,\cite{Chen:2011im} that the differences in the SI form factors due to small deformations of the nuclei can be safely neglected. In fact, we have
explicitly checked that this is indeed the case when using realistic nuclear 
density profiles obtained from a state-of-the art mean field calculation.
Thus, throughout this paper we consider the form factor in Eq.\,(\ref{eqn:SI_FF}) with 
no associated uncertainty.

\subsection{Uncertainties in the SD form factors} 
\label{sec:nuclear_uncertainties}

On the other hand, the effect of uncertainties in the SD form factors has not 
been addressed in the literature. The SD contribution to the WIMP-nucleus 
differential cross-section in Eq.\,(\ref{eqn:SI_and_SD}) can be expanded as 
a function of the WIMP couplings to the matrix elements of the axial-vector 
currents in protons ($a_p$) and neutrons ($a_n$),
\begin{eqnarray}
	\left(\frac{d\sigma}{dE_R}\right)_{SD} & = & \frac{16\,G_F^2 m_N}{\pi v^2}
	\frac{(J+1)}{J} \nonumber\\
	& & \left(a_p \langle S_p \rangle + a_n \langle S_n \rangle \right)^2  
	F^2_{SD}(E_R)\,, 
	\label{eqn:SD_cross_section}
\end{eqnarray}
where $J$ is the total spin of the nucleus and $\langle S_p \rangle$ 
($\langle S_n \rangle$) is the proton (neutron) spin averaged over the 
nucleus. The SD form factor $F^2_{SD}(E_R)=S(E_R)/S(0)$, is commonly expressed 
as a decomposition into isoscalar ($a_0=a_p+a_n$) and isovector ($a_1=a_p-a_n$) 
couplings,
\begin{equation}
	S(q)=a_0^2 S_{00}(q)+a_0a_1S_{01}(q)+a_1^2S_{11}(q),
\end{equation}
where $q$ is the momentum transfer. The quantities $S_{00}(q)$, $S_{11}(q)$ and 
$S_{01}(q)$ are the spin-dependent structure functions (SDSFs), 
and are computed using nuclear physics models, whereas the couplings 
$a_p$ and $a_n$ (and consequently $a_0$ and $a_1$) are specific of the particle 
physics model for DM and are computed from the diagrams describing the 
WIMP-nucleon interaction. 
In order to continue with a model independent approach we assume a 
specific relation between $a_p$ and $a_n$, and consider the 
cases\footnote{This is equivalent to reducing by one the dimensionality of our 
parameter space, assuming a relation between $\sigsdp$ and $\sigsdn$. Our analysis can easily be 
extended to consider the full four-dimensional parameter space 
$(\mwimp,\,\sigsi,\,\sigsdp,\,\sigsdn)$, but this renders the discussion more 
cumbersome. Furthermore, particle models for DM generally predict 
$|\sigsdn|\approx|\sigsdp|$.}
$a_p/a_n=\pm 1$.
Under this assumption, Eq.\,(\ref{eqn:SD_cross_section}) reduces to
\begin{equation}
	\left(\frac{d\sigma}{dE_R}\right)_{SD}=\left\{ 
	\begin{array}{lcr}
	\frac{64\,G_F^2 m_N}{v^2 (2J+1)}
	a_p^2\,S_{00}(q) 
	&;
	&\frac{a_p}{a_n}=1\,, \\
	\frac{64\,G_F^2 m_N}{v^2 (2J+1)} 
	a_p^2\,S_{11}(q) 
	\rule{0pt}{4ex}
	&;
	&\frac{a_p}{a_n}=-1\,. \\
	\end{array}
	\right.
\label{eqn:s11}
\end{equation}

\begin{figure*}
	\includegraphics[width=0.5\textwidth]{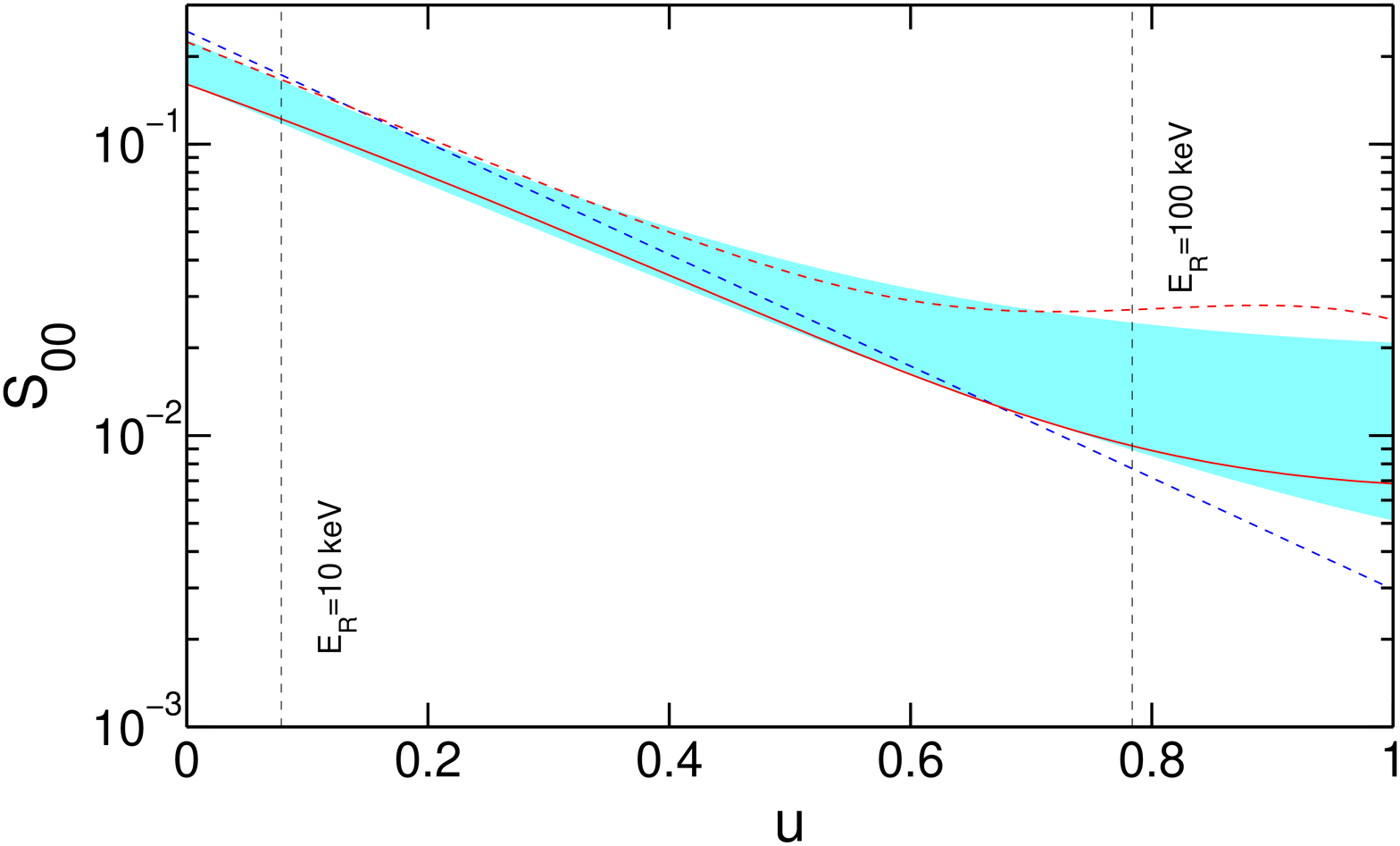}\hspace*{-0.62cm}
	\includegraphics[width=0.5\textwidth]{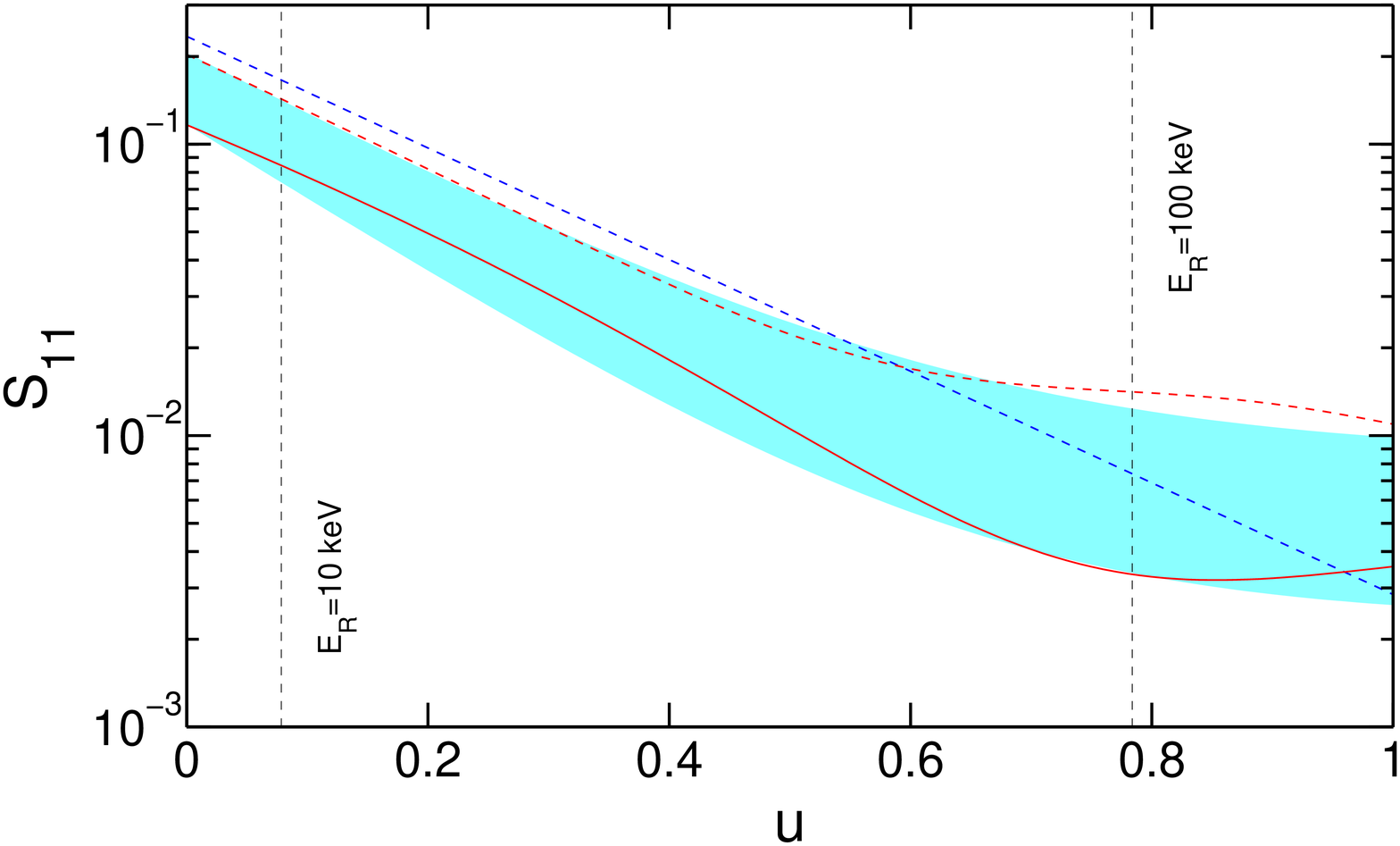}\\[-4ex]
\caption{\label{fig:FFs} Spin-dependent structure functions as a function of $u$, in the case of $a_p/a_n=1$ (left panel) or $a_p/a_n=-1$ (right panel). The solid (dashed) red lines correspond to the D-model \cite{Ressell:1993qm} (R-model \cite{Dimitrov:1994gc}) and the dotted blue line indicates the gaussian approximation of Eq.\,(\ref{eqn:gaussian}). The blue region covers the area spanned by the family of curves in Eq.\,(\ref{eqn:family}). The vertical black dashed lines indicate the WIMP search window used in the analysis.}
\end{figure*}

The SDSFs $S_{00}(q)$ and $S_{11}(q)$ can be calculated using a shell-model (ShM) description of the atomic nucleus, where the nuclear spin properties are obtained by the wave functions of a few valence nucleons, those which do not cancel out the spin of the nucleus in pairs. In particular, $S_{00}(q)$ and $S_{11}(q)$ are related to the transverse electric and longitudinal projections of the axial current. To calculate these quantities in the ShM, the nucleons are placed in energy levels according to the exclusion principle, assuming a particular interaction between nucleons (typically a harmonic oscillator potential) and including as many excited states as possible, making this kind of calculation very difficult.

ShM calculations are generally more reliable for heavy nuclei than for light ones. The same holds for nuclei close to magic numbers, elements featuring closed shells being more easily modeled. 
An example is $^{19}$F, that has 9 protons and 10 neutrons, thus only one proton above a magic number. On the other hand, the nucleus of $^{73}$Ge is much more difficult to model since it has 32 protons and 41 neutrons, the nearest closed shell being the one with 28 nucleons. 
In this case, deviations of the real nucleus from the ShM should be expected, as well as differences in the results when different ShMs are used. In the first part of the paper we consider the case of germanium, for which the only natural isotope that contributes to the SD cross-section is $^{73}$Ge.

In the case of $^{73}$Ge, various ShM calculations are available in the literature. We consider two different, commonly used parametrizations, from Ressel et al. \cite{Ressell:1993qm} and Dimitrov et al. \cite{Dimitrov:1994gc}, to which we refer as R- and D-models, respectively. 
They differ in the methodology and in the choice of the nuclear interaction potential, but both reproduce the value of the magnetic momentum of $^{73}$Ge. 
The SDSFs in both cases can be expressed as a function of the adimensional quantity $u$, related to the momentum transfer as $u=(qb)^2/2$, where $b$ is the oscillator size parameter, $b=A^{1/6}$.

The SDSFs for the R-and D-models are plotted as a function of $u$ in Fig.\,\ref{fig:FFs} by means of red dashed and solid red lines, respectively. The left (right) panel refers to the case $a_p/a_n=1$ ($a_p/a_n=-1$). 
The vertical, black dashed lines indicate the values of $u$ that correspond to the WIMP search window that we use in our analysis, from a threshold energy of 10 keV, to an energy of 100 keV (as currently done in CDMS-II).
The dotted blue lines indicate a gaussian approximation (see Eq.\,(\ref{eqn:gaussian}) below). Finally, the blue areas represent the regions spanned by a family of curves, obtained by a parametrization which interpolates between the R- and D-models that will be introduced in Sec.\,\ref{sec:parametrization}.

The two SDSFs differ in the zero momentum value (the R-model being larger for the whole energy range of interest for direct detection), and also in the shape at large energies. They both start as decreasing power-laws at low-energy flattening out as $u$ increases. However the transition happens sooner for the R-model (around $u=0.5$) than for the D-model. The slope for the D-model is also slightly steeper than for the R-model, especially in $S_{11}(q)$. As we will see in Sec. \ref{sec:results} these differences play an important role when determining the DM parameters.

There are finally some nuclei for which ShM computations of their form factors are not available. In these cases an approximation was introduced in Ref.\,\cite{Belanger:2008sj} that works well in the low momentum transfer
regime, but fails towards larger values of $q$,
\begin{equation}
	S_{ij}(q)=S(0)\,e^{-\frac{q^2R^2}{4}}\,,
\label{eqn:gaussian}
\end{equation} 
where $R$, is an effective radius, measured in fm, which can be written 
as,
\begin{equation}
	R  = 0.92\,A^{1/3} + 2.68 
	 - 0.78 \sqrt{(A^{1/3}-3.8)^2+0.2}\,. 
\end{equation}

\section{Determination of WIMP properties}
\label{sec:Bayesian}

We consider a set of benchmark scenarios (BM1, BM2 and BM3) listed in Table\,\ref{tab:benchmarks}, that define the phenomenological DM parameters $(\mwimp,\,\sigsi,\,\sigsd)$. 
These benchmarks are consistent with possible particle physics models for DM\footnote{In particular, the three benchmarks can be obtained within the context of neutralino DM in the general Minimal Supersymmetric Standard Model.}. 
We then assume the observation of a DM signal in a given direct detection experiment. 
The differential rate is computed for each benchmark point following Eq.\,(\ref{eqn:diff_rate}), and used to derive the total number of events $\lambda$.

\begin{table}
\begin{center}
\begin{tabular}{|c|ccc|ccc|}
	\hline
	& $\mwimp$ [GeV] & $\sigsi$ [pb] & $\sigsd$ [pb] &
	$\lambda$ & $\lambda^{SI}$ & $\lambda^{SD}$ \\
	\hline
	BM1 & 100 & $10^{-9}$ & $10^{-5}$	& 37.2 	& 36.4 	& 0.8\\
	BM2 & 50 & $10^{-9}$ & $10^{-5}$ 	& 42.1	& 41.2	& 0.9\\
	BM3 & 100 & $10^{-9}$ & $10^{-3}$	& 79.6	& 36.4 	& 43.2 \\
	\hline
\end{tabular}
\end{center}
\caption{\label{tab:benchmarks} Phenomenological parameters defining the three benchmark models. We include the predicted total number of recoil events, $\lambda$, as well as the number of events (calculated using the R-model) $\lambda^{SI}$ ($\lambda^{SD}$) due to SI (SD) interactions, for the experimental setup described in the text.}
\end{table}

We first particularize our analysis for the case of a germanium detector with a 
total exposure of $\epsilon=300$~kg yr. This could, e.g., correspond to the 
1 Ton phase of SuperCDMS, operating for a whole year with an efficiency of 
30\%. 
We define the energy window for WIMP searches in the range $E_T=10$~keV and $E_{max}=100$~keV, and calculate the 
number of events $\{ \lambda_i \}$ in a series of energy bins 
$\{E_i,E_i+\Delta E\}$ with $\Delta E=5$~keV. 
We also include a background with a rate of $4 \times 10^{-8}$ days$^{-1}$ kg$^{-1}$ keV$^{-1}$, which is of the order of the background expected for the SuperCDMS experiment in SNOLAB \cite{talk}. 
For the considered exposure this 
means a total of $0.02$ background events in each of the energy bins considered (i.e., we are almost dealing with a background free experiment). We assume that this background is flat (energy independent).

\begin{figure*}
	\hspace*{-0.4cm}
	\includegraphics[width=0.35\textwidth]{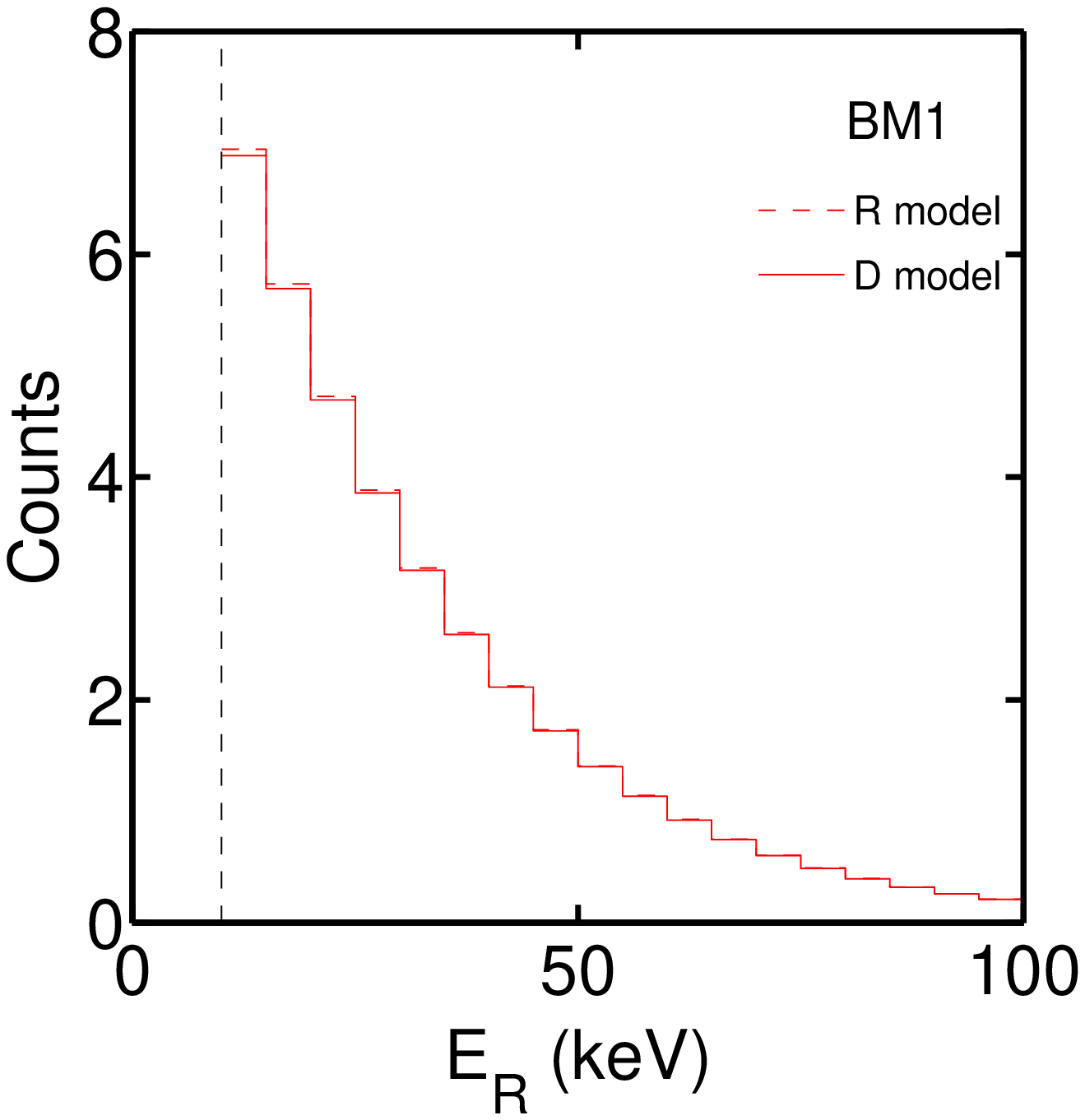}\hspace*{-0.4cm}
	\includegraphics[width=0.35\textwidth]{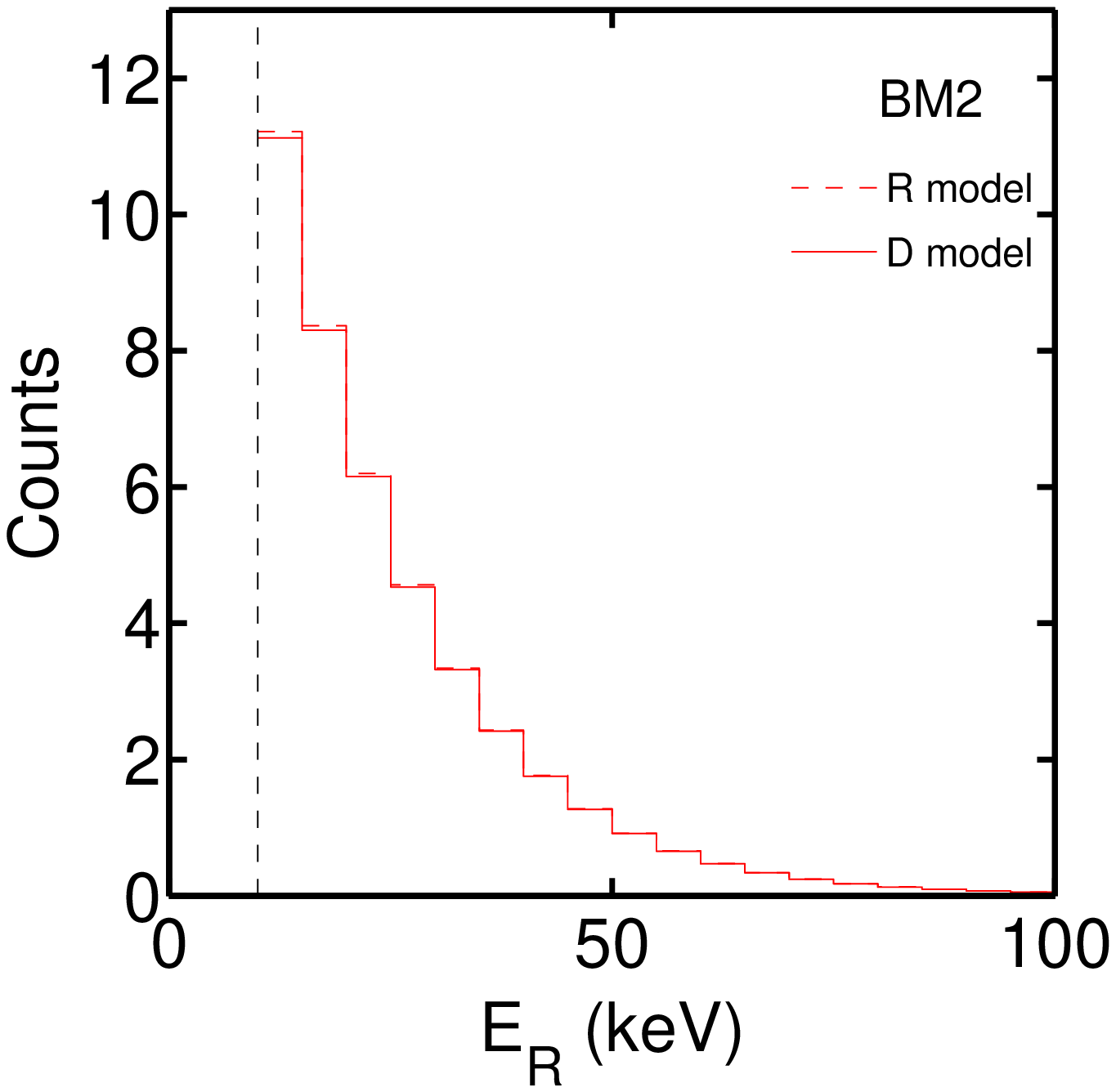}\hspace*{-0.4cm}
	\includegraphics[width=0.35\textwidth]{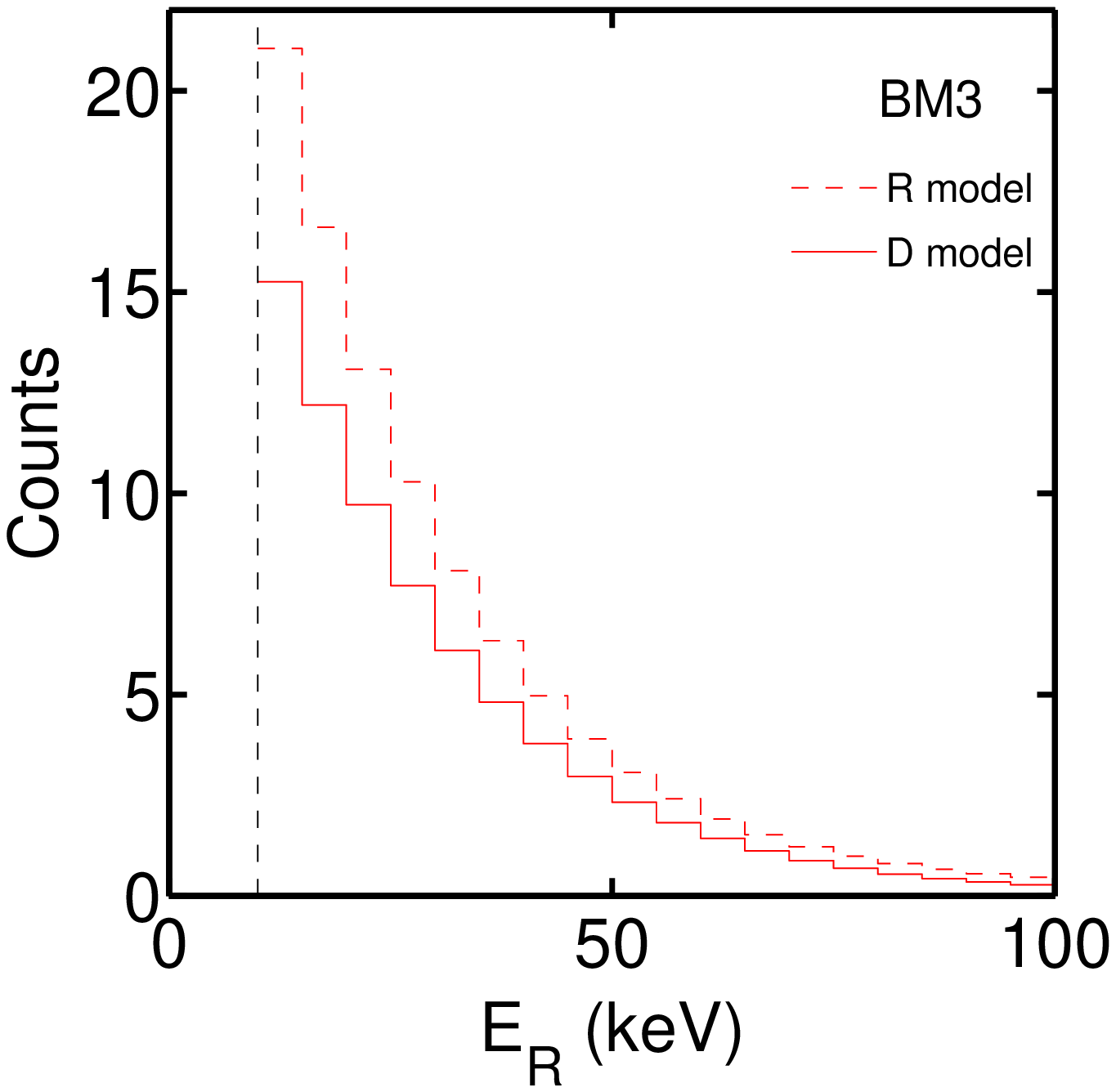}\\[-4ex]
\caption{\label{fig:benchmark_rate} Predicted DM spectra for benchmarks BM1, BM2 and BM3 (from left to right) for the experimental setup described in the text. The solid and dashed red lines correspond to the predictions using the R- and D-model for the SDSF, respectively. The vertical dashed line indicates the energy threshold $E_T$.}
\end{figure*}

The simulated energy spectra for the three benchmark points can be seen in Fig.\,\ref{fig:benchmark_rate}, where the solid red line corresponds to the results when the D-model is used and the dashed red line is obtained for the R-model. 
Practically no difference is observed for benchmarks BM1 and BM2.
This happens because for these two benchmark points the main contribution to the total detection rate is due to SI interactions. On the contrary, in BM3 the SD contribution is important and we observe how the predicted rate is significantly higher for the R-model than for the D-model. This is a consequence of the higher value of $S_{11}$ for the R-model in the whole energy range (see Fig.\,\ref{fig:FFs}).

We treat the quantities $\{ \lambda_i \}$ as the experimental information from which DM parameters have to be reconstructed. Our analysis is based on the Bayes theorem, which determines the posterior probability distribution (pdf) $p(\mathbf{\Theta}|\mathbf{D})$ of a set of parameters $\mathbf{\Theta}$ (for which a prior probability is assumed $p(\mathbf{\Theta})$) from a set of experimental data $\mathbf{D}$, encoded in the likelihood function $p(\mathbf{D}|\mathbf{\Theta})$ (or $\mathcal{L}(\mathbf{\Theta}$)),
\begin{equation}
	p(\mathbf{\Theta}|\mathbf{D}) = 
	\frac{p(\mathbf{D}|\mathbf{\Theta}) p(\mathbf{\Theta})}{p({\mathbf{D})}}\,.
	\label{eqn:Bayes}
\end{equation}

The evidence $p(\mathbf{D})$ in the denominator of Eq.\,(\ref{eqn:Bayes}) is a function of only the experimental data. For our purposes it works as a normalization factor and can therefore be ignored. 
The pdf in Eq.\,(\ref{eqn:Bayes}), in principle, depends on the priors $p(\mathbf{\Theta})$ and different choices of priors can affect the shape of the final pdf. 
However, should this happen, it would mean that the experimental data are not constraining enough and do not dominate the final probability distribution. Residual prior dependence can be seen, e.g., in Refs.\,\cite{Bertone:2011nj,Feroz:2011bj,deAustri:2006pe}.
Our scans are performed with MultiNest 2.9 \cite{Feroz:2008xx,Feroz:2007kg} interfaced with our own code for the computation of the number of recoil events and the likelihood. Scans are performed with 20000 live points and a tolerance of 0.0001.

In our case, for a given benchmark point the experimental data is $\mathbf{D}=(\{\lambda_i\})$ and a scan of the parameter space $\mathbf{\Theta}=(\mwimp,\,\sigsi,\,\sigsd)$ is performed. The ranges considered are $\mwimp=1-10^5$~GeV, $\sigsi=10^{-12}-10^{-6}$~pb, and $\sigsd=10^{-8}-1$~pb. Logarithmic priors are assumed for the three variables since the range scanned is quite large spanning up to eight orders of magnitude.

The likelihood $\mathcal{L}(\mathbf{\Theta})$ is calculated for each point in the scan, computing 
the number of recoil events $N_i$ in the $i$-th bin, and comparing it with the prediction of the benchmark model in the same bin, $\lambda_i$,
assuming that each experimental data follows an independent Poissonian distribution,
\begin{equation}
	\mathcal{L}(\Theta) = 
        \prod_i \frac{N_i(\mathbf{\Theta})^{\lambda_i} 
          e^{N_i(\mathbf{\Theta})}}{\lambda_i!}\,.
	\label{eqn:likelihood}
\end{equation}
The number of recoil events $N_i$ in the $i$-th bin are obtained integrating Eq.\,(\ref{eqn:diff_rate}) between $E_i$ and $E_i+\Delta E$, and including a certain number of background events $b_i$. The latter is included as a nuisance parameter in our scans, following a Poissonian distribution function with a mean value of 0.02.

The results of our scans are plotted in the next sections by means of one- or two-dimensional plots. When the probability for a subset of the original $\mathbf{\Theta}$ is considered, one can account for the presence of the hidden parameters in two different ways:
\begin{itemize}
\item by marginalizing over them, obtaining the pdf for the $j$-th parameter integrating over all the others
\begin{equation}
	p(\Theta_j|\mathbf{D}) = \int p(\mathbf{\Theta}|\mathbf{D}) \,
	d\Theta_1 ... \, d\Theta_{j-1} \, d\Theta_{j+1} \, d\Theta_n\,;
	\label{eqn:marginalization}
\end{equation}
\item by maximizing over them, obtaining the so-called profile likelihood
\begin{equation}
	\mathcal{L}(\Theta_j)= \max_{\Theta_1, ..., \Theta_{j-1},\Theta_{j+1},\Theta_n}
	\mathcal{L}(\mathbf{\Theta})\,.
\end{equation}
\end{itemize}

The profile likelihood is usually more sensitive to small fine-tuned regions with large likelihood, while the integration implemented for the pdf accounts for volume effects. Thus, a parameter space characterized by a complicated likelihood function may result in different pdf and profile likelihood distributions for the same parameter.
In the following we will present plots for both the the pdf and the profile likelihood since they have different statistical meanings and they provide complementary information.

\begin{figure*}
	\includegraphics[width=0.35\textwidth]{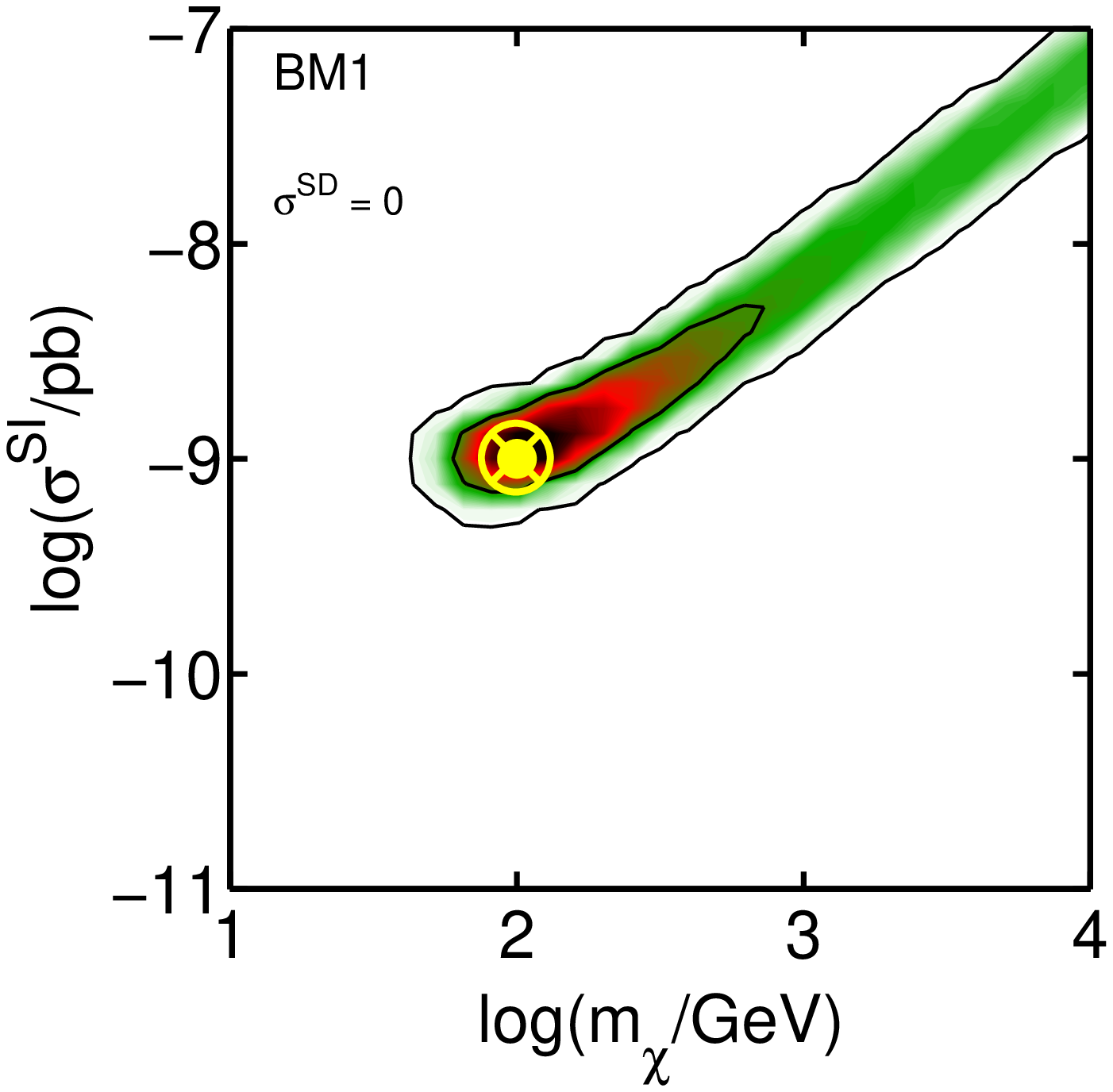}\hspace*{-0.62cm}
	\includegraphics[width=0.35\textwidth]{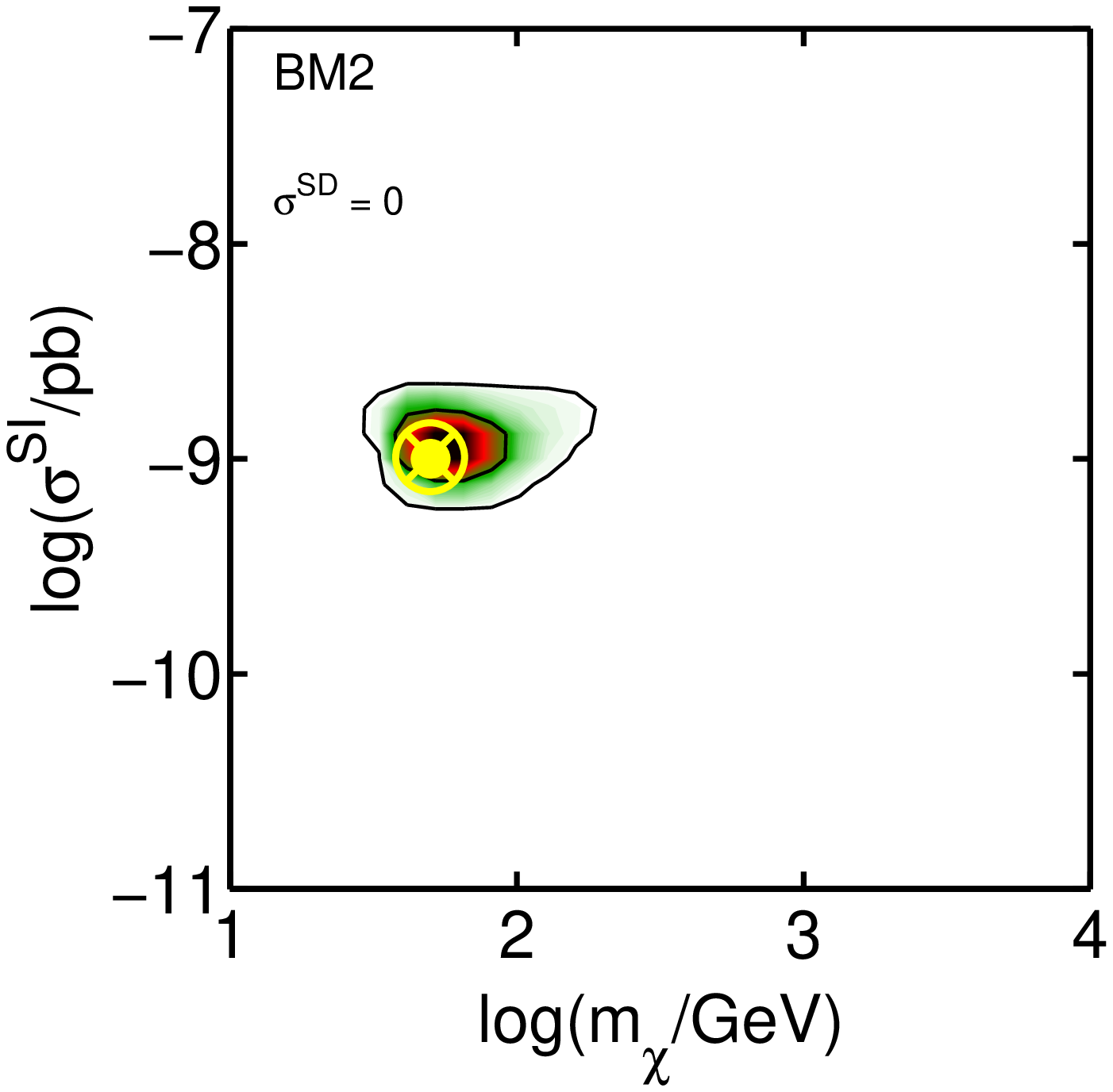}\hspace*{-0.62cm}
	\includegraphics[width=0.35\textwidth]{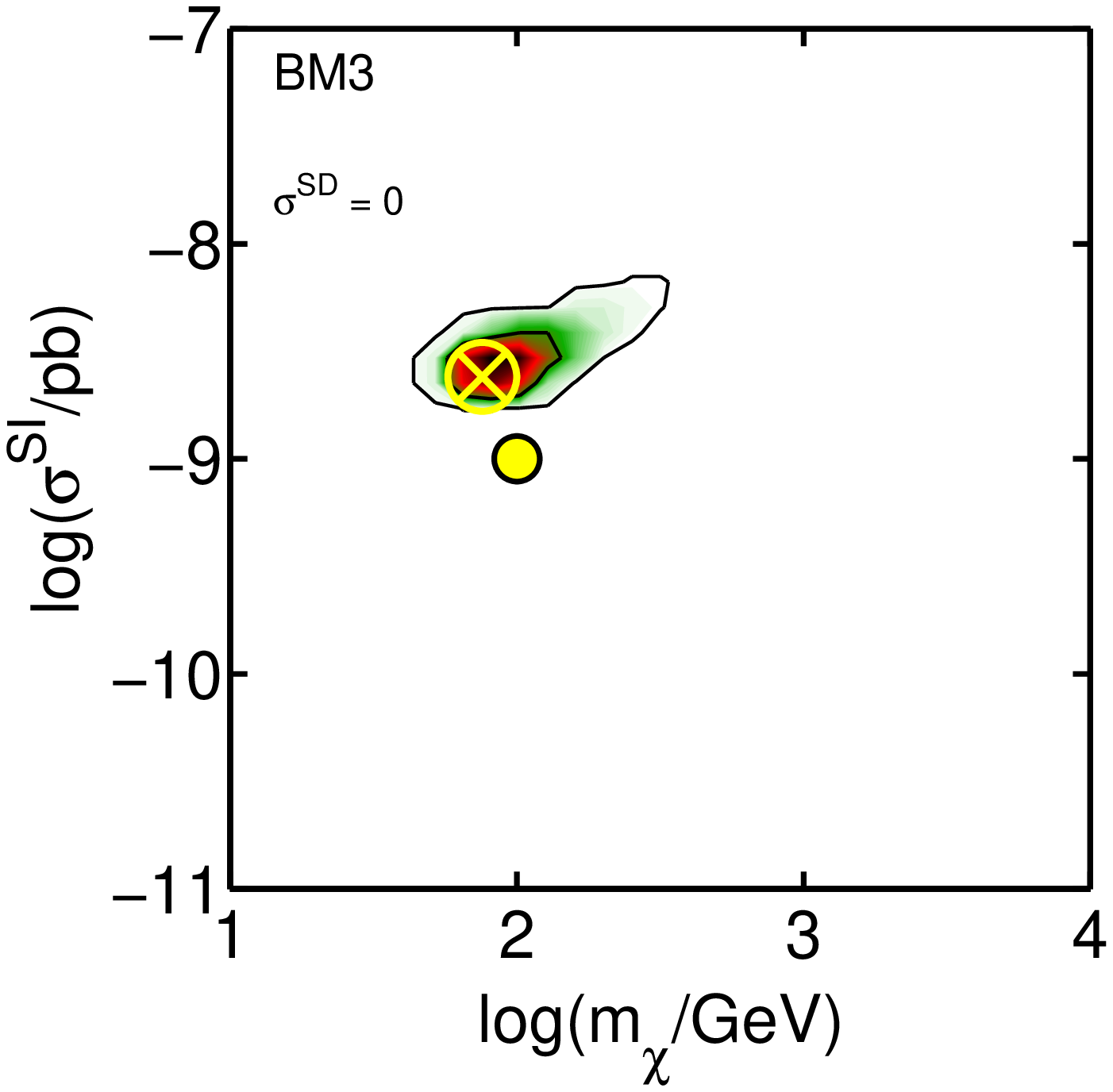}\\[-4ex]
\caption{\label{fig:noSD} Two-dimensional profile likelihood in the plane $(\mwimp,\, \sigsi$) for benchmarks BM1, BM2 and BM3, from left to right, assuming in the reconstruction that $\sigsd=0$. The inner and outer contours are 68\% and 99\% confidence level regions, respectively. The yellow dot indicates the benchmark values in each case, and the encircled yellow cross marks the positions of the best-fit point.}
\end{figure*}

It was recently pointed out in Ref.\,\cite{Strege:2012kv} that the method outlined here is affected by an intrinsic statistical limitation associated with the fact that only one set of simulated data is obtained for each benchmark point. In our work we do not incorporate this effect, since we want to isolate the variations due to nuclear uncertainties. Furthermore, our benchmark points are in regions with good coverage (see Figs.\,2 or 4 in Ref.\,\cite{Strege:2012kv}) and the number of events is relatively large.

\section{Results}
\label{sec:results}

We begin by considering the three benchmarks BM1, BM2 and BM3, with the total rate and energy spectrum of nuclear recoils as given in Table\,\ref{tab:benchmarks}. For concreteness we consider explicitly the case $a_p/a_n=-1$, which implies looking at only the $S_{11}$ component of the SDSF, according to Eq.(\ref{eqn:s11}). Other choices of $a_p/a_n$ lead to qualitatively similar results (but then a linear combination with the other components, $S_{00}$ and $S_{01}$ also appears). 
We emphasize at this point in that a full analysis can be done without fixing this ratio and including a fourth parameter in the scan, however this makes it more difficult to isolate the effects of uncertainties in the SDSF.

As a first exercise in parameter reconstruction, we assume (as it is often done) that the SD contribution is vanishing and we attempt to reconstruct the WIMP mass and SI cross-section from the experimental data (of course, this assumption is not made when preparing the simulated data from the benchmark points).

The resulting two-dimensional profile likelihood for these quantities\footnote{Note that, since we are scanning over only two parameters, there is no need of marginalization or maximization.} are given in Fig.\,\ref{fig:noSD}. 
An obvious thing to observe is that the reconstruction of these two parameters is good for benchmarks BM1 and BM2, since in these cases the SI contribution is the dominant term in the detection rate. 
This is obviously not the case in BM3, where SD interactions play a more important role. For this benchmark point, ignoring the SD contribution term leads to an overestimation of the SI independent cross-section of approximately a factor two (in order to account for the total detected rate).

Another feature that can be observed, and is consistent with the existing literature, is that the goodness of the reconstruction is very dependent on the mass of the DM candidate \cite{Green:2007rb,Green:2008rd} (see also Refs.\,\cite{Pato:2010zk,Peter:2011eu,Strege:2012kv}). In particular, we can see how in benchmark BM1 the 99\% confidence level contours are open for heavy WIMPs, whereas this is not the case for BM2 and BM3. 
In principle, increasing the DM mass makes the recoil energy spectrum flatter, as a consequence of the dependence of $v_{min}$ which enters through the reduced mass.
Thus one expects to produce a worse fit to the recoil spectrum as we scan more massive DM candidates. In benchmarks BM2 and BM3 this is the reason why heavy masses are disfavoured, however in BM1 the number of events in each energy bin is too low to pick up this tendency and very massive DM candidates can still produce good fit to the data. 
The presence of a flat background also enhances this effect.
On the other hand, the contours do not extend to low WIMP masses because particles with masses below $\mwimp\sim 30-40$~GeV produce a much steeper spectrum. 
Notice finally that the assumption $\sigsd=0$ leads to a lower limit for the SI cross-section that allows us to reconstruct the value of $\sigsi$ up to approximately a factor 5 (for a fixed value of DM mass).

An unbiased reconstruction of DM parameters, however, has to include the possibility that $\sigsd\neq 0$. In fact, when we allow for a non-negligible SD contribution to the WIMP scattering cross-section, we find that a new degeneracy in the parameter space arises: the same detected rate can be explained by a DM with either pure SD or pure SI interactions or, in general, a given combination of both as we see from Eqs.\,(\ref{eqn:diff_rate}) and (\ref{eqn:SI_and_SD}). 
This implies that the closed contours in Fig.\,\ref{fig:noSD} can extend towards arbitrarily small values of $\sigsi$.
It is when we acknowledge this possibility that uncertainties in the SDSF play a non-trivial role, as they affect the total rate and energy spectrum of WIMP recoils.

Choosing germanium as a case study, we consider the two calculations, R- and D-models, for the SDSF of $^{73}$Ge (the isotope that contributes to the SD cross-section) that were introduced in Sec.\,\ref{sec:nuclear_uncertainties}, performing the Bayesian inference for both SDSFs. Strictly speaking, we have to select one SDSF from which the simulated data for a given benchmark is generated and, then, one SDSF for the computation of the likelihood in Eq.\,(\ref{eqn:likelihood}) at the moment of the scan. 
The two choices are independent, leaving us with four possibilities of combining the two SDSFs. In particular, we can either use the same (R- or D-) model for the generation of the simulated experimental data and for the parameter reconstruction or we can generate the data with one model and perform the reconstruction with the other one. 
This last possibility gives an idea on how sizable the mis-reconstruction of DM parameters can be if we use an incorrect model for the SDSF, this is, one different to the real one.
We perform this computation for the three benchmark points.

\begin{figure*}
\includegraphics[width=0.35\textwidth]{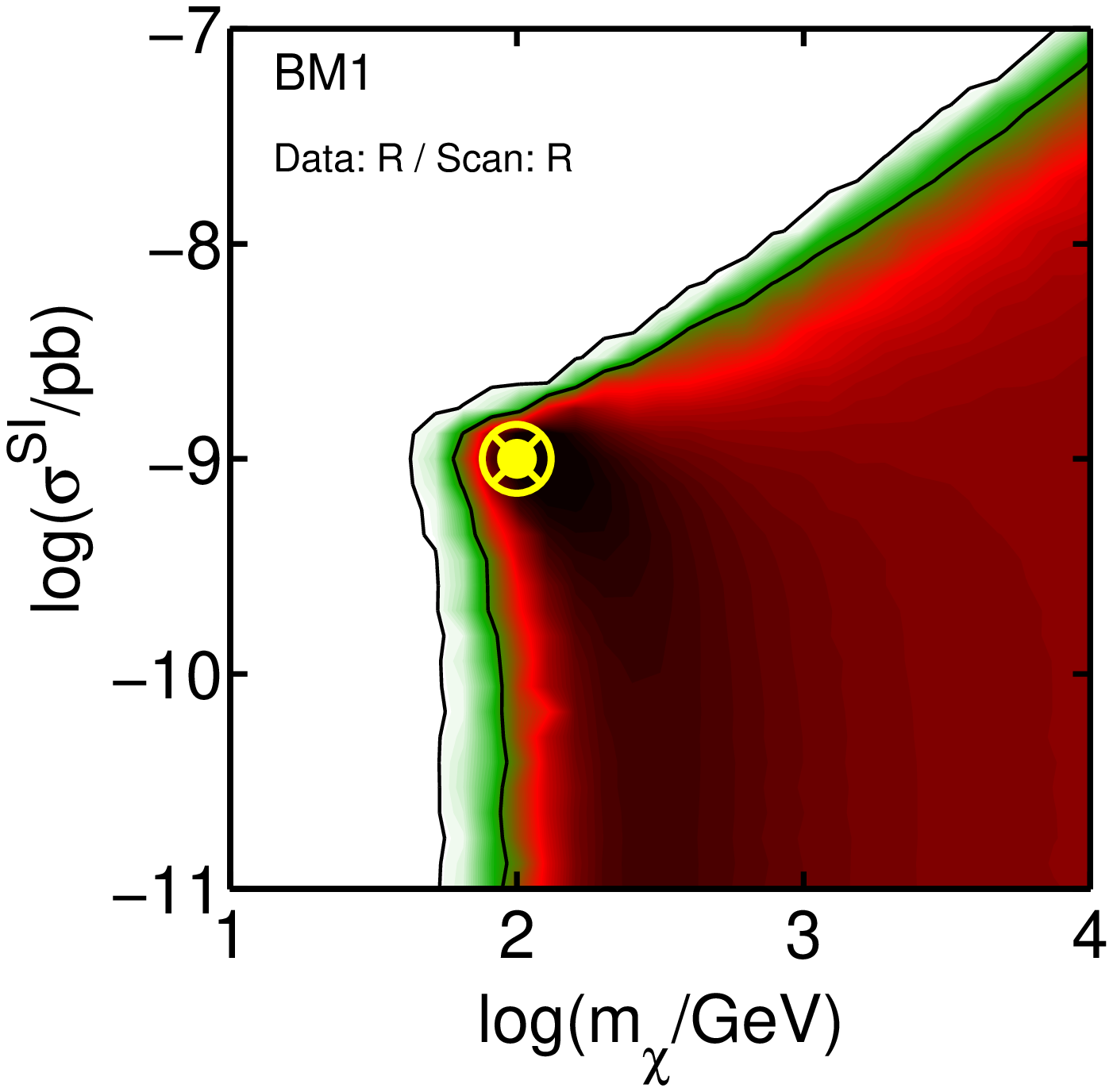}\hspace*{-0.62cm}
\includegraphics[width=0.35\textwidth]{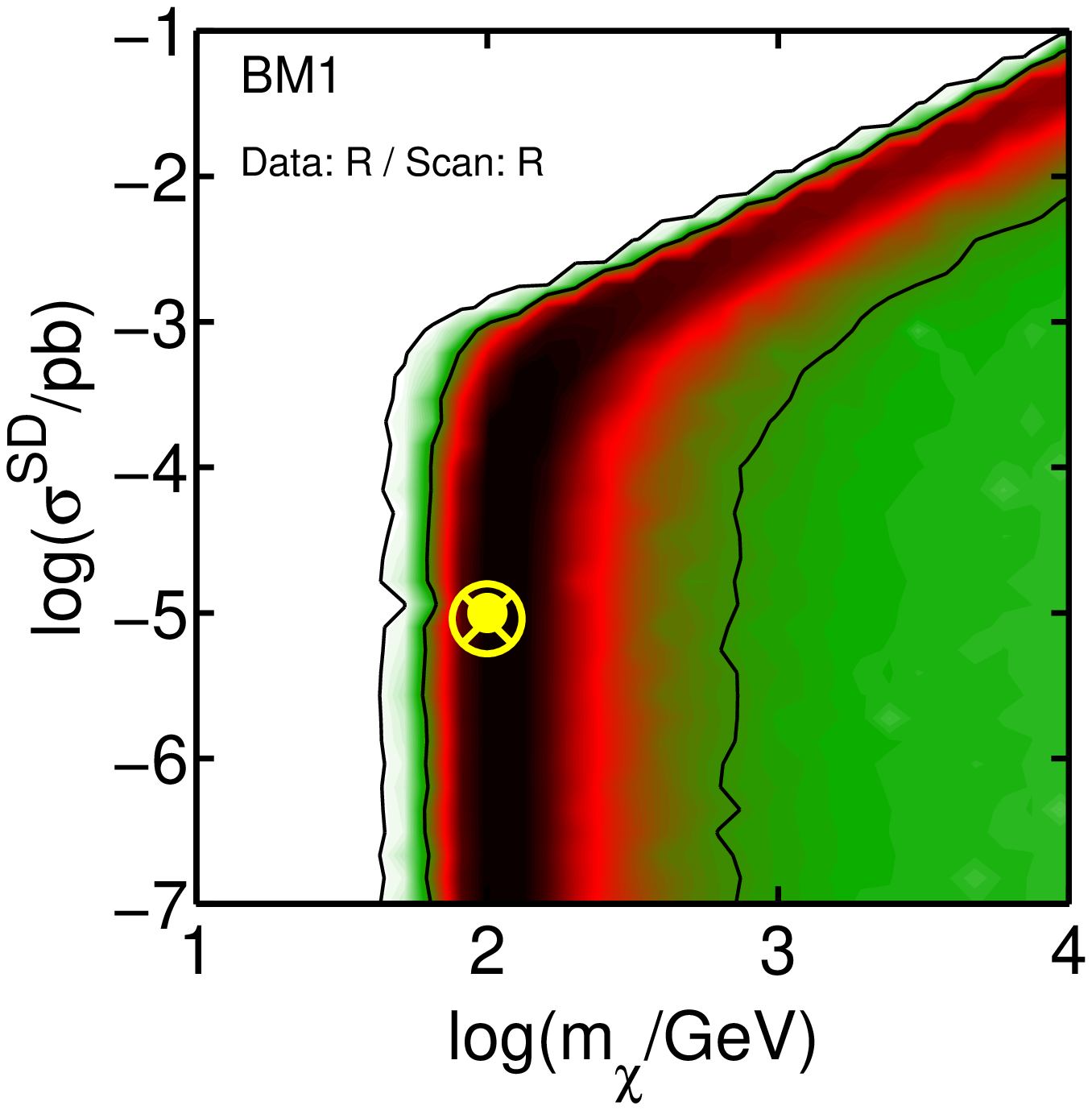}\hspace*{-0.62cm}
\includegraphics[width=0.35\textwidth]{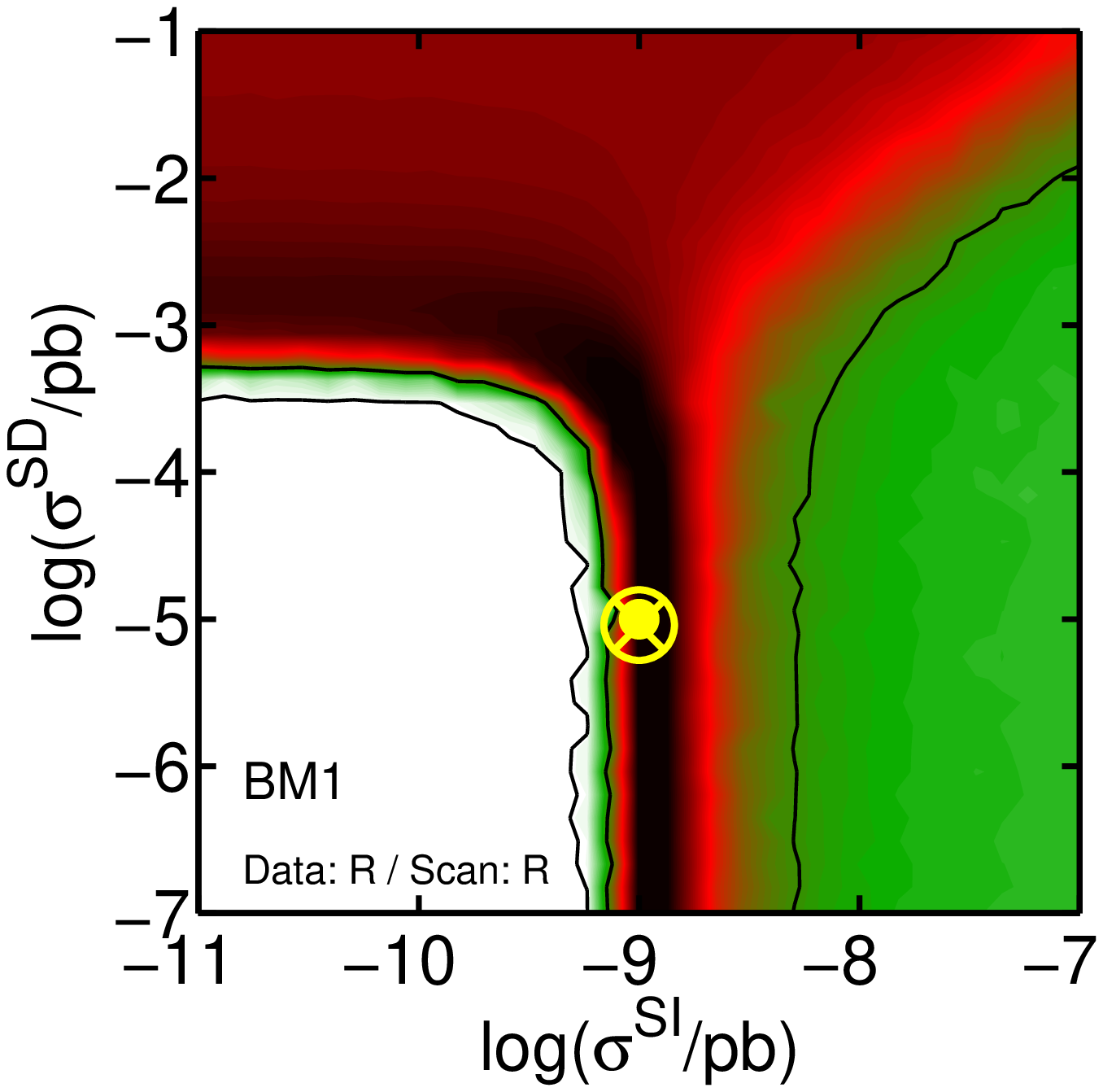}\\[-4ex]
\includegraphics[width=0.35\textwidth]{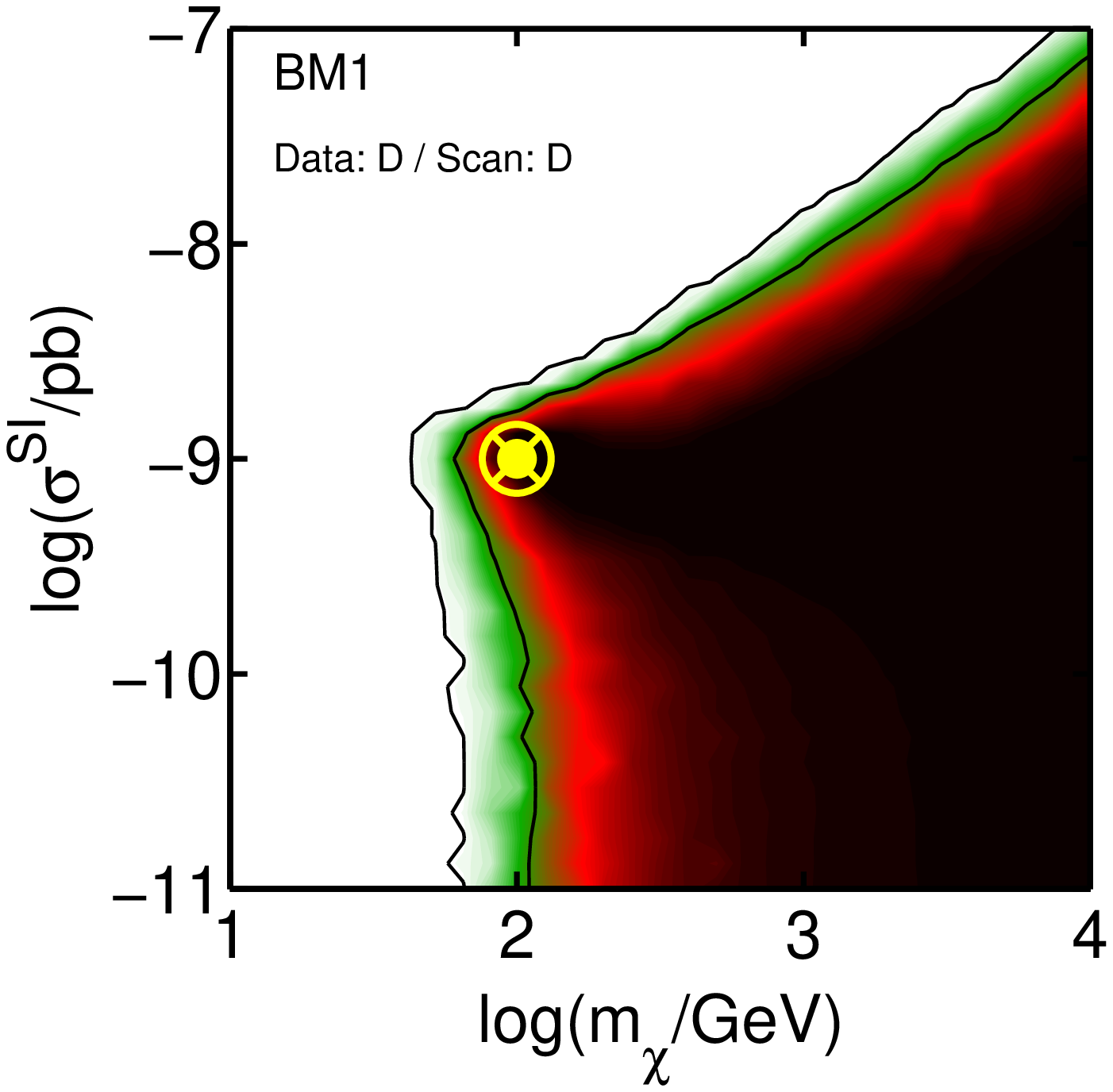}\hspace*{-0.62cm}
\includegraphics[width=0.35\textwidth]{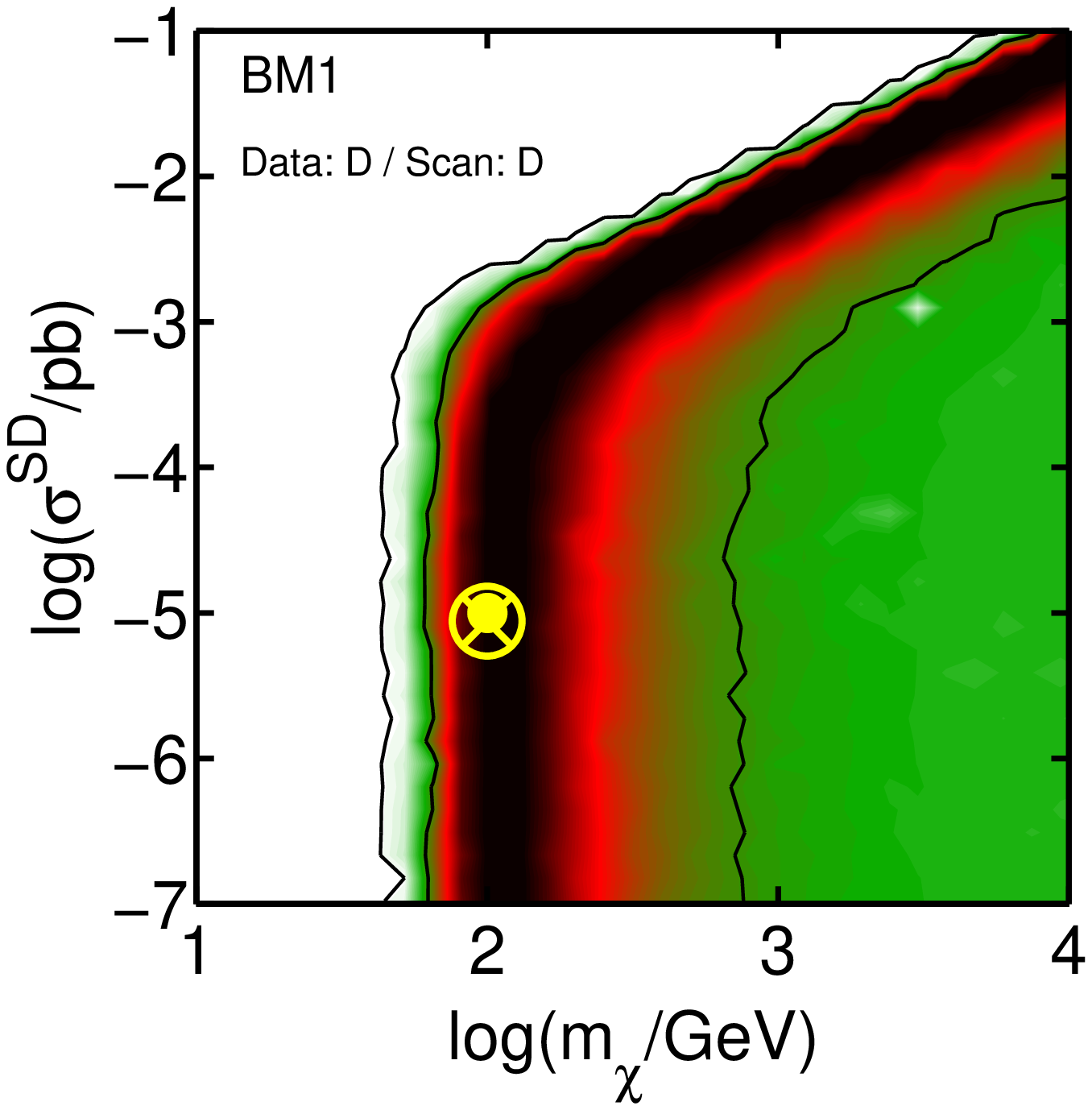}\hspace*{-0.62cm}
\includegraphics[width=0.35\textwidth]{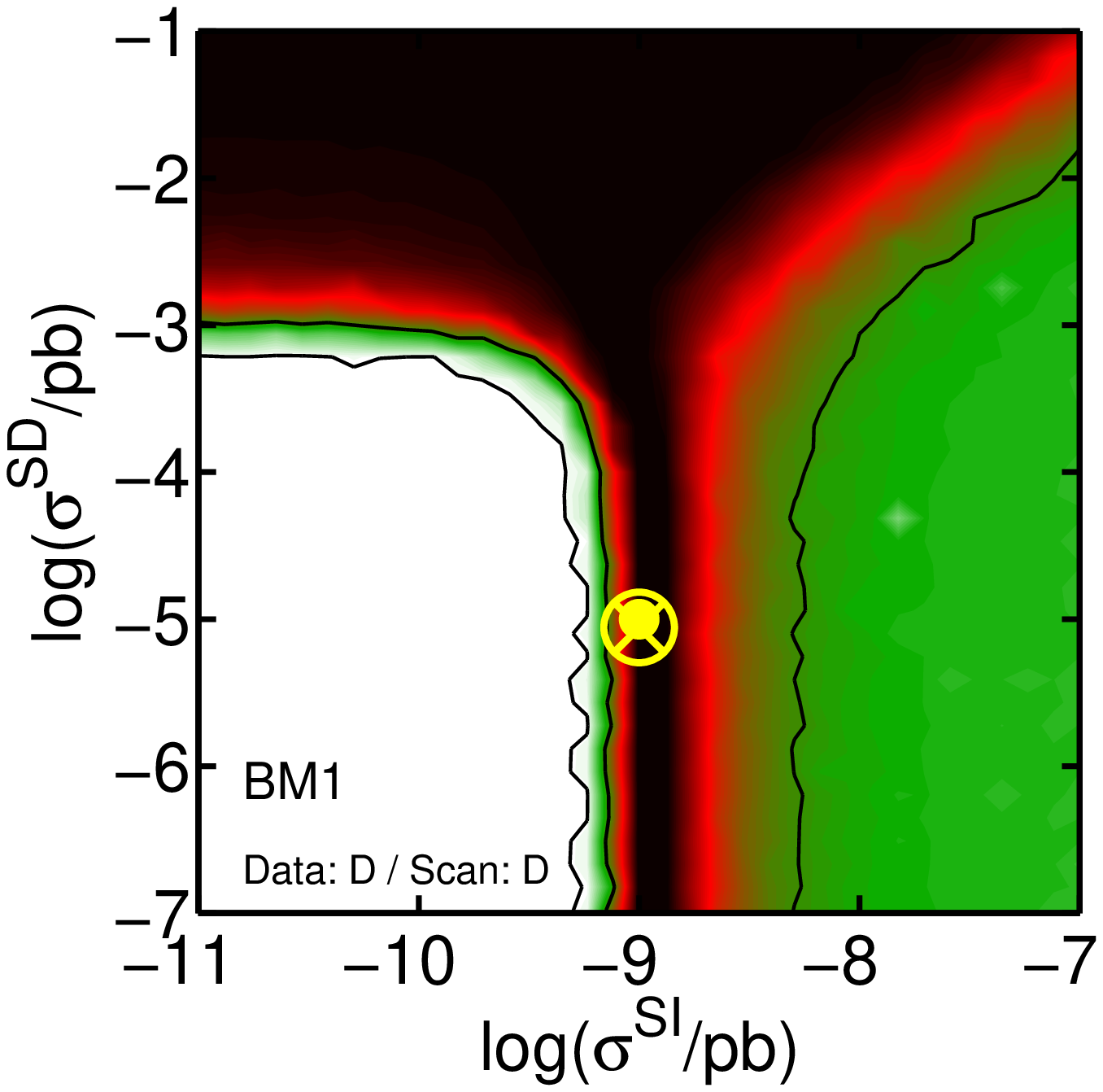}\\[-4ex]
\includegraphics[width=0.35\textwidth]{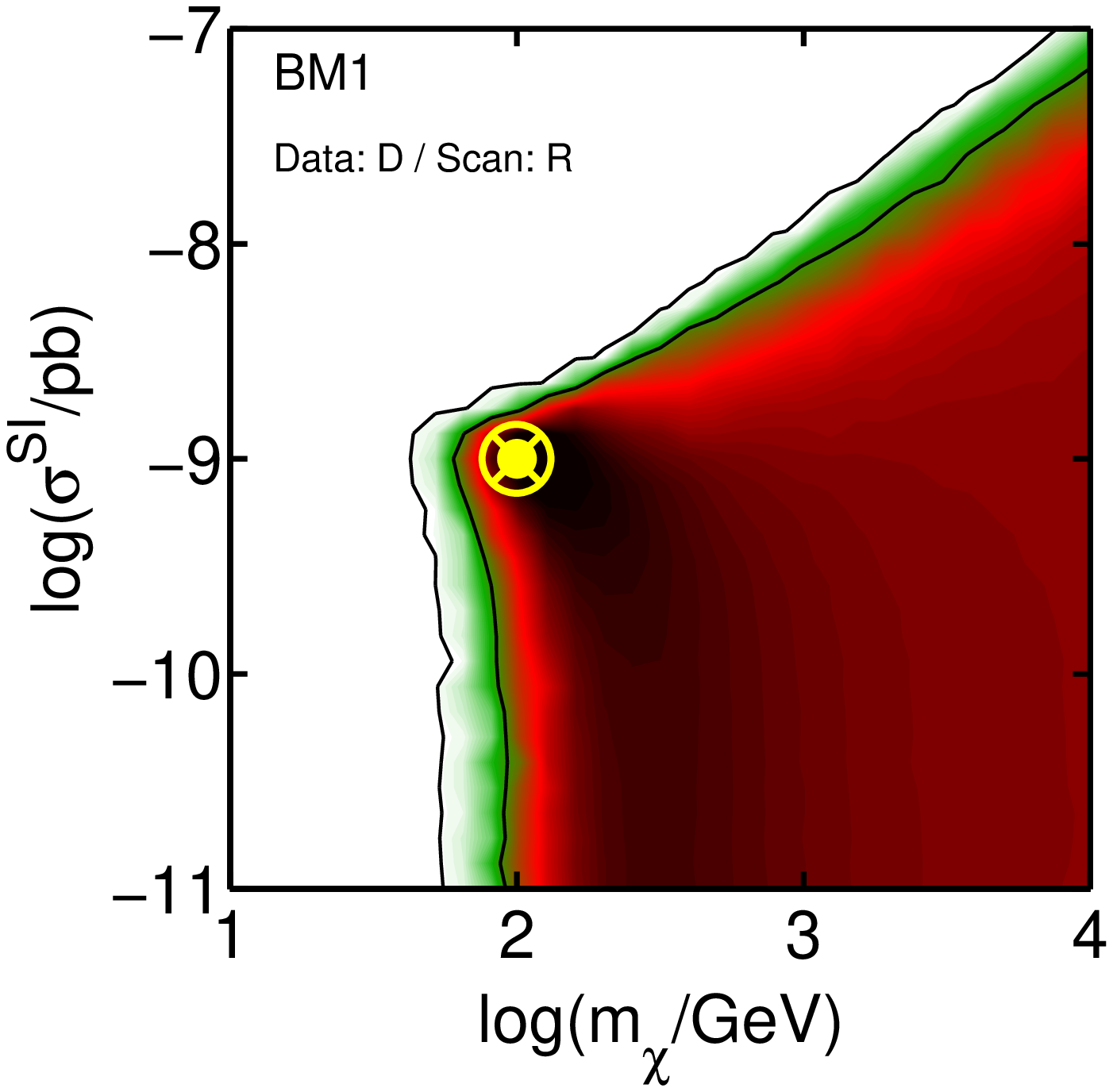}\hspace*{-0.62cm}
\includegraphics[width=0.35\textwidth]{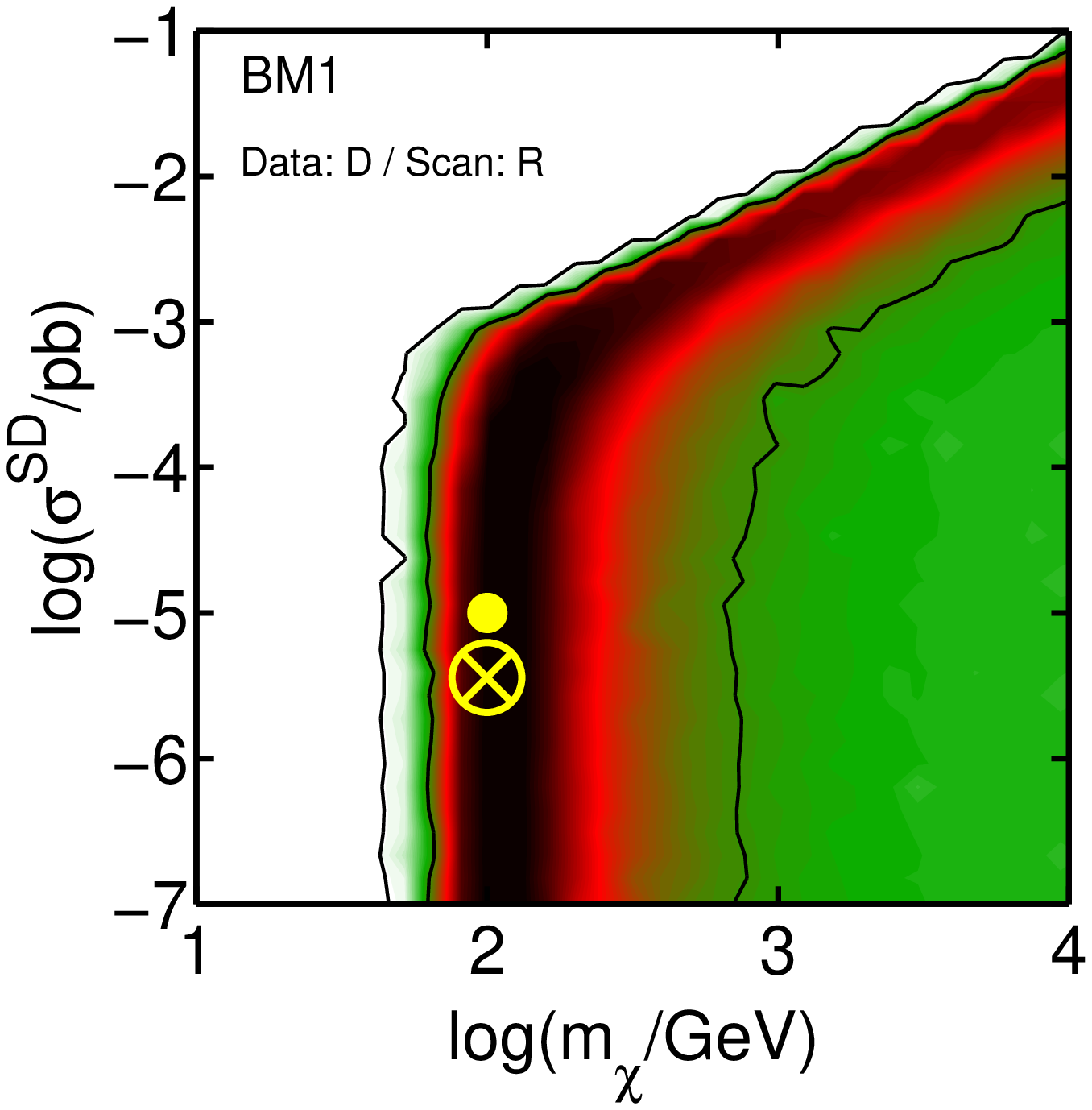}\hspace*{-0.62cm}
\includegraphics[width=0.35\textwidth]{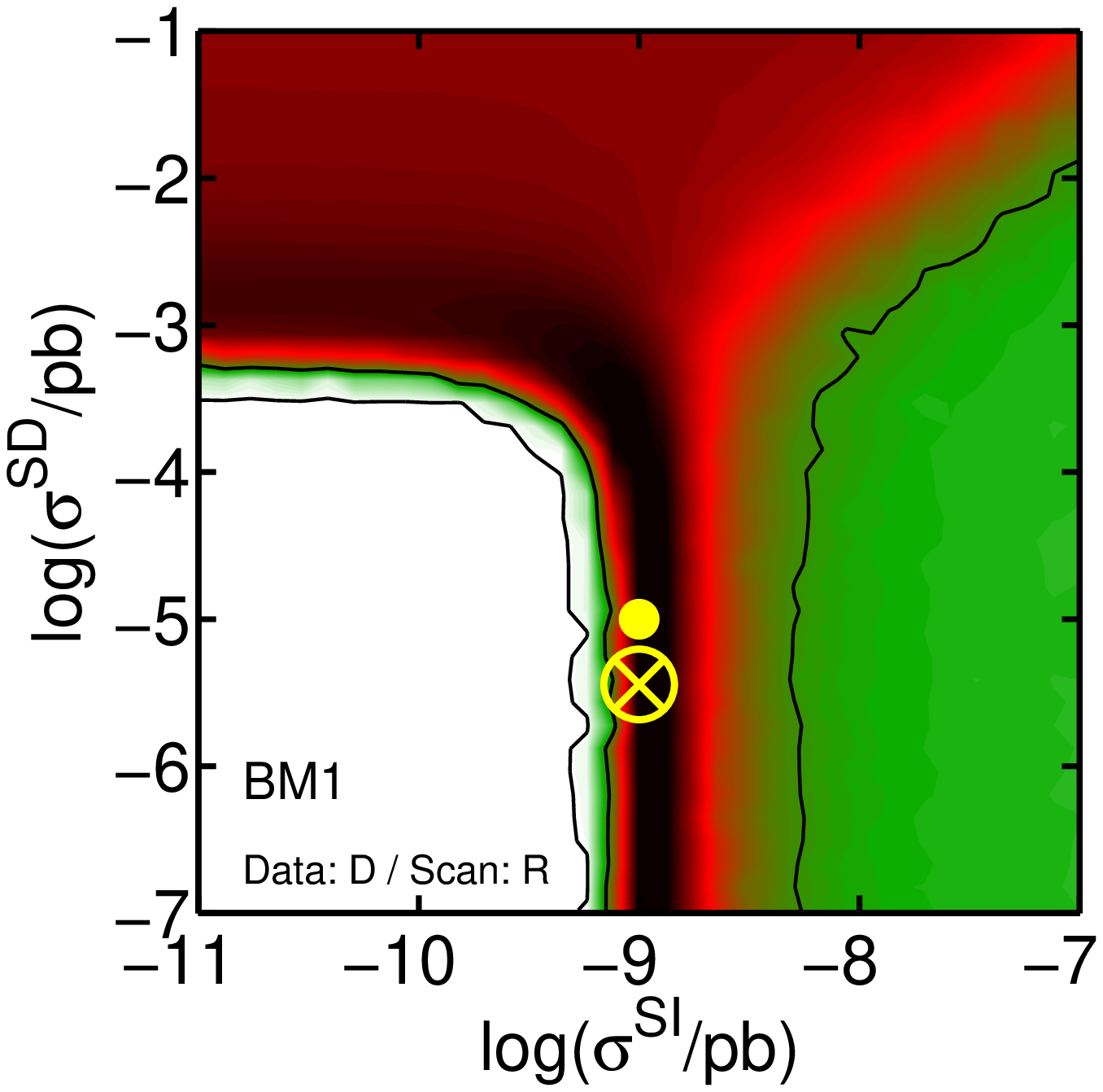}\\[-4ex]
\includegraphics[width=0.35\textwidth]{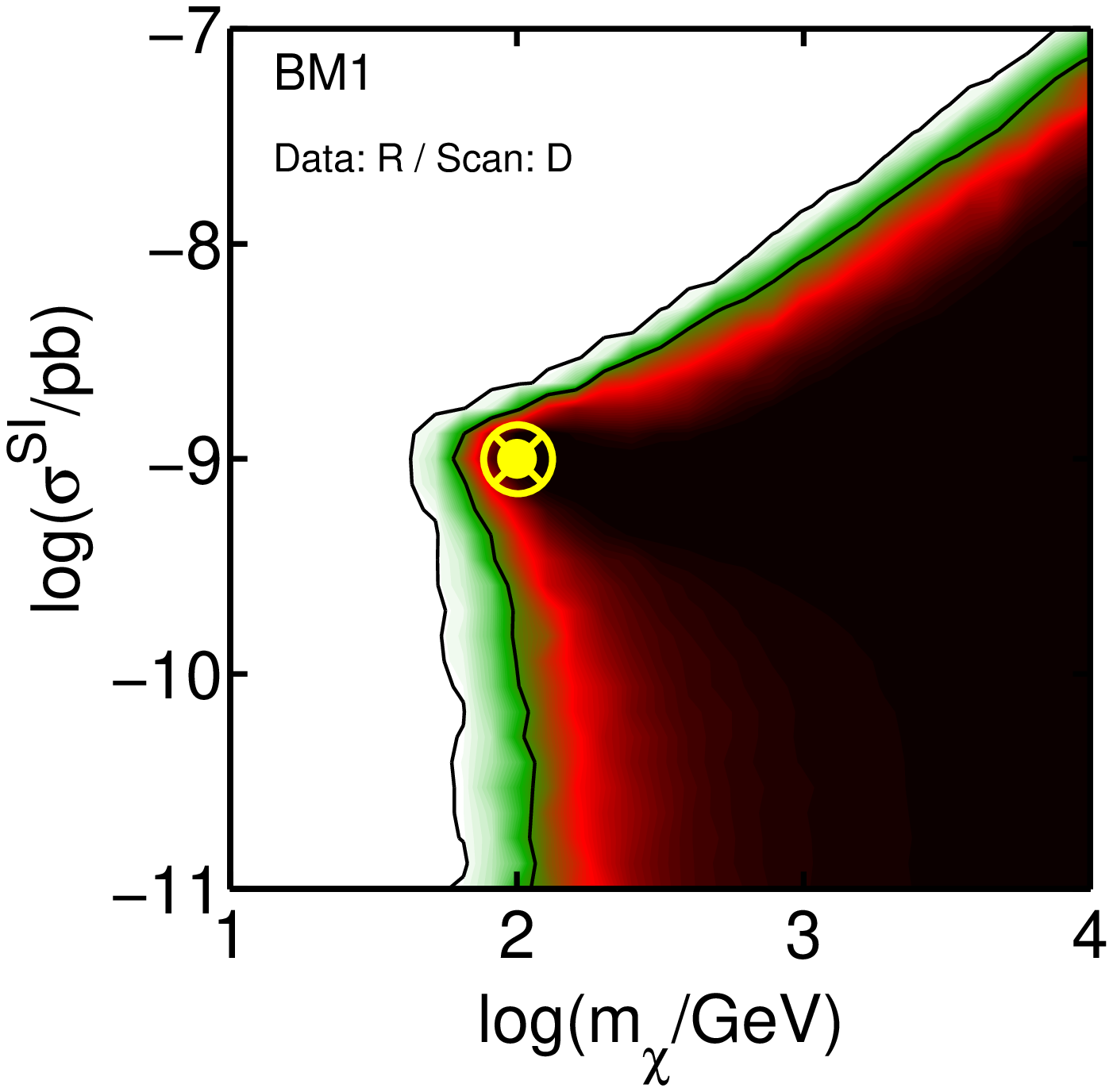}\hspace*{-0.62cm}
\includegraphics[width=0.35\textwidth]{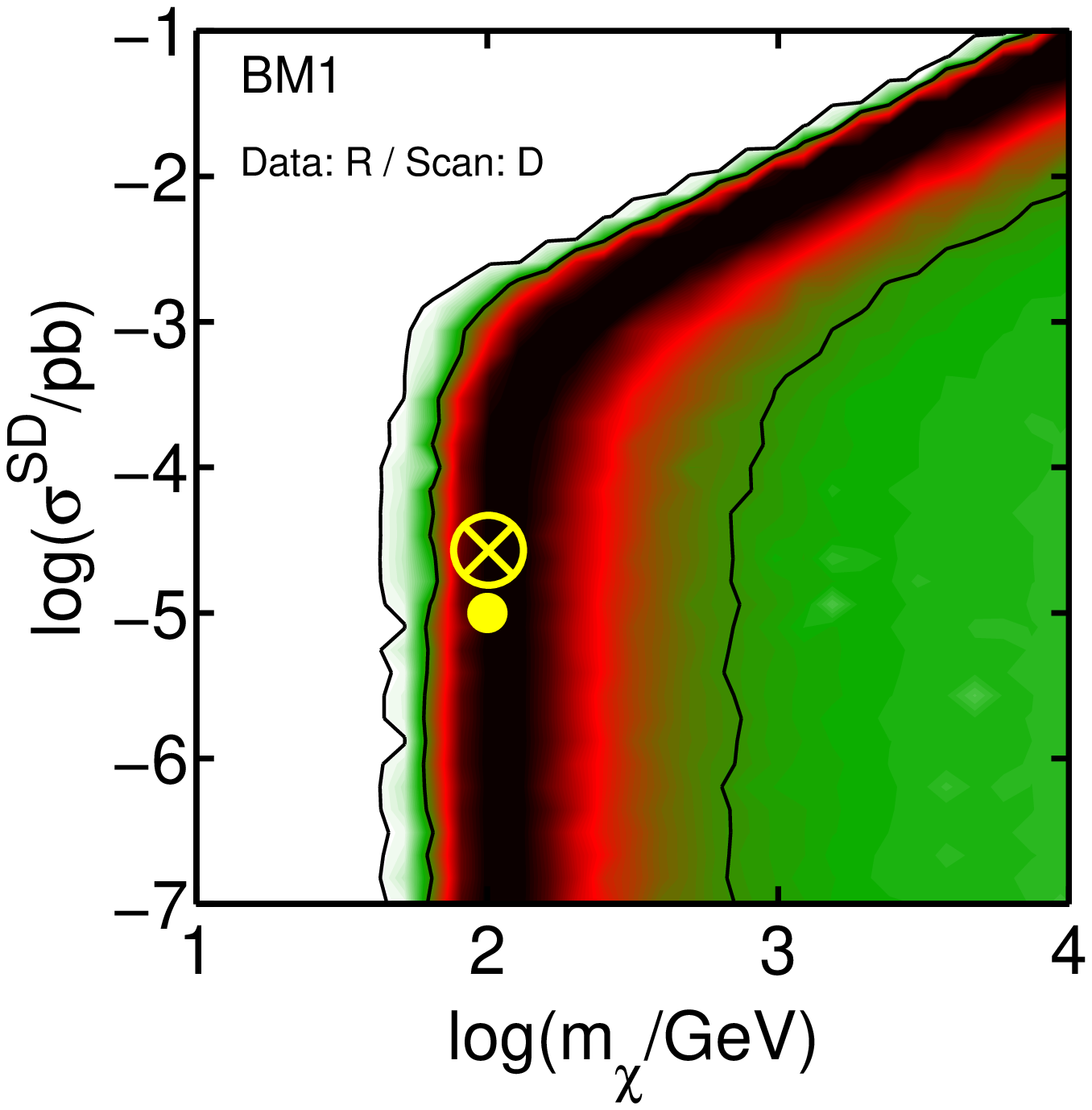}\hspace*{-0.62cm}
\includegraphics[width=0.35\textwidth]{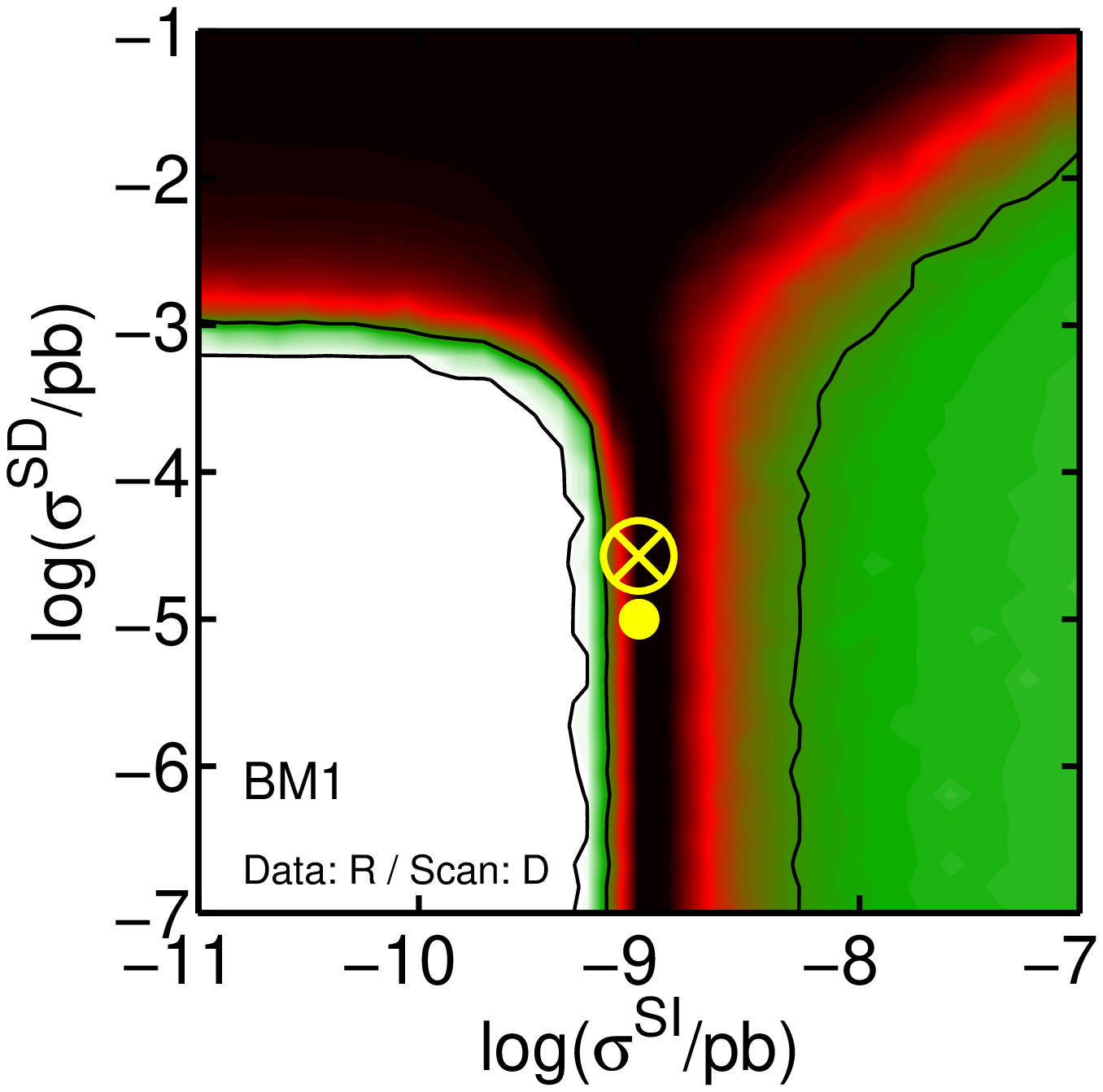}\\[-4ex]
\caption{\label{fig:BM1_profl} Two-dimensional profile likelihood for the reconstructed parameter space $(\mwimp,\,\sigsi,\,\sigsd)$ in benchmark model BM1. In the first and second rows the model used for the SDSF in the simulated experimental data and in the scan for parameter reconstruction is the same (R-model in the first row and D-model in the second). In the third row the D-model is used for the simulated data and the R-model for the parameter reconstruction and the reverse is done in the fourth row (see the captions of the different panels). The inner and outer contours are 68\% and 99\% confidence levels, respectively. The yellow dot indicates the benchmark value of the parameters, while the yellow encircled cross denotes the position of the best-fit values.}
\end{figure*}

\begin{figure*}
\includegraphics[width=0.35\textwidth]{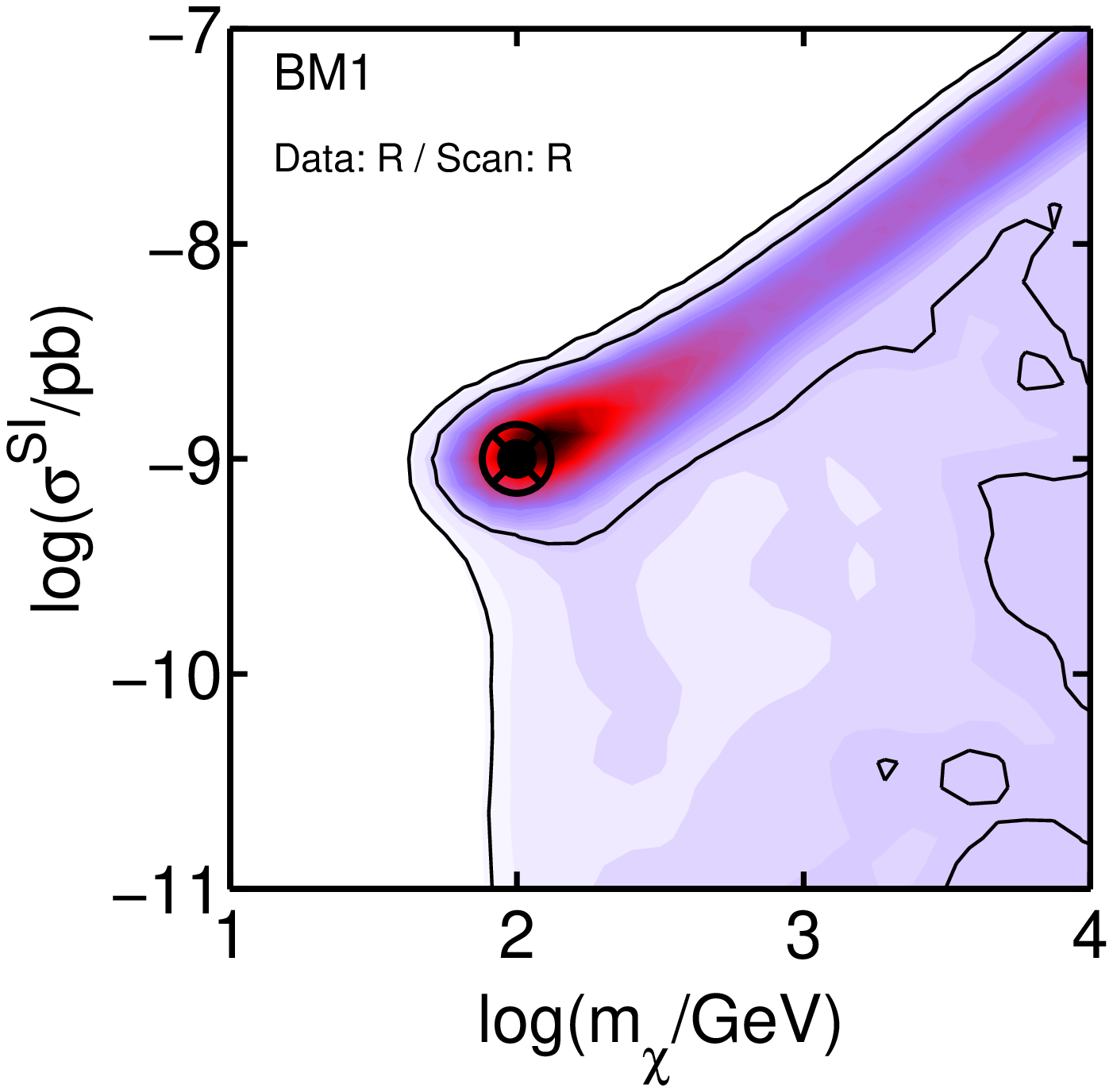}\hspace*{-0.62cm}
\includegraphics[width=0.35\textwidth]{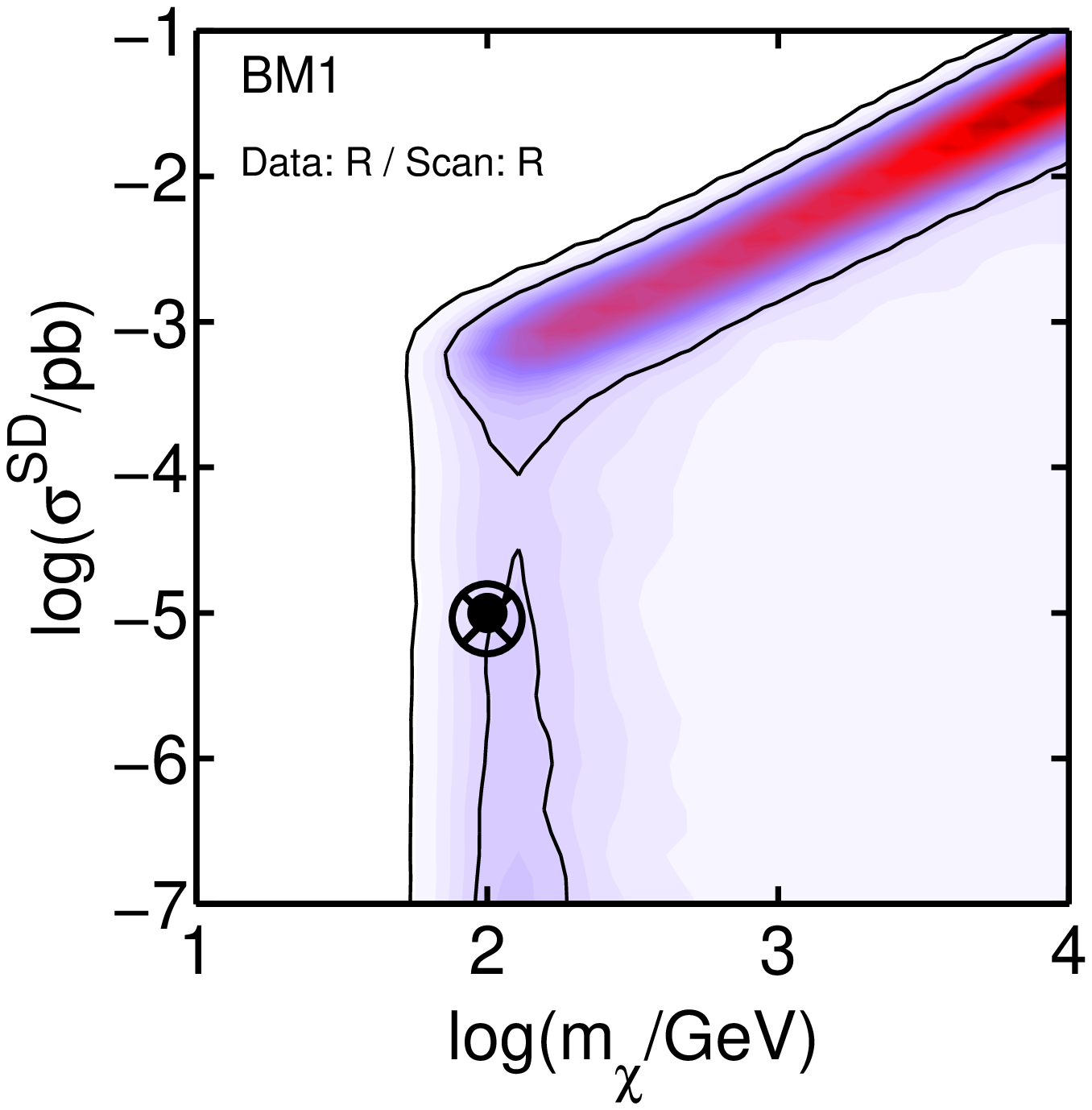}\hspace*{-0.62cm}
\includegraphics[width=0.35\textwidth]{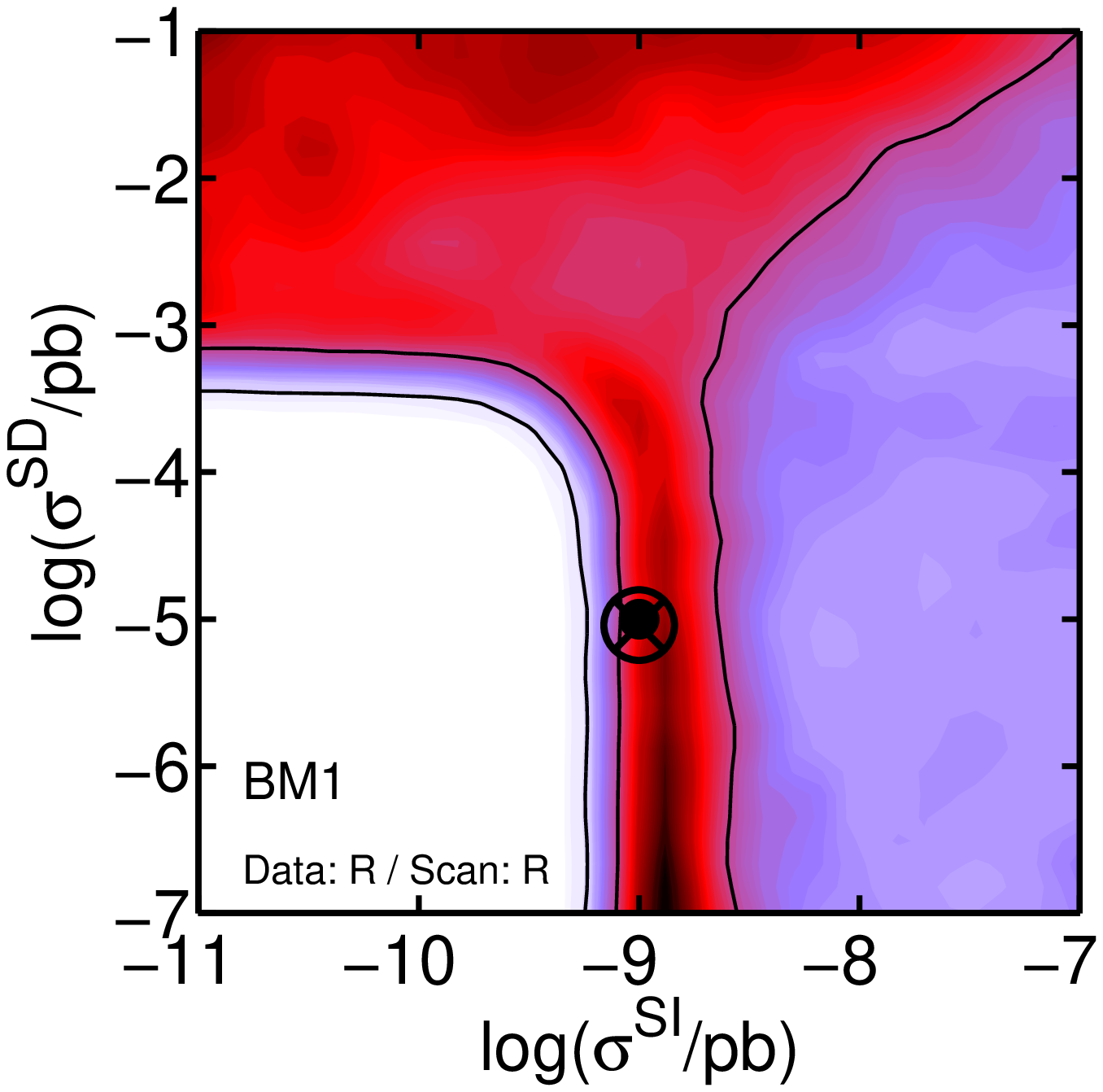}\\[-4ex]
\includegraphics[width=0.35\textwidth]{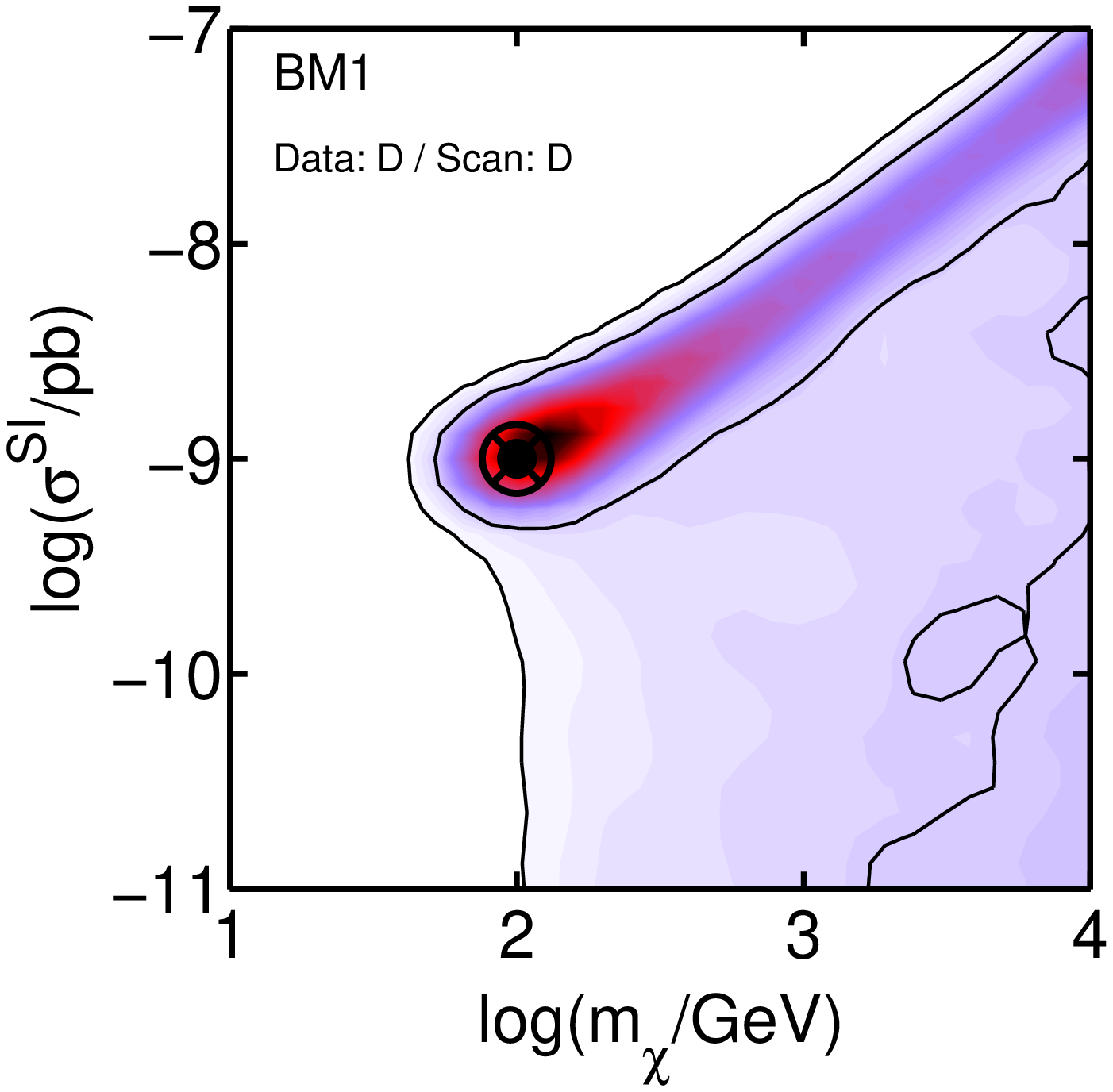}\hspace*{-0.62cm}
\includegraphics[width=0.35\textwidth]{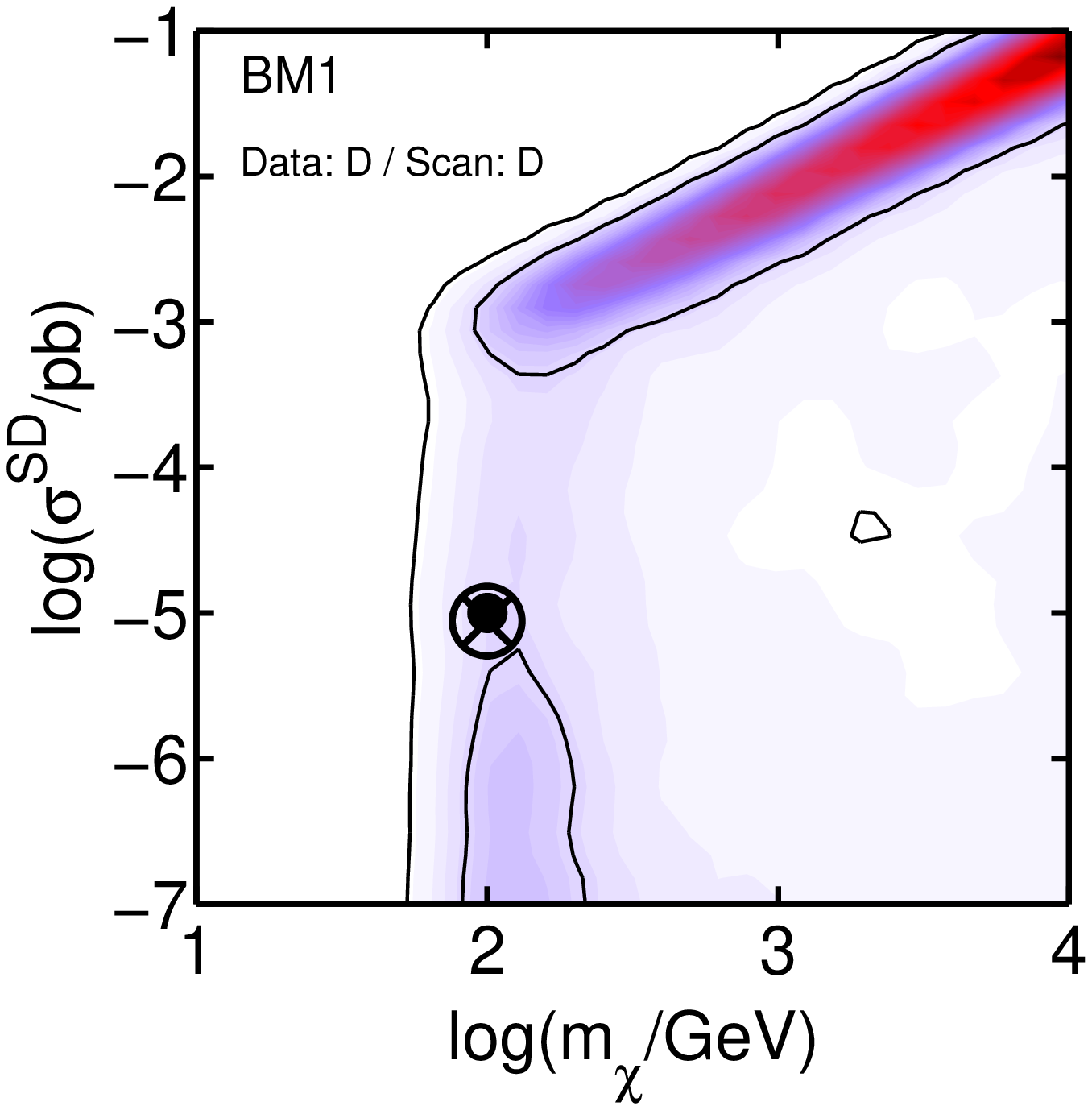}\hspace*{-0.62cm}
\includegraphics[width=0.35\textwidth]{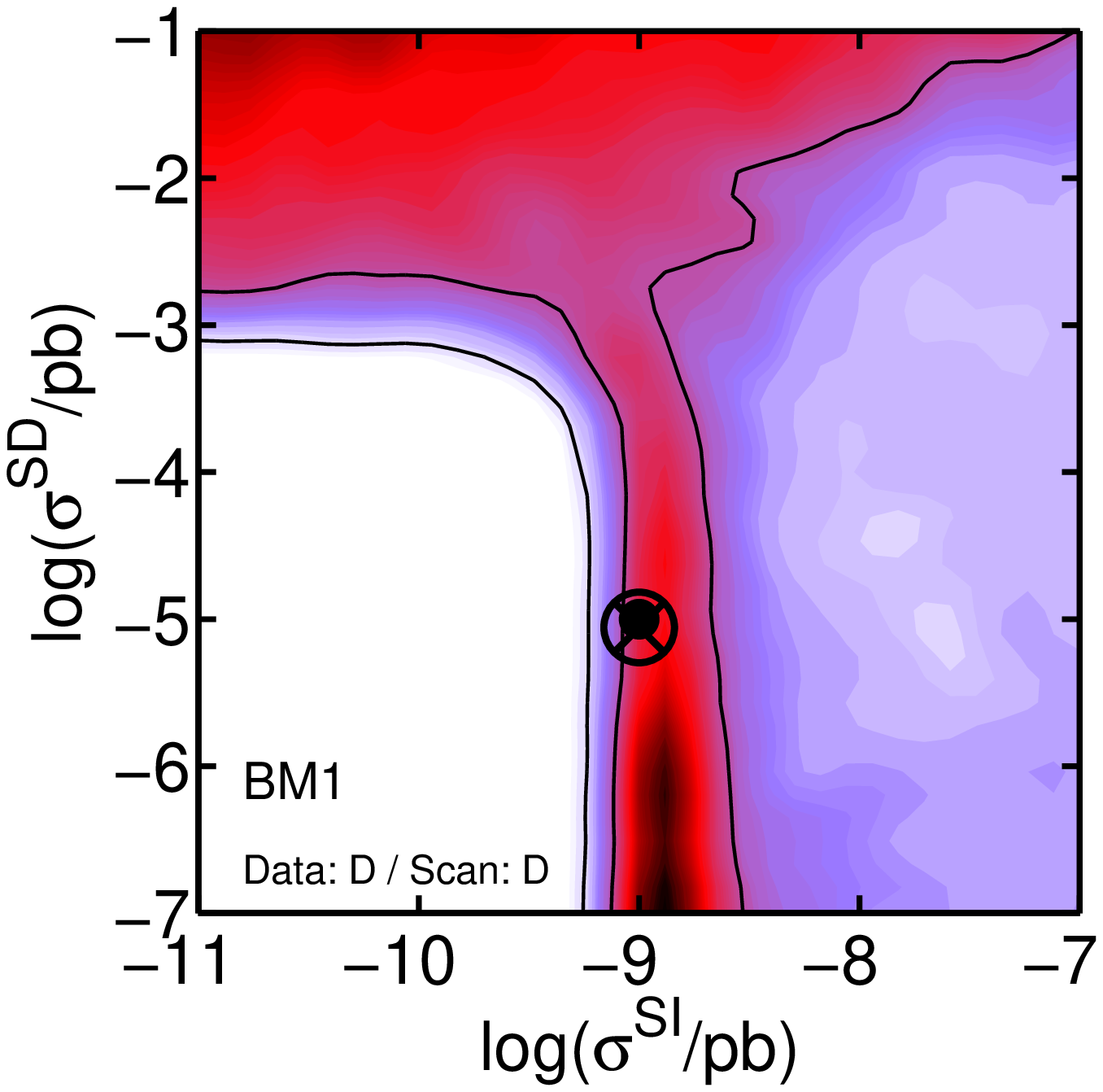}\\[-4ex]
\includegraphics[width=0.35\textwidth]{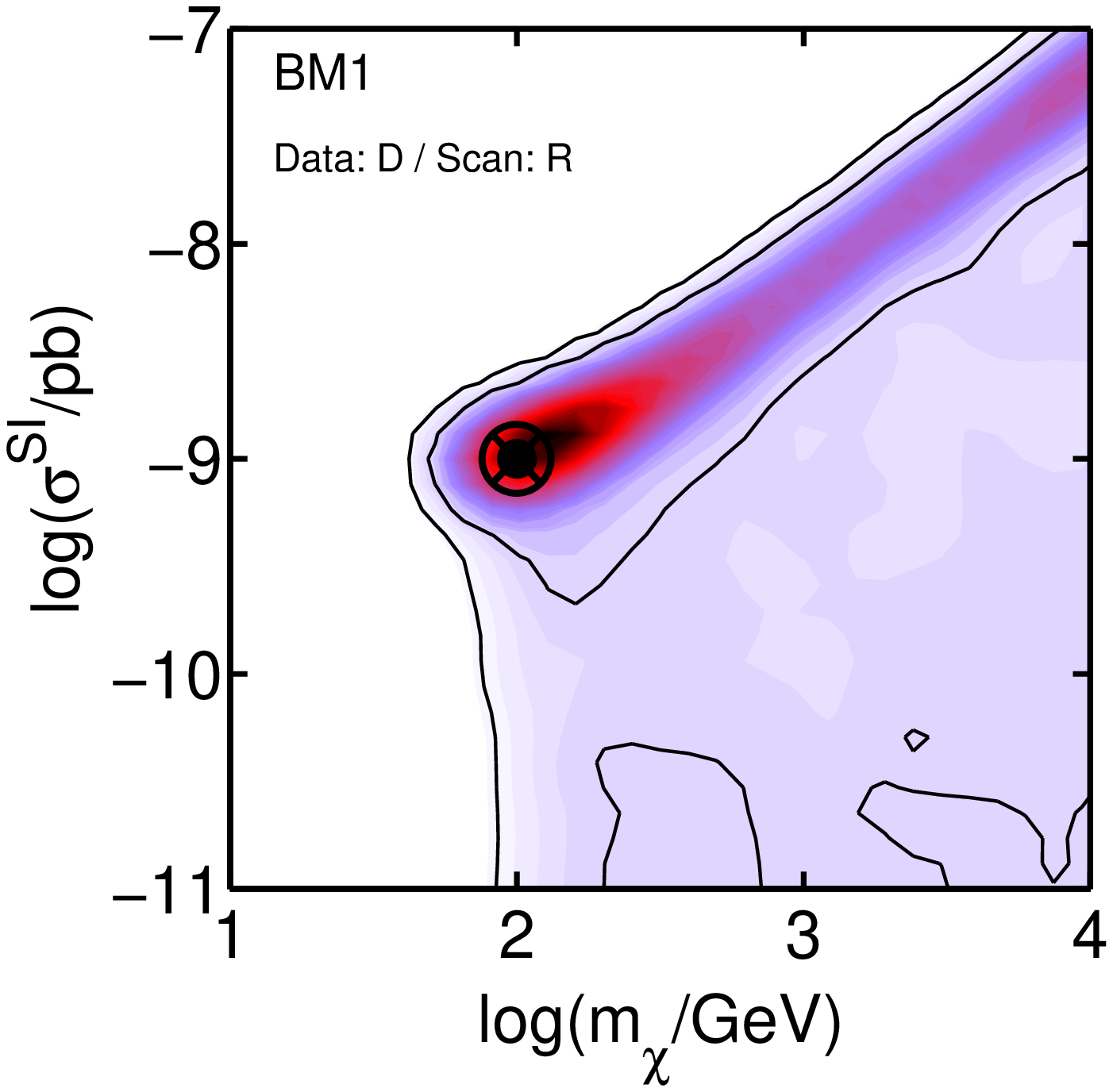}\hspace*{-0.62cm}
\includegraphics[width=0.35\textwidth]{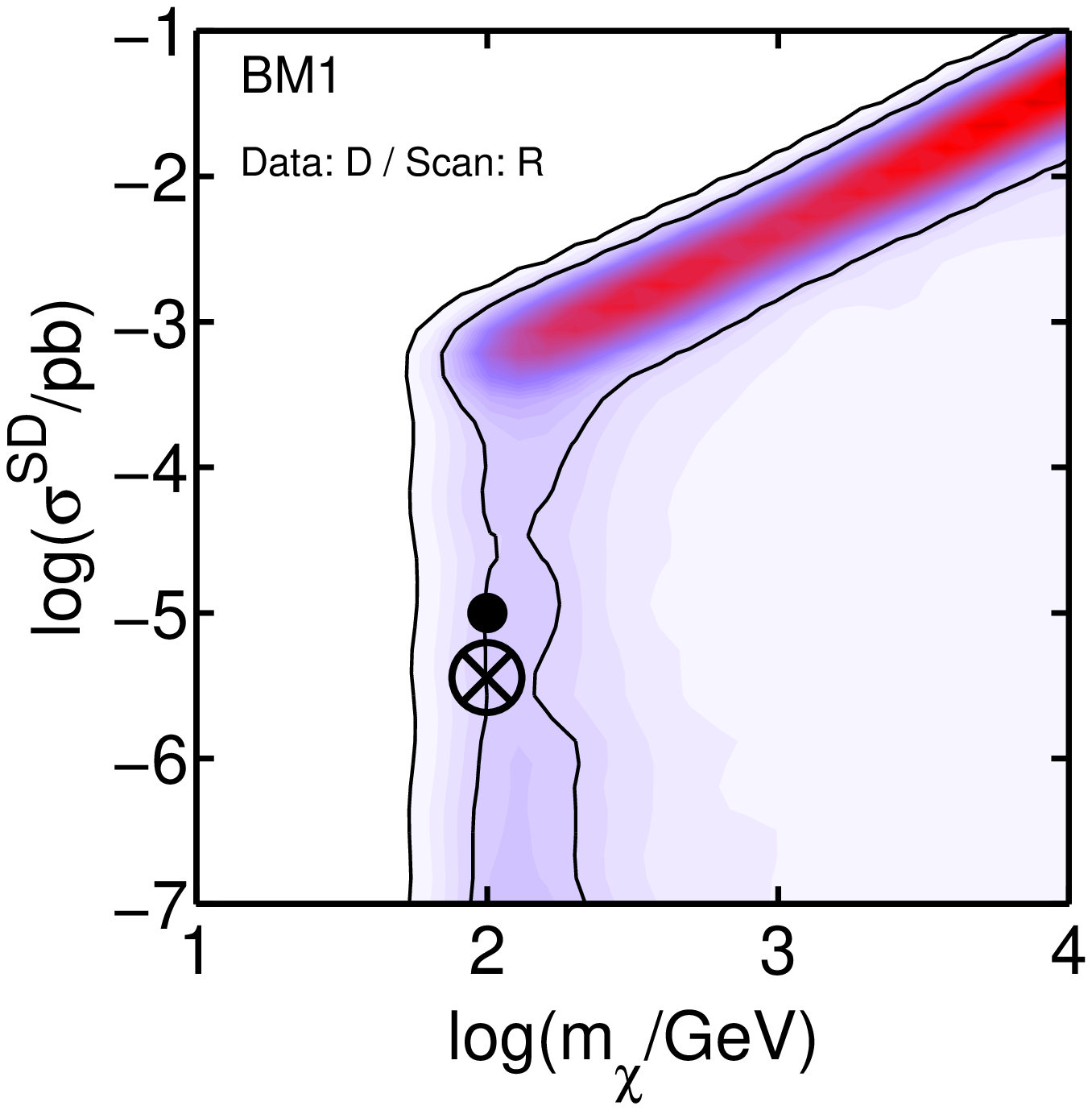}\hspace*{-0.62cm}
\includegraphics[width=0.35\textwidth]{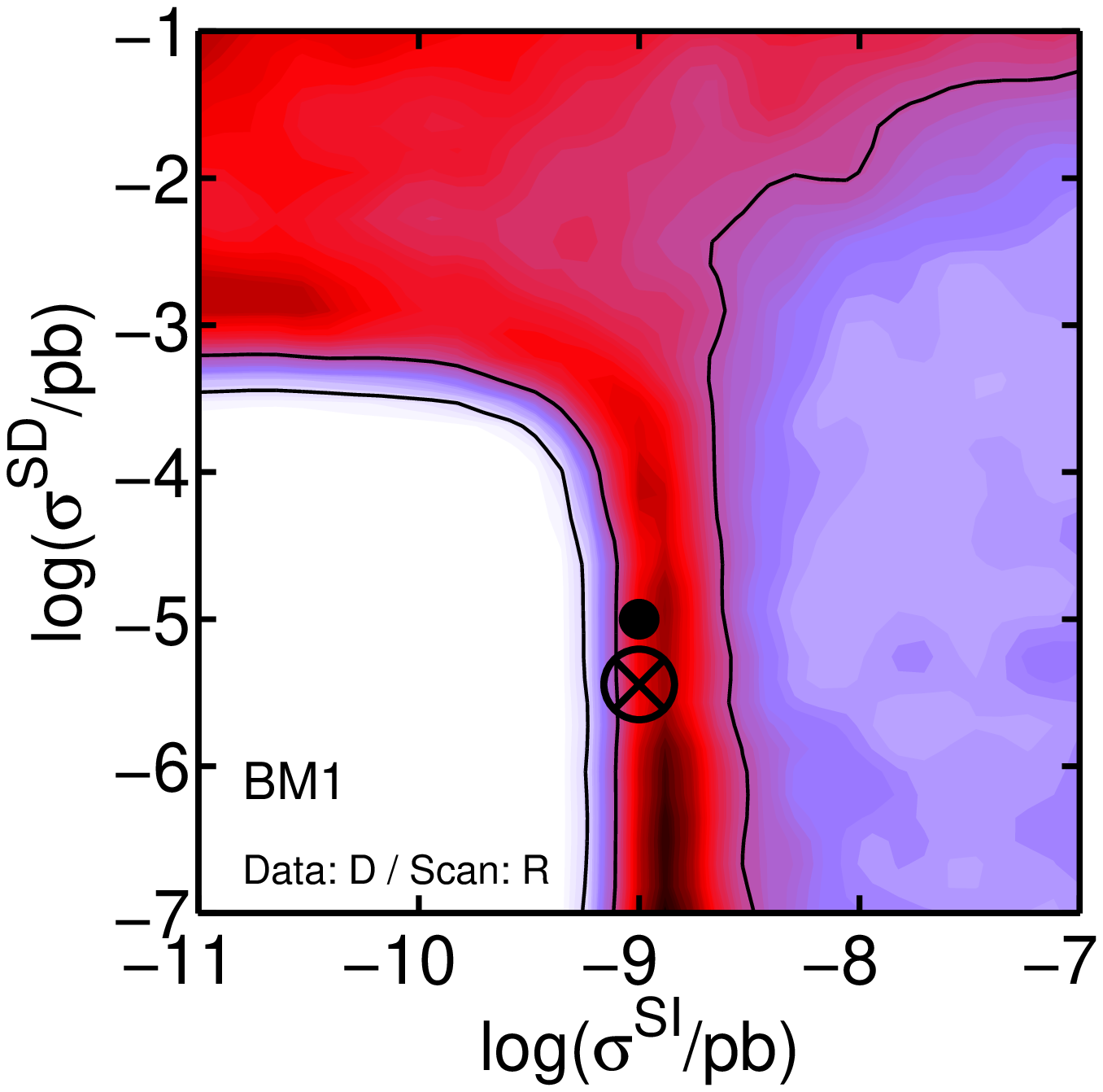}\\[-4ex]
\includegraphics[width=0.35\textwidth]{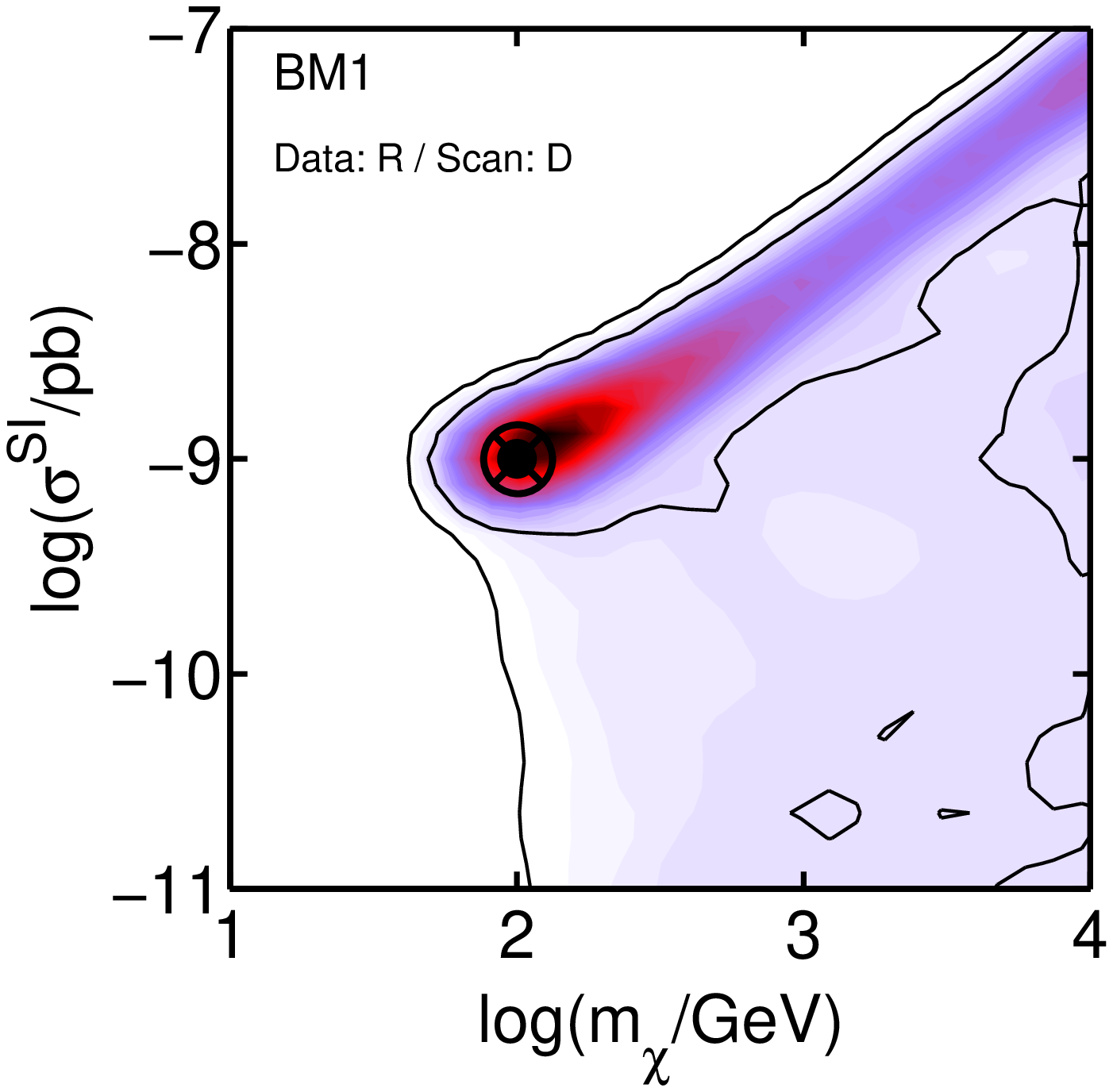}\hspace*{-0.62cm}
\includegraphics[width=0.35\textwidth]{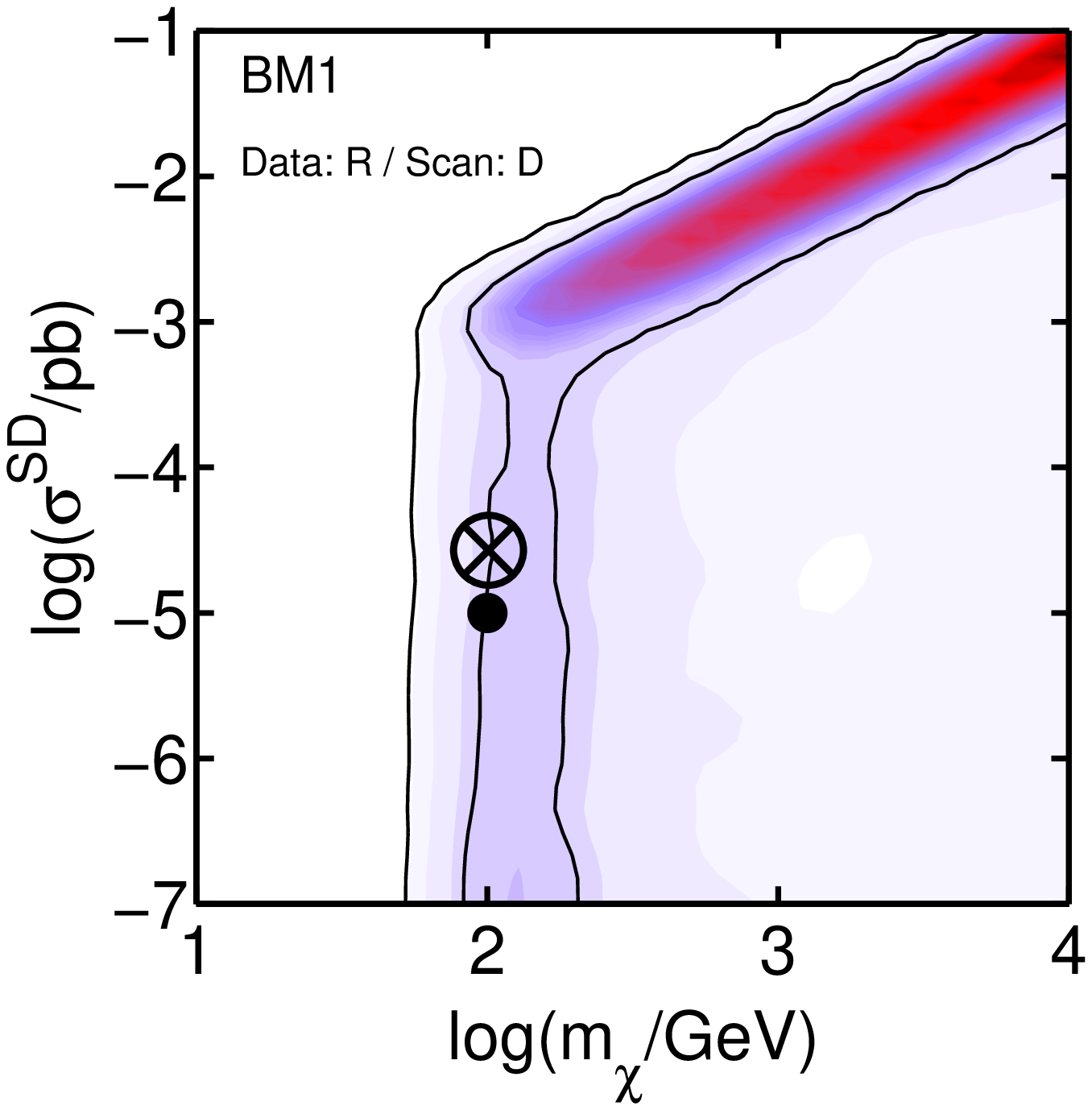}\hspace*{-0.62cm}
\includegraphics[width=0.35\textwidth]{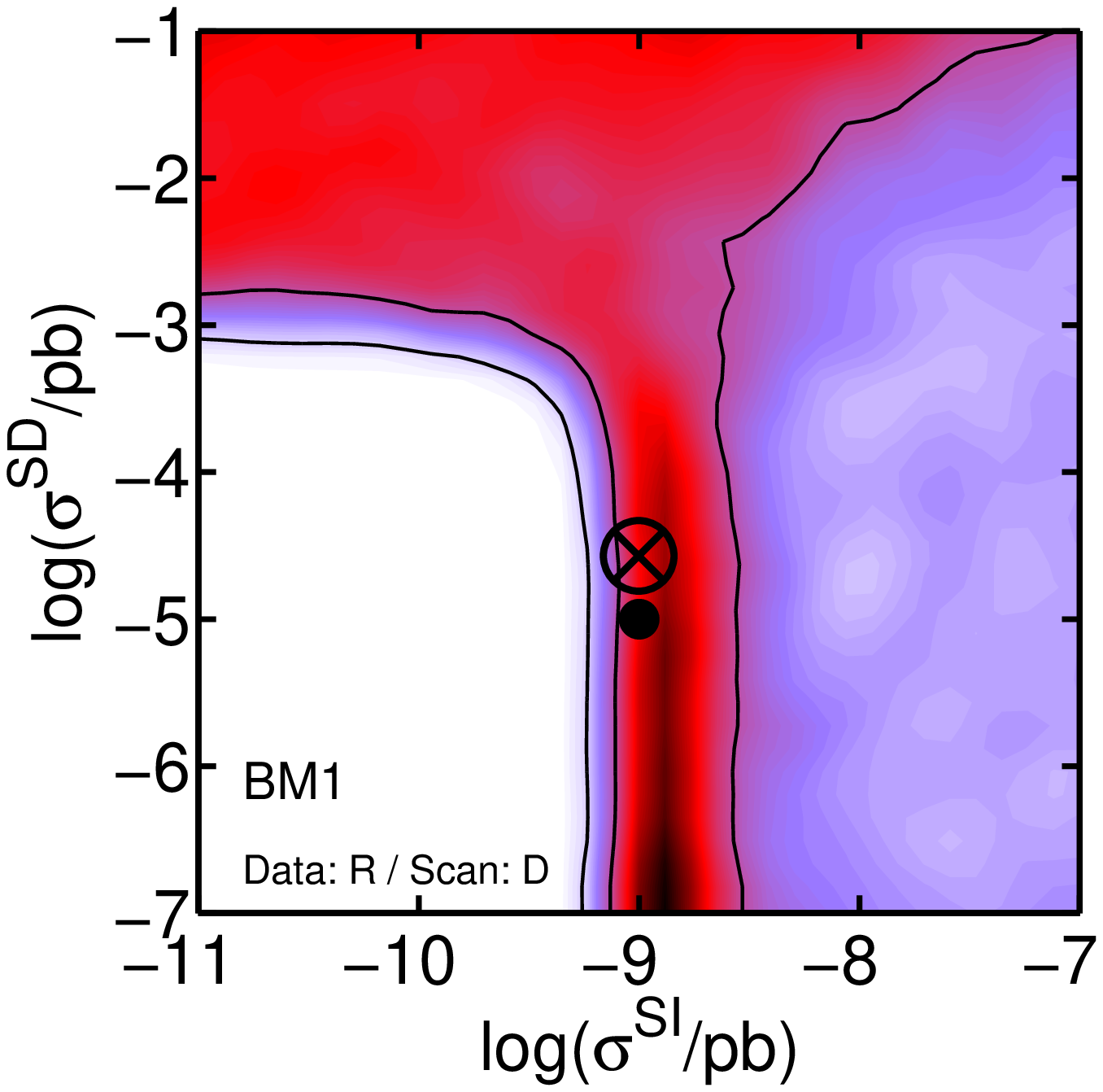}\\[-4ex]
\caption{\label{fig:BM1_pdf} The same as in Fig.\,\ref{fig:BM1_profl} but for the marginalized pdf.}
\end{figure*}

The results for the first benchmark, BM1, are displayed in Fig.\,\ref{fig:BM1_profl} (profile likelihood) and Fig.\,\ref{fig:BM1_pdf} (pdf). The four rows correspond to the possible combinations of SDSFs and we indicate which model is used in the generation of data and in the calculation of the likelihood in the scan.
Comparing the distribution of the profile likelihood with that of the pdf, we can observe the effect of maximization versus marginalization. 
There are regions of the parameter space that are contained in the 99\% confidence level contour of the profile likelihood which are however left out of the credible interval contours of the pdf.
This happens because the good agreement with the data is produced only in a small volume of the three-dimensional parameter space and the integration in the third dimension decreases the corresponding value for the pdf \footnote{
The contours in the two-dimensional plots for the profile likelihood look smoother than for the pdf. This occurs, in particular, when integrating over regions with an almost flat likelihood, where it is difficult to obtain uniform sampling. The resulting pdf can present unphysical structures. The fact that the background is included as a free parameter in the scan introduces additional fluctuations.
}.

The plots in the first column of Figs.\,\ref{fig:BM1_profl} and \ref{fig:BM1_pdf} represent the reconstruction of the WIMP mass and SI cross-section, and they can therefore be compared with the leftmost plot of Fig.\,\ref{fig:noSD}. We observe that $\sigsi$ can now be arbitrarily small as long as the $\sigsd$ is large enough to reproduce the observed DM rate  
and that the assumption $\sigsd=0$ leads to over constrained contours.
The SD form factor (that results for both R-and D-models of the SDSF) is steeper than the SI form factor. 
Thus, in principle using the information from the energy spectrum it would not be possible to properly fit BM1 data with only SD interactions (large values of the WIMP mass provide in this case a better fit, since they would correspond to a flatter energy spectrum).
However, the number of recoils events in the high energy bins is too small to be sensitive to those differences.

Similarly, there is no lower bound for the SD cross-section. This is evidenced in the second column of both Figs.\,\ref{fig:BM1_profl} and \ref{fig:BM1_pdf}, where $\sigsd$ is plotted as a function of the WIMP mass. 
As commented above, when trying to fit the energy spectrum of a point dominated by SI interactions (such as BM1) in terms of axial interactions, we get better results for large WIMP masses, due to the SDSF being steeper. The trend is evident in the pdf plots where larger values of masses are associated to a brighter pdf than lower masses, but the small number of events prevents this tendency to have any significant effect on the shape of the contours.

Finally, the degeneracy in the reconstruction of the SI and SD contributions for a given set of experimental data is clearly evidenced in the third column of Figs.\,\ref{fig:BM1_profl} and \ref{fig:BM1_pdf}, where $\sigsd$ is plotted versus $\sigsi$ and the resulting compatible regions show an ``inverse L'' pattern.
The lower left corner of the plot is empty since both the SD and SI cross-section are too small to produce the simulated number of nuclear recoils, however, as stressed in the previous paragraphs, both $\sigsi$ and $\sigsd$ can be sizable if the WIMP mass is also large.
These plots also show that for this particular benchmark point SD interactions provide the dominant contribution to the WIMP rate for $\sigsd\gsim 10^{-3}$ pb.

The interpretation of the results for the different rows allows us to determine to what extent the uncertainties in the SDSF affect the reconstruction of DM parameters. 
We stress again that for BM1 the differential event rate is dominated by SI interactions, thus we do not observe significant differences when changing the SDSF in the computation of the simulated recoils (see the left panel in Fig.\,\ref{fig:benchmark_rate}).
As a consequence, the plots on the first line are indistinguishable from those in the third line in Figs.\,\ref{fig:BM1_profl} and \ref{fig:BM1_pdf}. The same happens between the second and fourth lines.
On the other hand, small differences arise when different SDSF are used in the computation of the likelihood.
As already pointed out, the R- and D-models differ in the zero-momentum value, as well as in the slope.
Indeed, we find that when the R-model is used in the scan to reconstruct the DM parameters (rows one and three) the resulting  $\sigsd$ can be smaller than when the D-model is used (rows two and four). This happens because the SDSF of the R-model is always larger than in the D-model, so the correct number of recoils is reproduced with a slightly smaller $\sigsd$.
Notice in particular how, although the best-fit value for $\sigsd$ is correctly reconstructed when the same SDSF is used for generating and reconstructing the points (first two rows), there is a mismatch when different models are used. For example, if data are generated with the R-model and scanned using the D-model (third row), the best-fit value for $\sigsd$ is lower (by about a factor two) than the actual one. Of course, the contrary occurs when data are generated with the D-model and scanned with the R-model (fourth row). This behaviour can be observed for the three benchmark points.
Regarding the reconstructed WIMP mass, the distribution is similar when either the R- or D-model is used, although the latter slightly favours heavier WIMPs to compensate for the steeper slope.

\begin{figure*}
\includegraphics[width=0.35\textwidth]{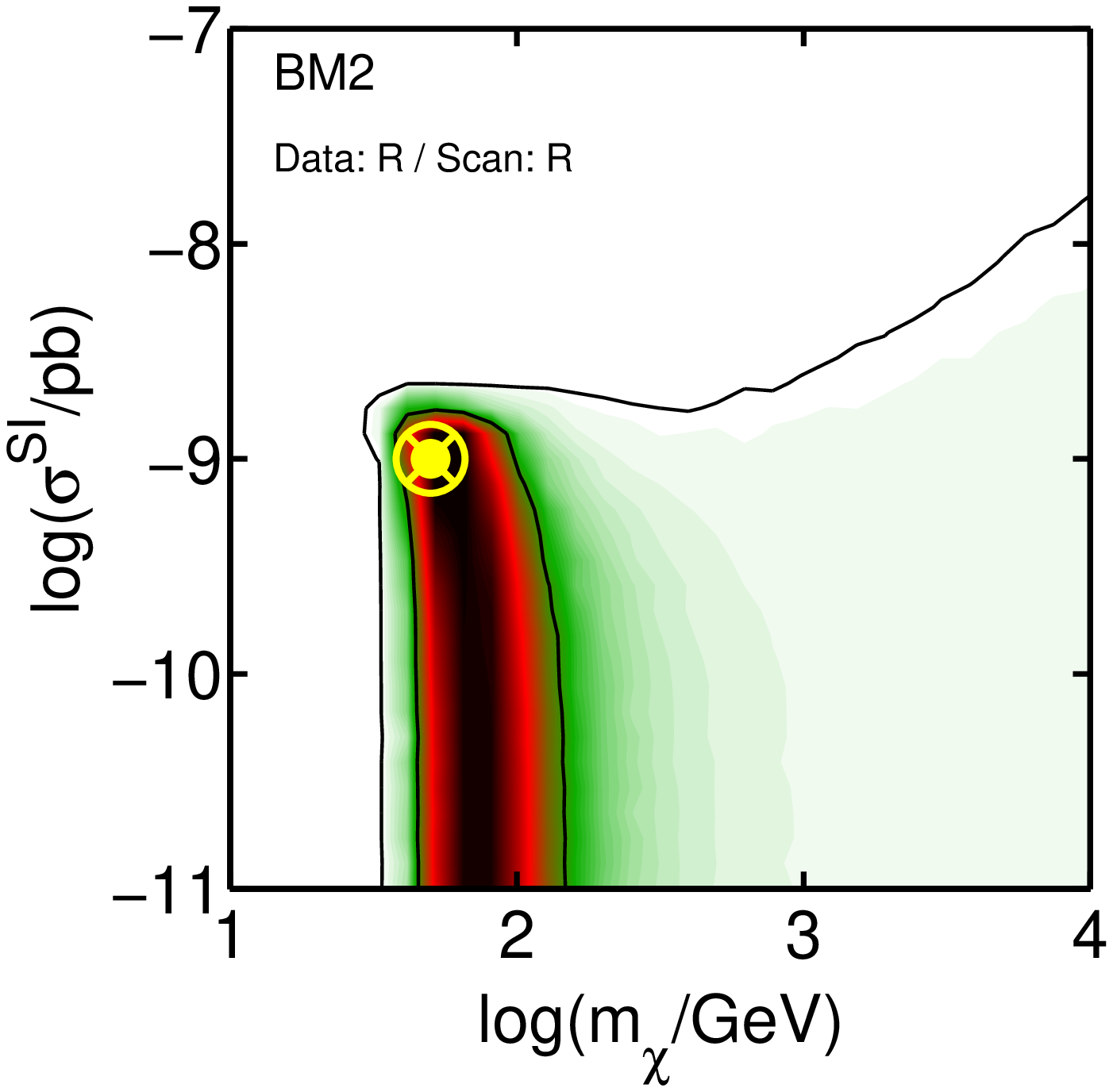}\hspace*{-0.62cm}
\includegraphics[width=0.35\textwidth]{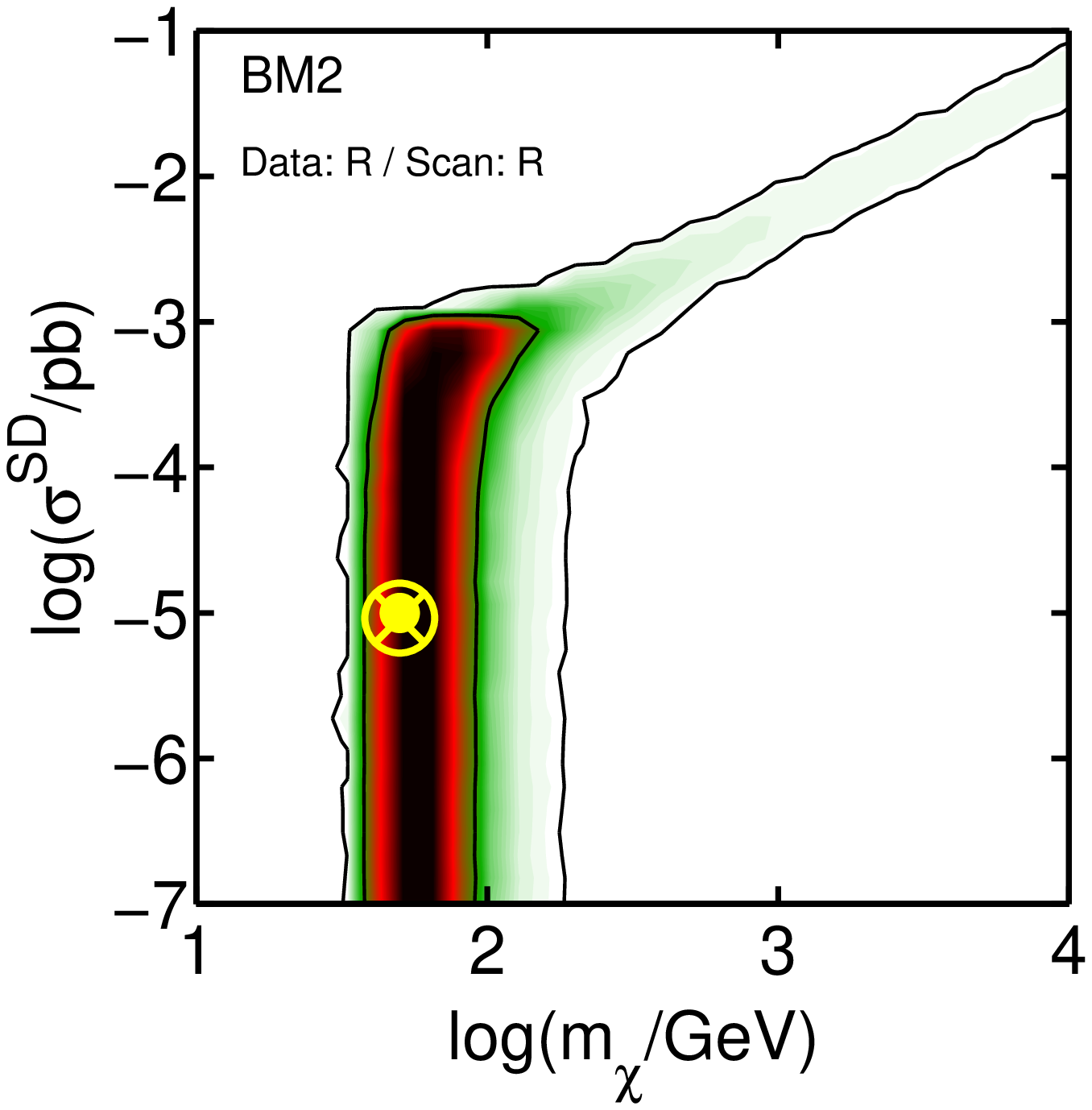}\hspace*{-0.62cm}
\includegraphics[width=0.35\textwidth]{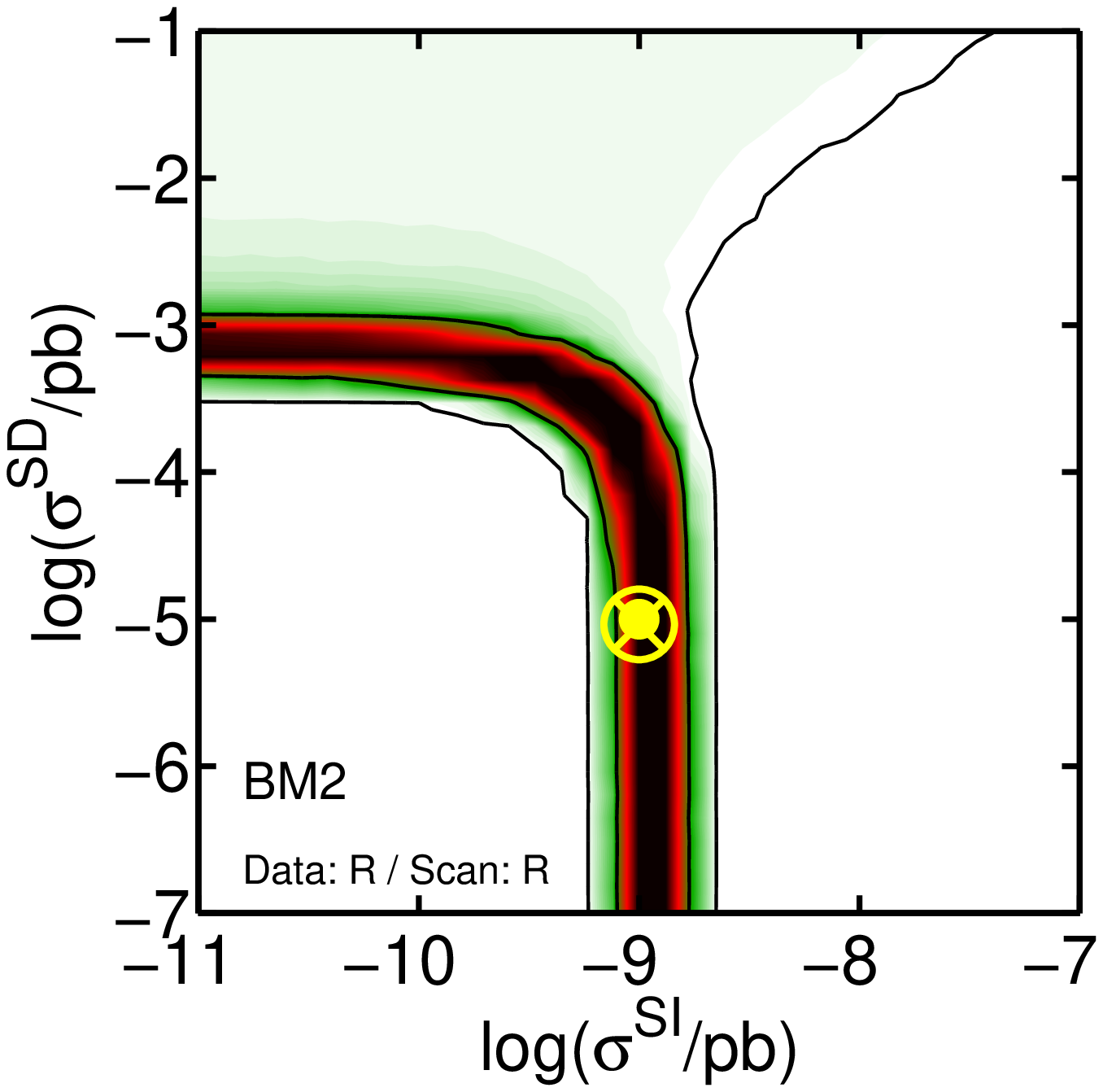}\\[-4ex]
\includegraphics[width=0.35\textwidth]{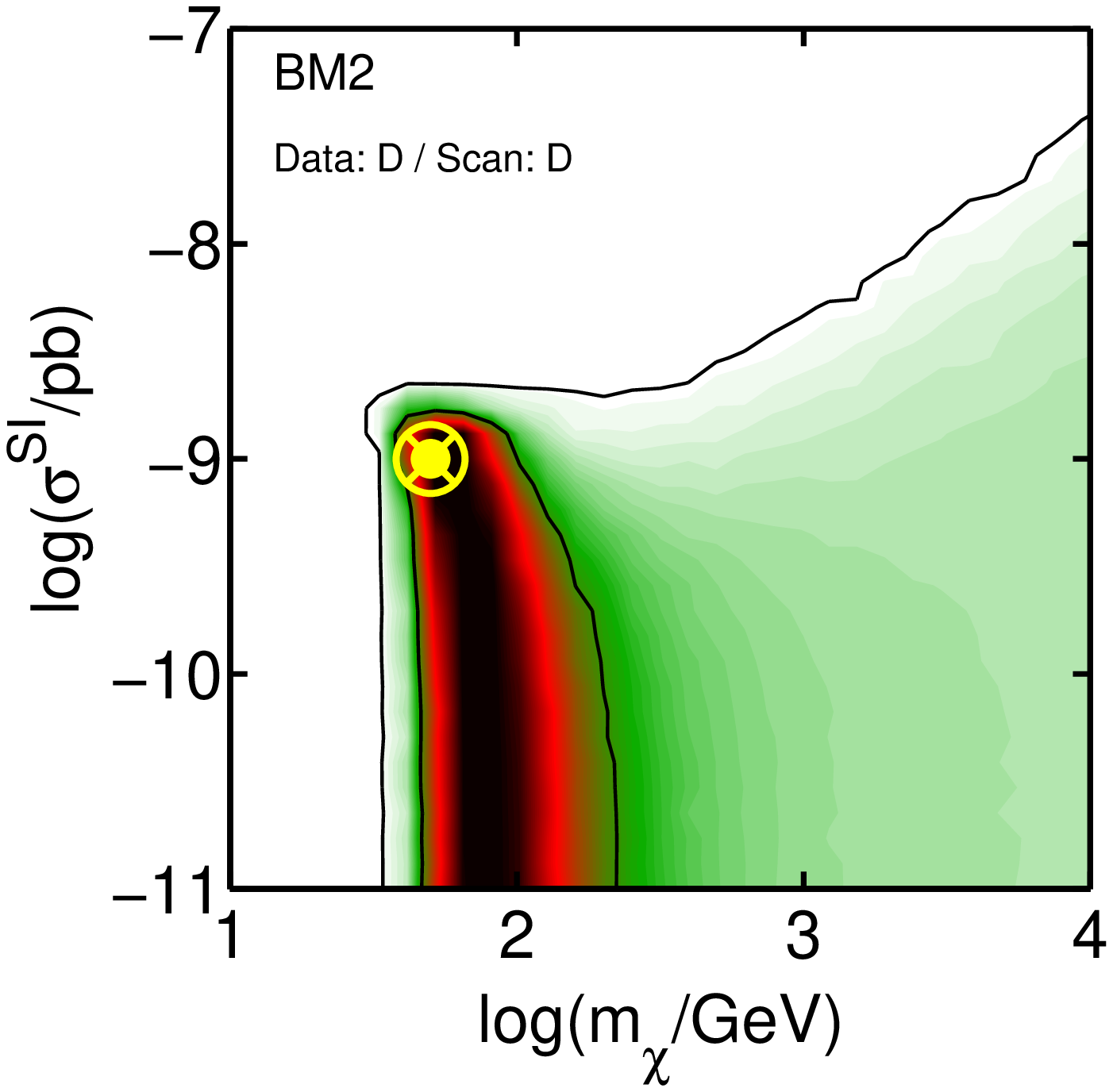}\hspace*{-0.62cm}
\includegraphics[width=0.35\textwidth]{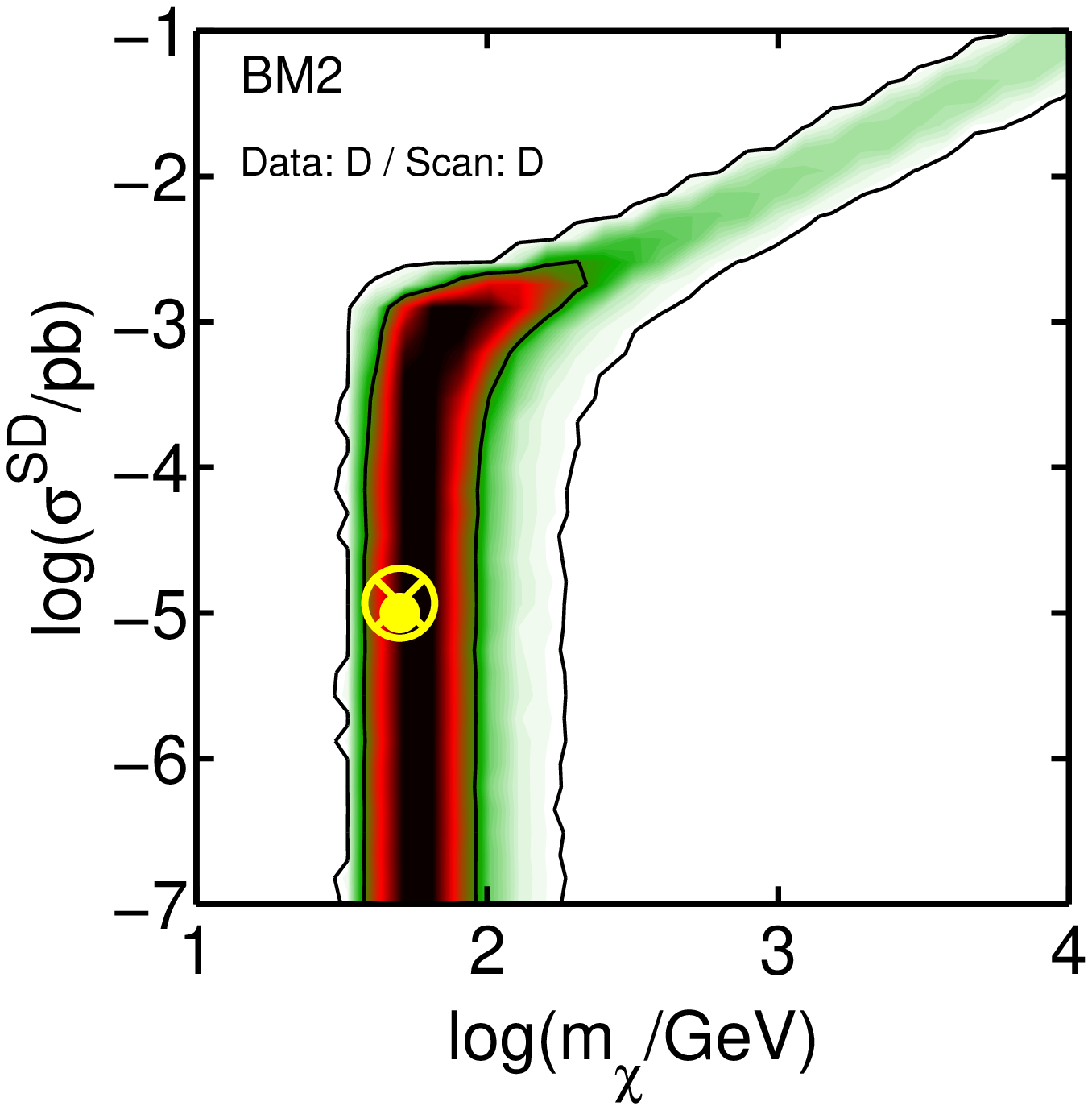}\hspace*{-0.62cm}
\includegraphics[width=0.35\textwidth]{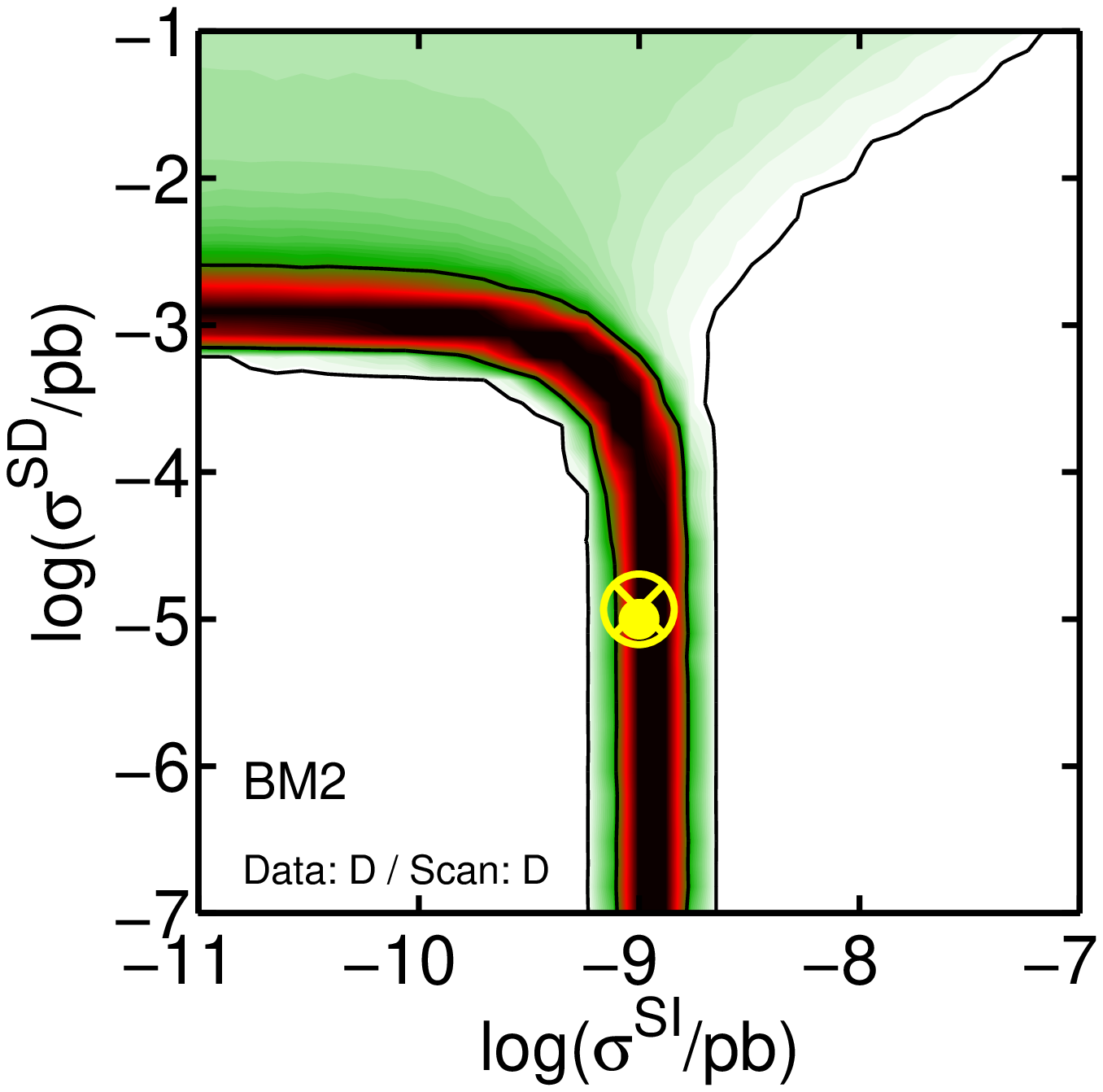}\\[-4ex]
\includegraphics[width=0.35\textwidth]{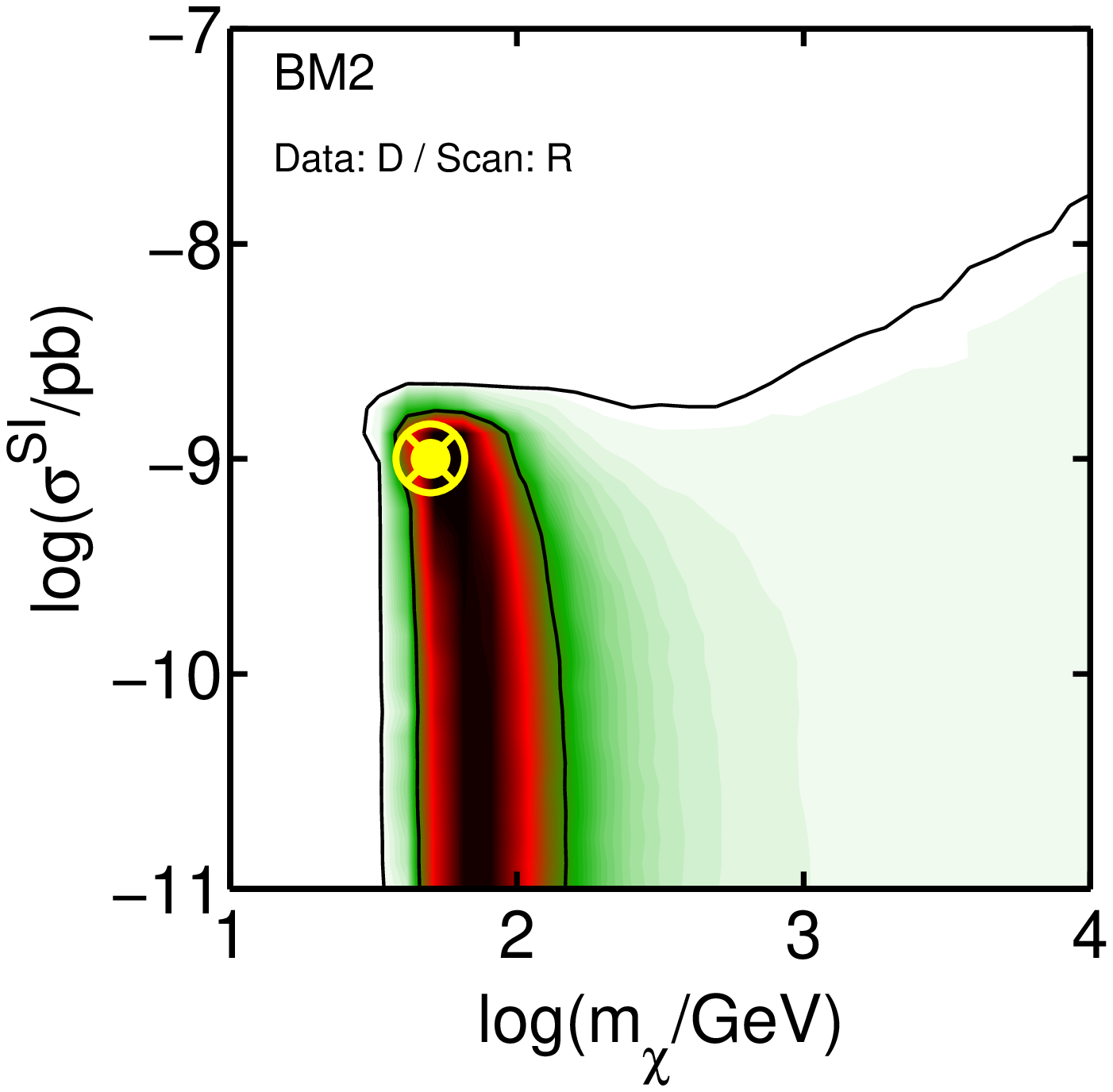}\hspace*{-0.62cm}
\includegraphics[width=0.35\textwidth]{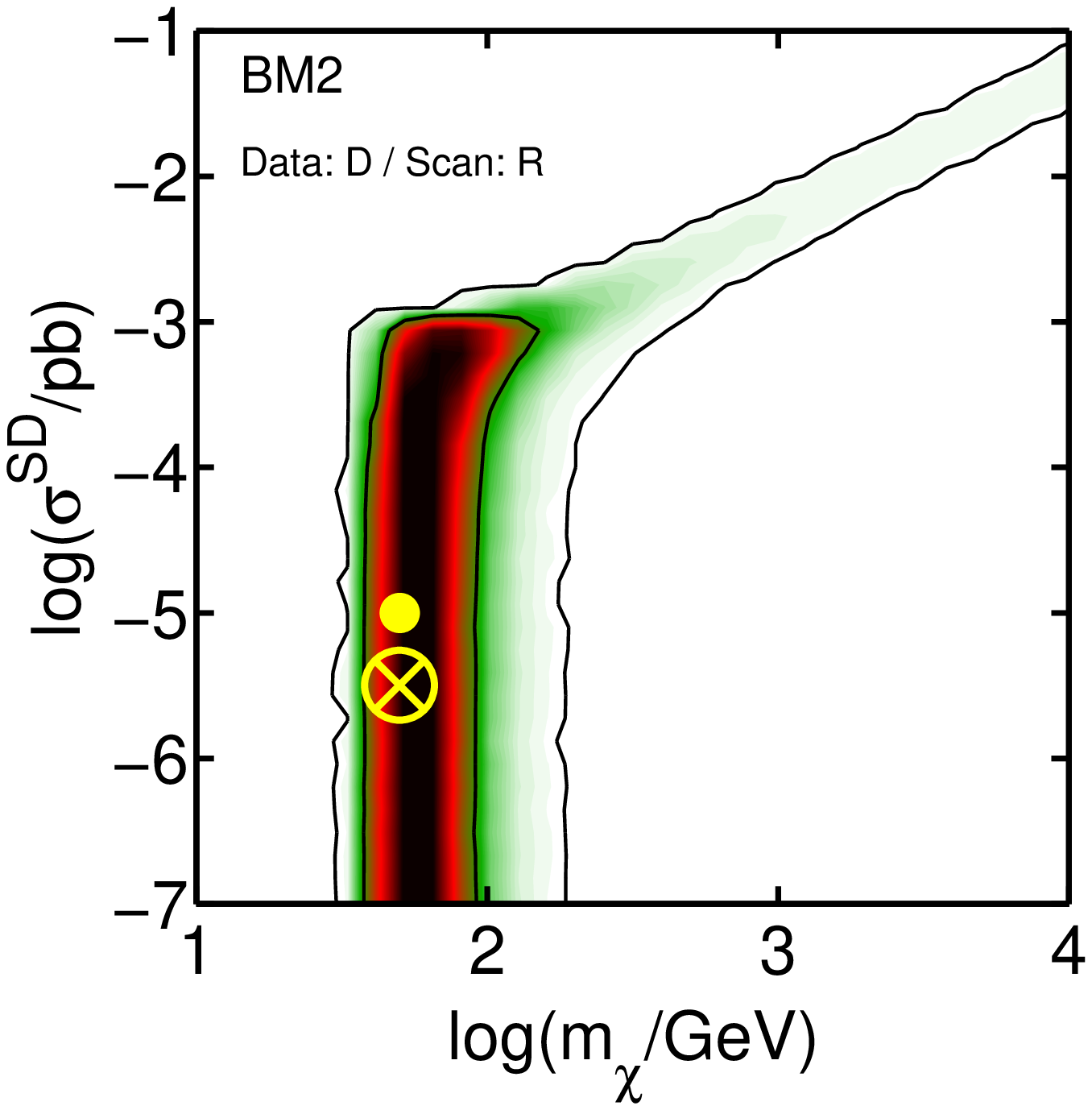}\hspace*{-0.62cm}
\includegraphics[width=0.35\textwidth]{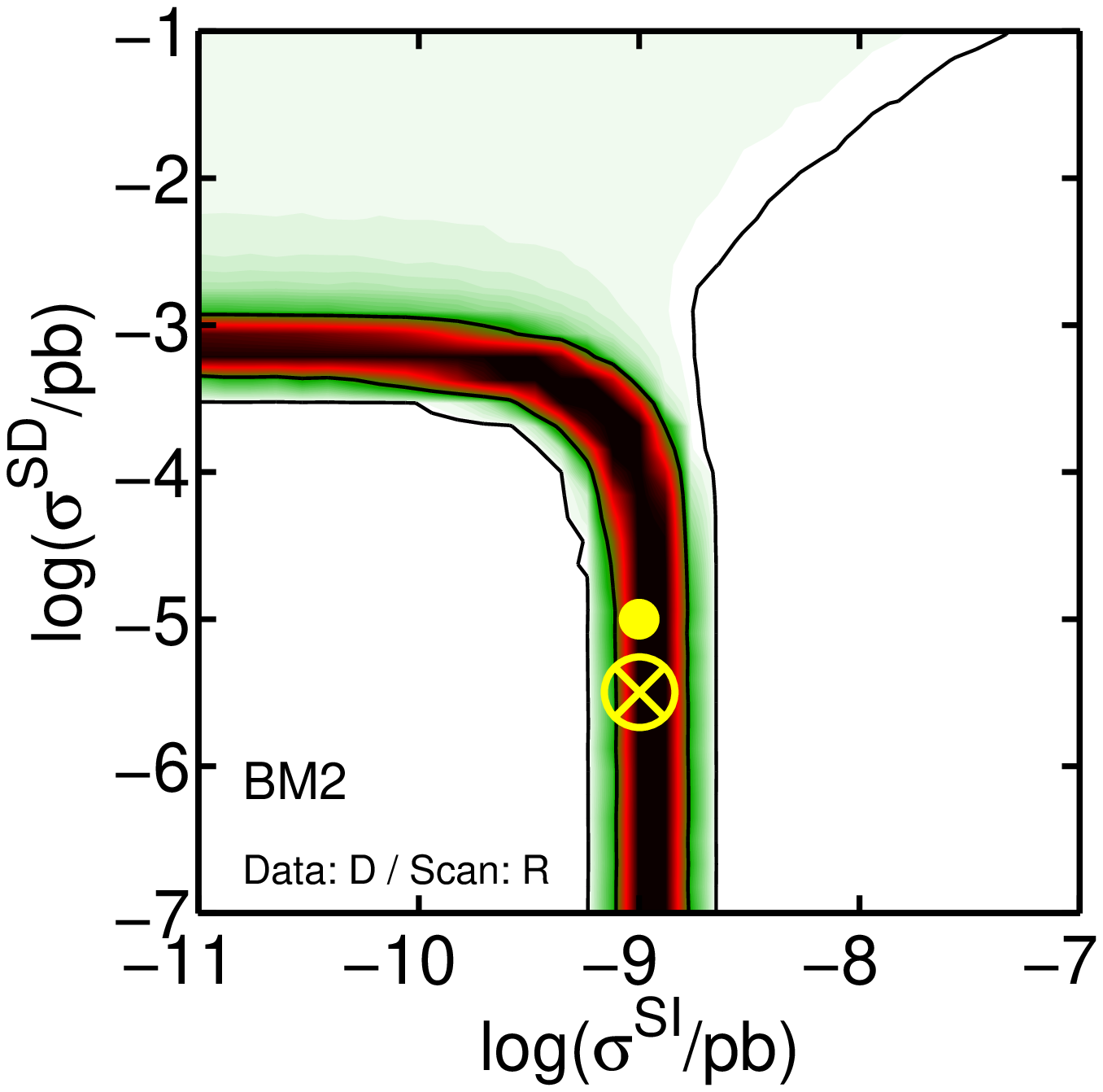}\\[-4ex]
\includegraphics[width=0.35\textwidth]{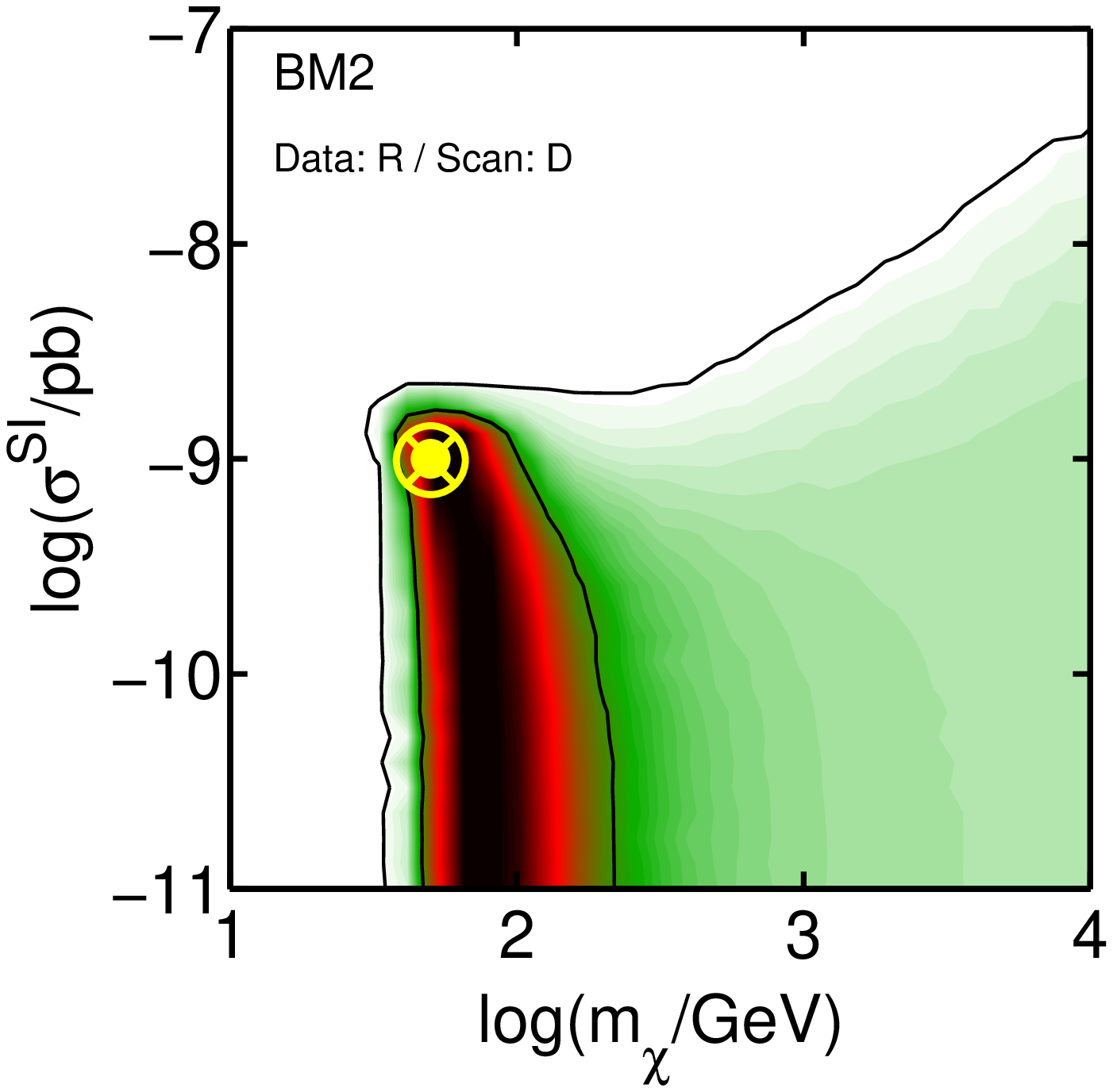}\hspace*{-0.62cm}
\includegraphics[width=0.35\textwidth]{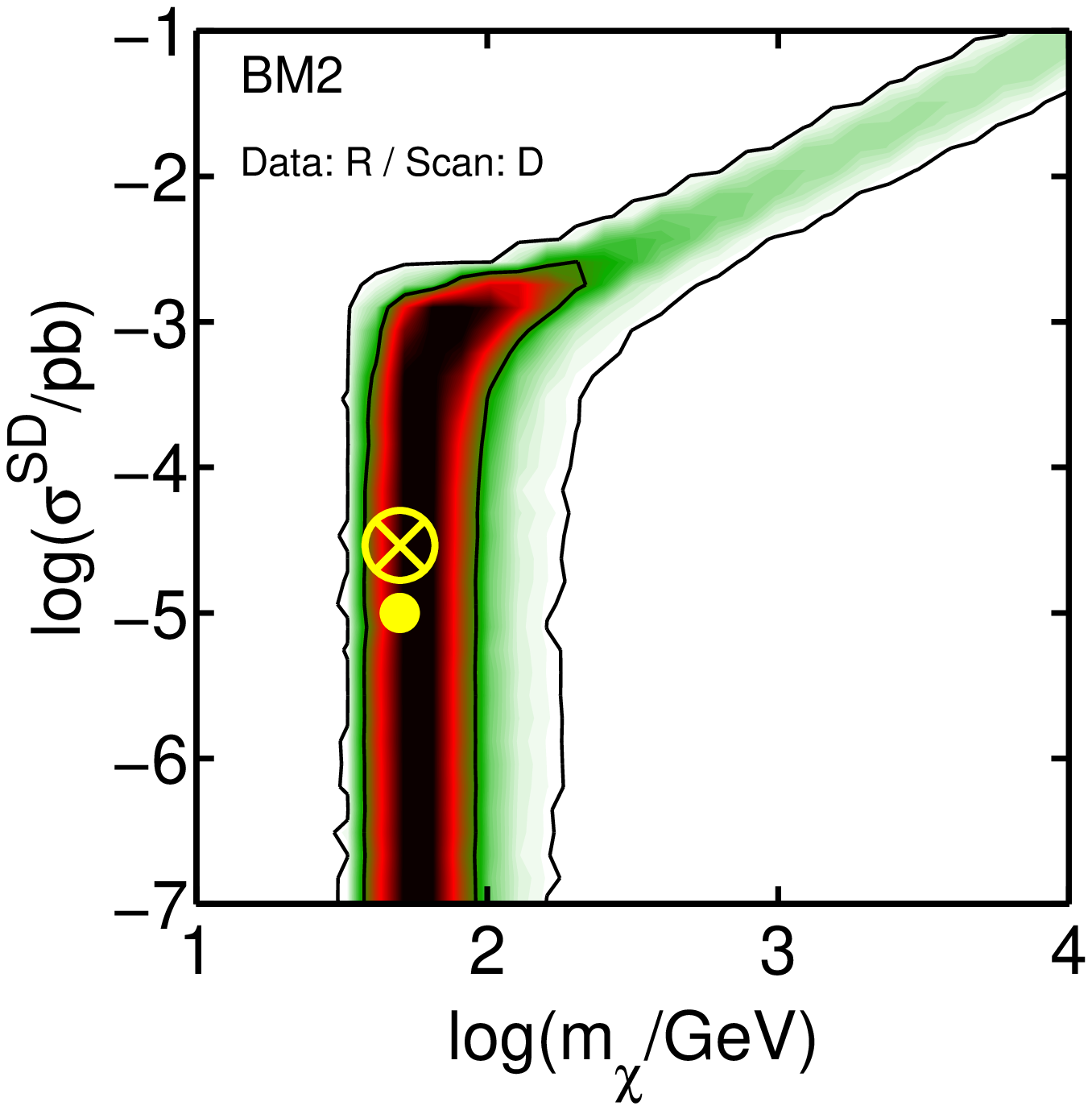}\hspace*{-0.62cm}
\includegraphics[width=0.35\textwidth]{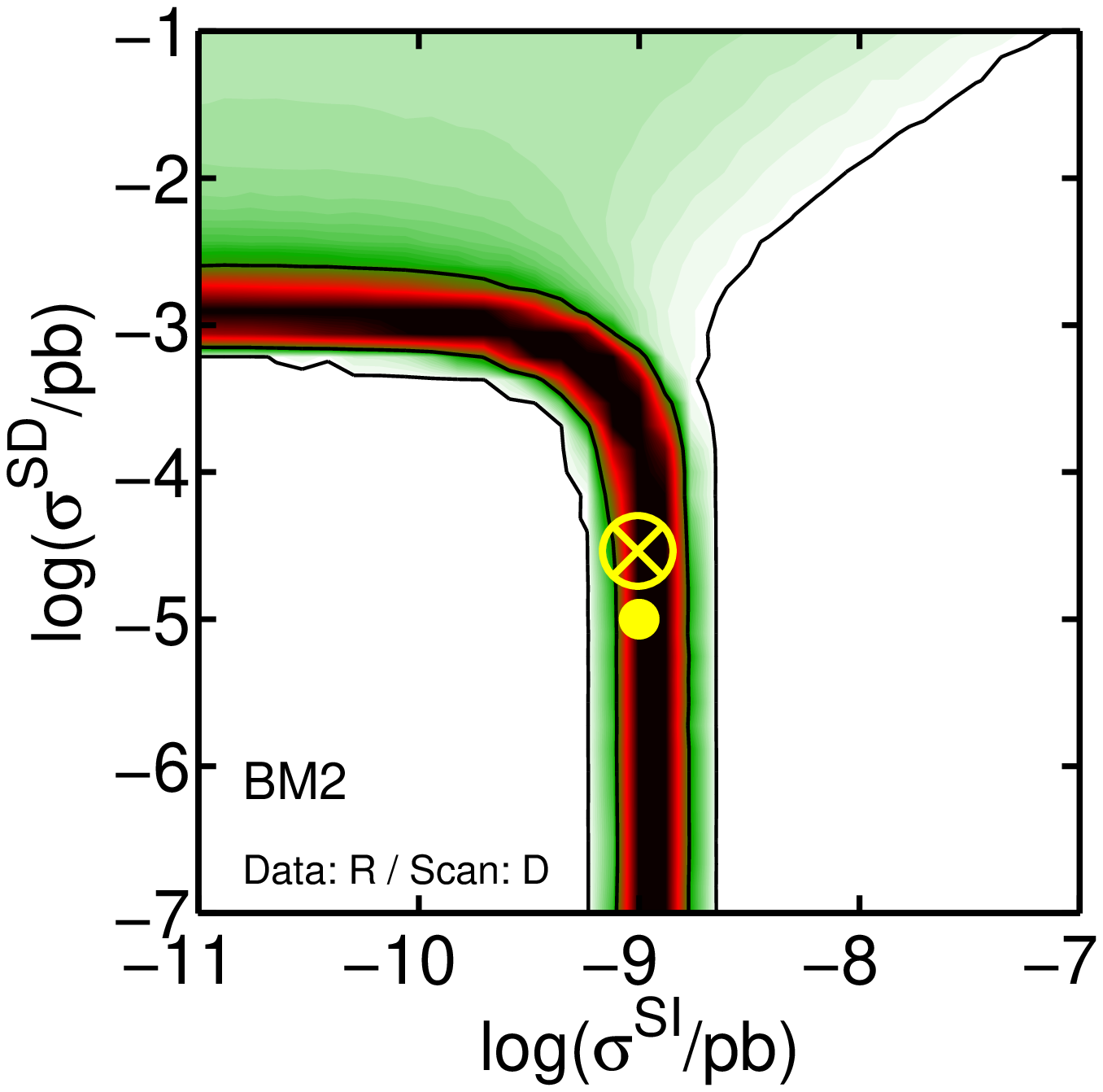}\\[-4ex]
\caption{\label{fig:BM2_profl} The same as in Fig.\,\ref{fig:BM1_profl} but for benchmark BM2.}
\end{figure*}

\begin{figure*}
\includegraphics[width=0.35\textwidth]{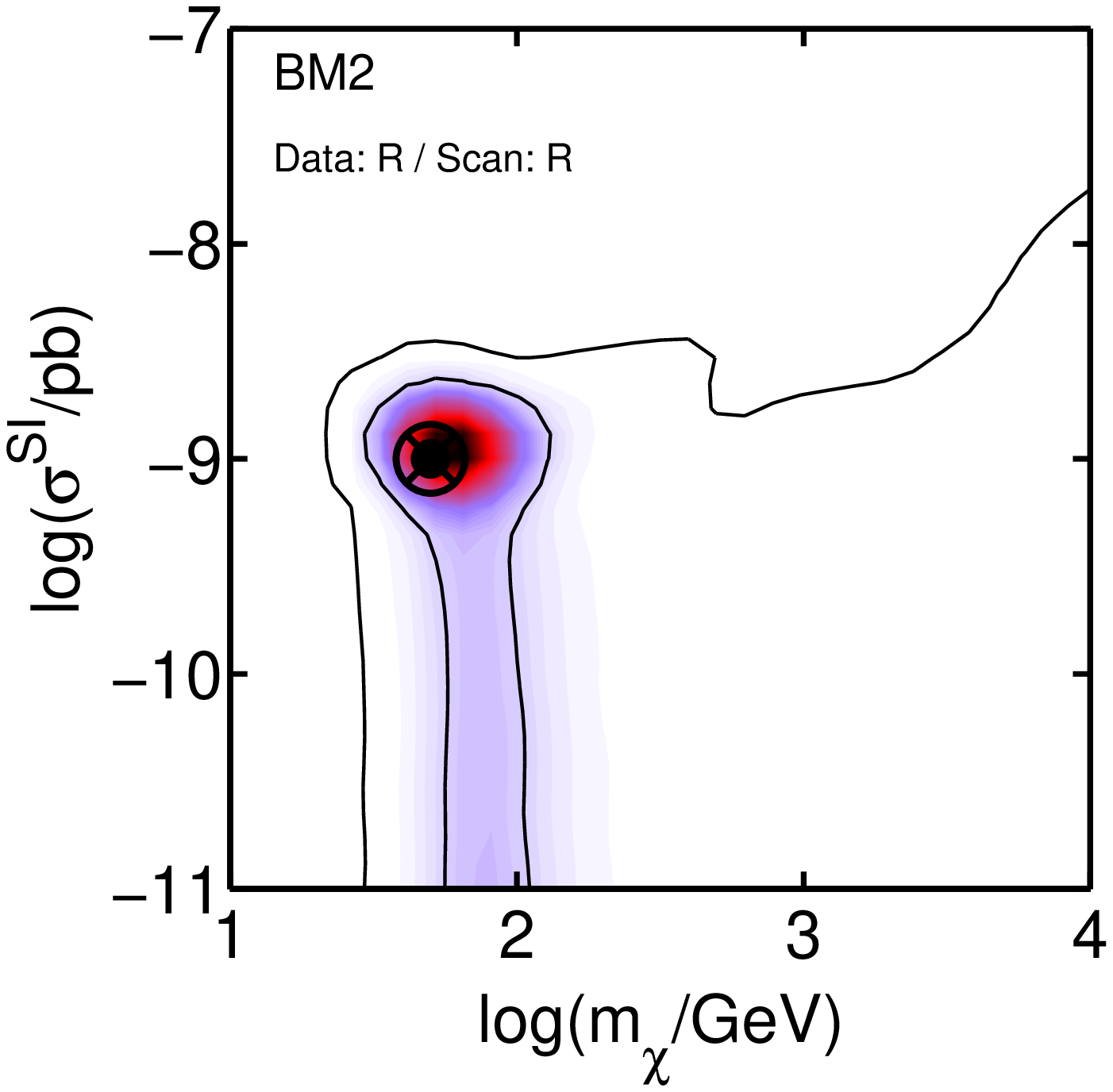}\hspace*{-0.62cm}
\includegraphics[width=0.35\textwidth]{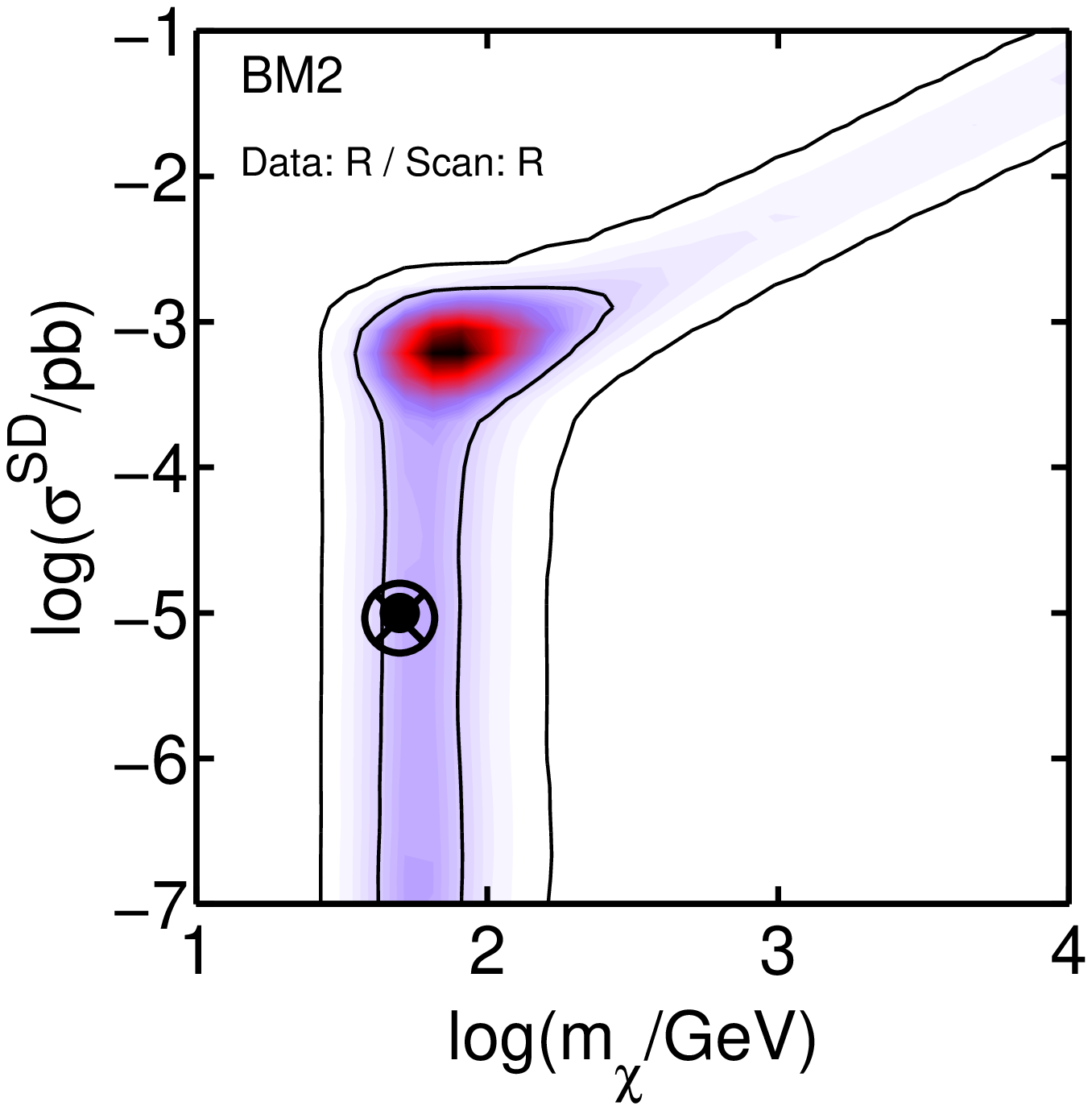}\hspace*{-0.62cm}
\includegraphics[width=0.35\textwidth]{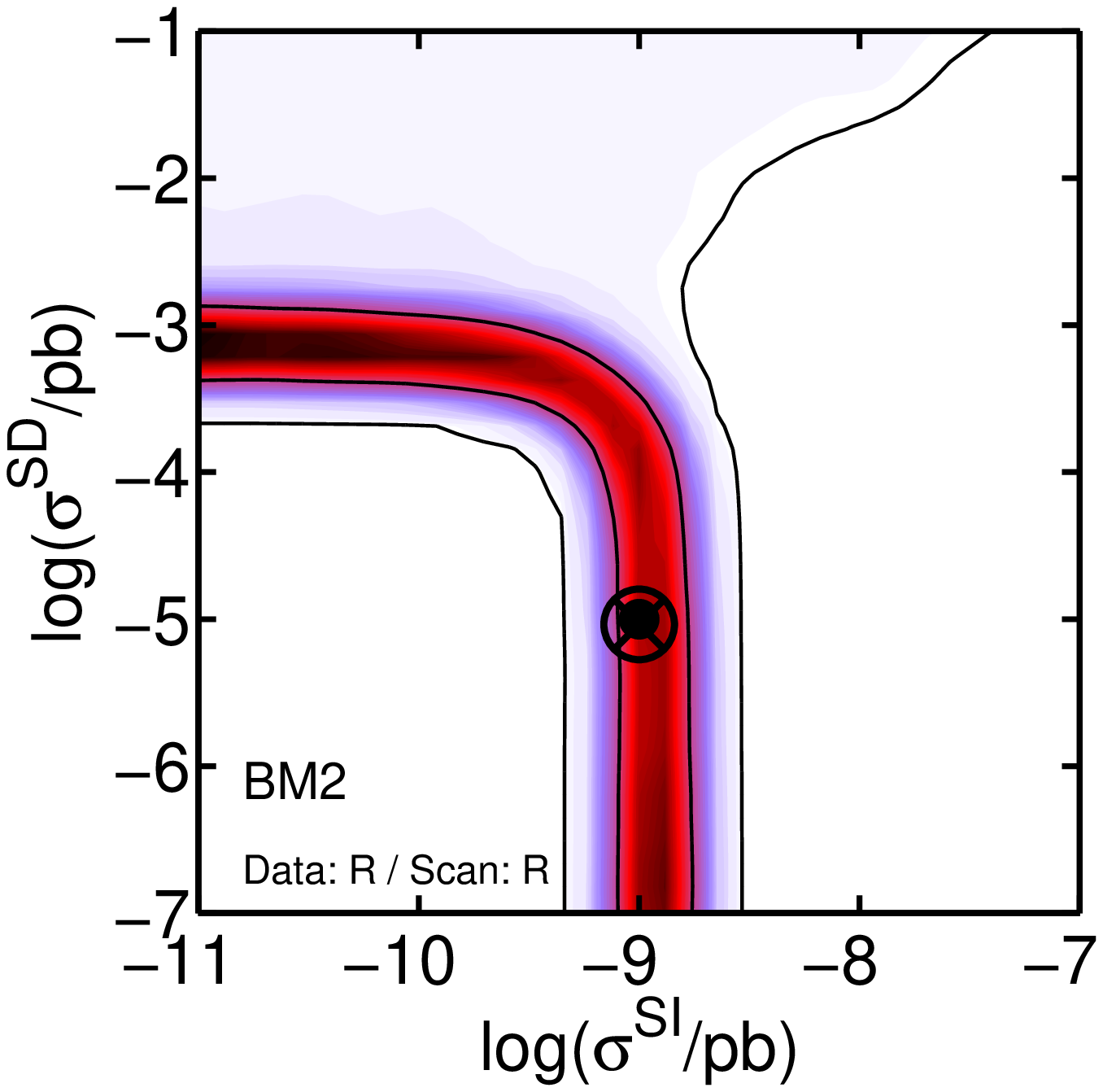}\\[-4ex]
\includegraphics[width=0.35\textwidth]{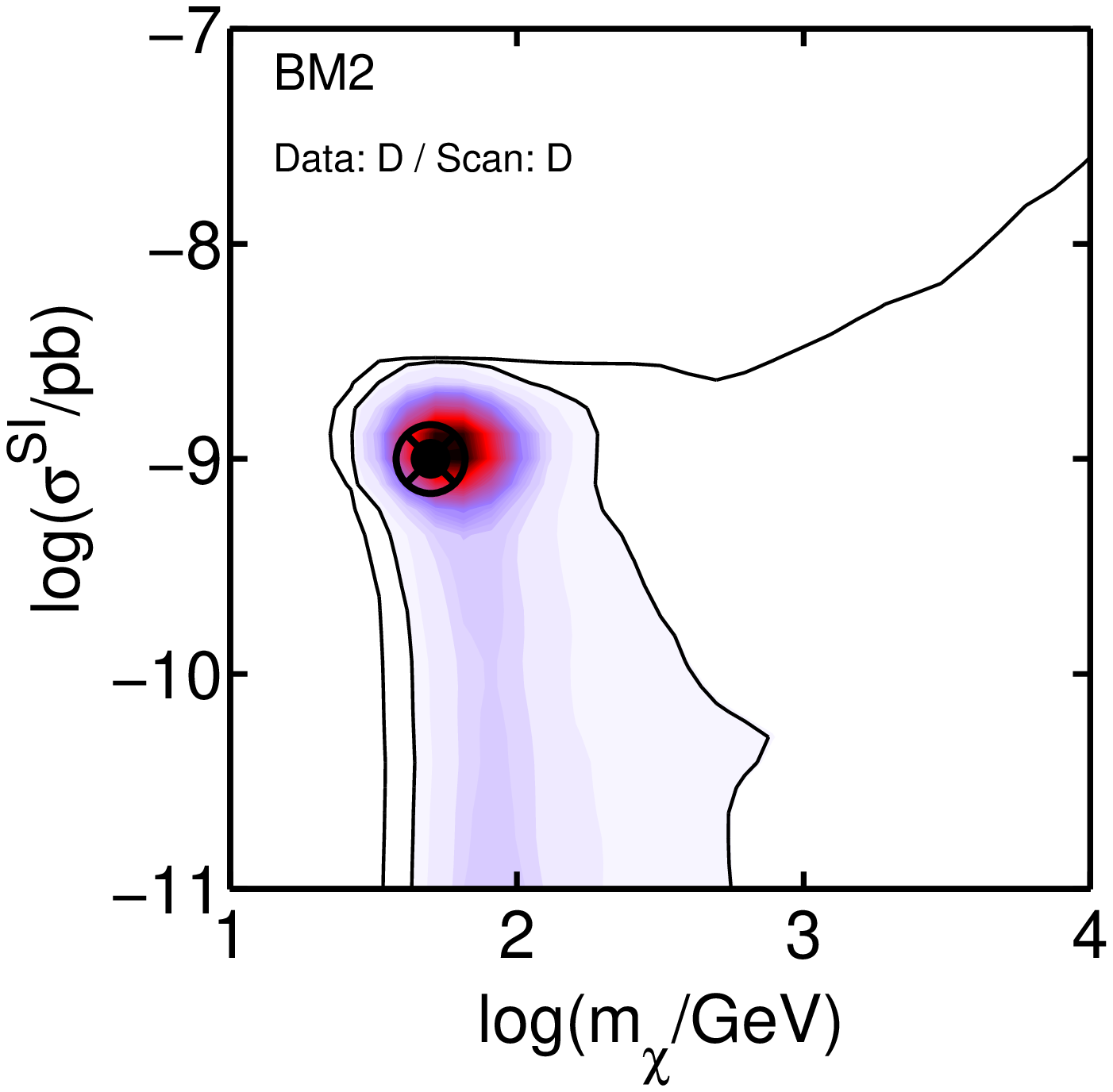}\hspace*{-0.62cm}
\includegraphics[width=0.35\textwidth]{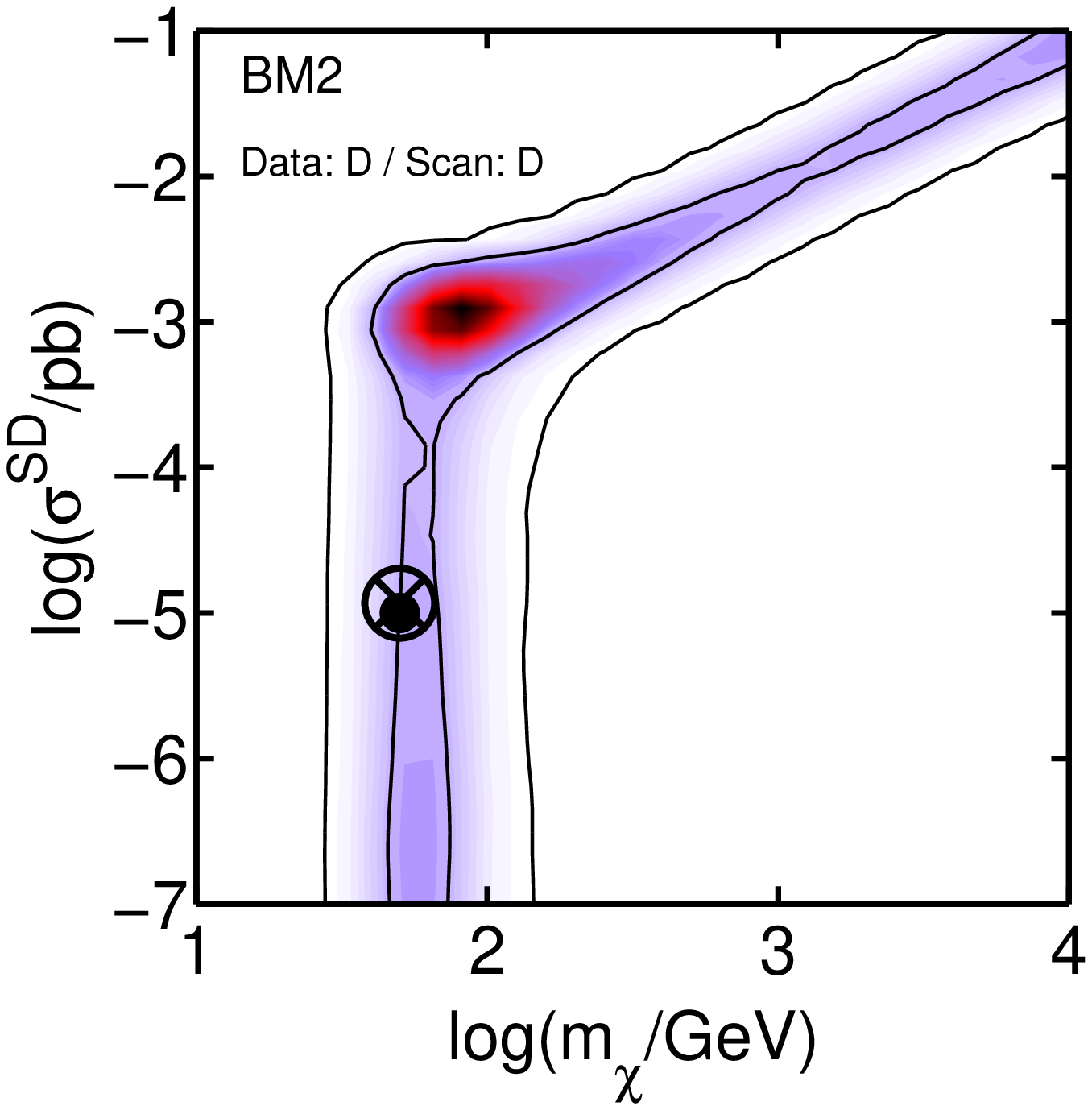}\hspace*{-0.62cm}
\includegraphics[width=0.35\textwidth]{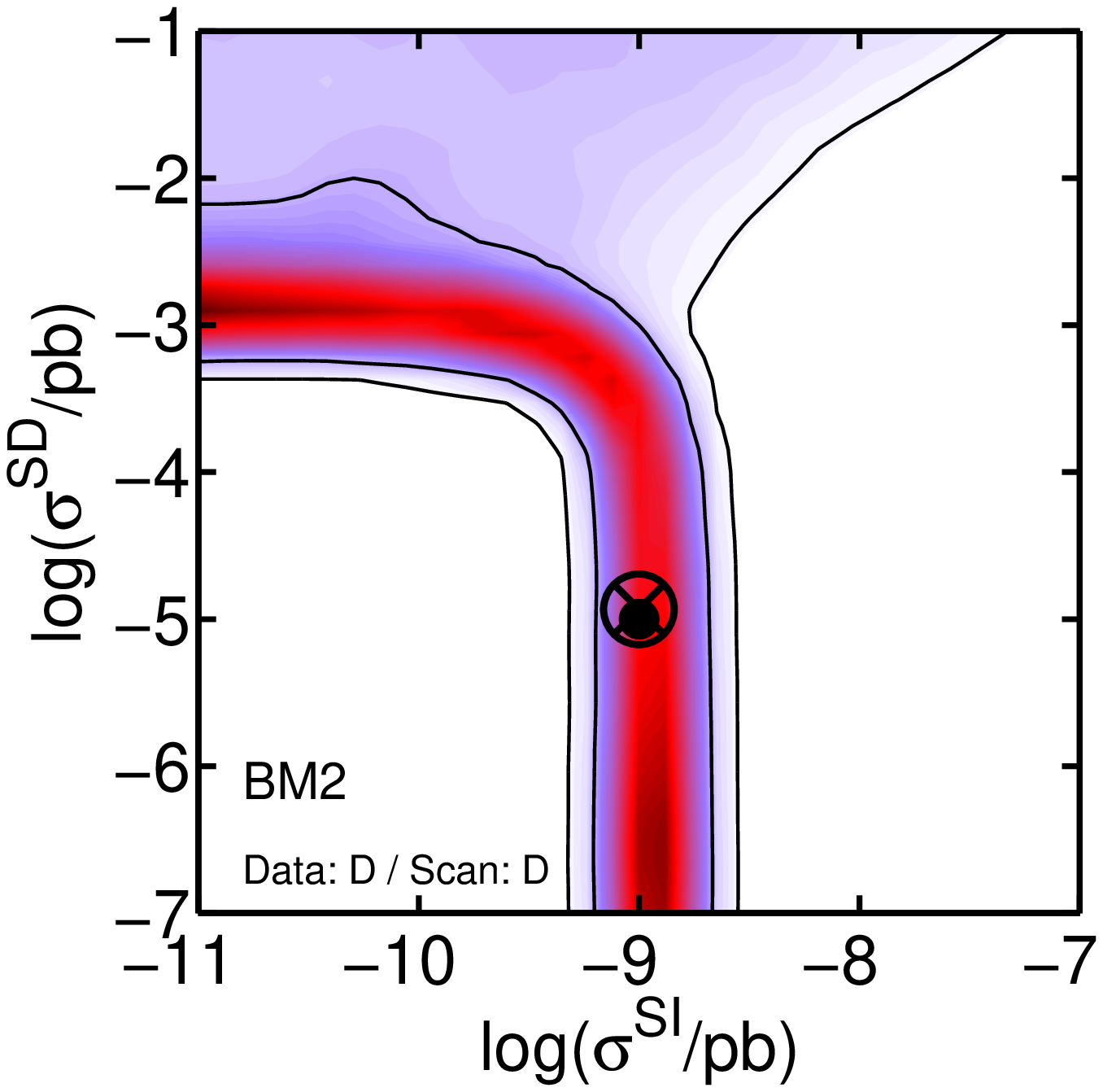}\\[-4ex]
\includegraphics[width=0.35\textwidth]{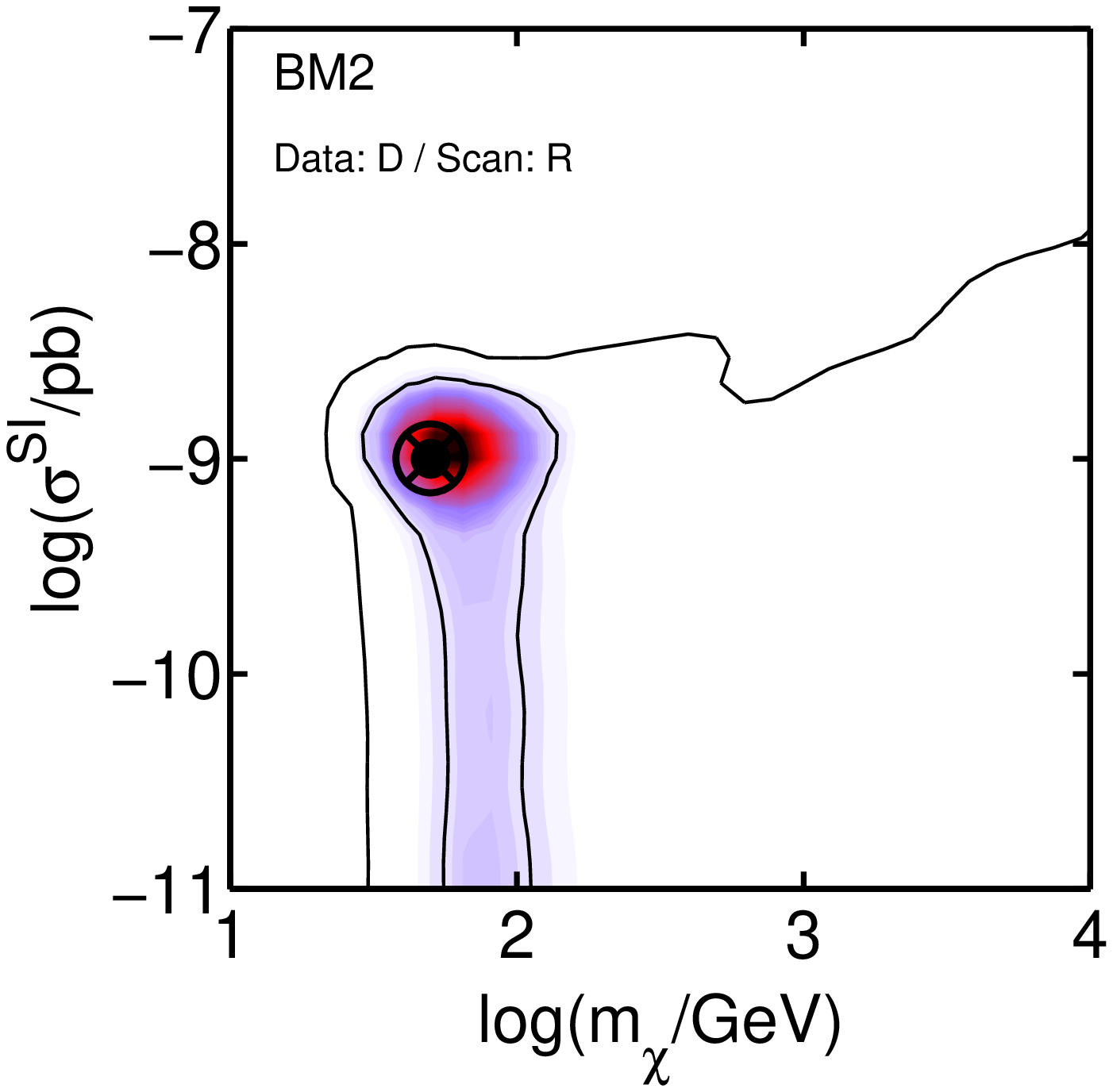}\hspace*{-0.62cm}
\includegraphics[width=0.35\textwidth]{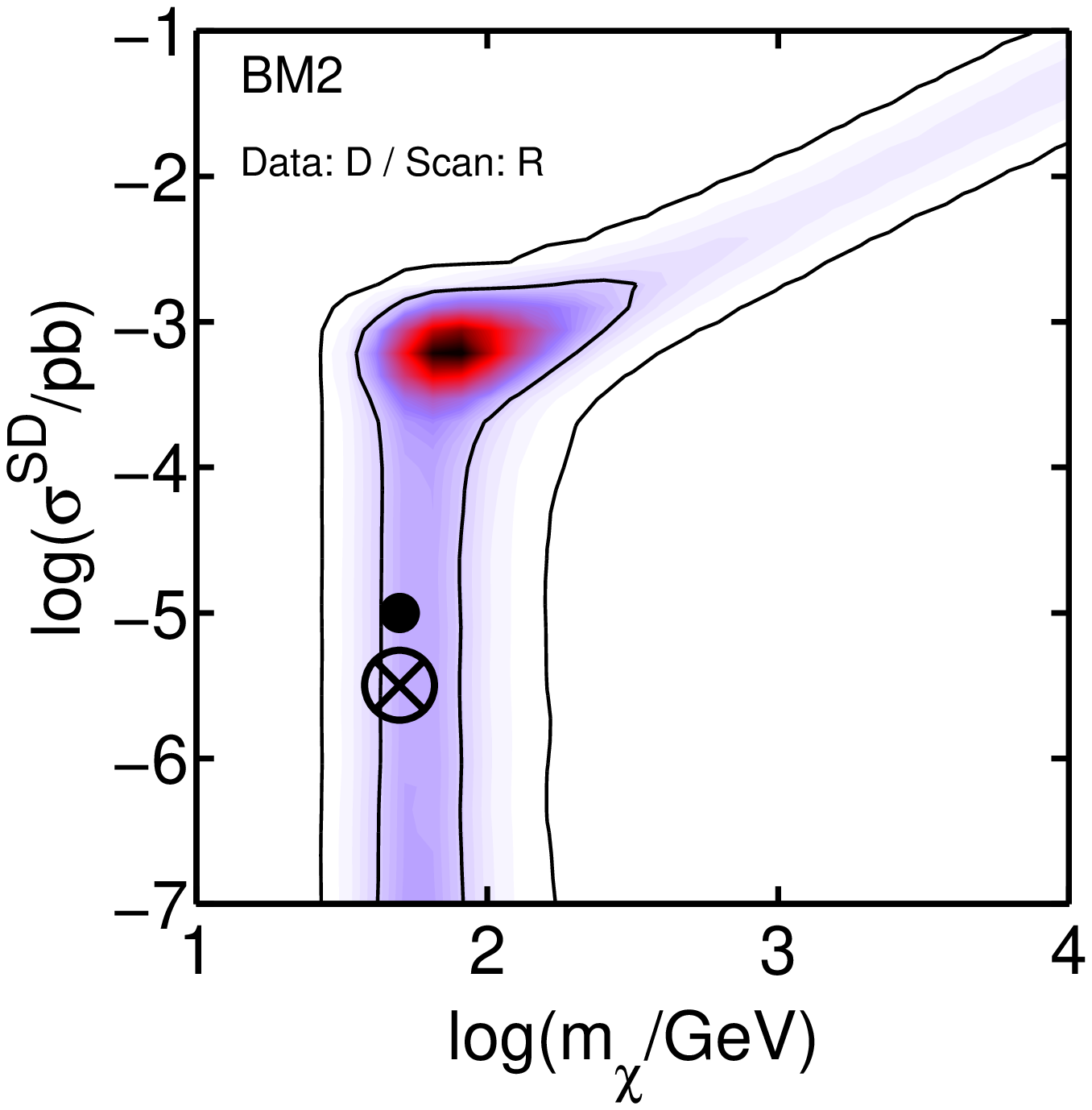}\hspace*{-0.62cm}
\includegraphics[width=0.35\textwidth]{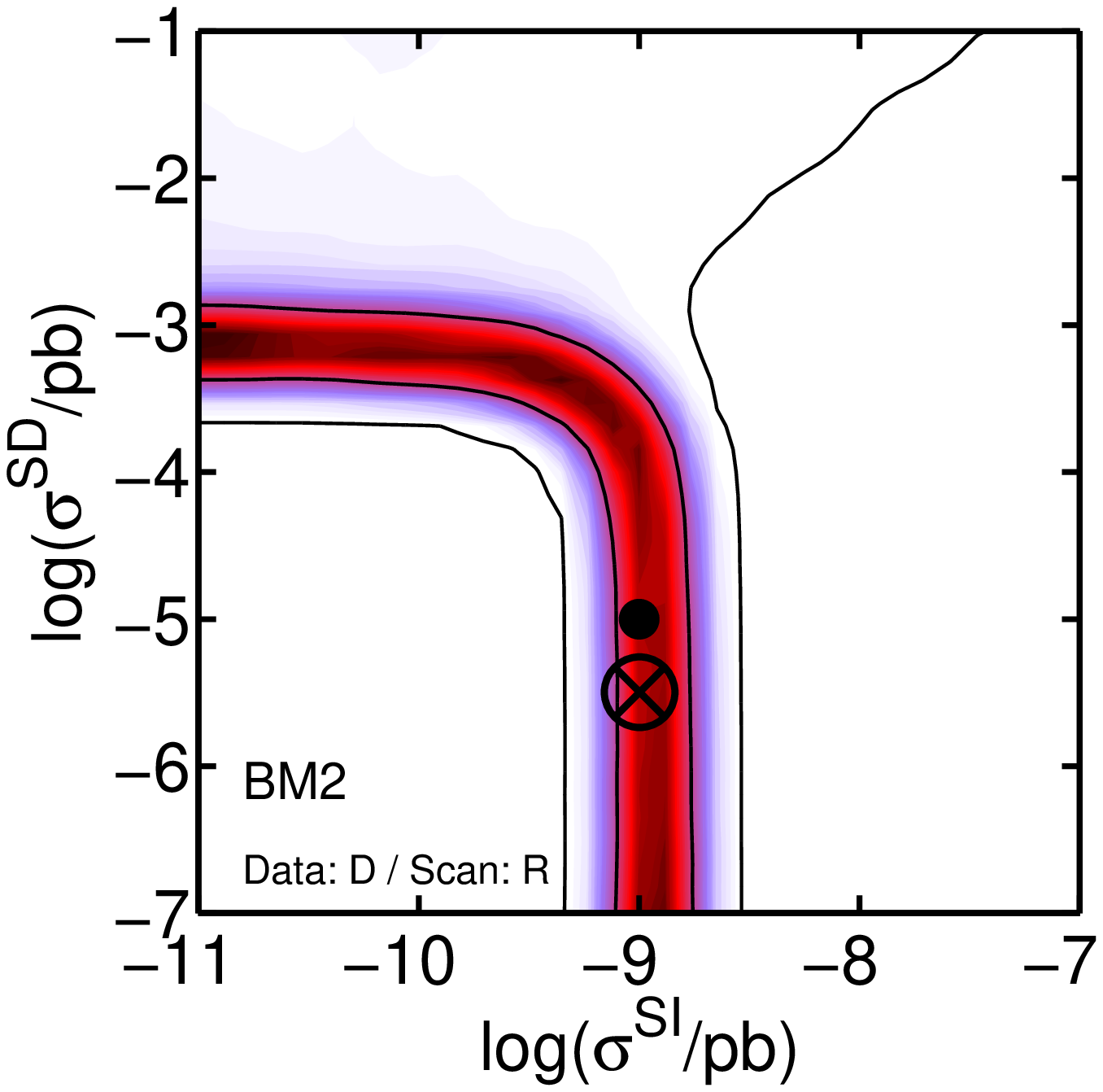}\\[-4ex]
\includegraphics[width=0.35\textwidth]{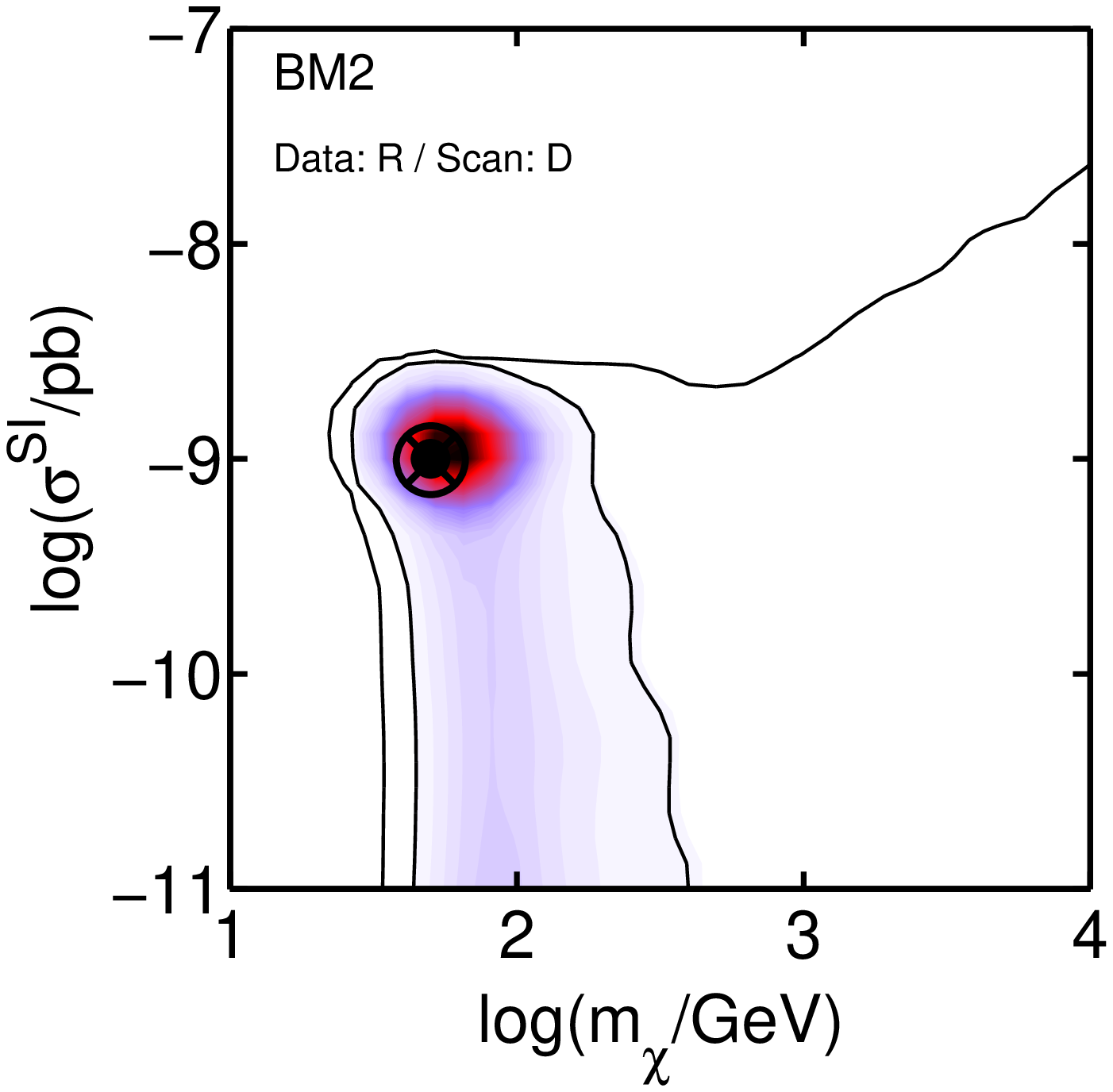}\hspace*{-0.62cm}
\includegraphics[width=0.35\textwidth]{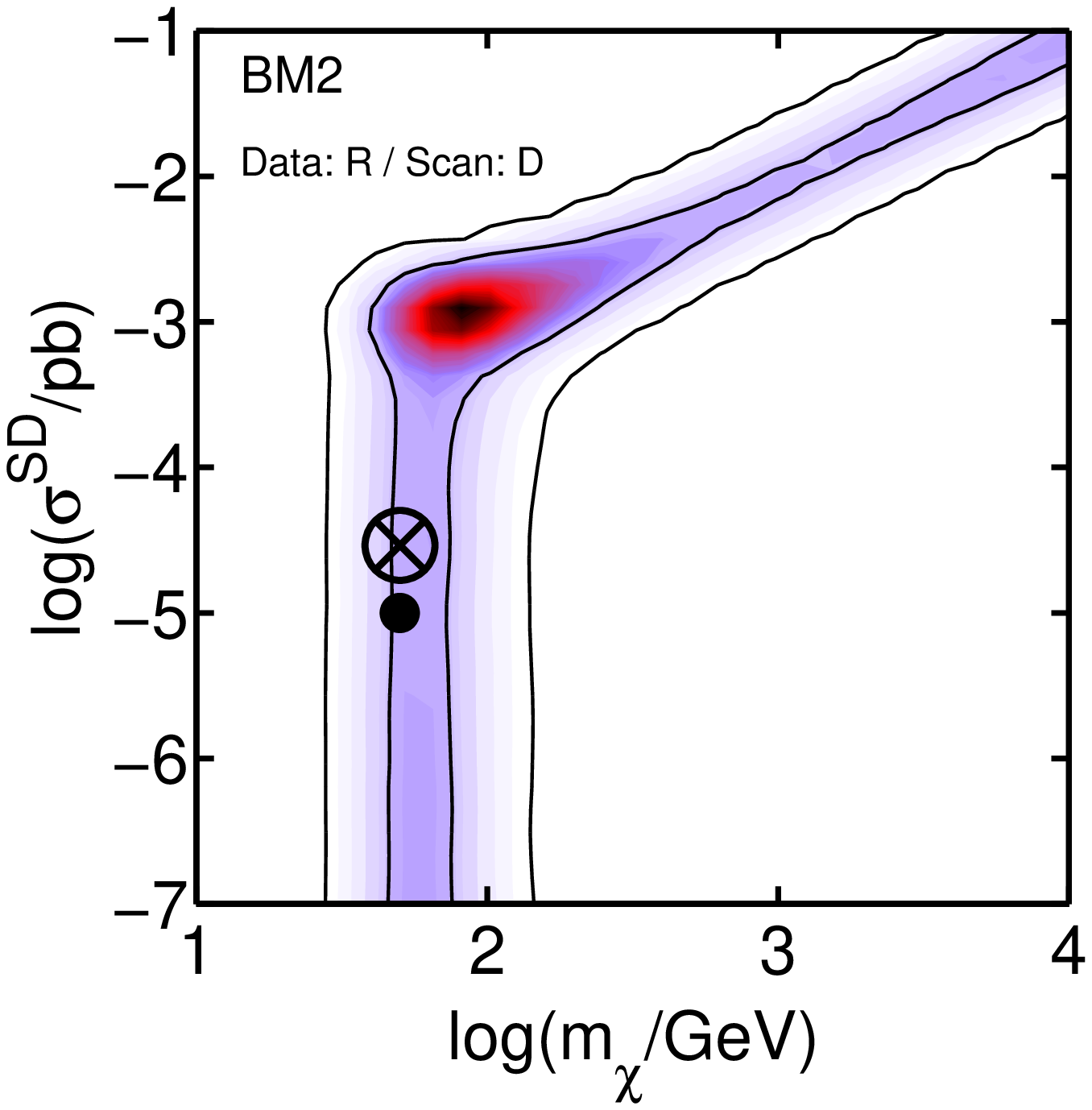}\hspace*{-0.62cm}
\includegraphics[width=0.35\textwidth]{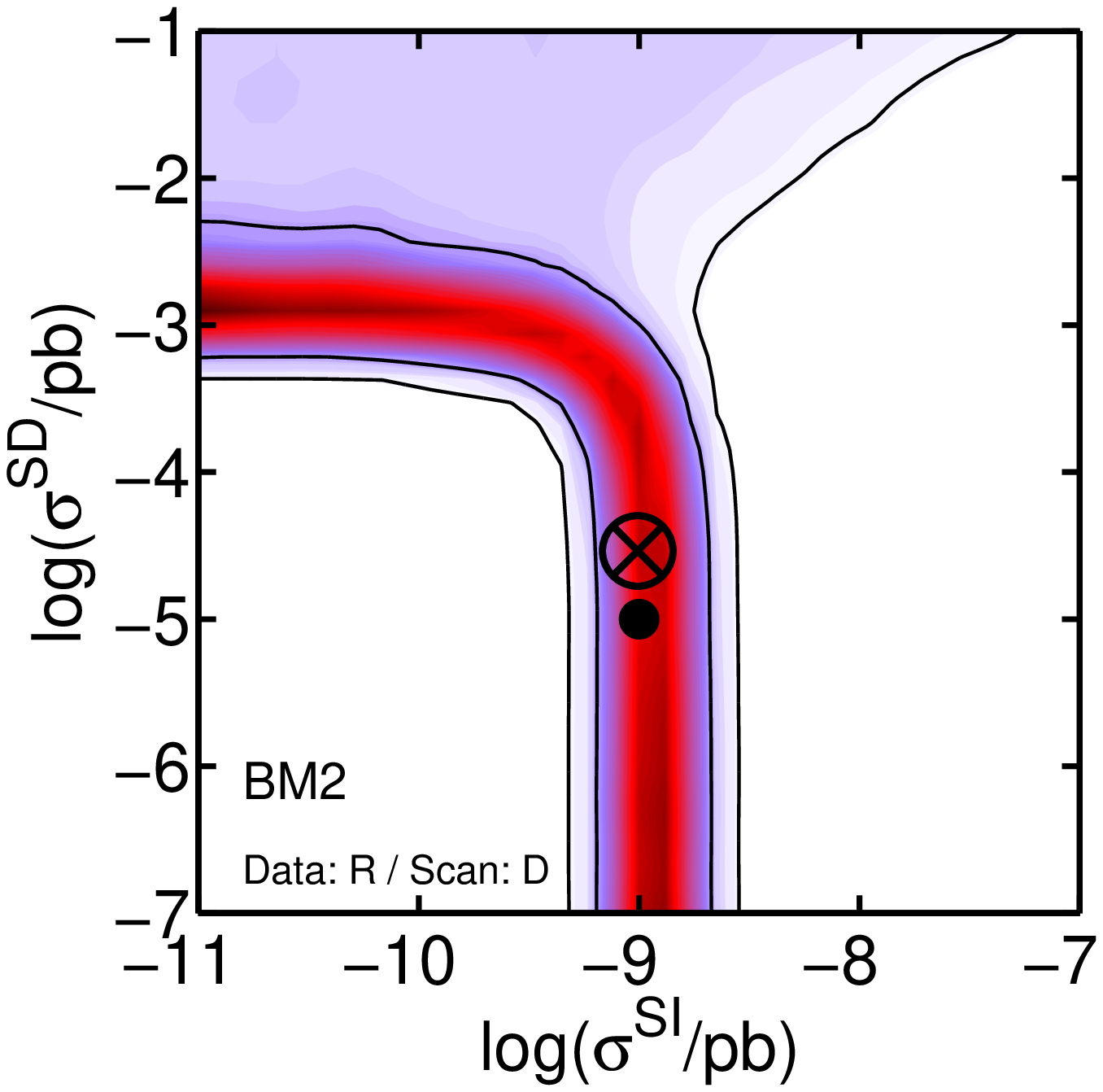}\\[-4ex]
\caption{\label{fig:BM2_pdf} The same as in Fig.\,\ref{fig:BM1_pdf} but for the benchmark BM2.}
\end{figure*}

%%%%%%%%%%%
%%%% BM2 %%%%
%%%%%%%%%%%

Let us now consider the second benchmark, BM2. We proceed as in the previous case and 
show in Figs.\,\ref{fig:BM2_profl} and \ref{fig:BM2_pdf} the corresponding reconstruction of the phenomenological parameters in terms of the profile likelihood and pdf, respectively.
The difference between profile likelihood and pdf (due to the volume 
effect) is now more striking, especially regarding the SD component and WIMP mass. As we see in Fig.\,\ref{fig:BM2_profl} the regions with a best likelihood lie around the correct mass but span many orders of magnitude in $\sigsd$. These regions, however, have a small volume and are disfavoured when the pdf is plotted. We should emphasize at this point that the information from both sources has a different statistical meaning and therefore this is no evidence of inconsistency.

As in the previous scenario, the detection rate in this benchmark point is due almost 
entirely to SI interactions, and there are no 
differences between the simulated data with either the R- or D- model for the SDSF (see the middle plot in Fig.\,\ref{fig:benchmark_rate}).
However, the number of events is now  
significantly larger and this allows a better determination of the slope of the recoil spectrum.
This has two effects: first, the WIMP mass can be more accurately predicted (points with a heavy WIMP being now more disfavoured than in the previous example), and second leads to larger differences in the reconstruction of $\sigsd$ when different models for the SDSF are used.
Notice, for example, how heavy WIMPs are a viable possibility only if the contribution from the SD cross-section is sufficiently large (otherwise the shape of the spectrum is not flat enough).
In spite of this, the degeneracy between $\sigsi$ and $\sigsd$ persists. 
The main difference in the reconstruction using the R- or D-model is again the value of the lowest $\sigsd$ compatible
with the data (when $\sigsi$ is negligible), which is smaller for the R-model.
Also the contours corresponding to the 68\% confidence level extend towards larger WIMP masses in the case of the reconstruction using the D-model, in order to compensate for its greater steepness.

%%%%%%%%%%%
%%%% BM3 %%%%
%%%%%%%%%%%

\begin{figure*}
\includegraphics[width=0.35\textwidth]{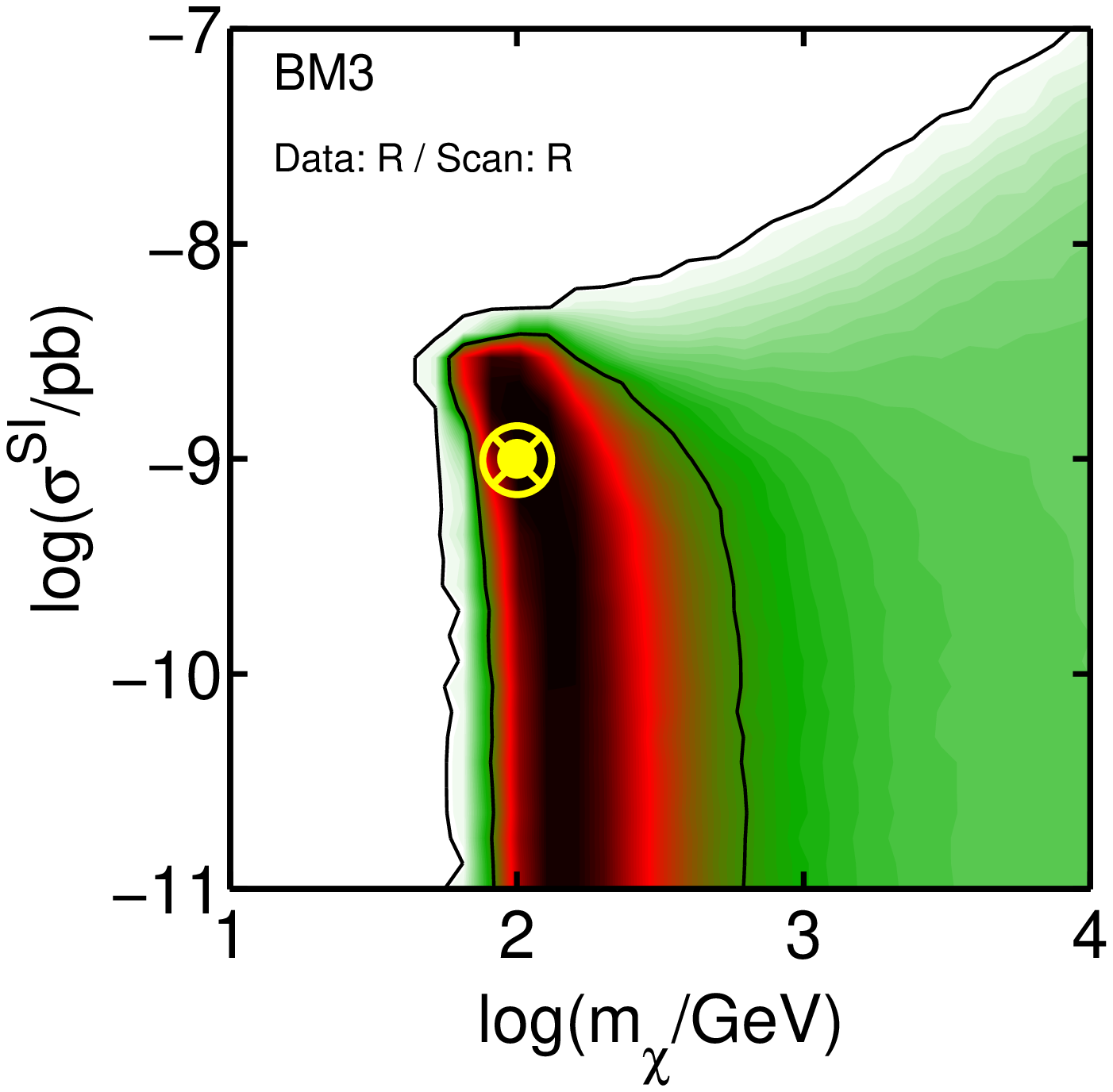}\hspace*{-0.62cm}
\includegraphics[width=0.35\textwidth]{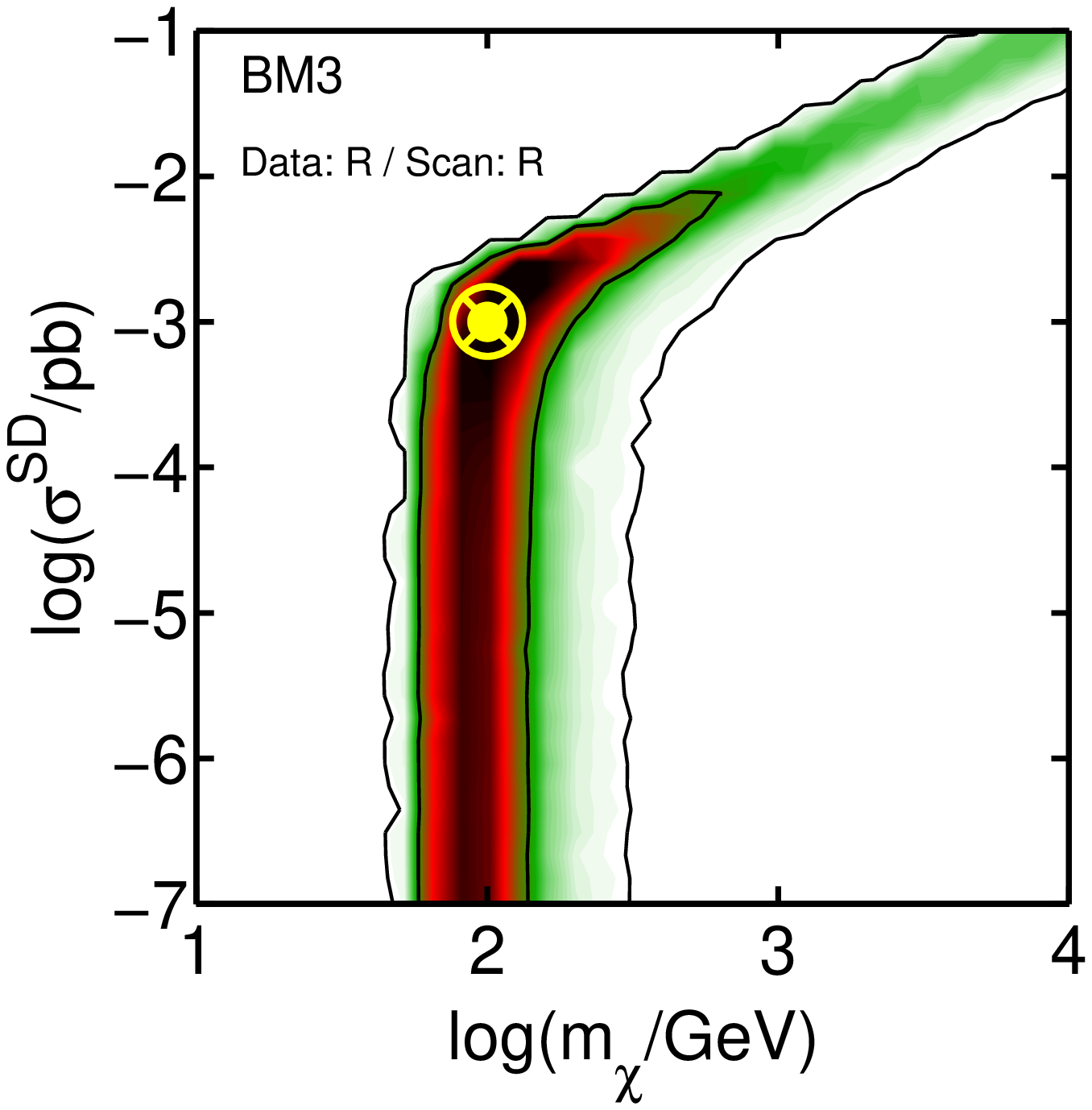}\hspace*{-0.62cm}
\includegraphics[width=0.35\textwidth]{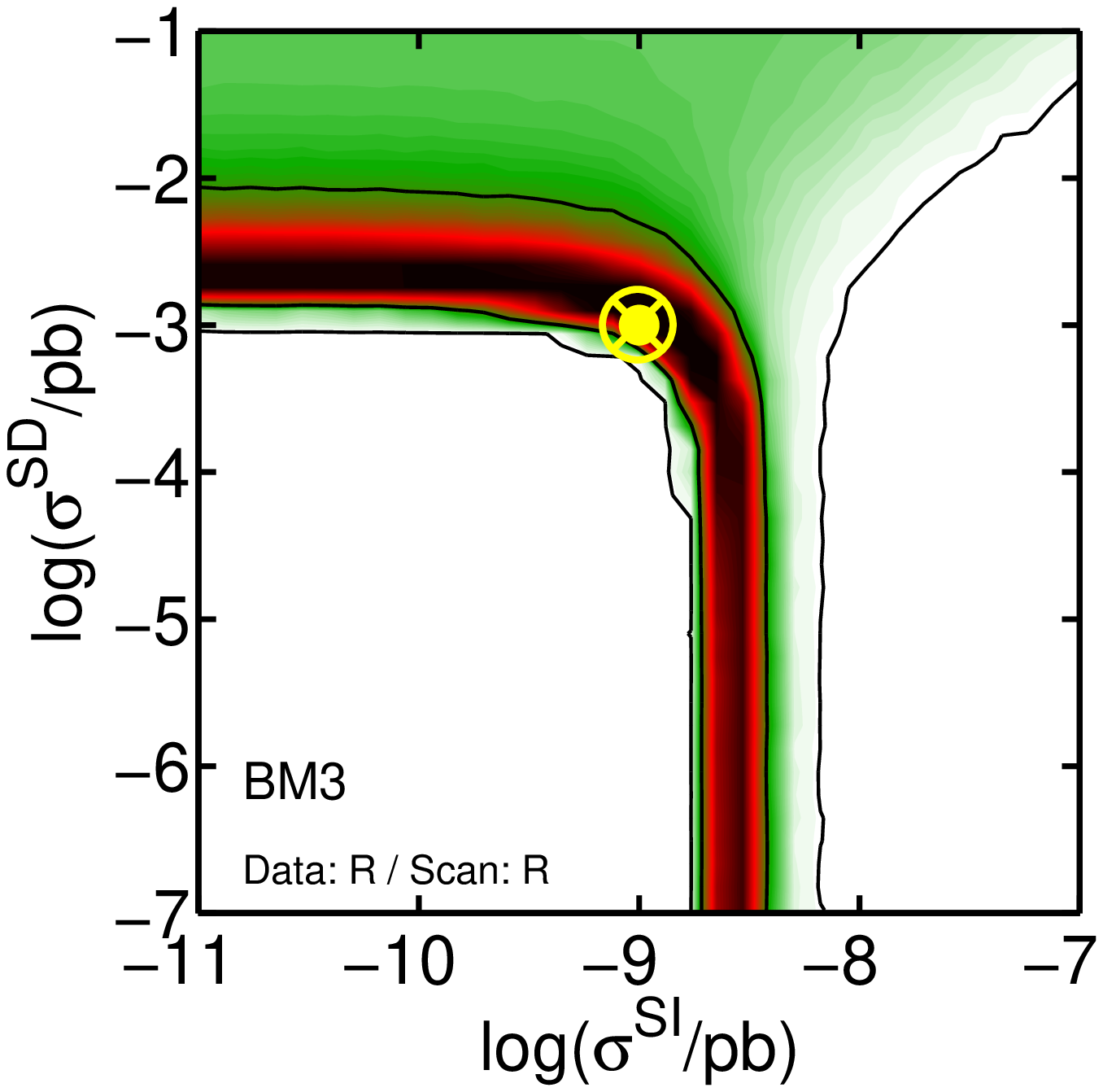}\\[-4ex]
\includegraphics[width=0.35\textwidth]{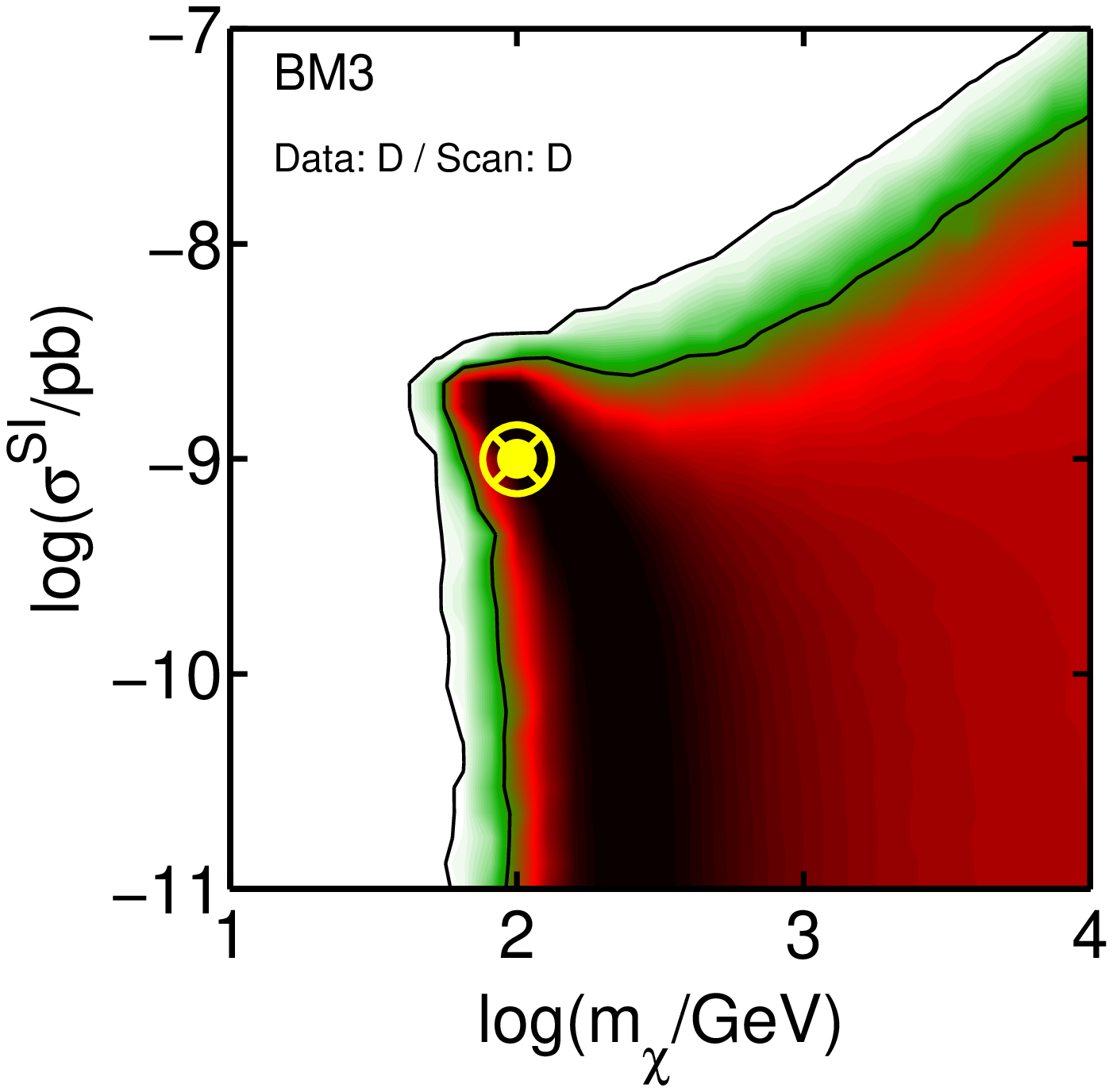}\hspace*{-0.62cm}
\includegraphics[width=0.35\textwidth]{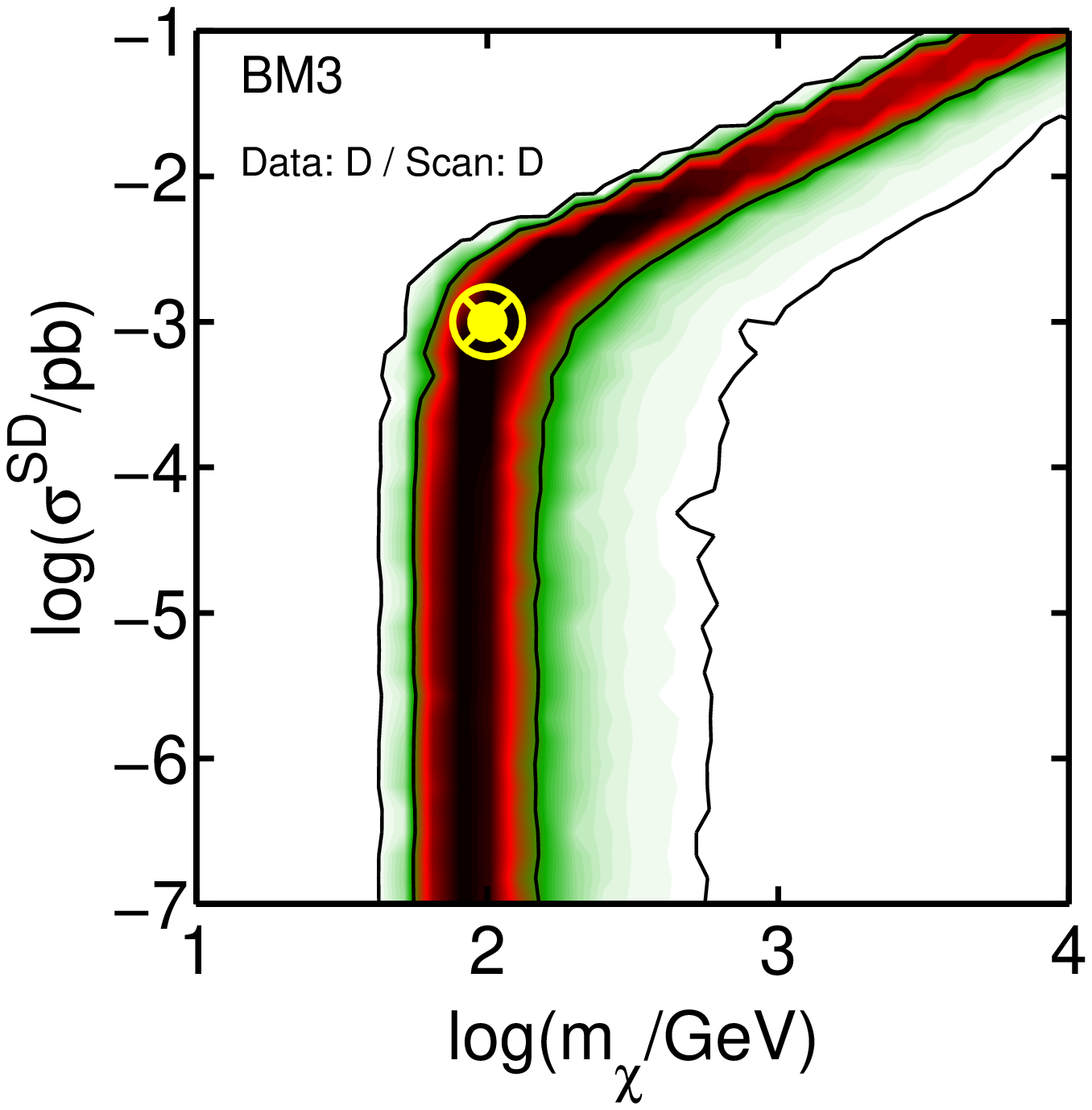}\hspace*{-0.62cm}
\includegraphics[width=0.35\textwidth]{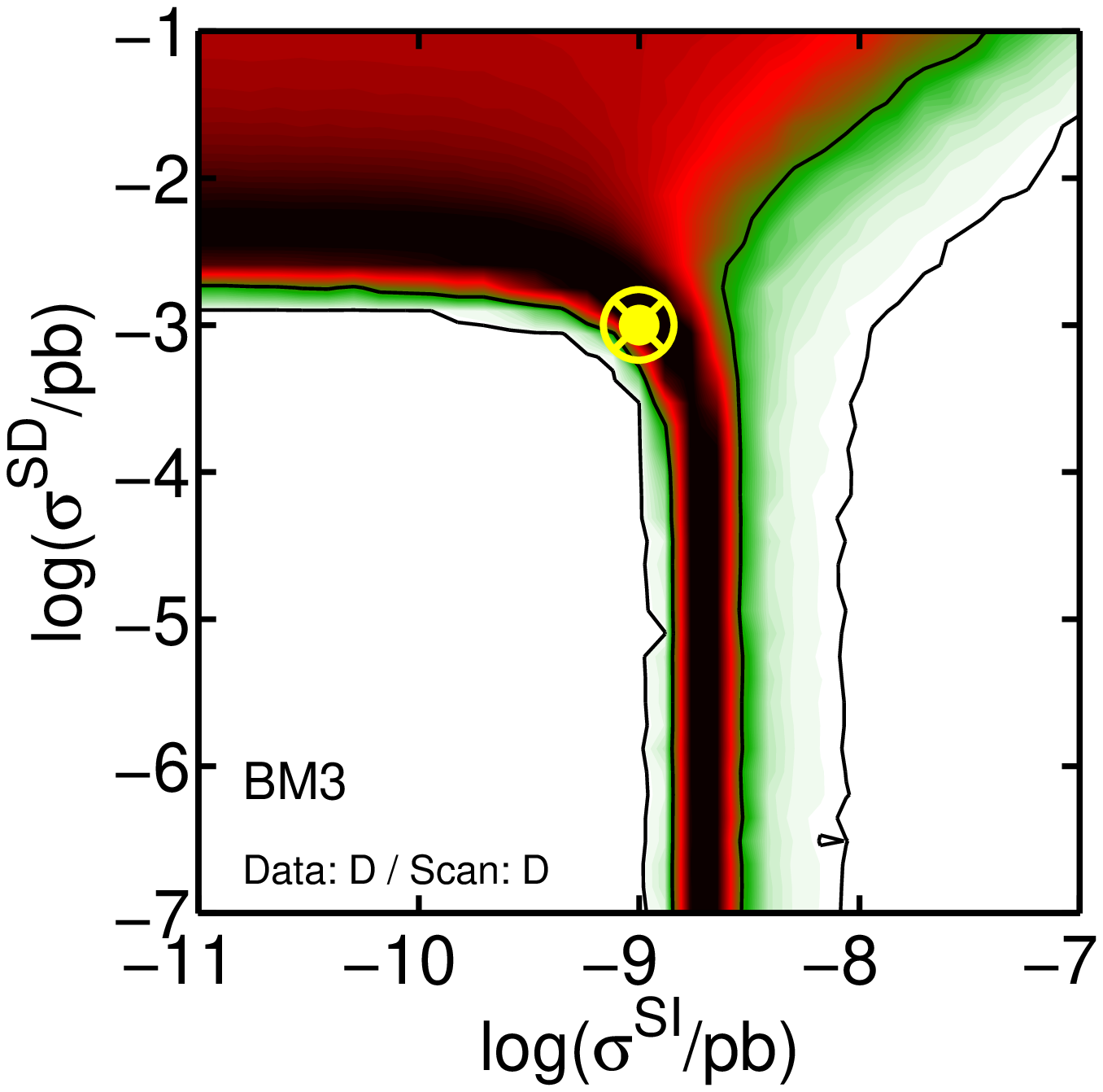}\\[-4ex]
\includegraphics[width=0.35\textwidth]{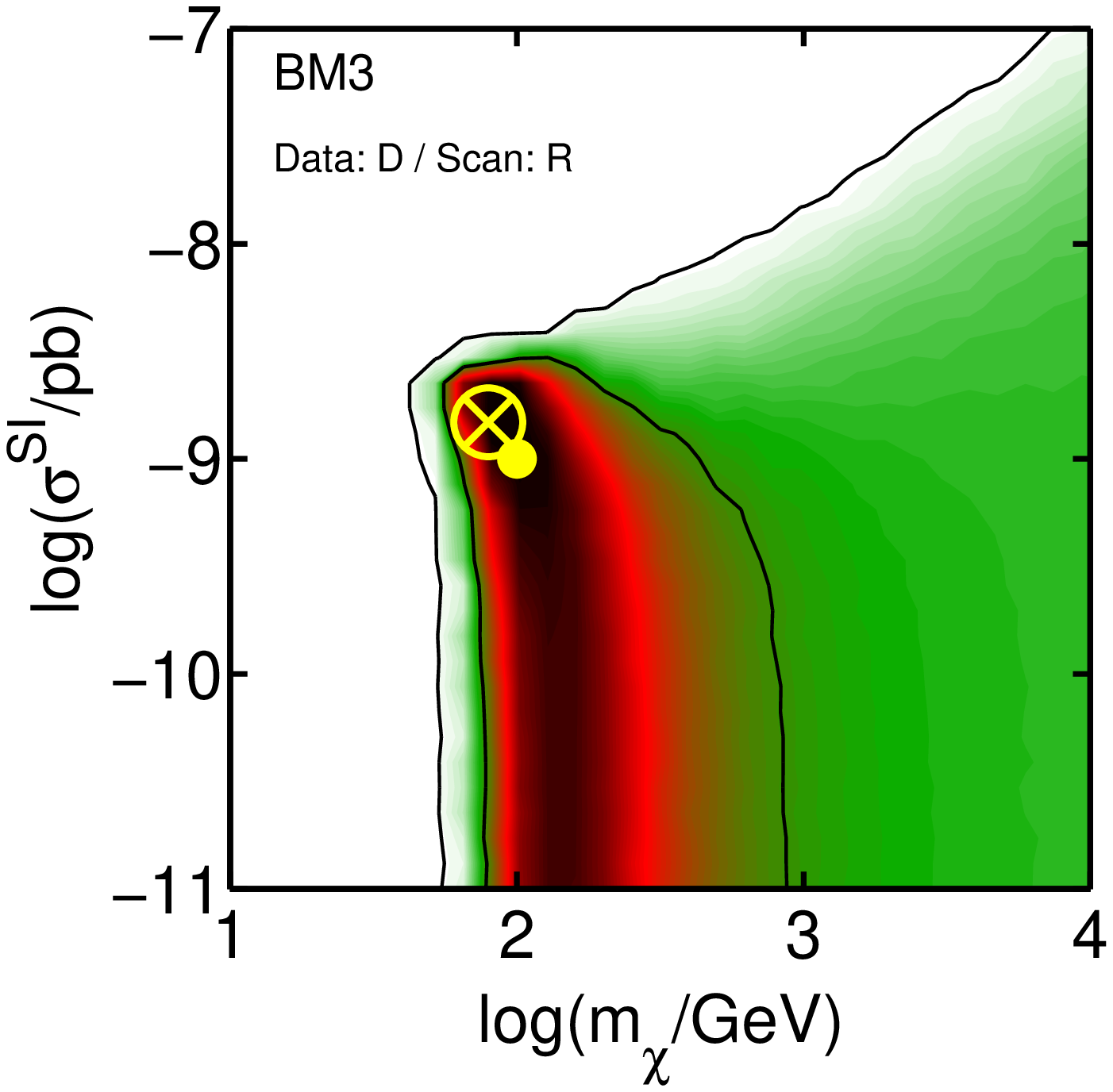}\hspace*{-0.62cm}
\includegraphics[width=0.35\textwidth]{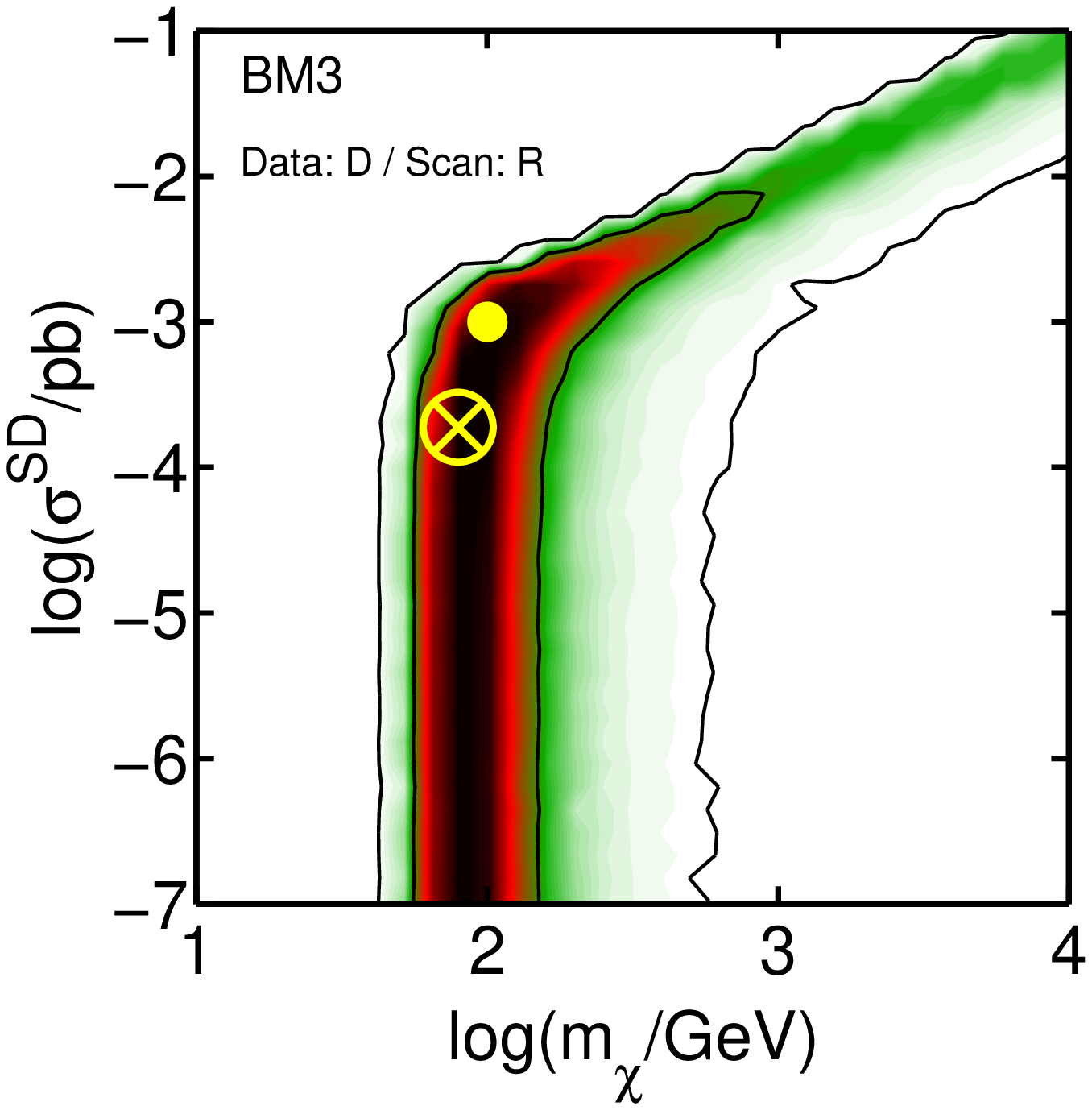}\hspace*{-0.62cm}
\includegraphics[width=0.35\textwidth]{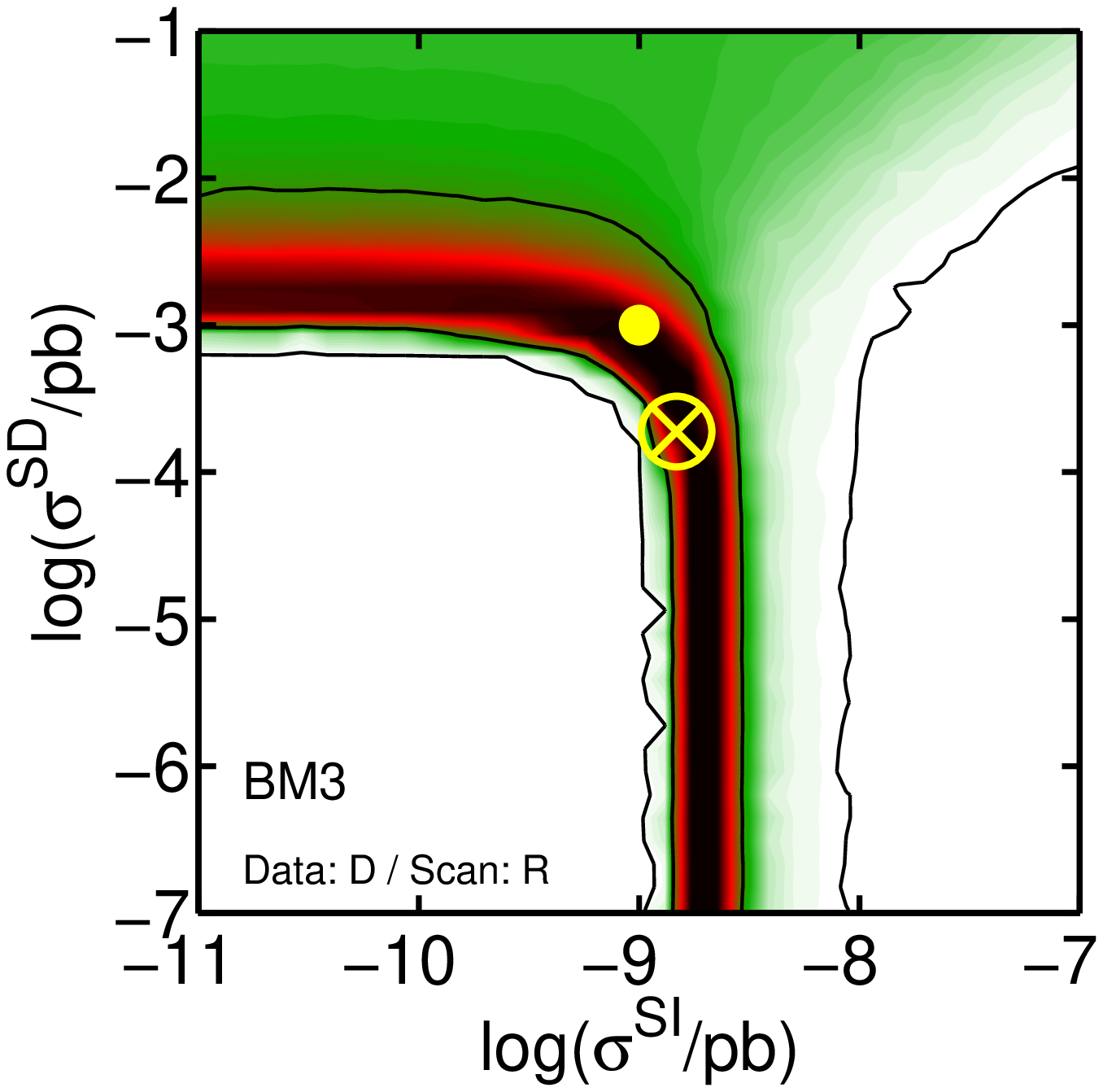}\\[-4ex]
\includegraphics[width=0.35\textwidth]{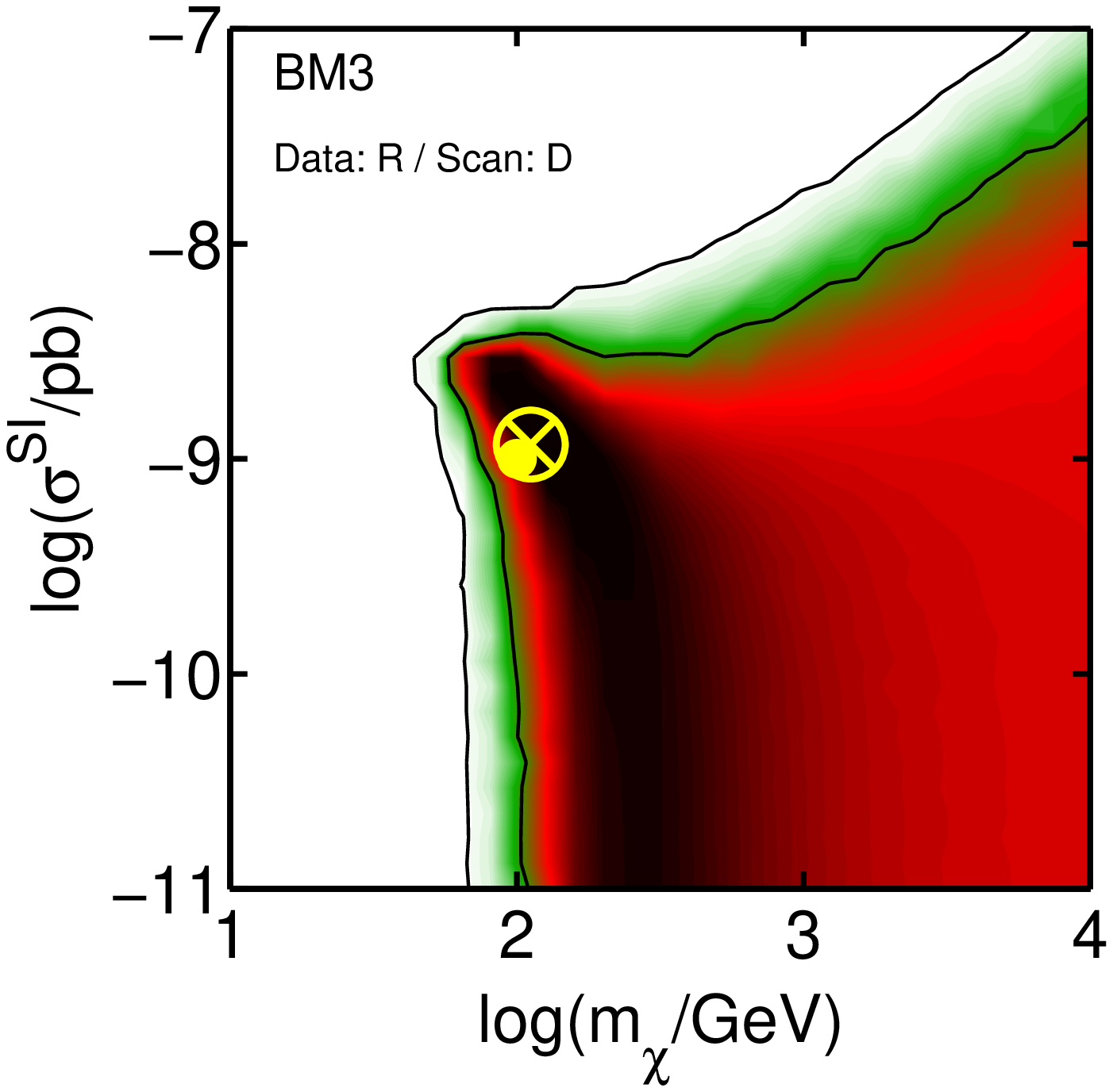}\hspace*{-0.62cm}
\includegraphics[width=0.35\textwidth]{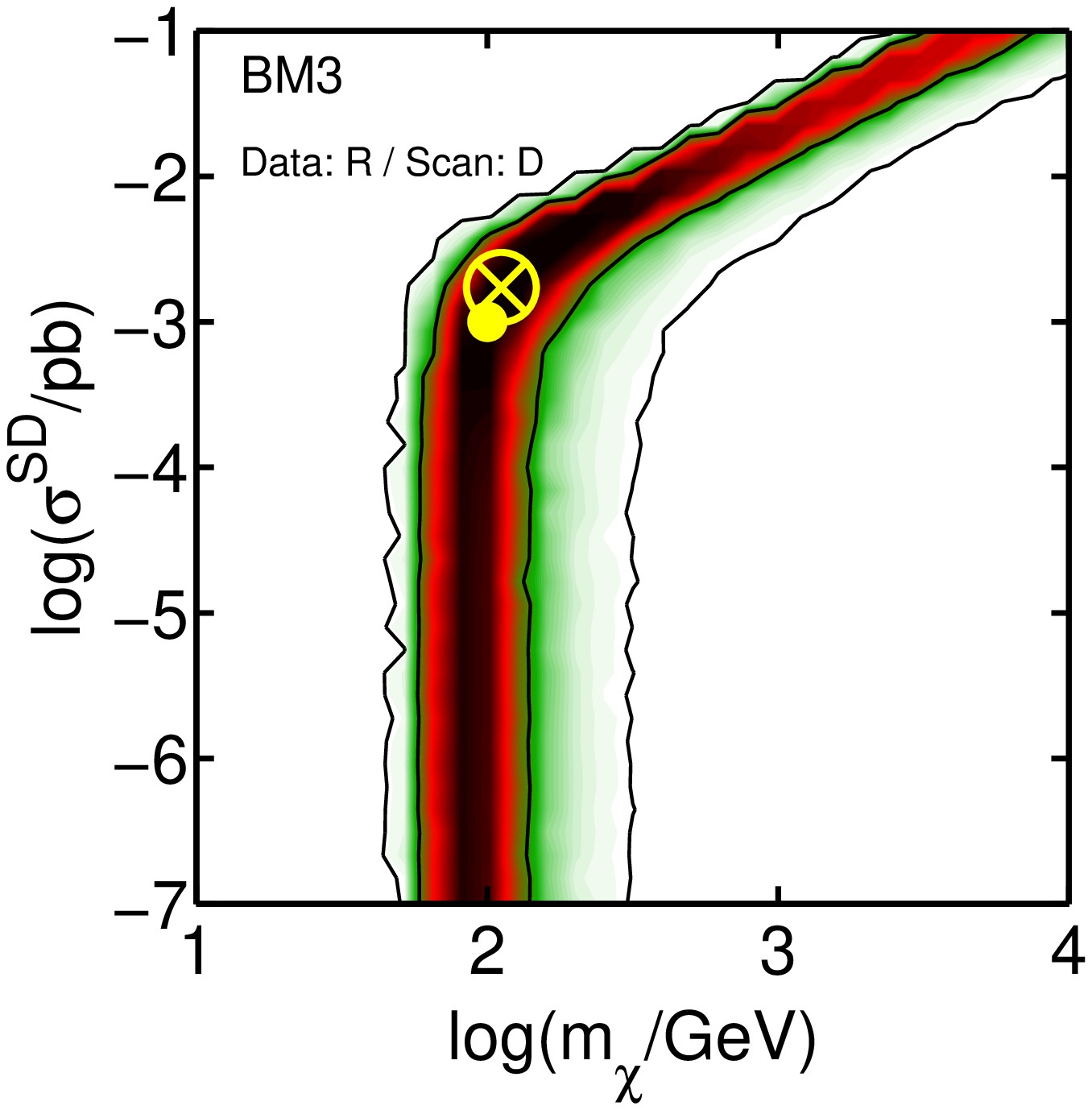}\hspace*{-0.62cm}
\includegraphics[width=0.35\textwidth]{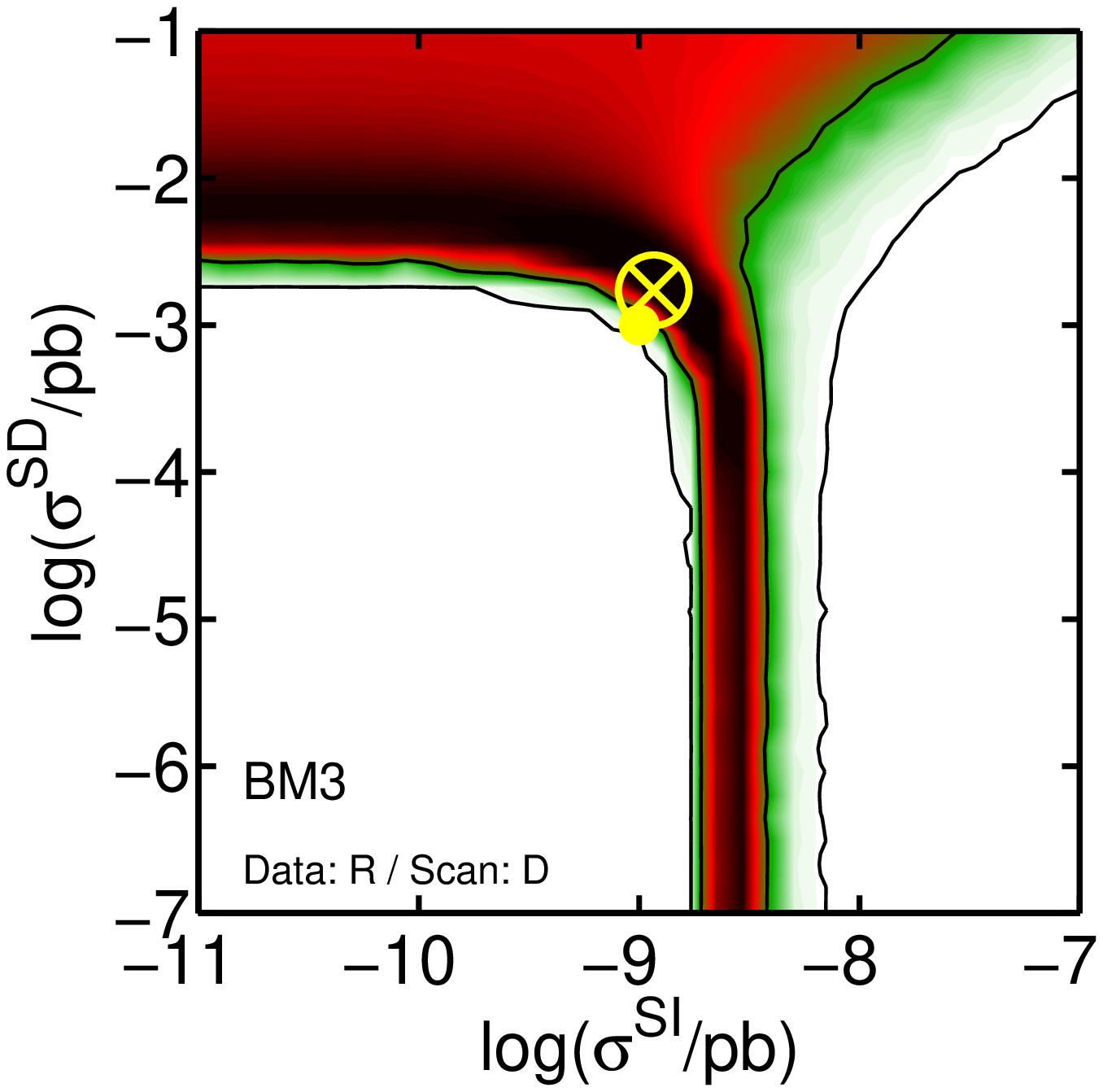}\\[-4ex]
\caption{\label{fig:BM3_profl} The same as in Fig.\,\ref{fig:BM1_profl} but for benchmark BM3.}
\end{figure*}

\begin{figure*}
\includegraphics[width=0.35\textwidth]{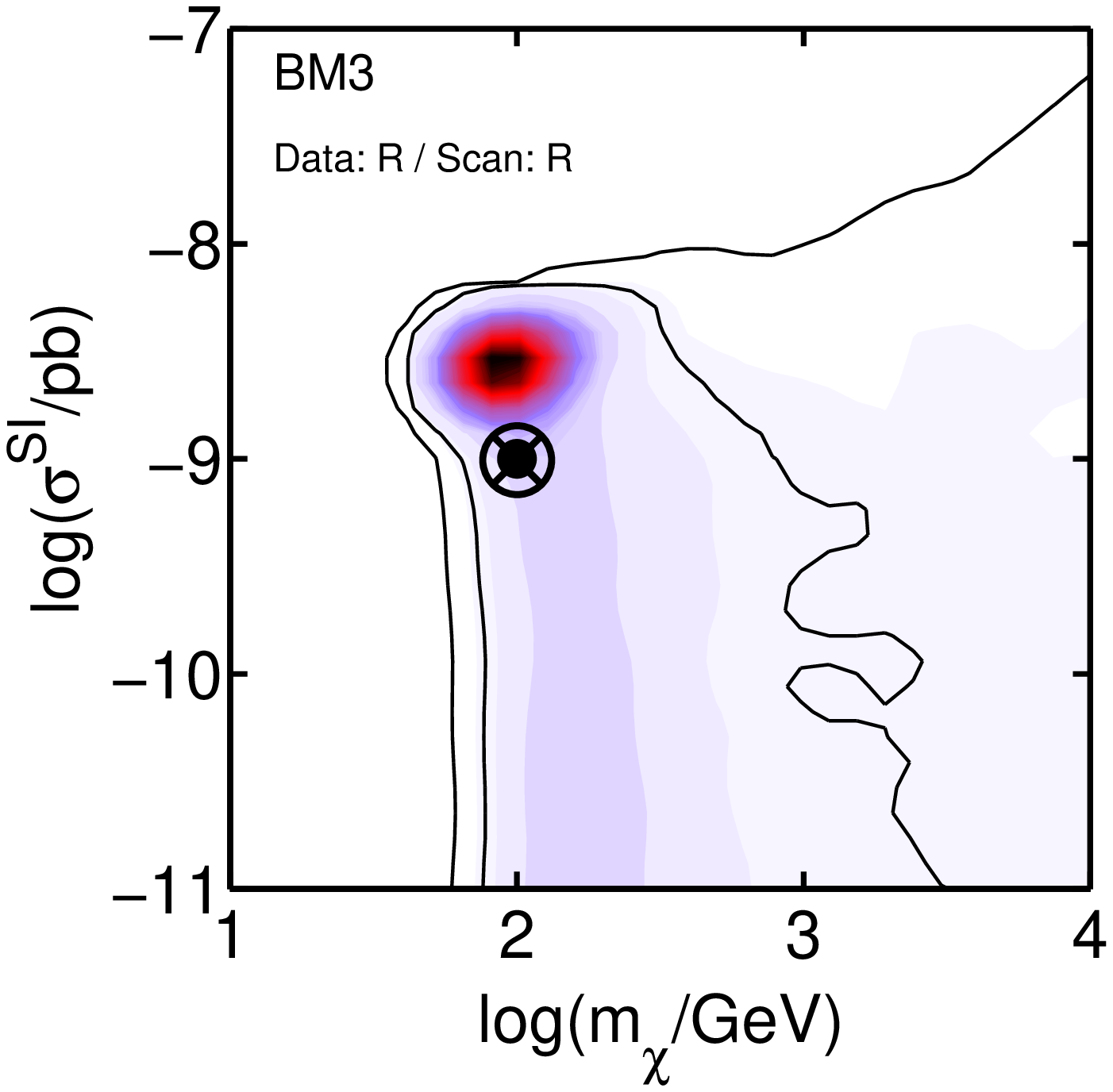}\hspace*{-0.62cm}
\includegraphics[width=0.35\textwidth]{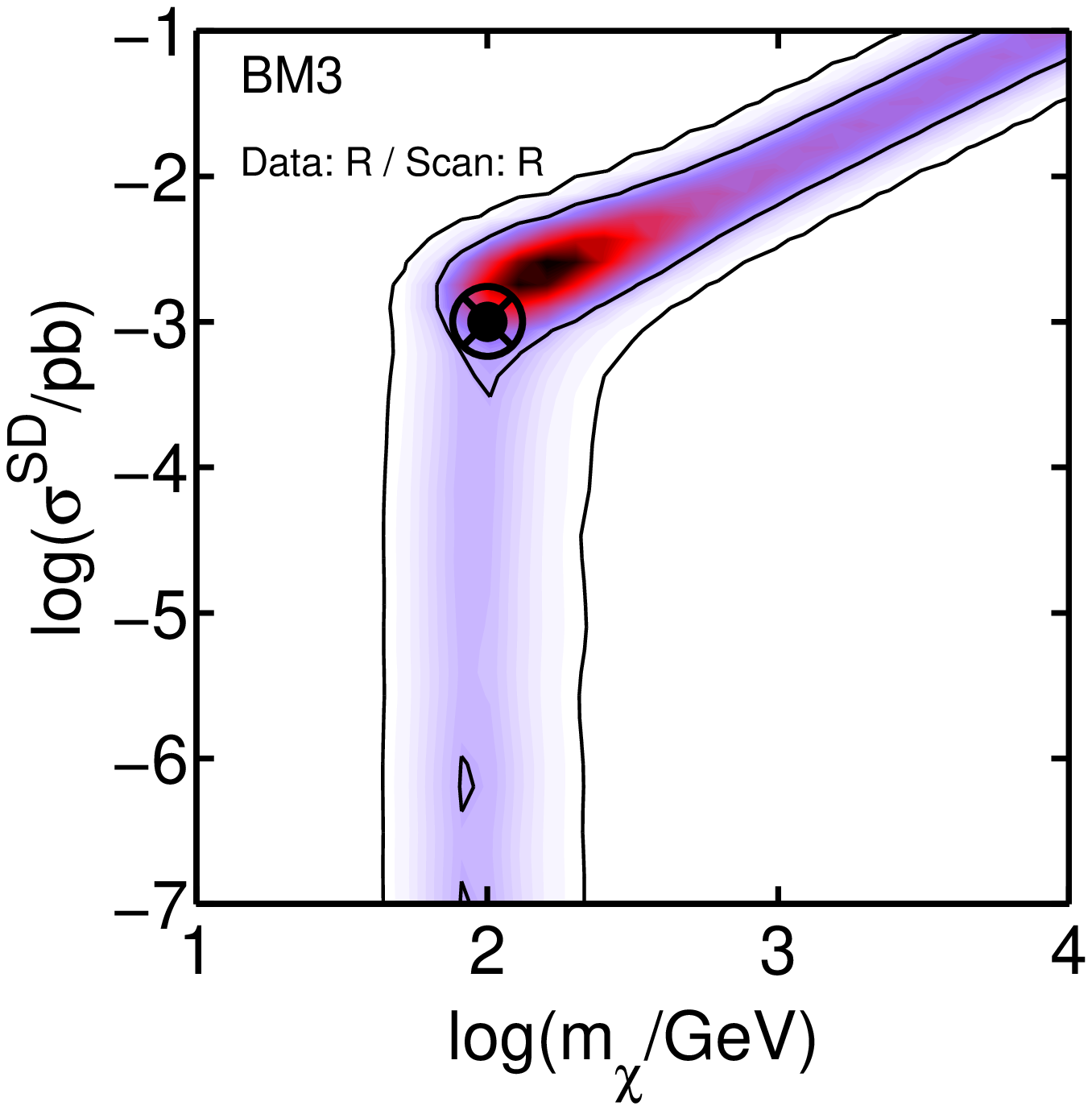}\hspace*{-0.62cm}
\includegraphics[width=0.35\textwidth]{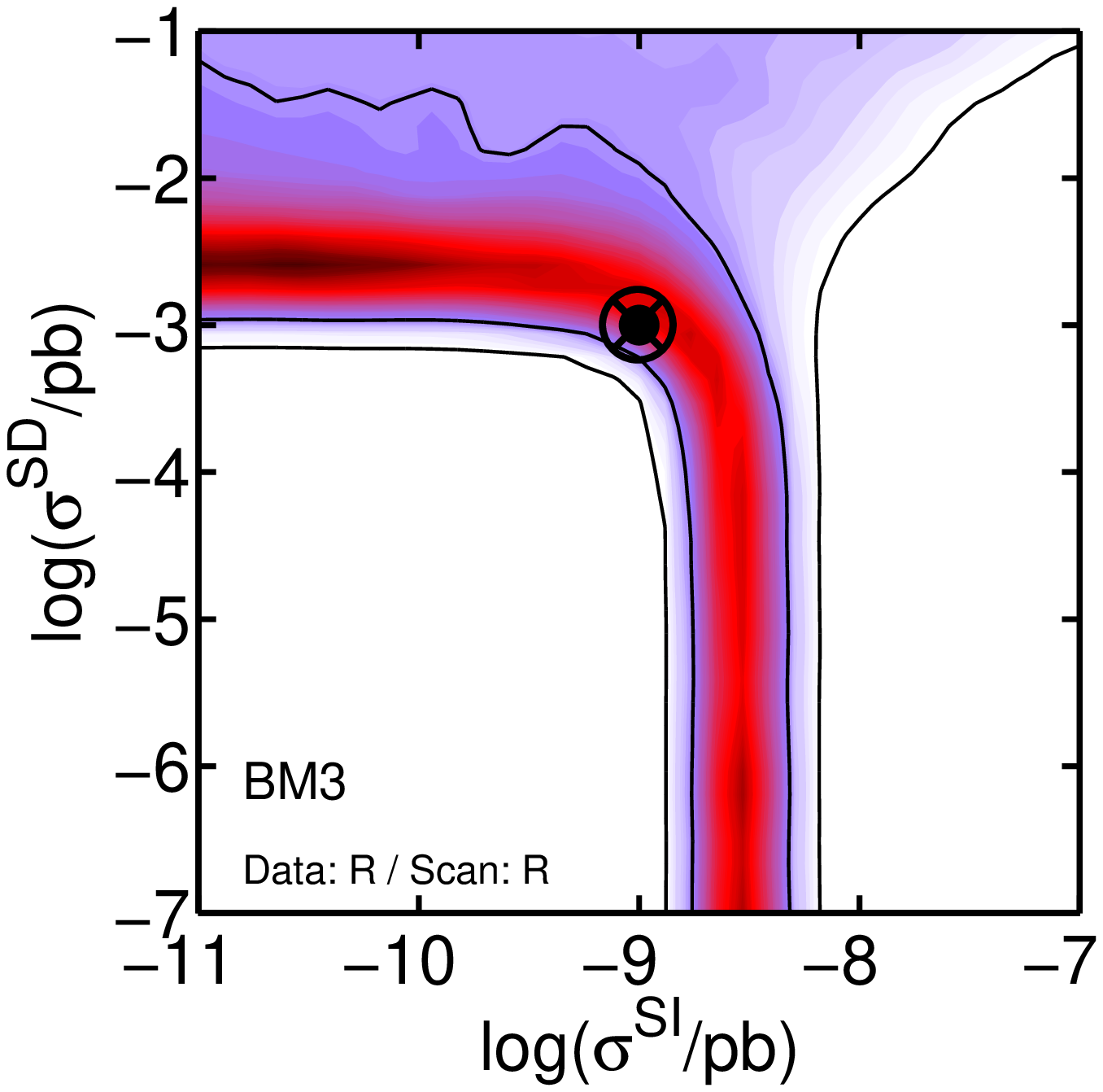}\\[-4ex]
\includegraphics[width=0.35\textwidth]{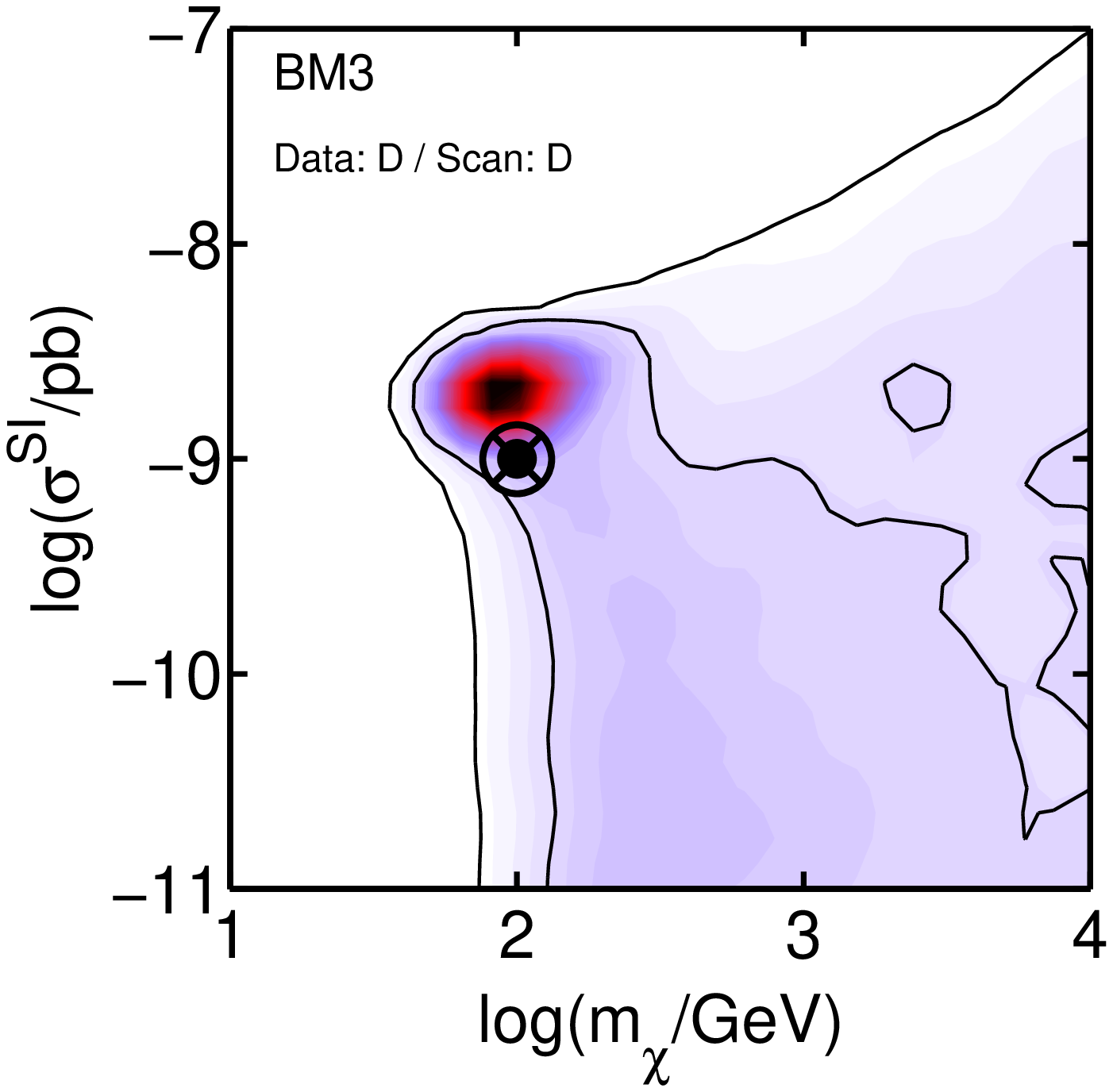}\hspace*{-0.62cm}
\includegraphics[width=0.35\textwidth]{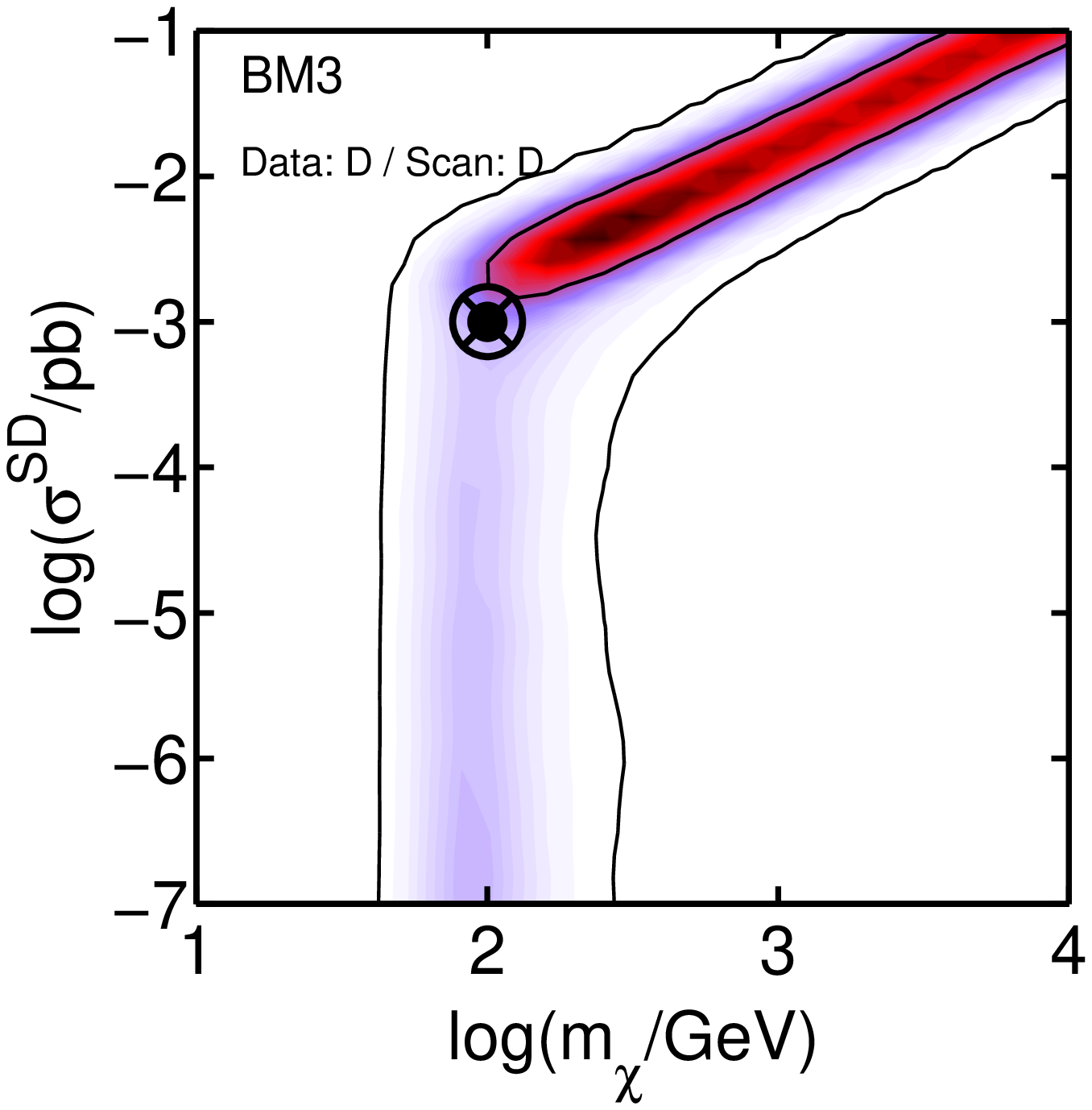}\hspace*{-0.62cm}
\includegraphics[width=0.35\textwidth]{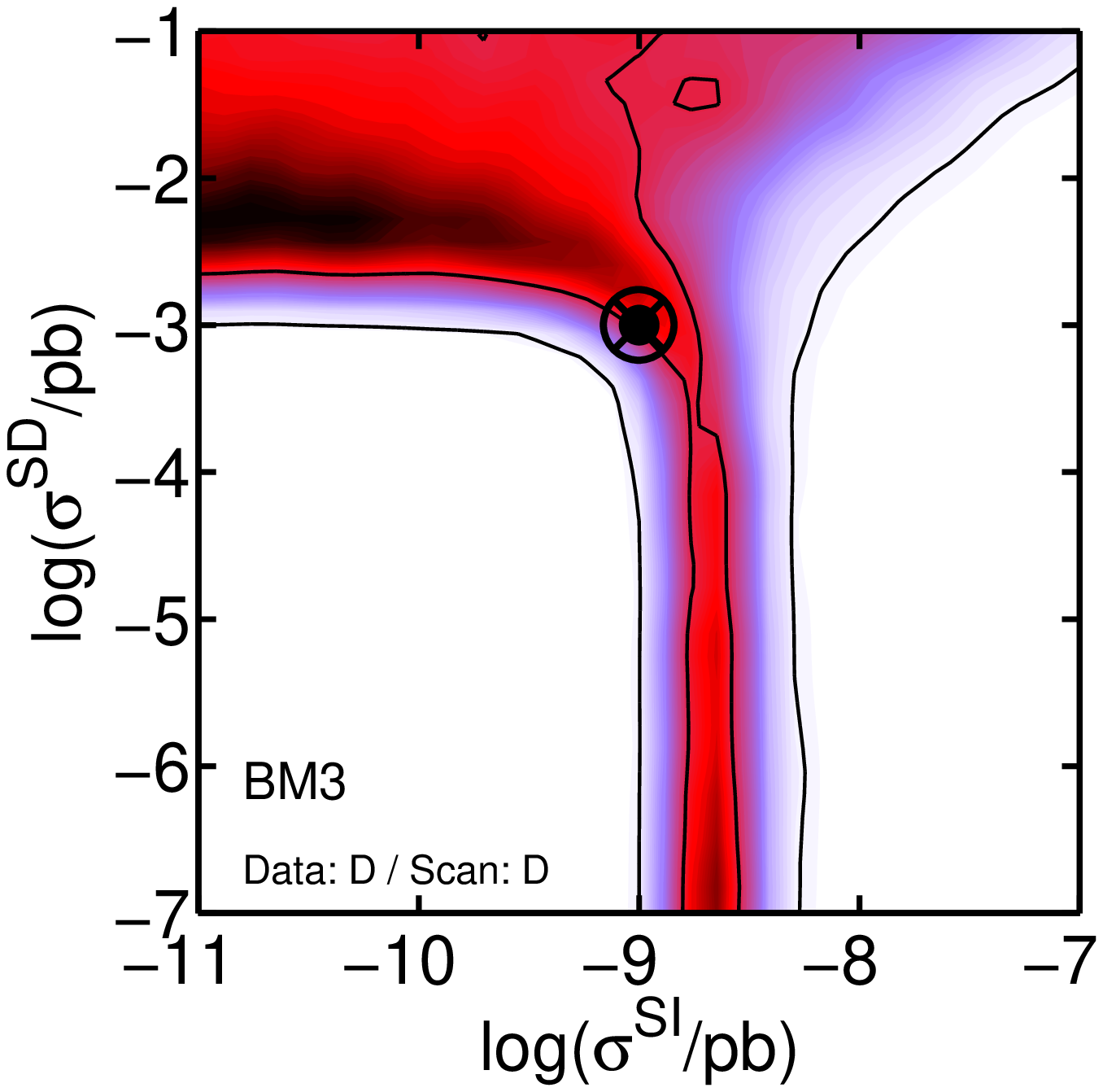}\\[-4ex]
\includegraphics[width=0.35\textwidth]{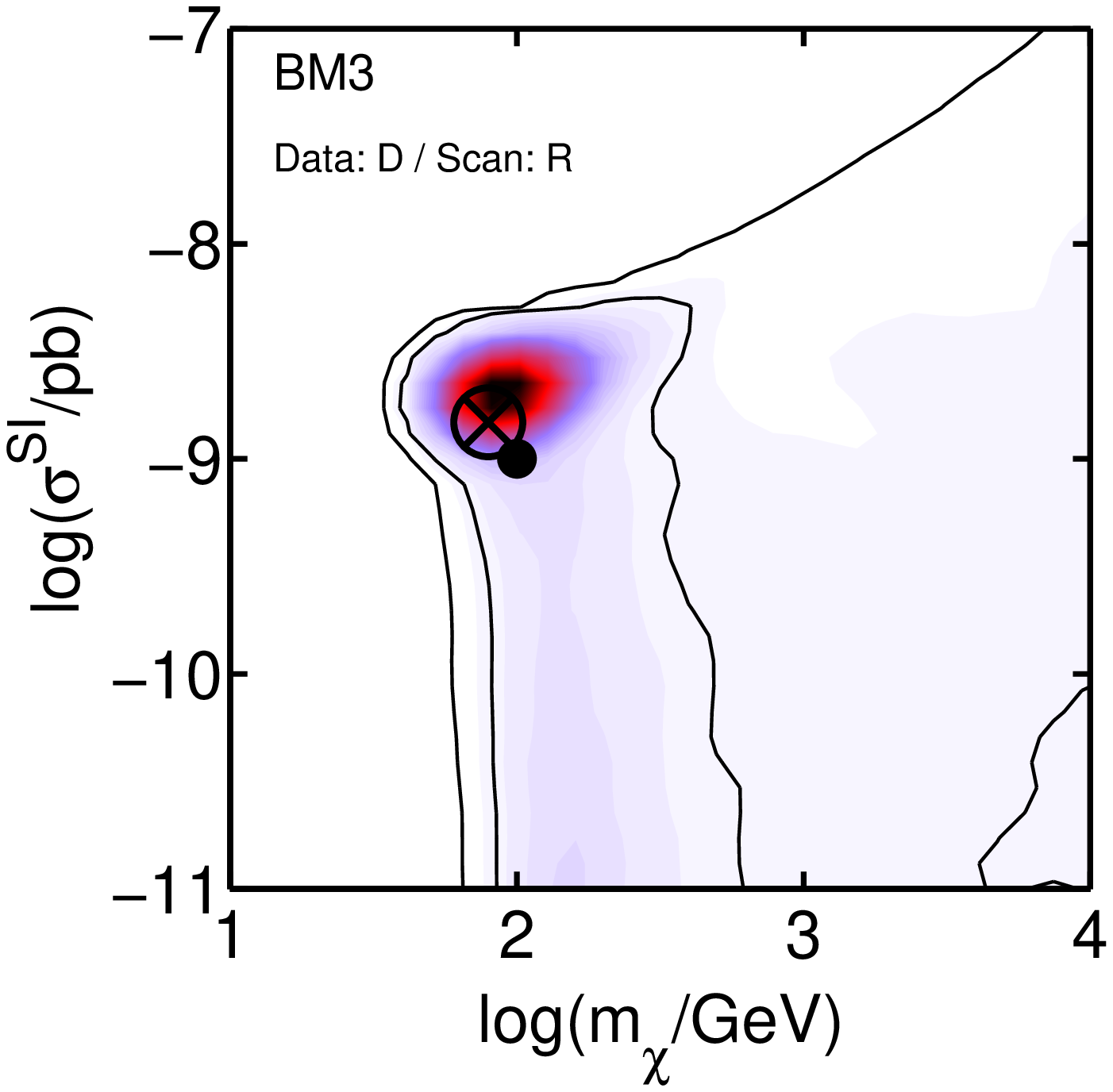}\hspace*{-0.62cm}
\includegraphics[width=0.35\textwidth]{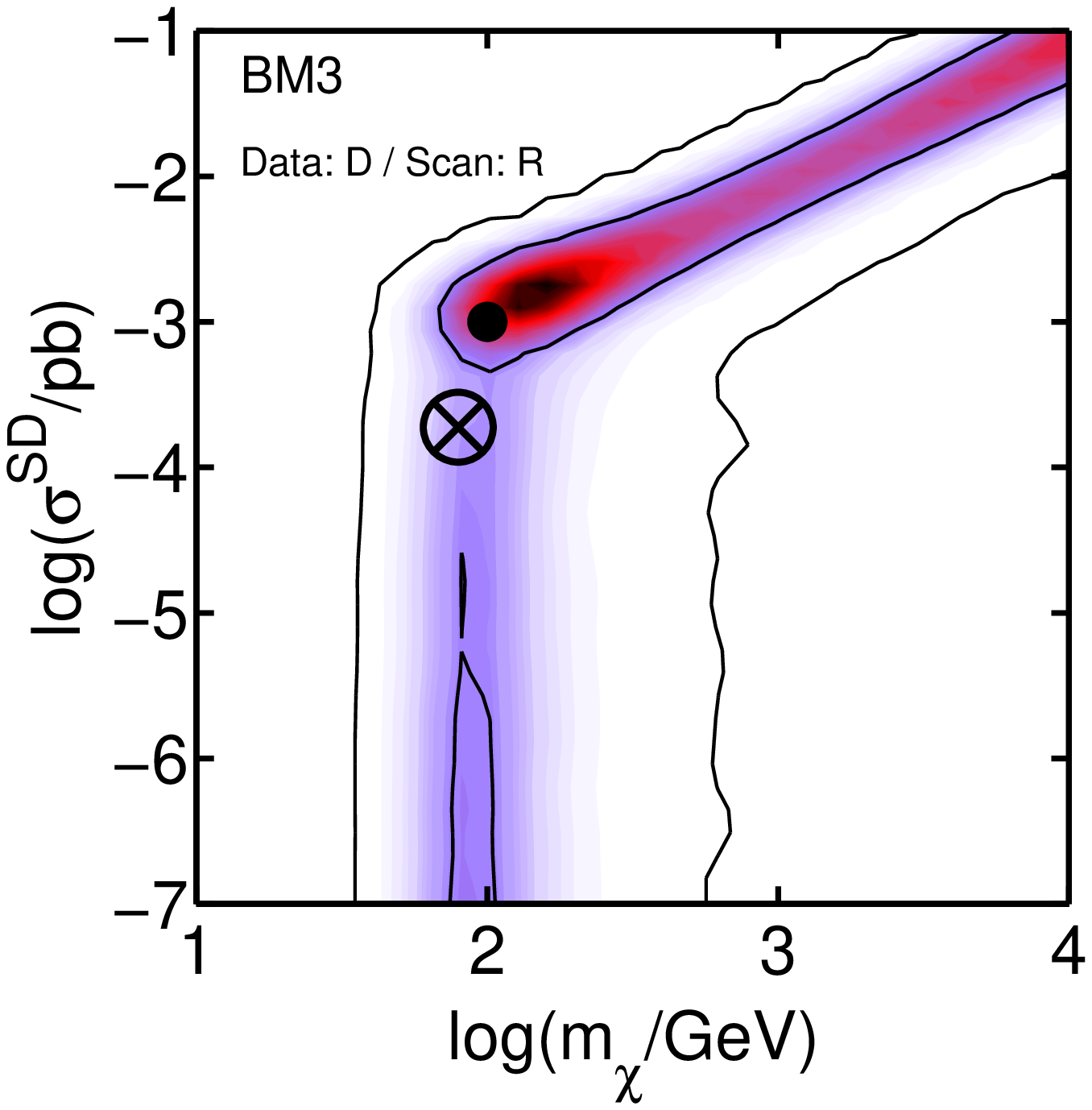}\hspace*{-0.62cm}
\includegraphics[width=0.35\textwidth]{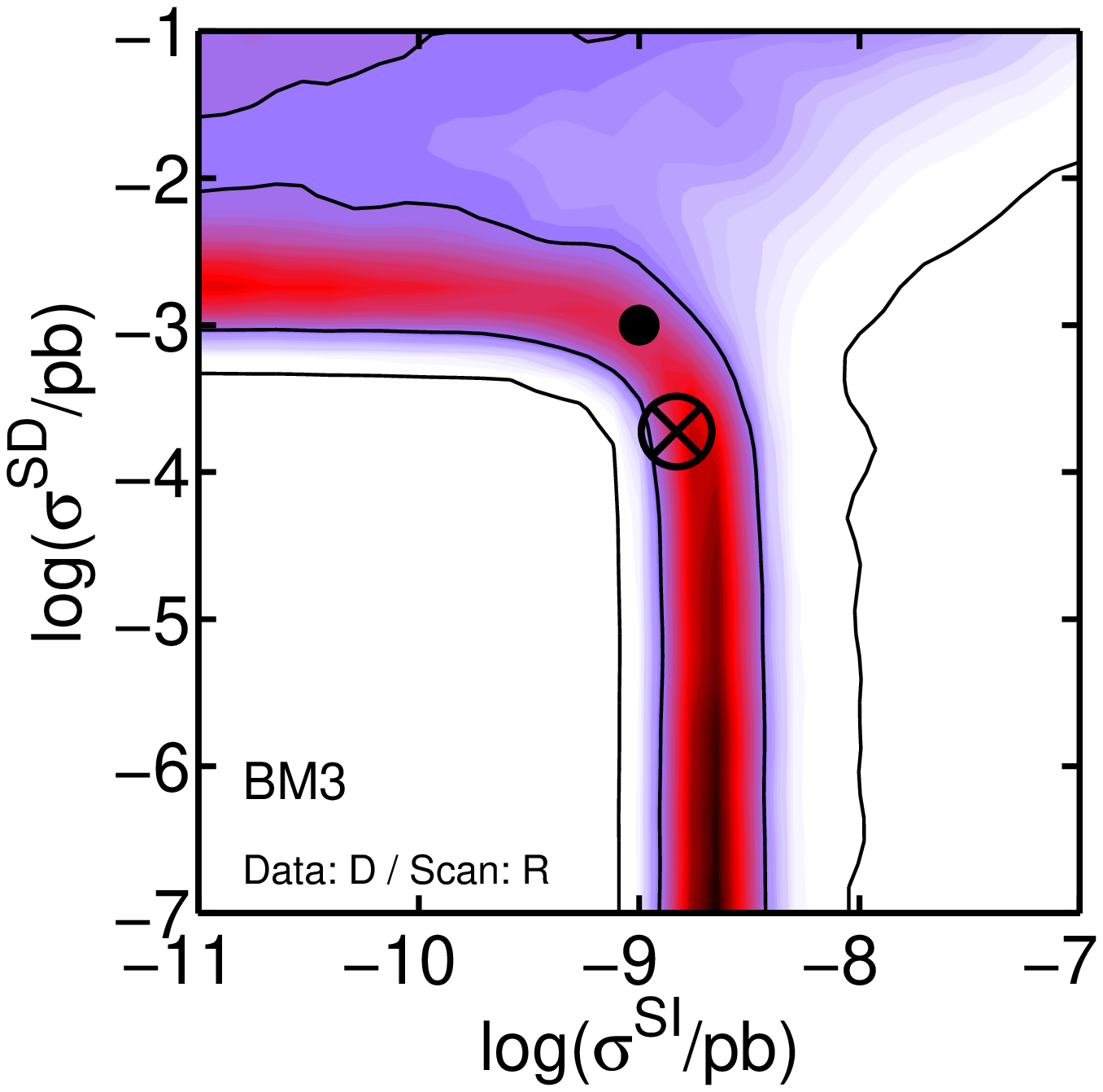}\\[-4ex]
\includegraphics[width=0.35\textwidth]{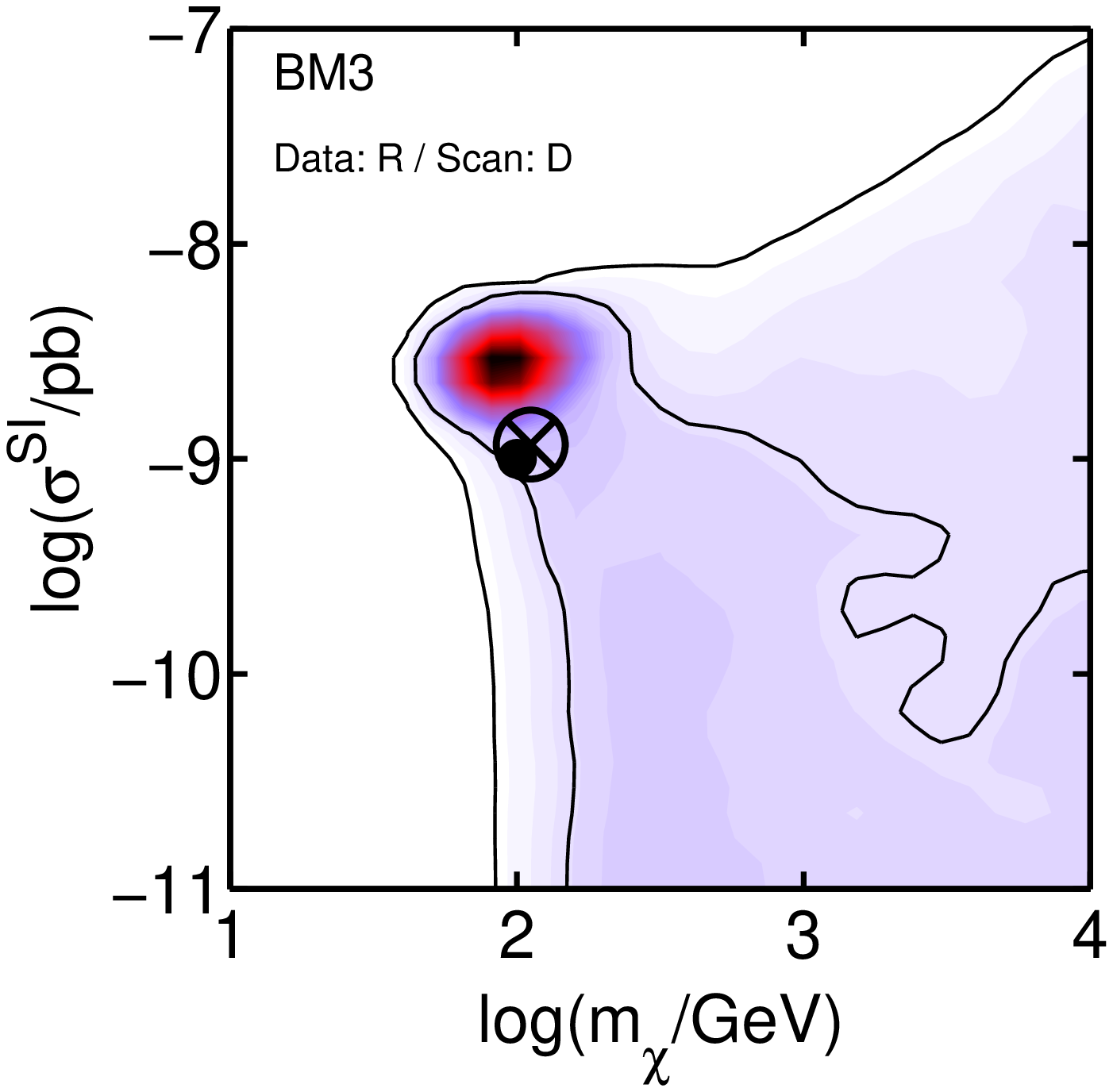}\hspace*{-0.62cm}
\includegraphics[width=0.35\textwidth]{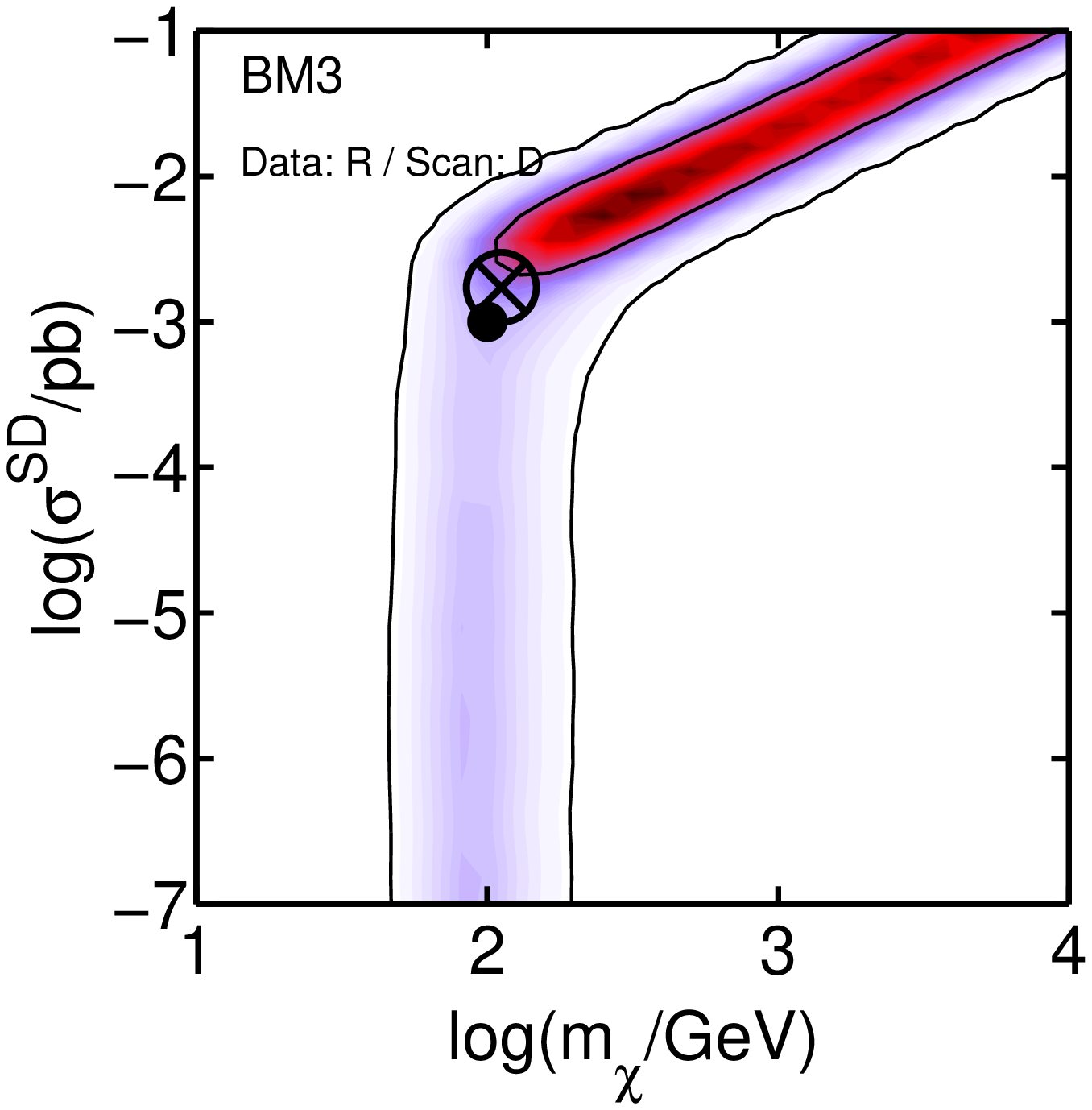}\hspace*{-0.62cm}
\includegraphics[width=0.35\textwidth]{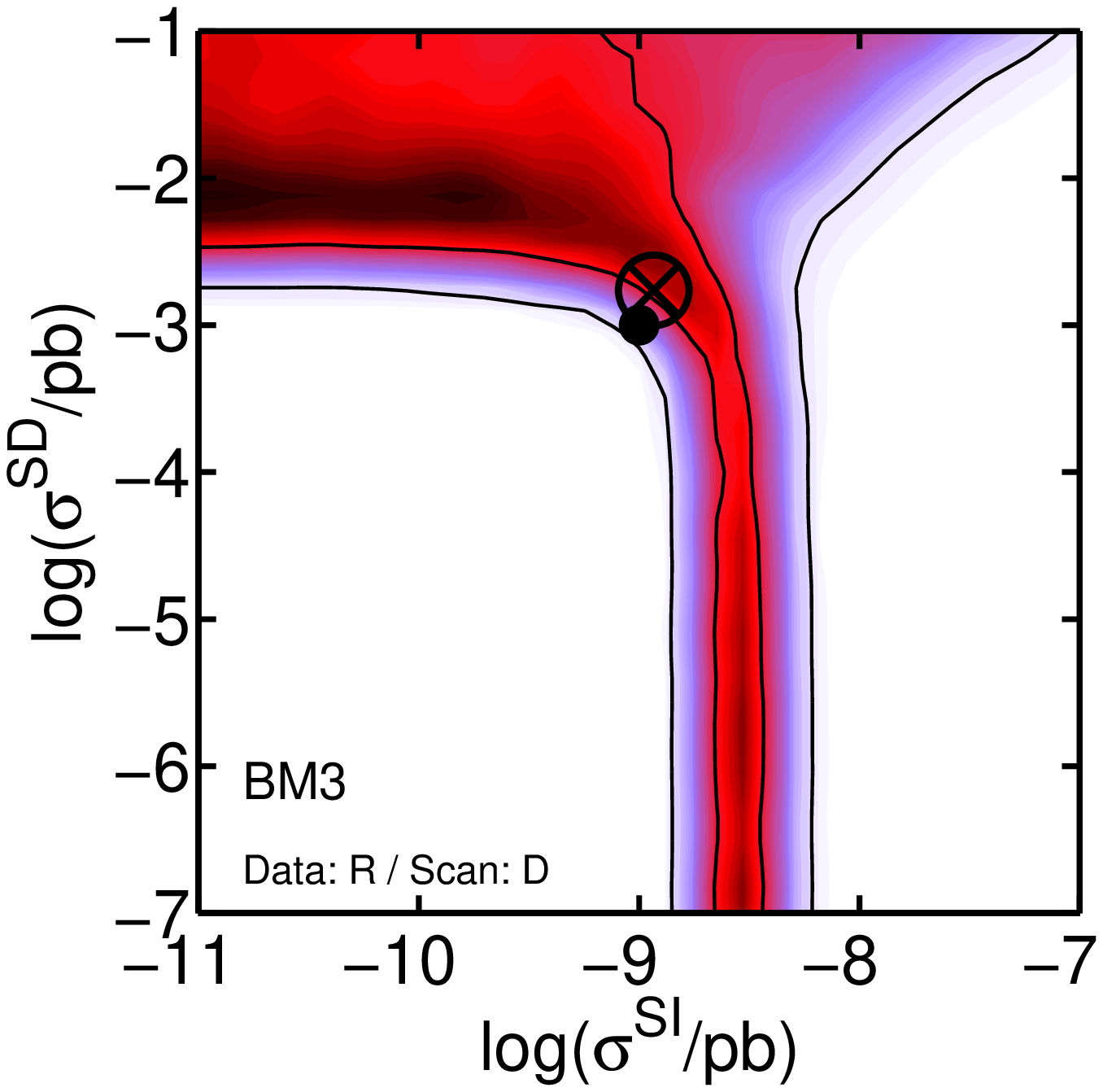}\\[-4ex]
\caption{\label{fig:BM3_pdf} The same as in Fig.\,\ref{fig:BM1_pdf} but for the benchmark BM3.}
\end{figure*}

Let us finally address the third choice of benchmark point, BM3. Contrary to the previous two cases, the SD cross-section is a significant contribution to the total event rate, as we can see in Table\,\ref{tab:benchmarks} (although not entirely dominant). 
Therefore we expect that variations in the SDSF play a more important role in the reconstruction of the DM  parameters. The results are displayed in Figs.\,\ref{fig:BM3_profl} (profile likelihood) and \ref{fig:BM3_pdf} (pdf).

Ignoring the contribution from the SD term is not a good approximation in this scenario. Notice that the reconstruction of $\sigsi$ in the limit when $\sigsd=0$ is larger (by approximately a factor 2) than the nominal value of the BM3 point. This can be appreciated on the lower parts of the plots in the third column of both Figs.\,\ref{fig:BM3_profl} and \ref{fig:BM3_pdf} (and is consistent with the results of Fig.\,\ref{fig:noSD}).
Moreover, the reconstructed value of $\sigsd$ also varies, depending on whether the scan is performed with the R- or D-model of the SDSF, once more due to the different prediction in the zero-momentum value.
As we already noted in the previous benchmark point, an effect in the reconstruction of the WIMP mass can also be appreciated between these two possibilities. The reconstruction performed with D-model favours heavier masses (in fact the 68\% confidence level contours of the pdf are open for heavy WIMPs) than those obtained for R-model, since the D-model for the SDSF is steeper and this can be compensated with a larger value of the WIMP mass, which flattens the spectrum.

\section{Parametrization of uncertainties in the spin-dependent structure functions}
\label{sec:parametrization}

In the previous section we have shown that the choice of model for the SDSF has an important effect in the reconstruction of DM parameters. 
So far our conclusions are based on the comparison of the results obtained using two different computations for the SDSF of $^{73}$Ge.
In order to consider these effects in a more systematic way, in this section we attempt to include uncertainties in the SDSFs as part of the scan.

To do this, a description of the structure functions has to be found in terms of a relatively small number of parameters. 
We propose the use of the following family of functions, which reproduces non-trivial features in the shape of SDSFs,
\begin{equation}
	S_{ij}(u)=N\left((1-\beta)e^{-\alpha u} + \beta\right)\,.
	\label{eqn:family}
\end{equation}
The parameter $N$ acts as an overall normalization that allows us to fit the value at zero-momentum, $\beta$ controls the height of a possible tail at large momentum and $\alpha$ provides the slope of the decreasing part in the the low-momentum regime\footnote{We have explicitly checked that although a five-parameters fit is able to reproduce better some features of the SDSF in certain nuclei (e.g., $^{129}$Xe and $^{131}$Xe), this has a negligible impact in the reconstruction of DM parameters.}.

\subsection{Germanium detectors}

In order to account for uncertainties in the SDSFs we have determined the maximum and minimum values of the three parameters $N$, $\alpha$ and $\beta$  in Eq.\,(\ref{eqn:family}) which define an area that contains the calculations of the R- and D-models. 
The range considered for $S_{11}(q)$ is the following: $N=[0.12,\,0.21]$, $\beta=[0.020,\,0.042]$, and $\alpha=[5.0,\,6.0]$. For illustrative purposes we display in Fig.\,\ref{fig:FFs} the area (in blue) spanned by the family of curves that can be obtained by varying the above parameters in the given ranges. As we see, the R- and D-models correspond approximately to the extremes of the above intervals.

\begin{figure*}
\includegraphics[width=0.35\textwidth]{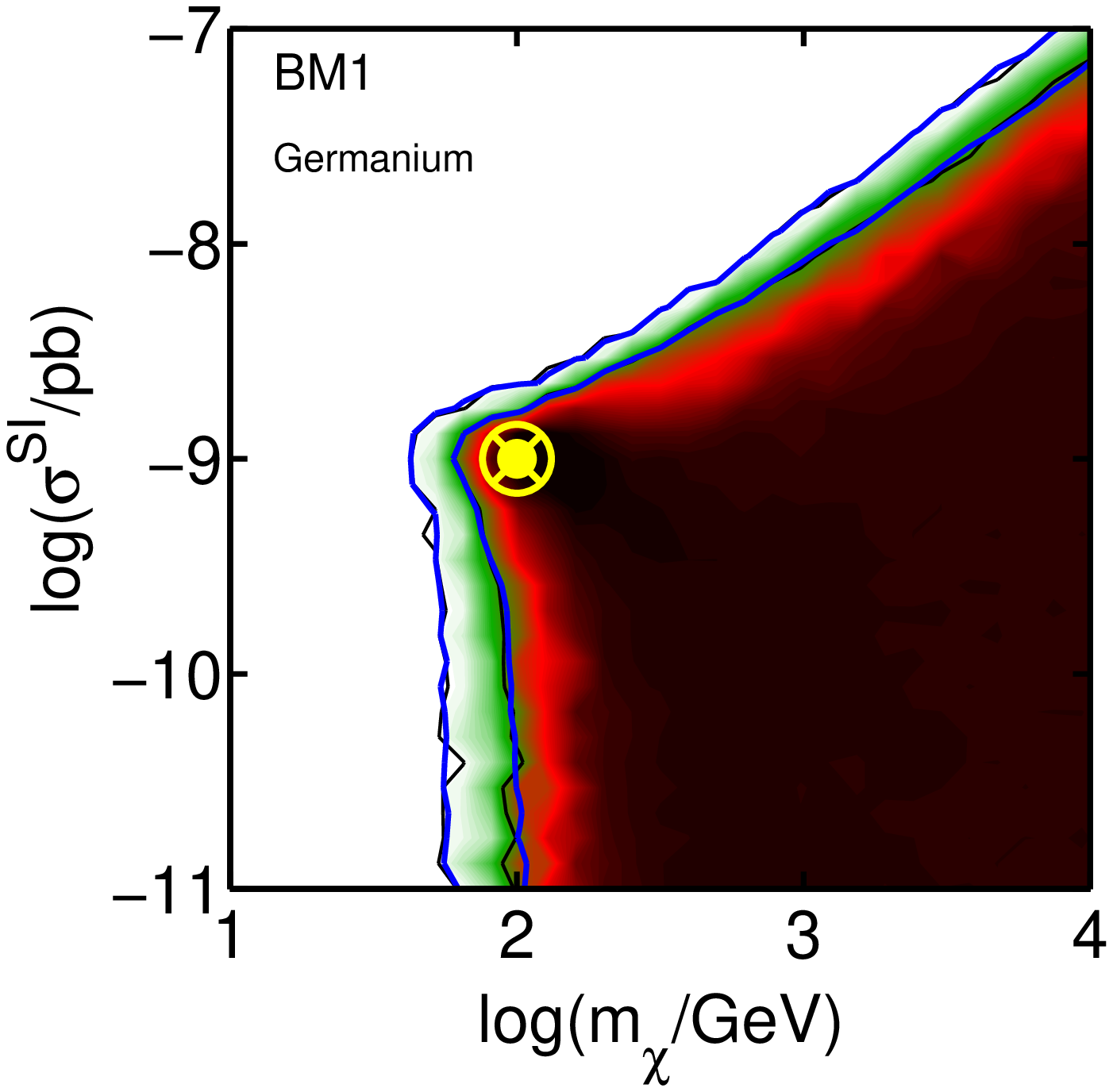}\hspace*{-0.62cm}
\includegraphics[width=0.35\textwidth]{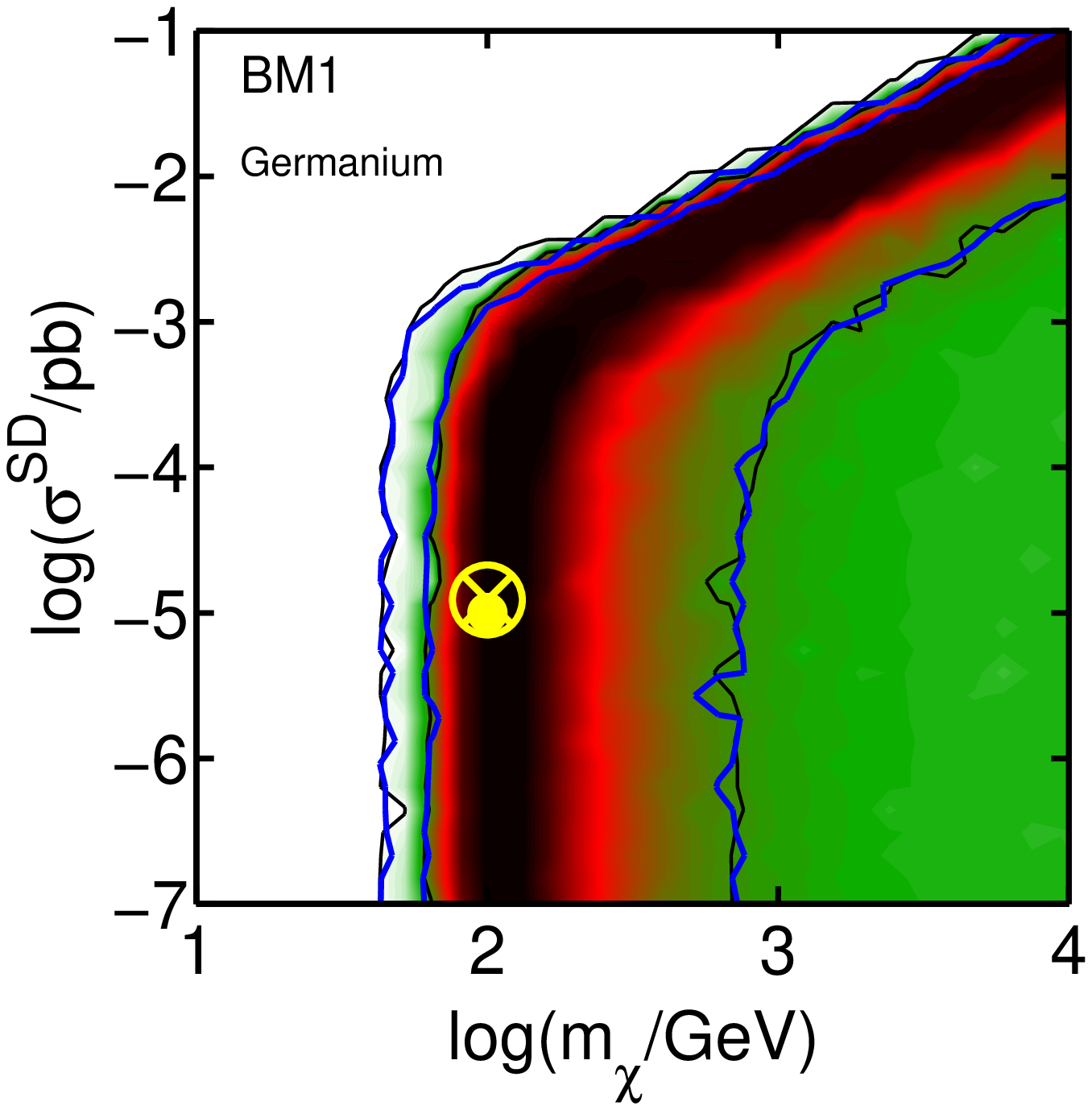}\hspace*{-0.62cm}
\includegraphics[width=0.35\textwidth]{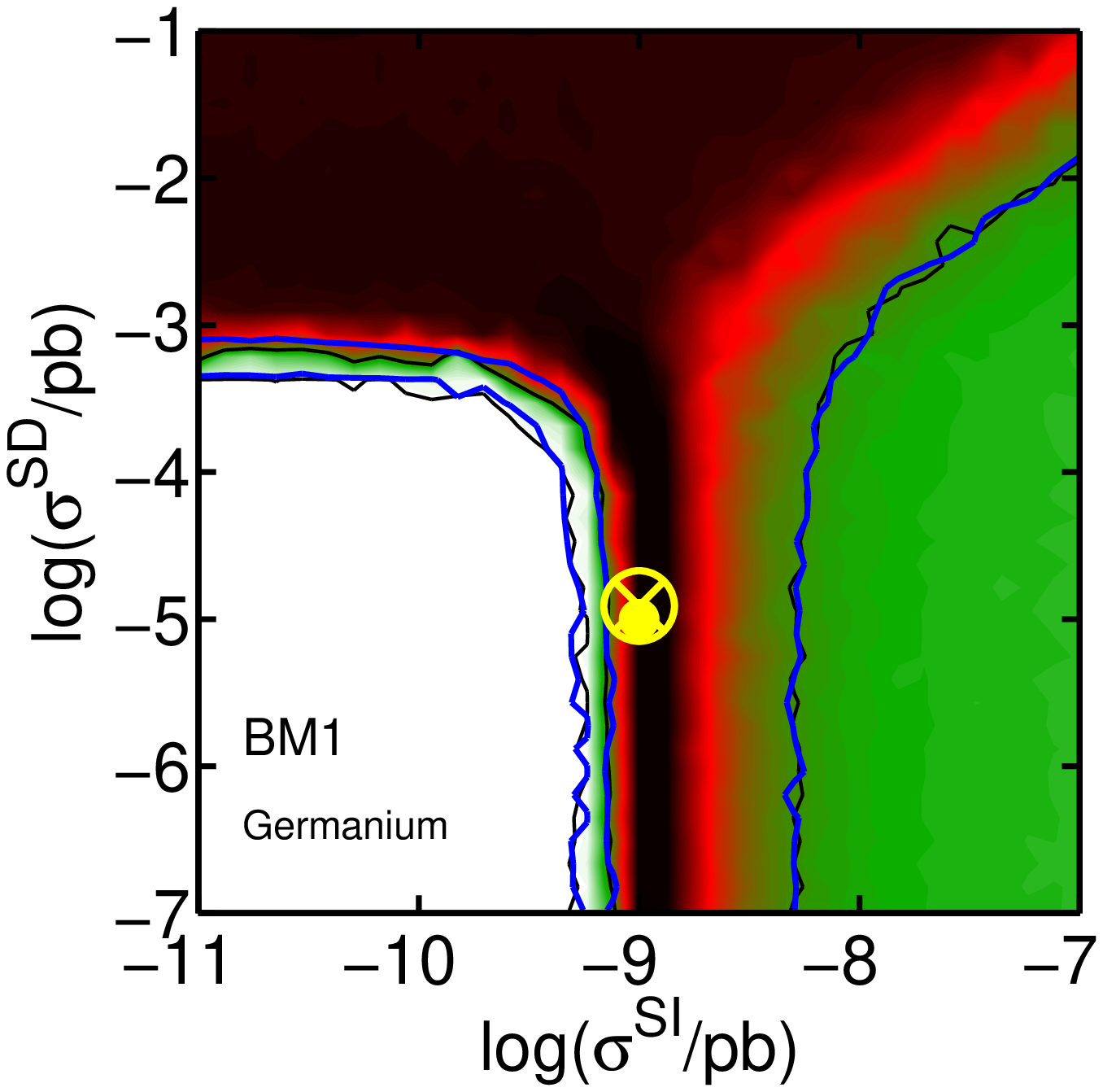}\\[-4ex]
\includegraphics[width=0.35\textwidth]{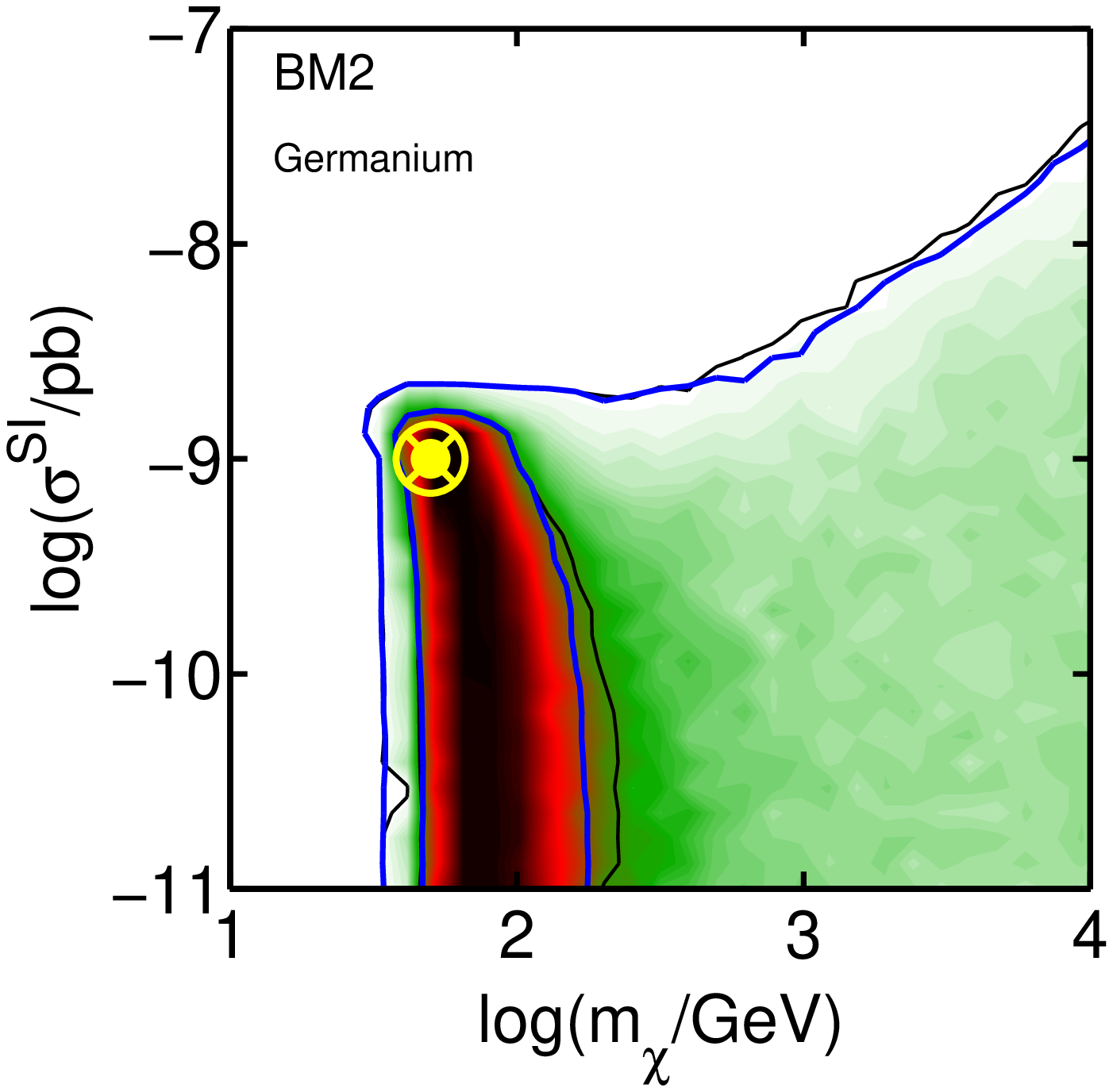}\hspace*{-0.62cm}
\includegraphics[width=0.35\textwidth]{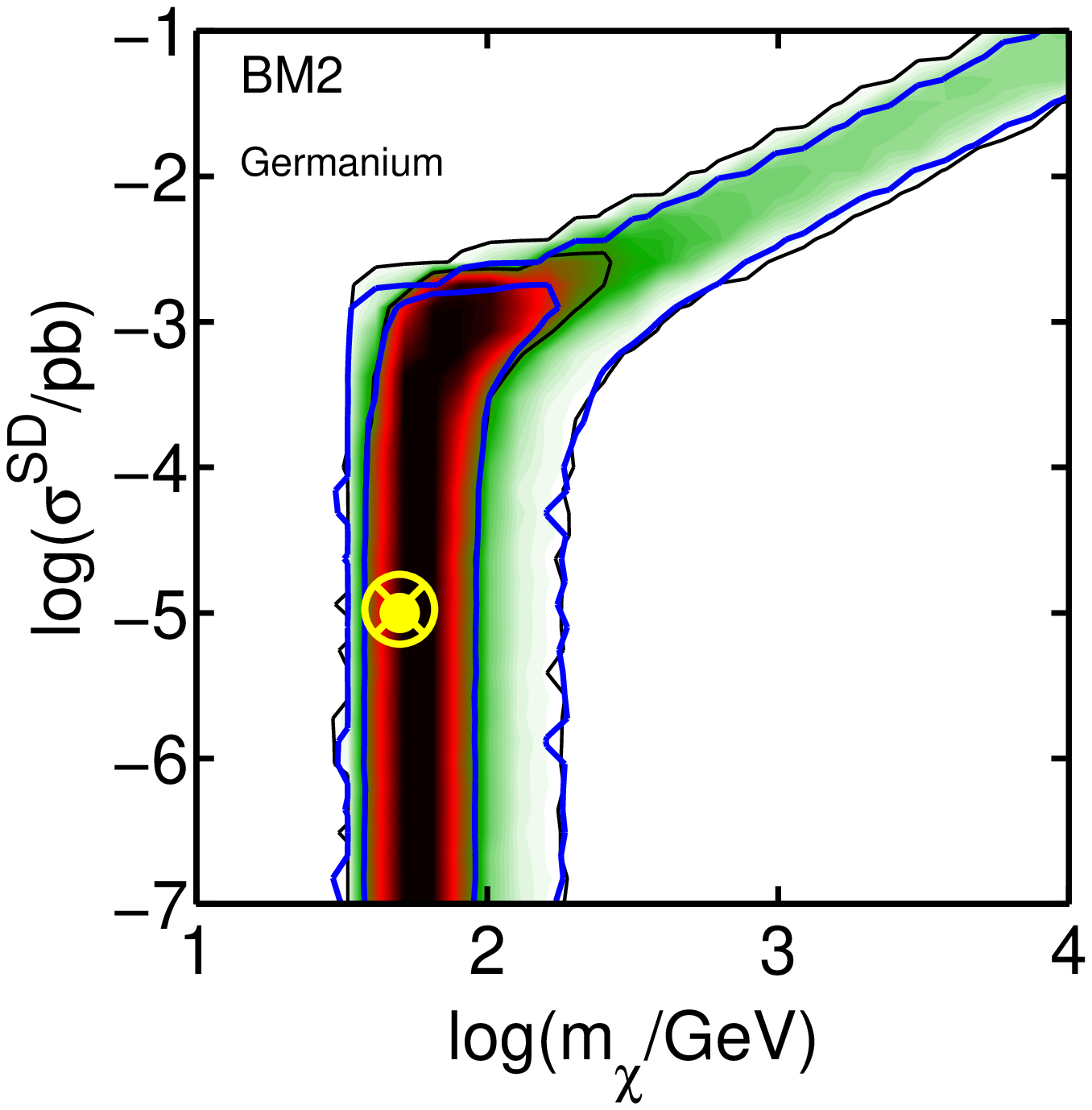}\hspace*{-0.62cm}
\includegraphics[width=0.35\textwidth]{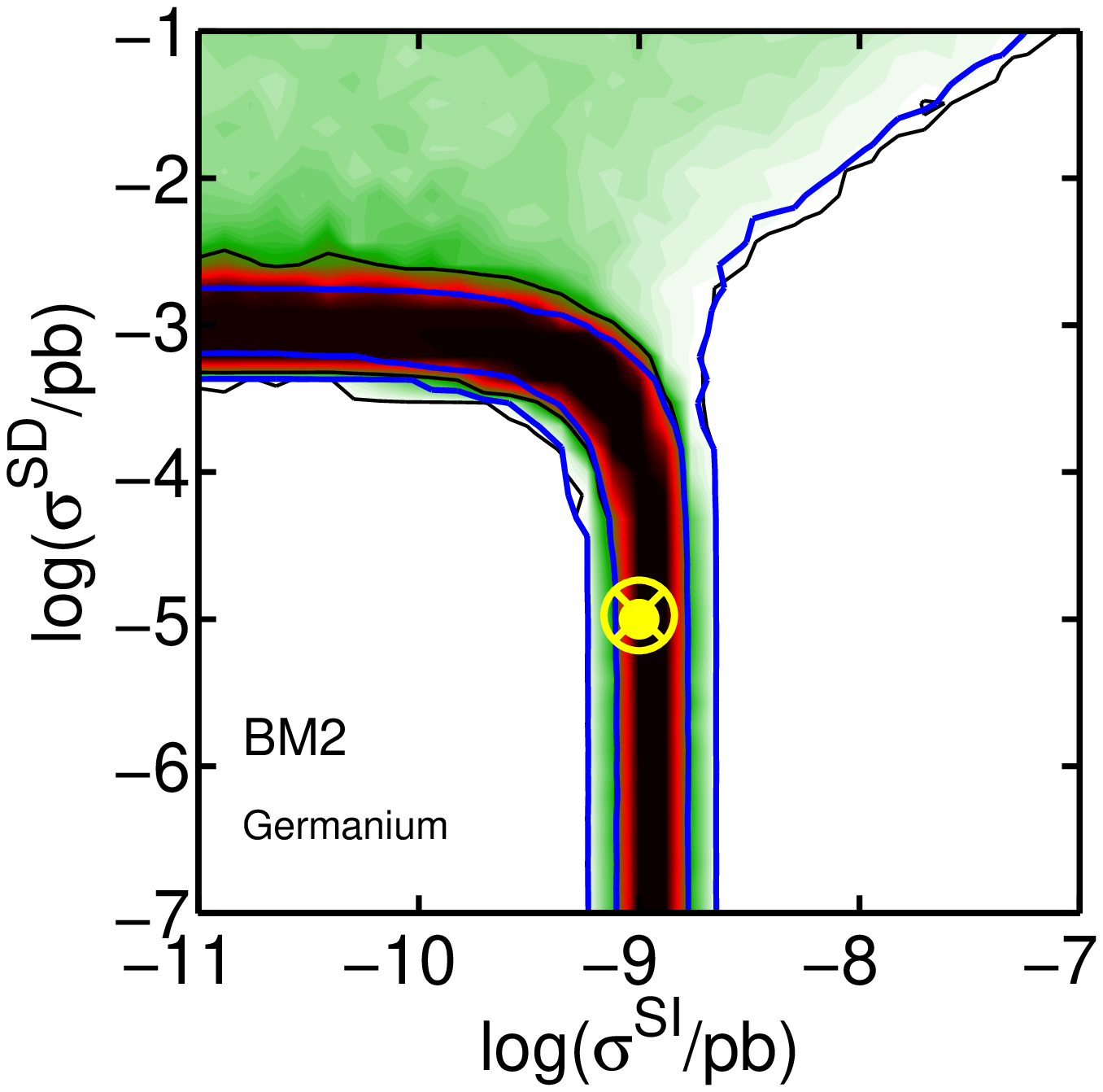}\\[-4ex]
\includegraphics[width=0.35\textwidth]{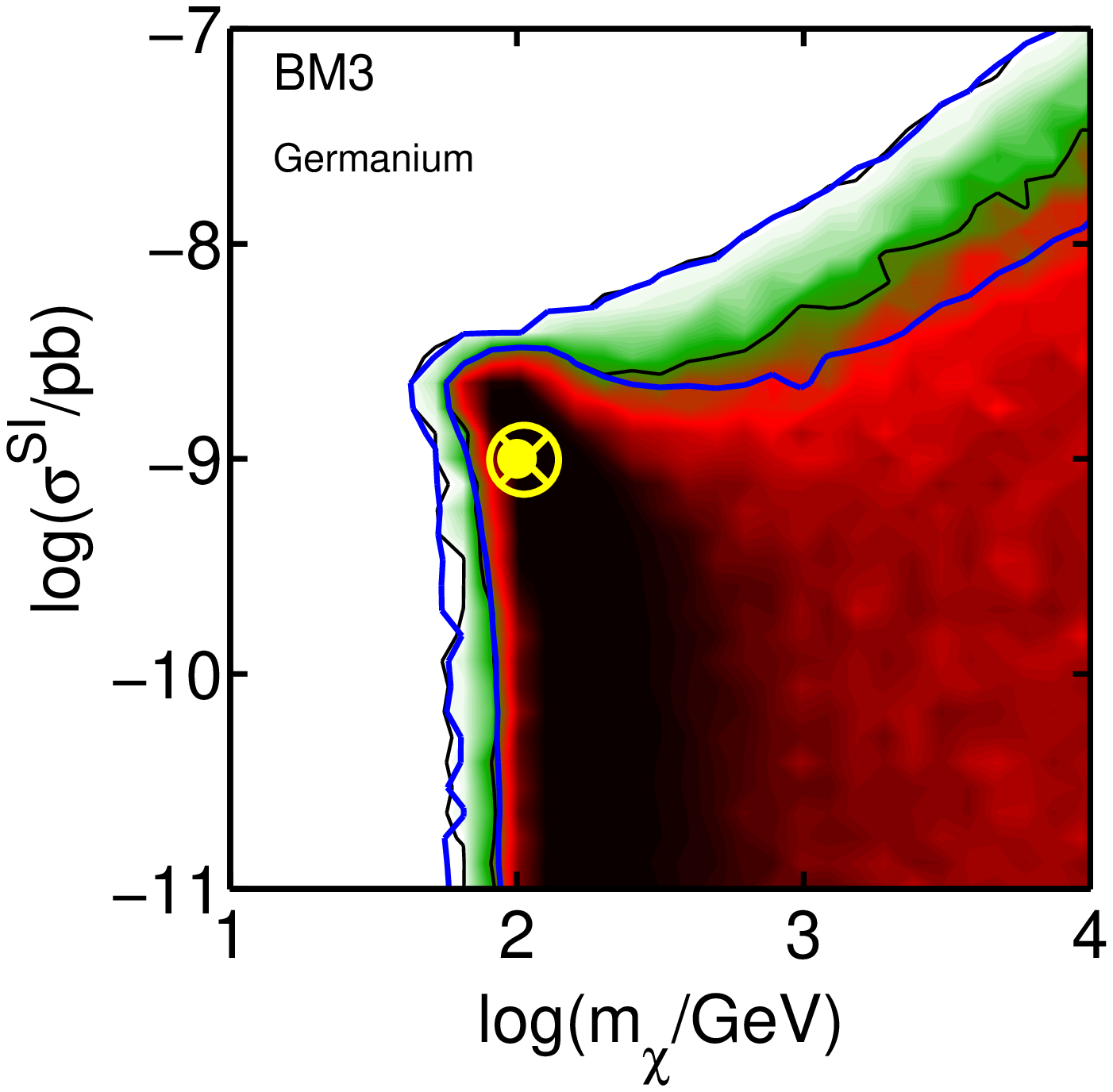}\hspace*{-0.62cm}
\includegraphics[width=0.35\textwidth]{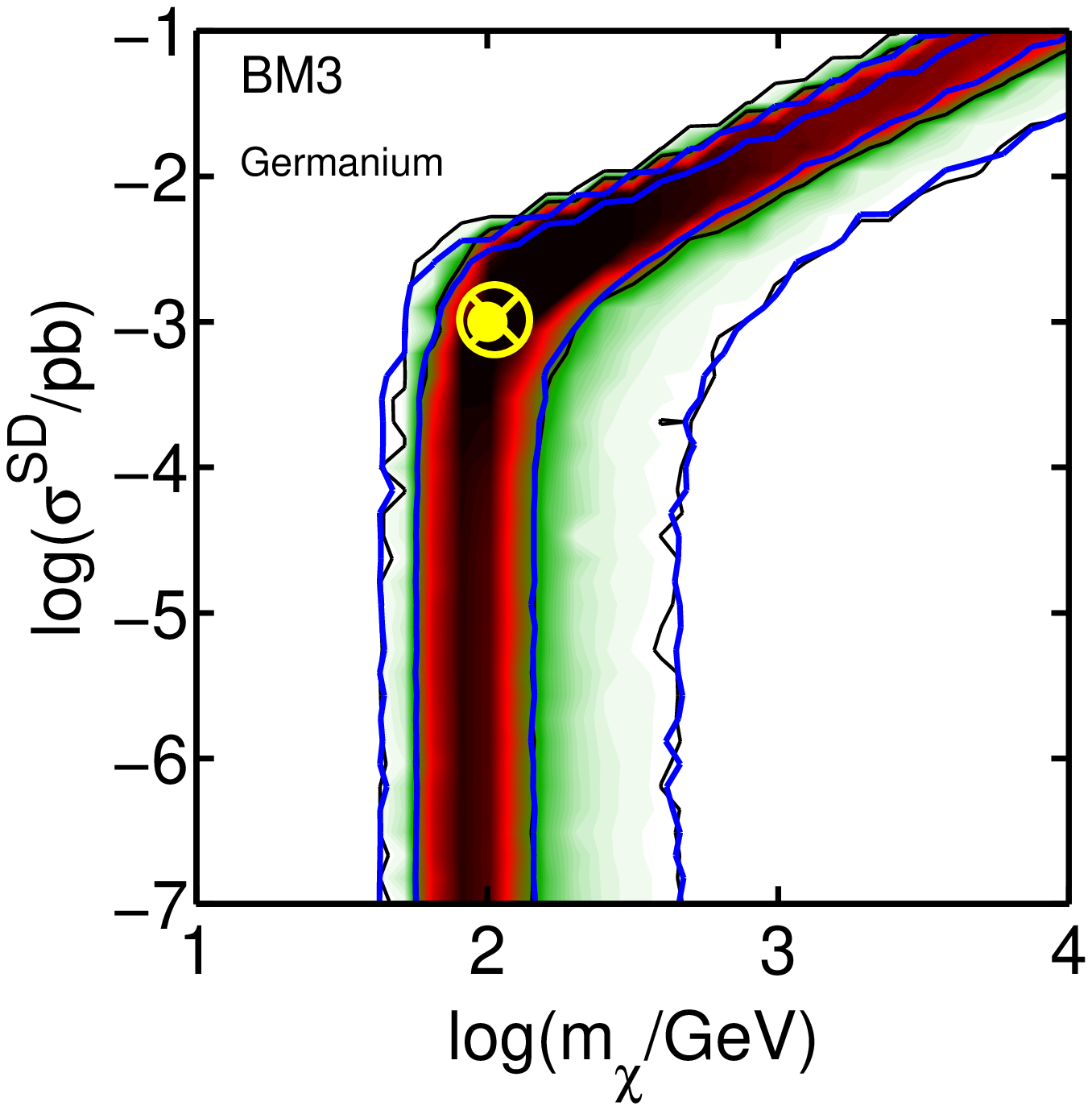}\hspace*{-0.62cm}
\includegraphics[width=0.35\textwidth]{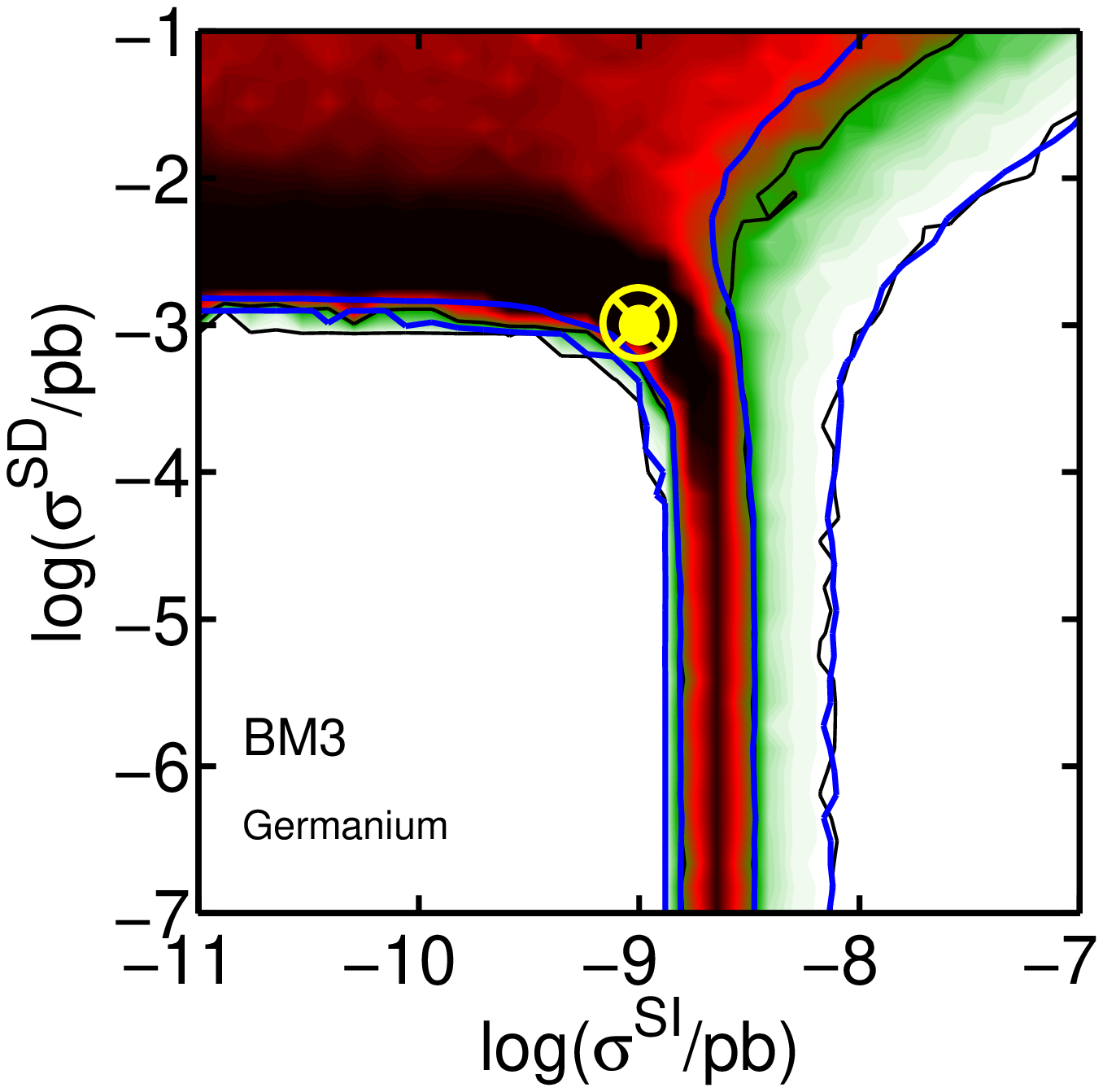}\\[-4ex]
\caption{Two-dimensional profile likelihood for the reconstructed parameter space $(\mwimp,\,\sigsi,\,\sigsd)$ in benchmark models BM1, BM2, and BM3 (from top to bottom), including nuclear uncertainties in the SDSF through the three-parameter model introduced in Eq.\,(\ref{eqn:family}). The inner and outer black contours are 68\% and 99\% confidence levels, respectively. The solid blue line corresponds to the case without uncertainties. The yellow dot indicates the benchmark value of the parameters, while the yellow encircled cross the position of the best-fit values.
\label{fig:SD_profl_nuisance}}
\end{figure*}

We repeat the scan for each benchmark extending the parameter space including $N$, $\alpha$ and $\beta$.
The number of events $\{\lambda_i\}$ of the simulated experimental data are obtained assuming a SDSF with $(N=0.16,\,\beta=0.031,\,\alpha=5.5)$ which is located in the center of the above-mentioned ranges. 
Fig.\,\ref{fig:SD_profl_nuisance} shows the resulting reconstructed contours in the profile likelihood of the DM properties in the three benchmark models. 
For comparison, we also indicate by means of blue lines the contours of the reconstructed DM parameters when nuclear uncertainties are not included and where the values of $N$, $\alpha$, and $\beta$ are fixed to their central values.

We observe that in the case of BM1 the differences with respect to the case with no uncertainties are very small. One can only observe a slight widening in the determination of $\sigsd$ when uncertainties in the SDSF are included, but otherwise the reconstructed regions in the parameter space show very little differences. This occurs because in BM1 the DM candidate interacts mainly through SI interactions and it is thus fairly independent of the details of the SD term.
Something similar occurs in the case of BM2, although the widening of the reconstruction of $\sigsd$ is more evident now. Also the 68\% confidence level curves corresponding to the WIMP mass extend to slightly larger values (notice that the logarithmic scale makes this effect more difficult to observe). 
Finally, it is in benchmark BM3 that the largest effects are found, since the SD contribution is larger. Once more, a widening in the determination of $\sigsd$ is observed, which is now more evident in the 68\% confidence level lines. Also the inclusion of uncertainties in the SDSF enlarge the contours for large WIMP masses.

\subsection{Xenon detectors}

\begin{figure*}
	\includegraphics[width=0.45\textwidth]{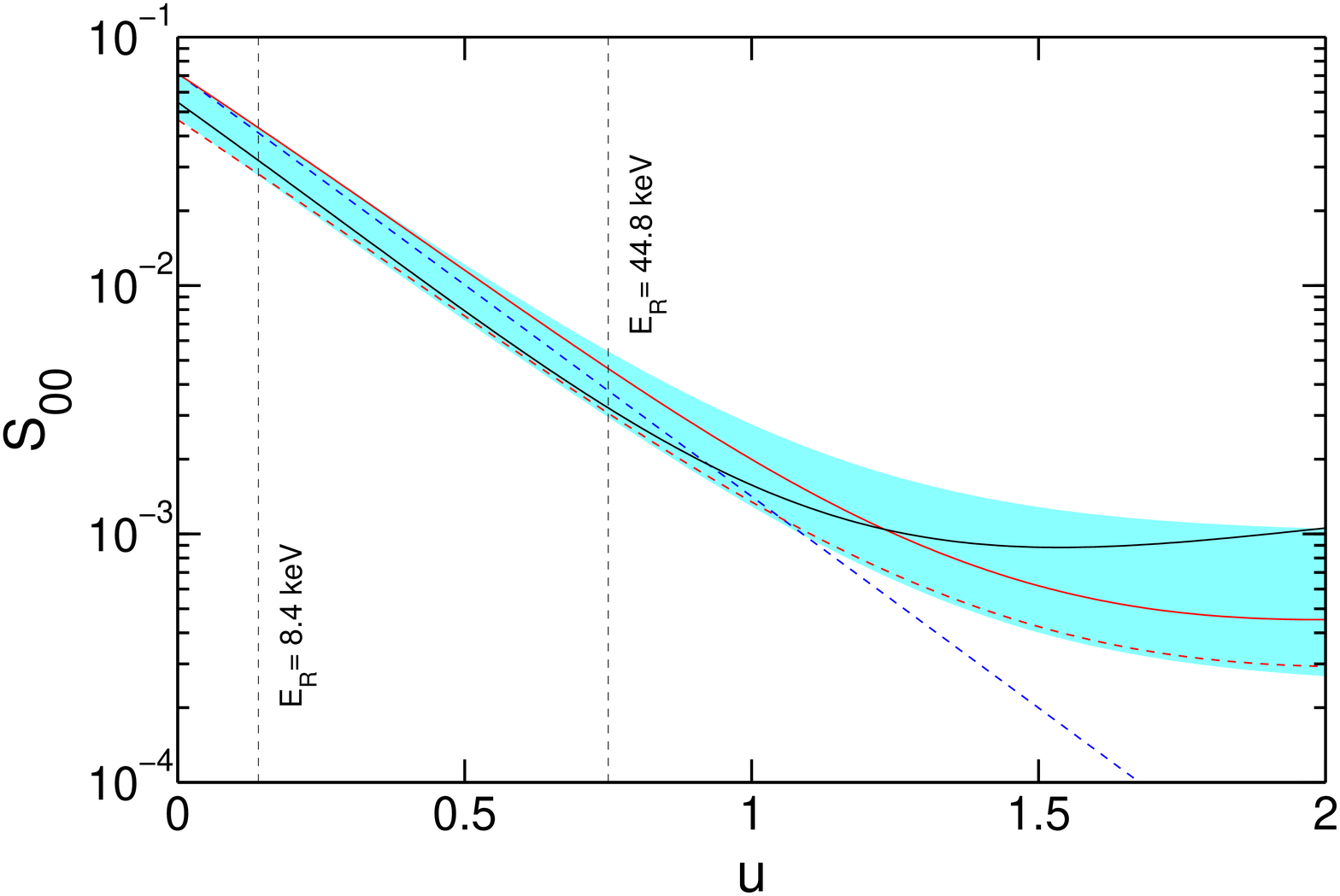}
	\includegraphics[width=0.45\textwidth]{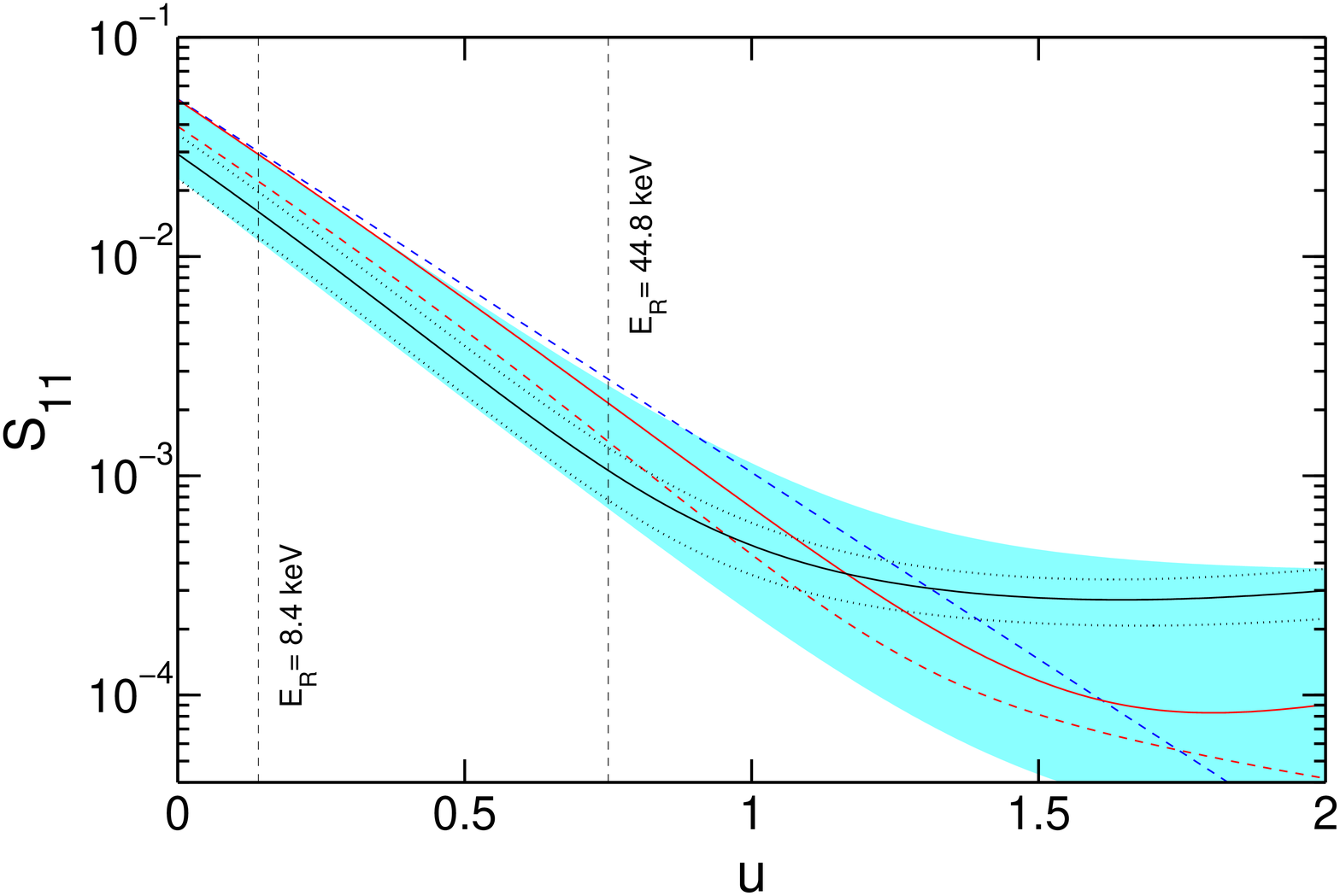}\\
	\includegraphics[width=0.45\textwidth]{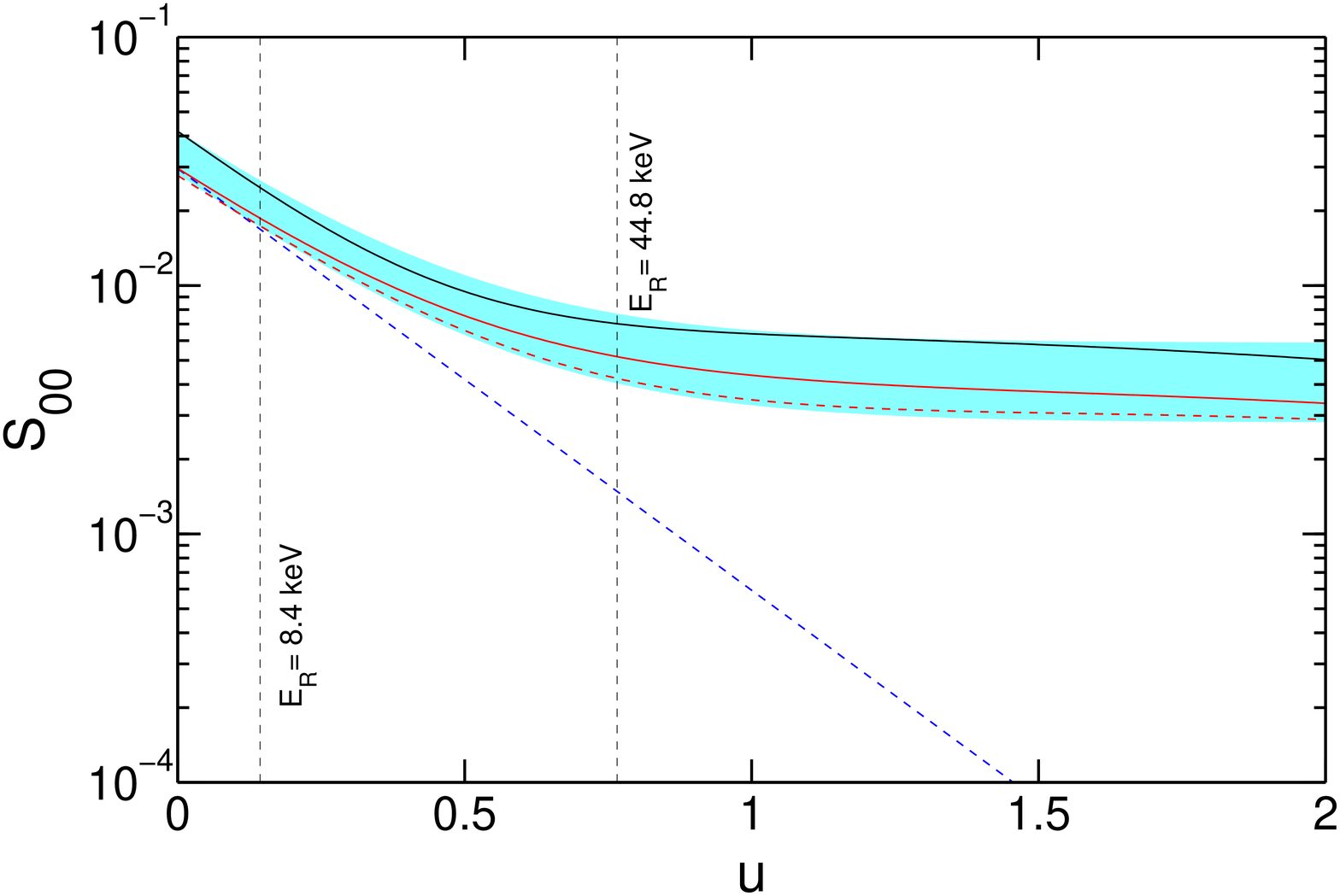}
	\includegraphics[width=0.45\textwidth]{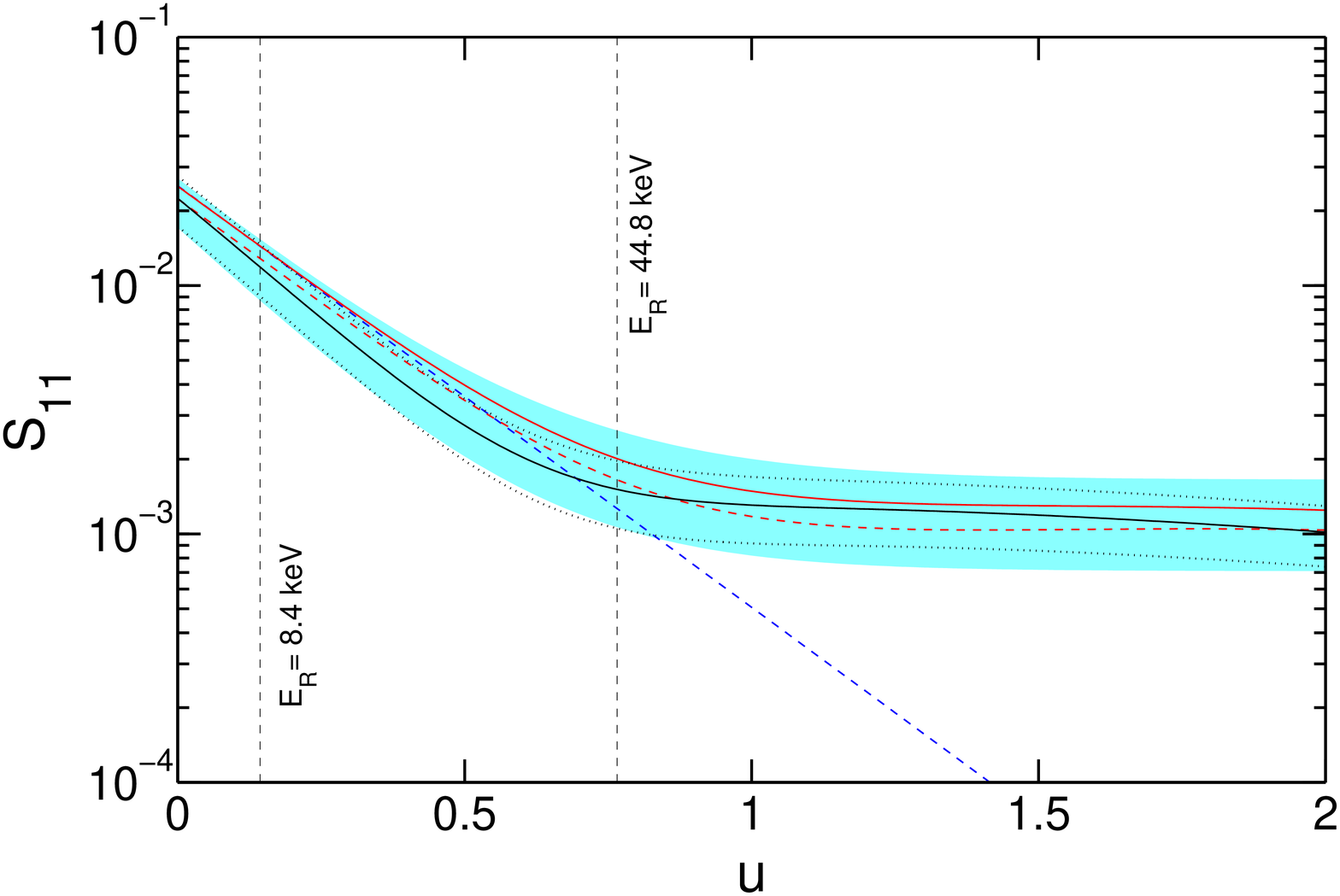}
\caption{\label{fig:FFs_Xe} The same as in Fig.\,\ref{fig:FFs} but for the case of $^{129}$Xe (top row) and $^{131}$Xe (bottom row). The solid (dashed) red lines correspond to the ShM calculation using the Bonn A (Nijmegen II) potential  \cite{Ressell:1997kx}. The solid black line corresponds to the determination of Ref.\,\cite{Menendez:2012tm} and the dotted black lines are the errors associated to it (the errors for $S_{00}$ are negligible and are not shown). The dotted blue line indicates the gaussian approximation of Eq.\,(\ref{eqn:gaussian}). The blue region covers the area spanned by the family of curves in Eq.\,(\ref{eqn:family}) with the parameters defined in the text. The vertical black dashed lines indicate the WIMP search window used in the analysis.}
\end{figure*}

The same procedure can be used for xenon detectors. Natural xenon contains 
two isotopes $^{129}$Xe (with a 26.4\% isotopic abundance) and $^{131}$Xe 
(21.29\%) which are sensitive to the SD component of the WIMP interaction (in 
particular to the SD cross-section of the WIMP with neutrons). 
As in the case of germanium, we consider various parametrizations of the SD form
factor for these nuclei from Ref.\,\cite{Ressell:1997kx}, in which the nuclear shell model was applied to two different potentials describing the nucleon-nucleon interaction, the Bonn A \cite{HjorthJensen:1995ap} and Nijmegen II \cite{Stoks:1994wp} potentials.
We also include a recent result from Ref.\,\cite{Menendez:2012tm} in which the so called gcn5082 interaction \cite{Caurier:2007wq} is used. 
Then we repeat
the analysis of the previous section modeling the uncertainties in the xenon
SDSF by means of the parametrizations in Eq.\,(\ref{eqn:family}),
changing the values of the $(N,\,\alpha,\,\beta)$ parameters to define the area that contains the above-mentioned models for the SDSFs. 
In particular, for the $S_{11}$ component in $^{129}$Xe we 
consider $N=[0.029,\,0.052]$, $\alpha=[4.2,\,4.7]$, and $\beta=[1.0\times10^{-3},7\times10^{-3}]$. Similarly, in $^{131}$Xe 
the ranges for $S_{11}$ are $N=[0.017,\,0.027]$, $\alpha=[4.3,\,5.0]$, and $\beta=[4.2\times10^{-2},6.1\times10^{-2}]$. 
The various models for the SDSFs are represented in Fig.\,\ref{fig:FFs_Xe}, together with the envelopes for $S_{00}$ and $S_{11}$ in both isotopes.
We consider the same exposure as in the previous case ($\epsilon=300$~kg yr) but the energy range of the WIMP detection window is now taken to be $E_R=[8.4,\,44.8]$~keV, mimicking that of the XENON100 experiment.

\begin{figure*}
\includegraphics[width=0.35\textwidth]{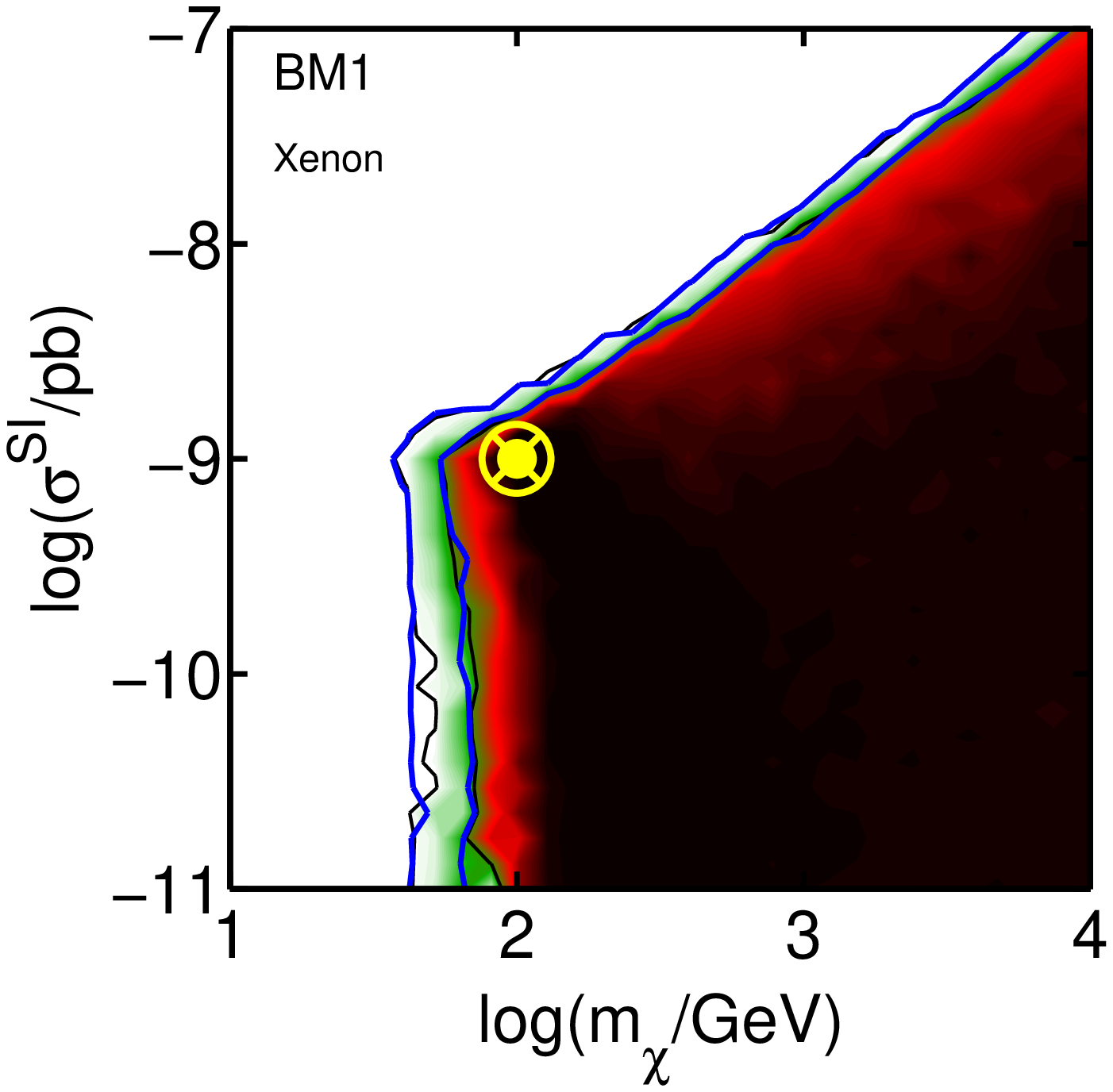}\hspace*{-0.62cm}
\includegraphics[width=0.35\textwidth]{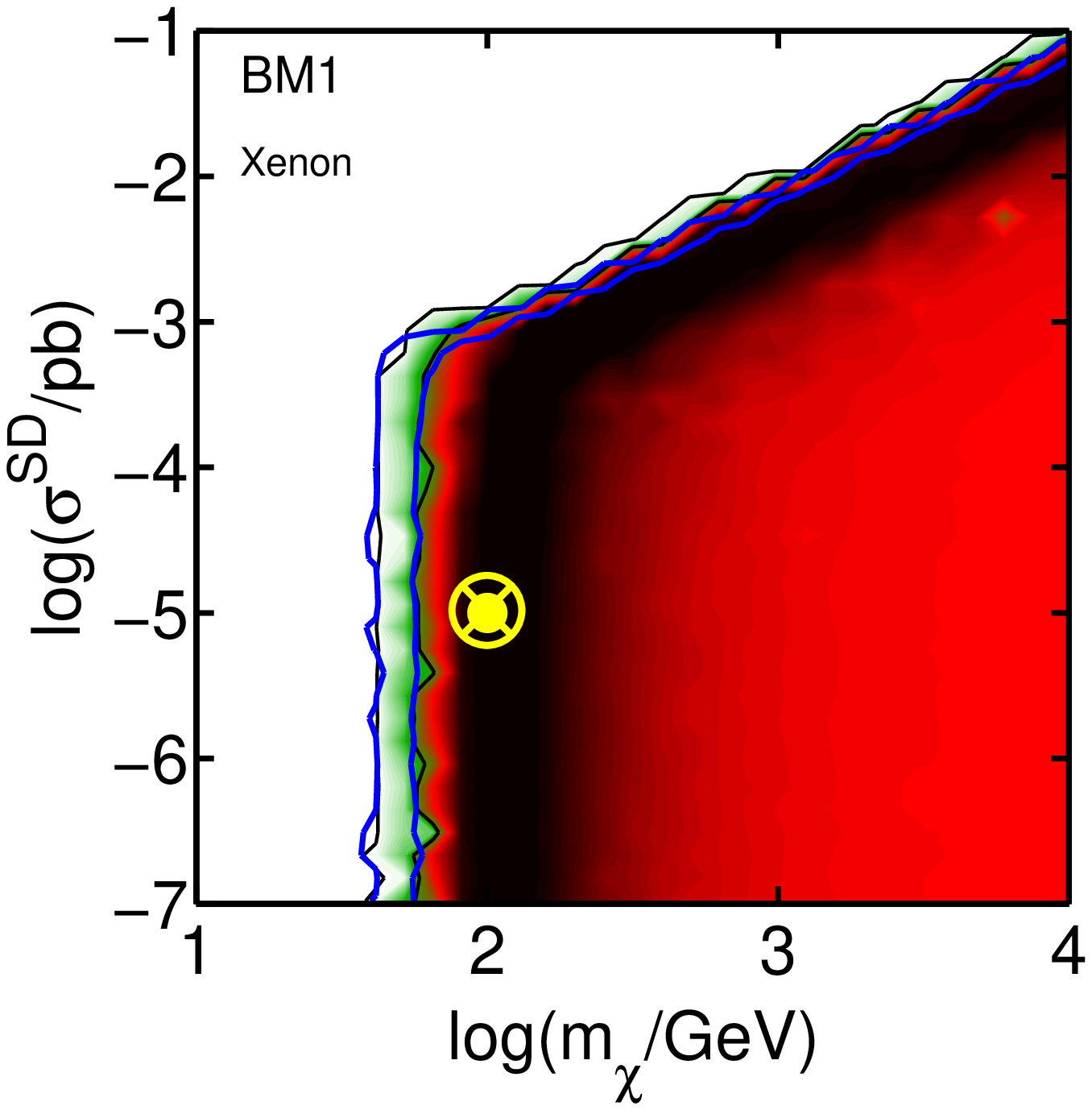}\hspace*{-0.62cm}
\includegraphics[width=0.35\textwidth]{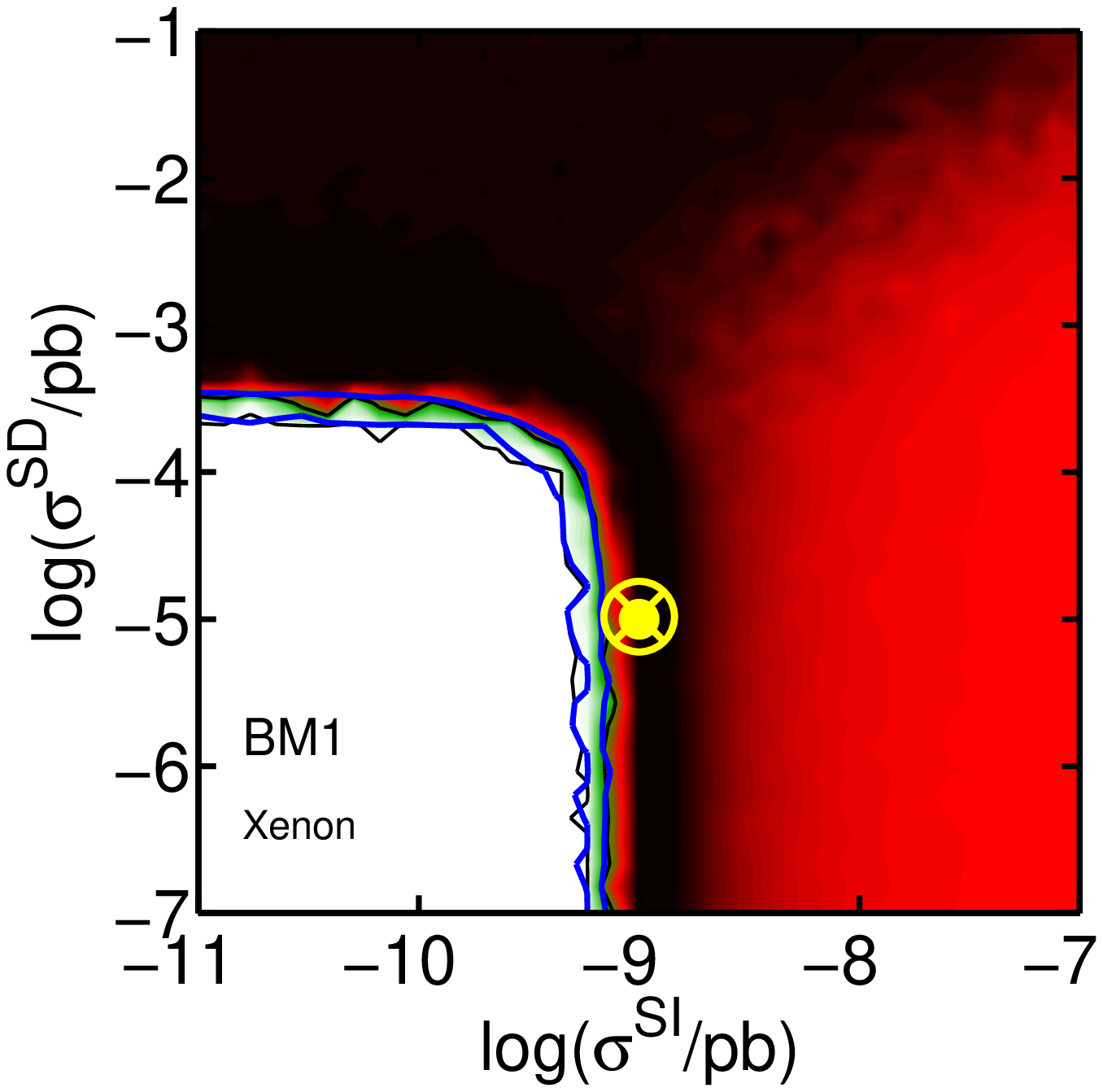}\\[-4ex]
\includegraphics[width=0.35\textwidth]{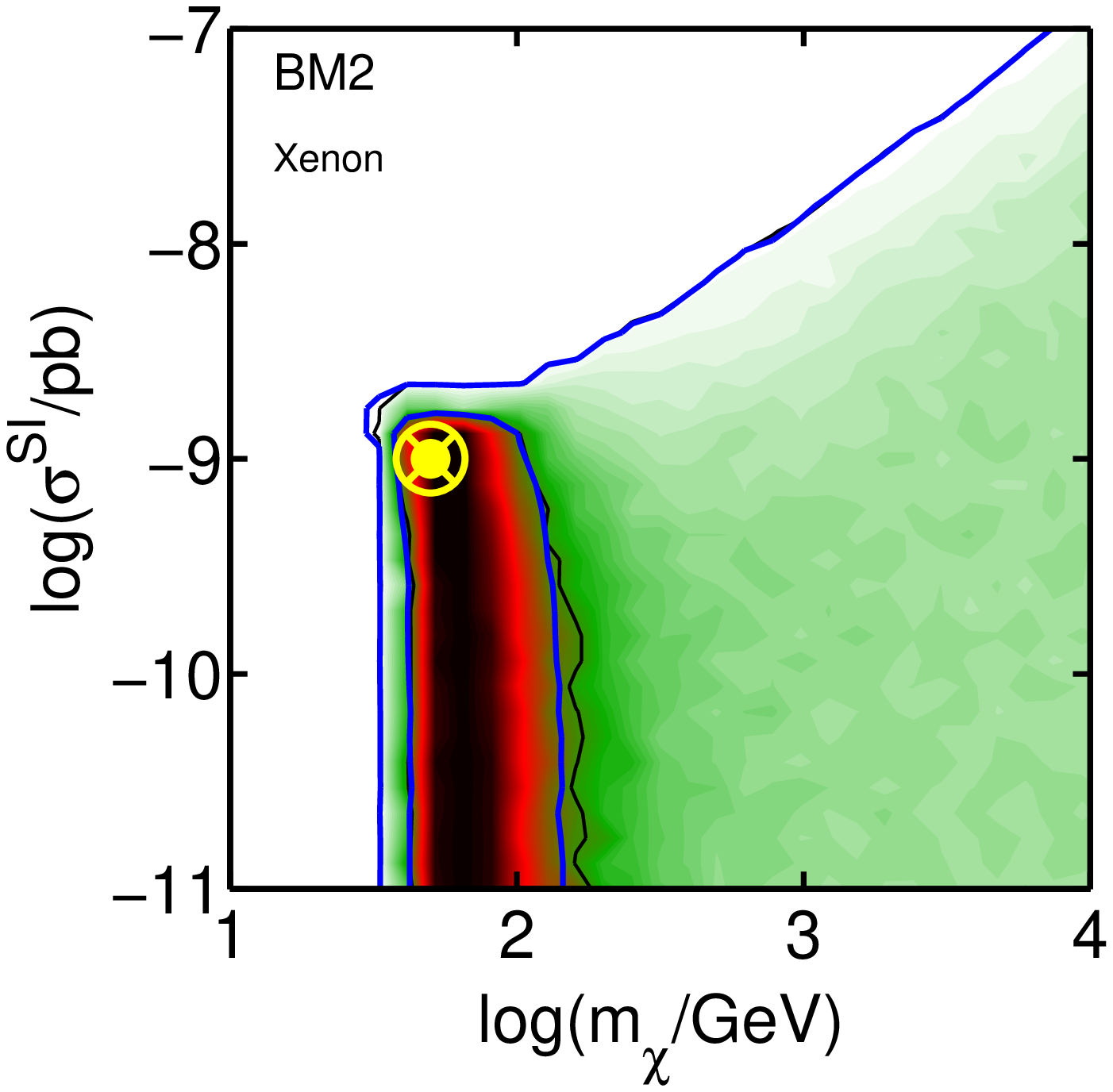}\hspace*{-0.62cm}
\includegraphics[width=0.35\textwidth]{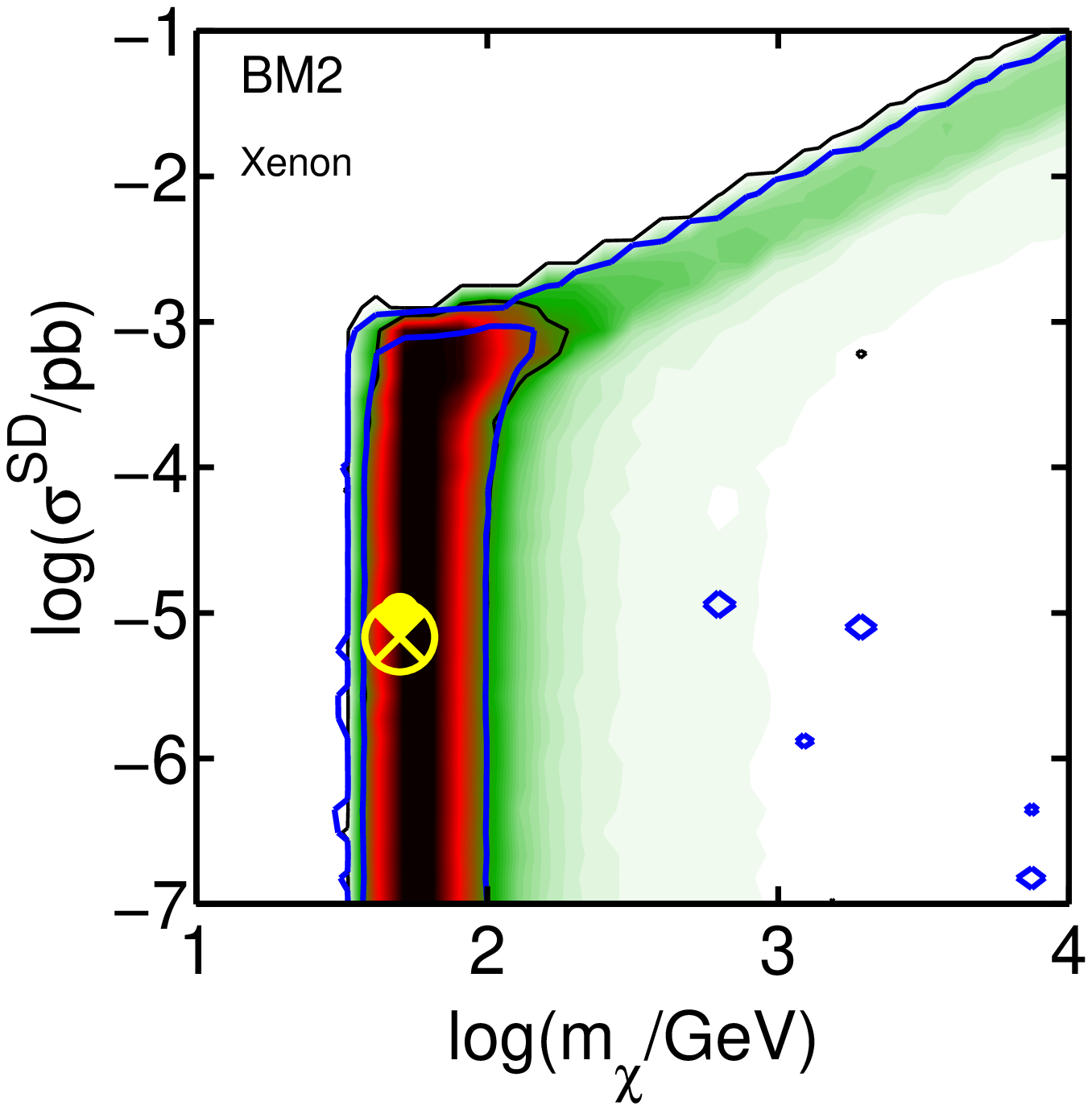}\hspace*{-0.62cm}
\includegraphics[width=0.35\textwidth]{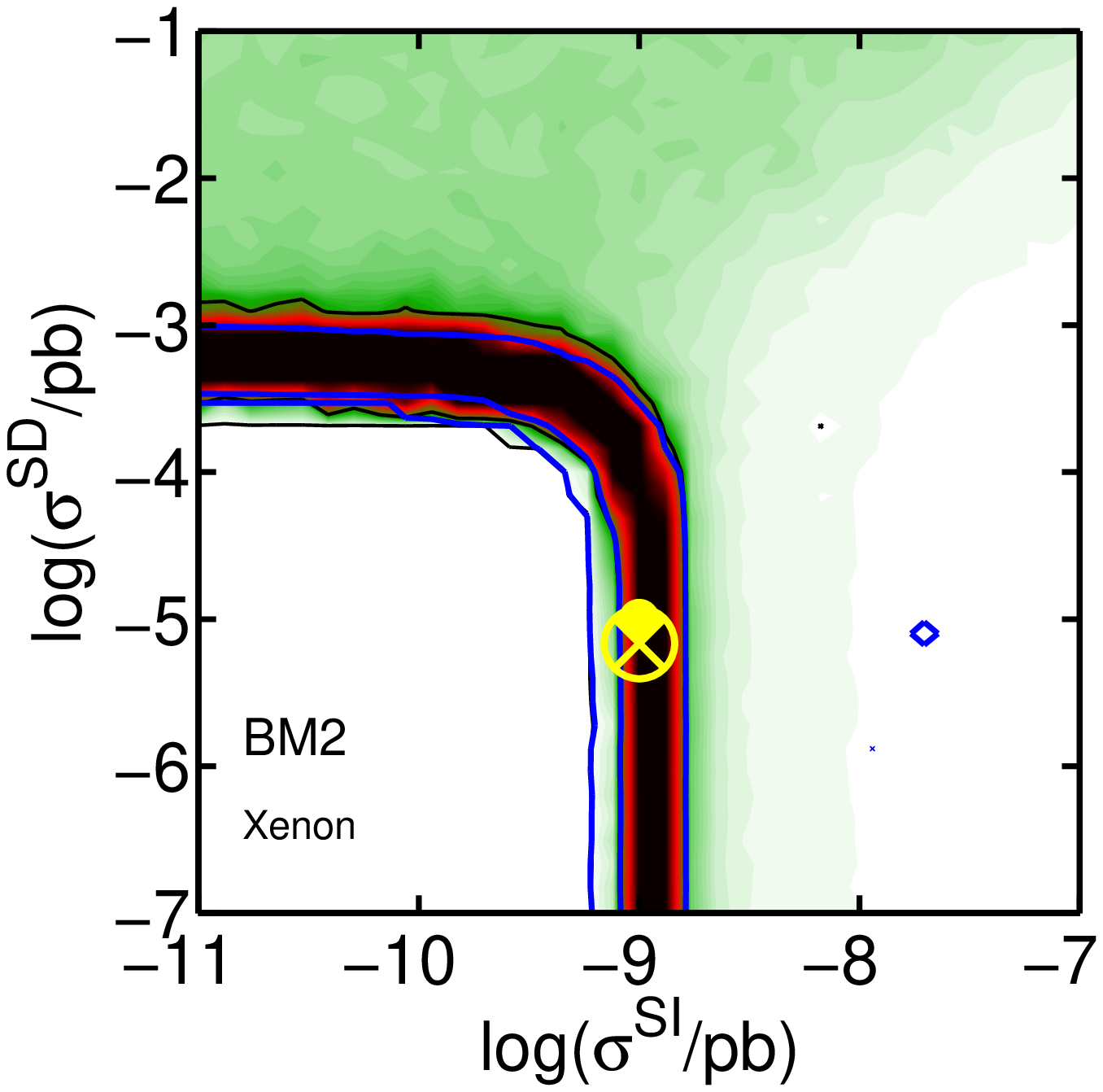}\\[-4ex]
\includegraphics[width=0.35\textwidth]{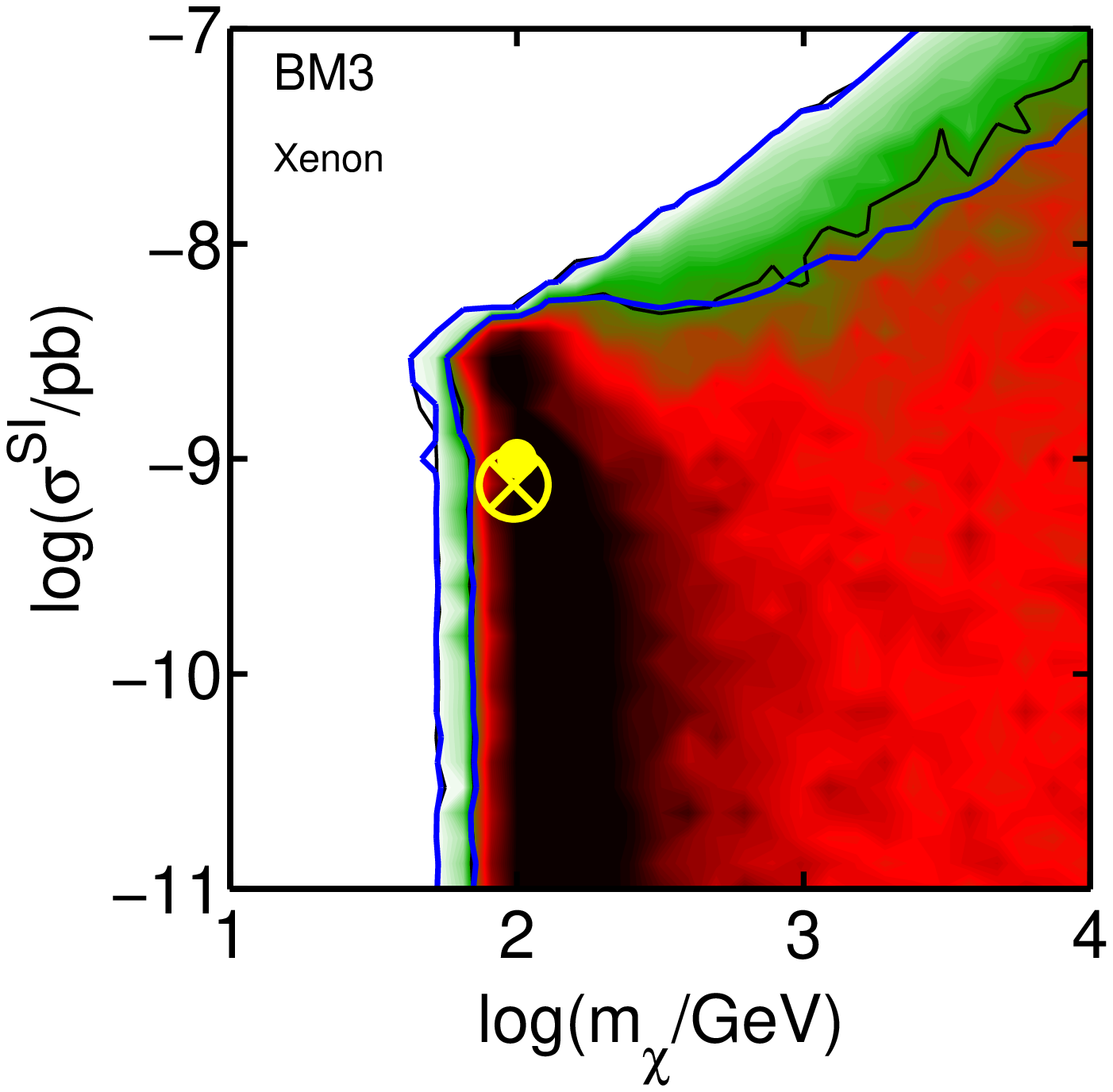}\hspace*{-0.62cm}
\includegraphics[width=0.35\textwidth]{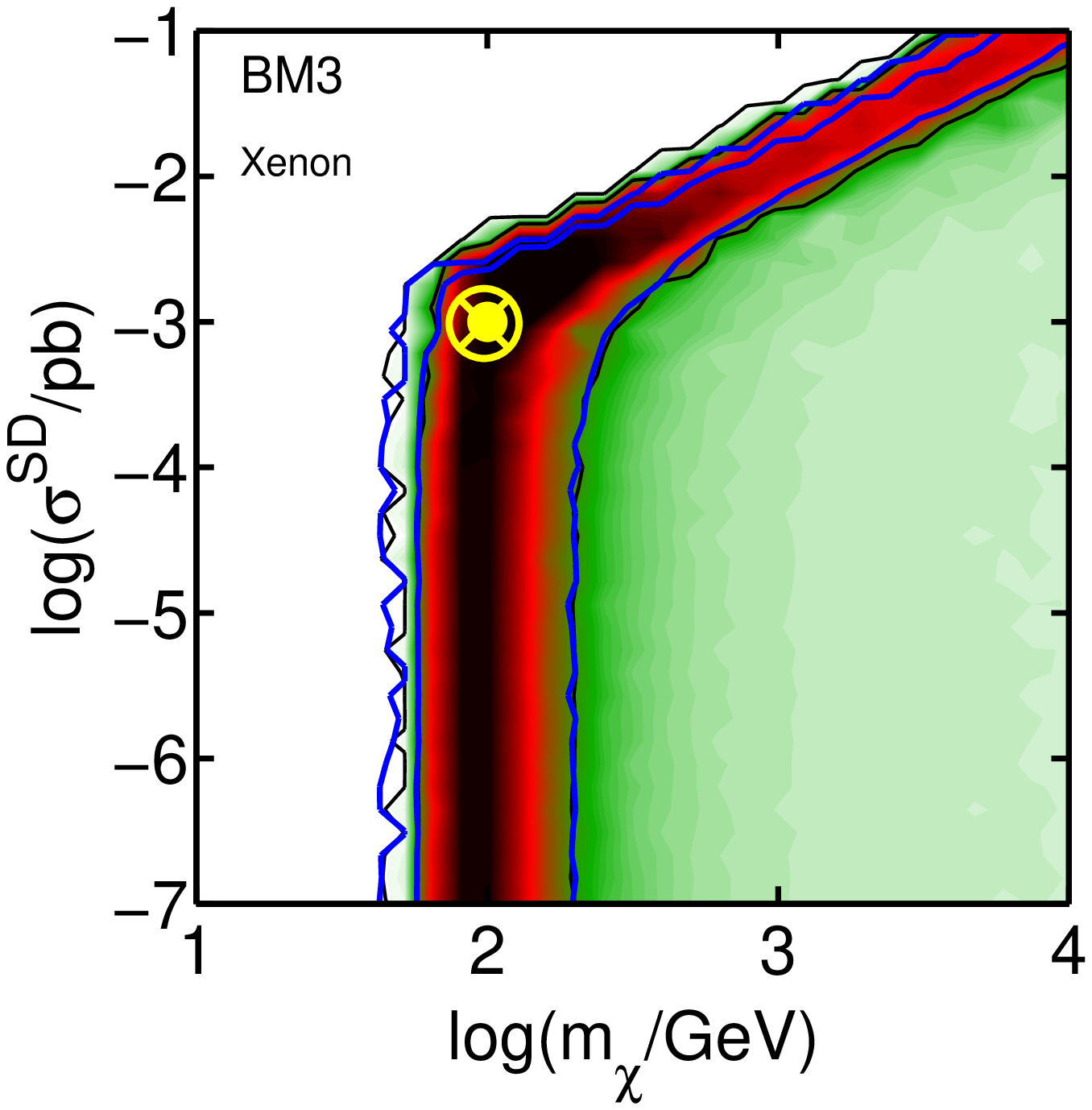}\hspace*{-0.62cm}
\includegraphics[width=0.35\textwidth]{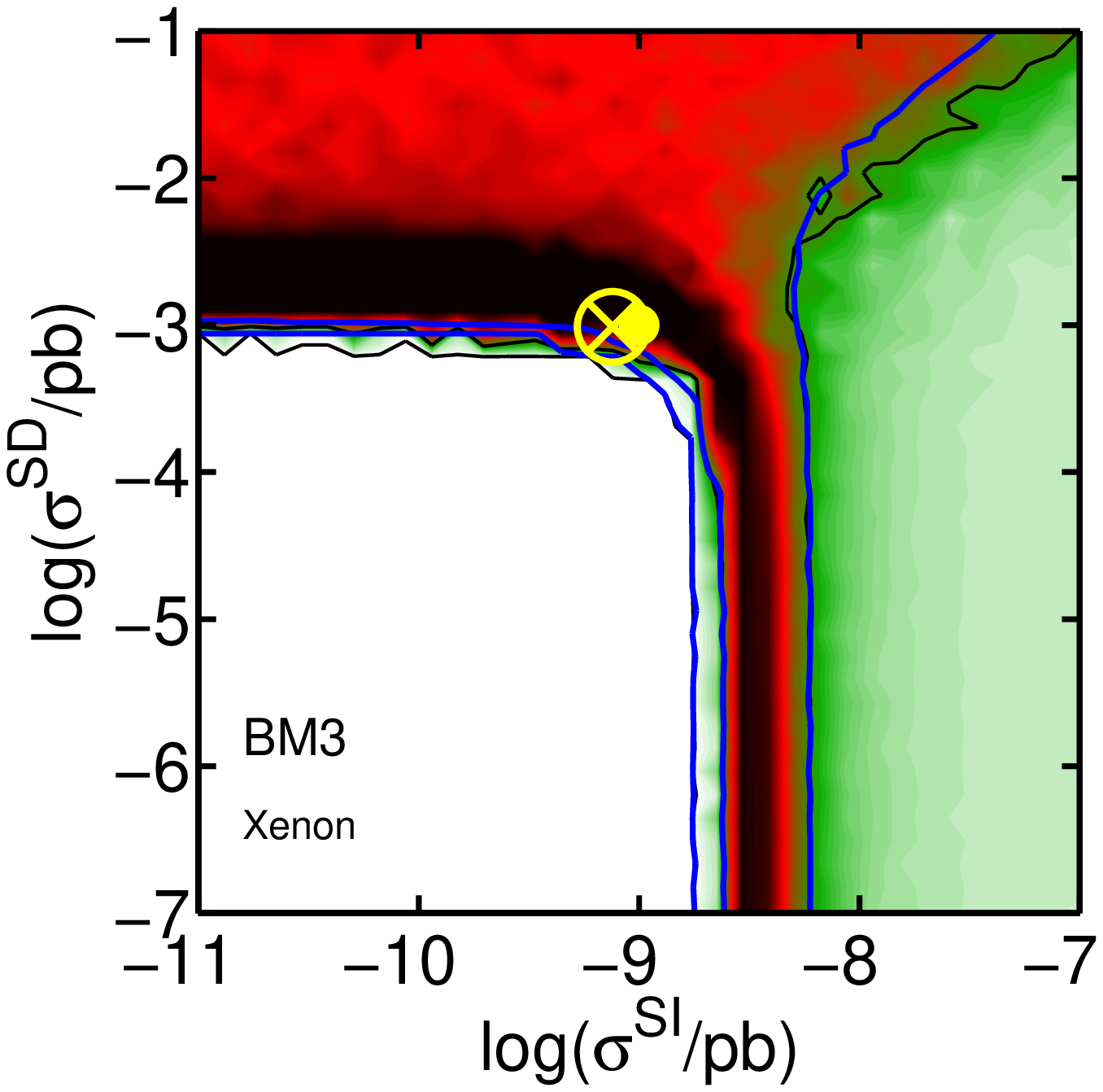}\\[-4ex]
\caption{The same as in Fig.\,\ref{fig:SD_profl_nuisance} but for the case of a xenon detector.
\label{fig:XeSD_profl_nuisance}}
\end{figure*}

Uncertainties in the SDSF for xenon have the same qualitative effect as in germanium. Namely, the predictions for the WIMP mass and the SD component of its scattering cross-section are affected. The resulting contours for the profile likelihood 
benchmarks BM1, BM2 and BM3 are displayed in 
Figs.\,\ref{fig:XeSD_profl_nuisance}.
We can observe that the effect is similar in magnitude to the case of germanium (despite being a heavier nucleus than germanium, the isotopic abundance of the elements sensitive to the SD coupling is larger in xenon).  
Once more, deviations are larger for BM2 and BM3 than in BM1.

The inclusion of uncertainties on SDSF through the 
parametrization in Eq.\,(\ref{eqn:family}) is a procedure that can be applied to 
other nuclei. In the case of germanium and xenon, the existence of different SDSF computations allowed us to 
define the ranges
in which the three parameters of Eq.\,(\ref{eqn:family}) are varied.

\subsection{Comparison with astrophysical uncertainties}
\label{sec:astro}

To put our results in context, we need to compare the effects of nuclear uncertainties in the SDSF that we just discussed with those originating from astrophysical uncertainties in the parameters of the DM halo. 
In order to introduce the latter, we have
considered a halo model motivated by $N$-body simulations, which differs from the 
standard halo model in a high-velocity tail 
\cite{Vogelsberger:2008qb,Ling:2009eh,Fairbairn:2008gz,Kuhlen:2009vh}.
The distribution function is taken from Ref.\,\cite{Lisanti:2010qx} and it is 
characterized by the presence of an additional parameter $k$ that controls the deviations of $F(v)$ from the standard halo model,
\begin{equation}
	F(v) = N_k^{-1} v^2 \left[ e^{-v^2/k v_0^2} - e^{-v_{esc}^2/k v_0^2} \right]^k
	\Theta(v_{esc}-v),
	\label{eqn:F(v)}
\end{equation}
where $N_k=v_0^3 e^{-y_e^2}\int_{0}^{y_e}$ $dy~y^2(e^{-(y^2-y_e^2)/k}-1)^k$ 
and $y_e=v_{esc}/v_0$. In the limit of vanishing $k$ the standard halo model is recovered.
We then consider that the three parameters that define the velocity distribution function vary in the ranges
 $v_{esc}\in[478, 610]$~km s$^{-1}$,  $v_0\in [170, 290]$~km s$^{-1}$, and $k\in[0.5, 3.5]$, and include them in our scan as nuisance parameters. 
The local DM density is also subject to observational uncertainties. Its value can be estimated from a set of experimental constraints
that fix the local gravitational potential of the Milky Way, with typical values
ranging from 0.2 to 0.6 GeV cm$^{-3}$ \cite{Catena:2009mf,Salucci:2010qr,Pato:2010yq,Iocco:2011jz}.

\begin{figure*}
\includegraphics[width=0.35\textwidth]{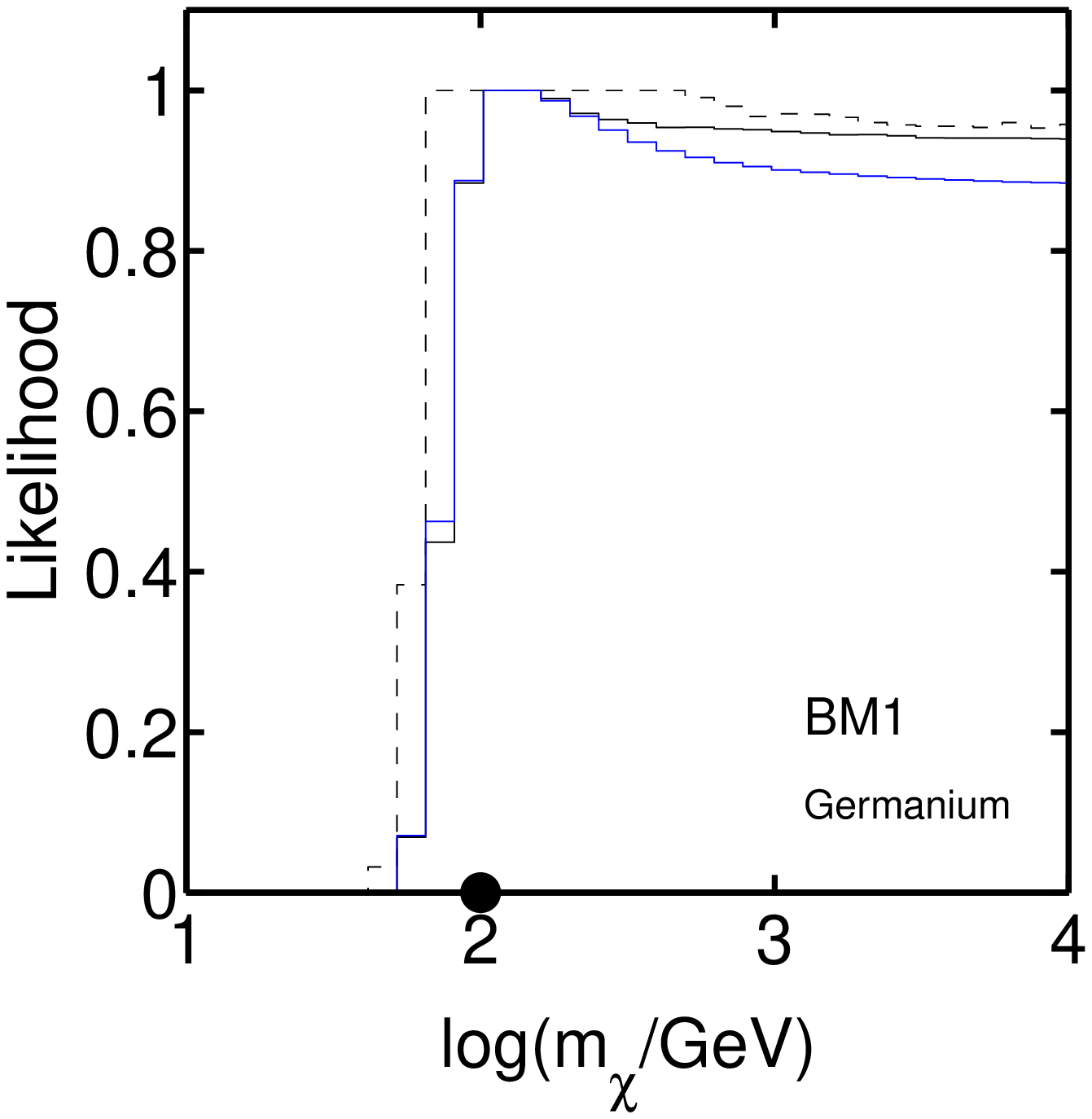}\hspace*{-0.62cm}
\includegraphics[width=0.35\textwidth]{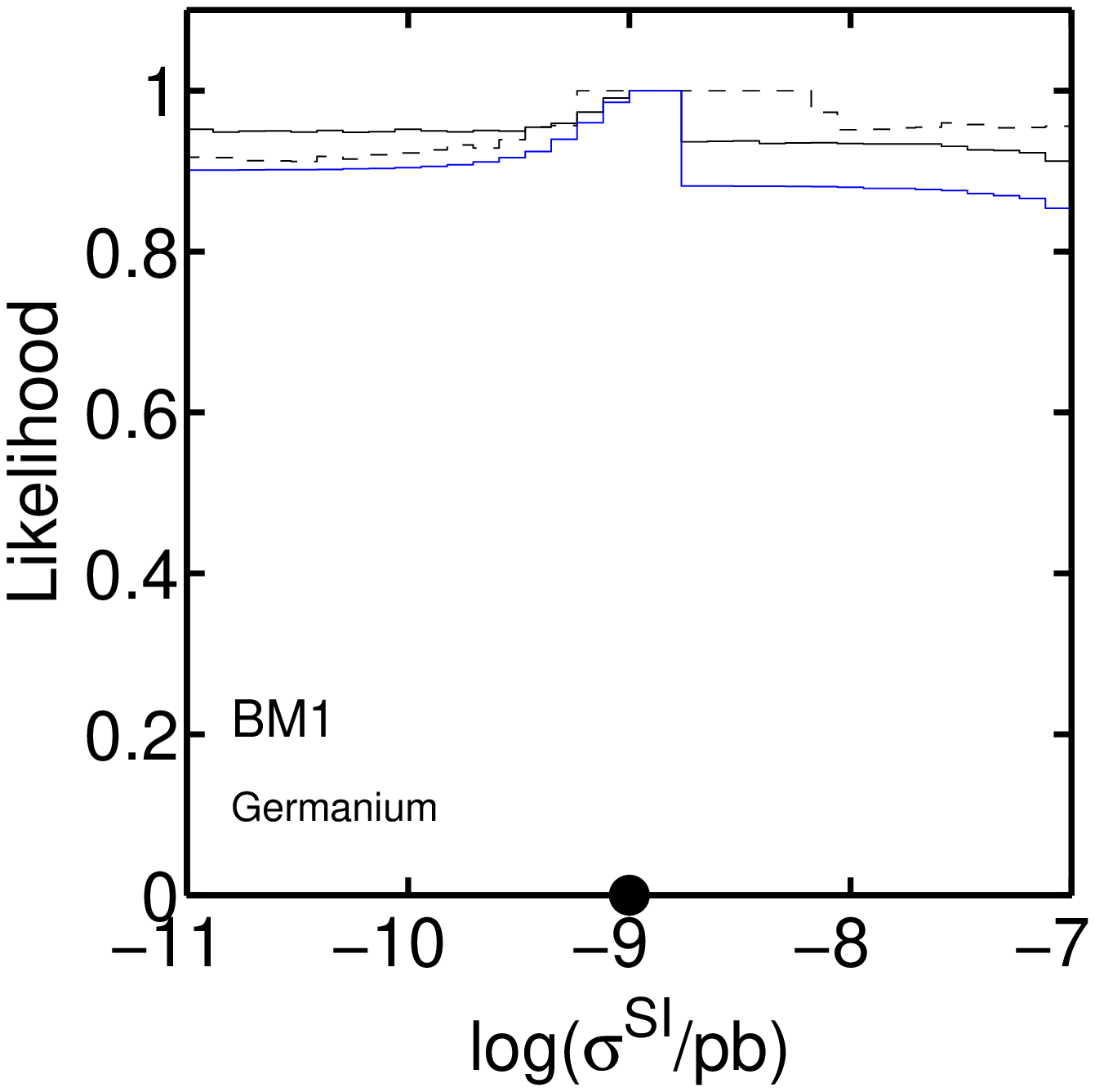}\hspace*{-0.62cm}
\includegraphics[width=0.35\textwidth]{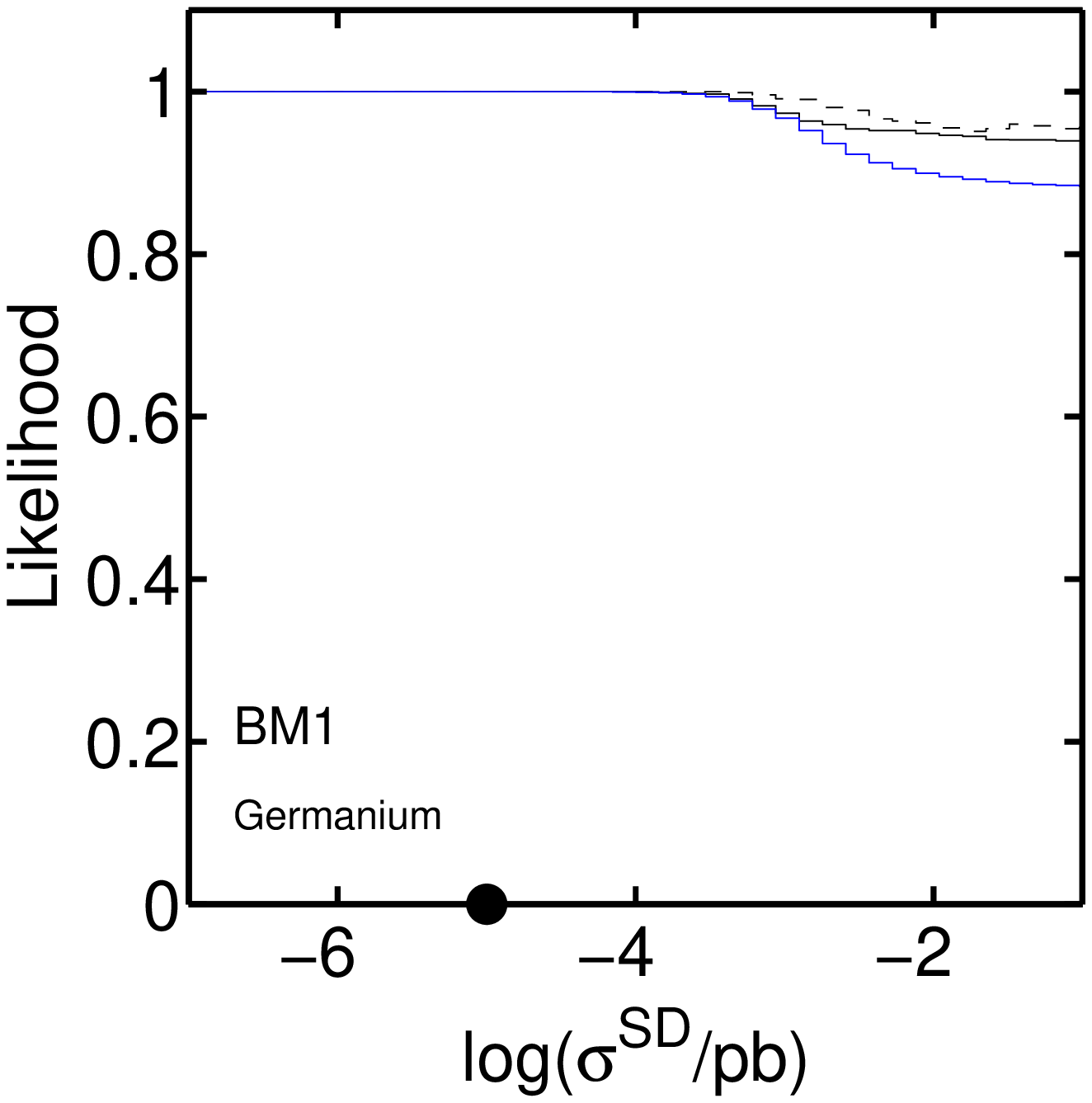}\\[-4.ex]
\includegraphics[width=0.35\textwidth]{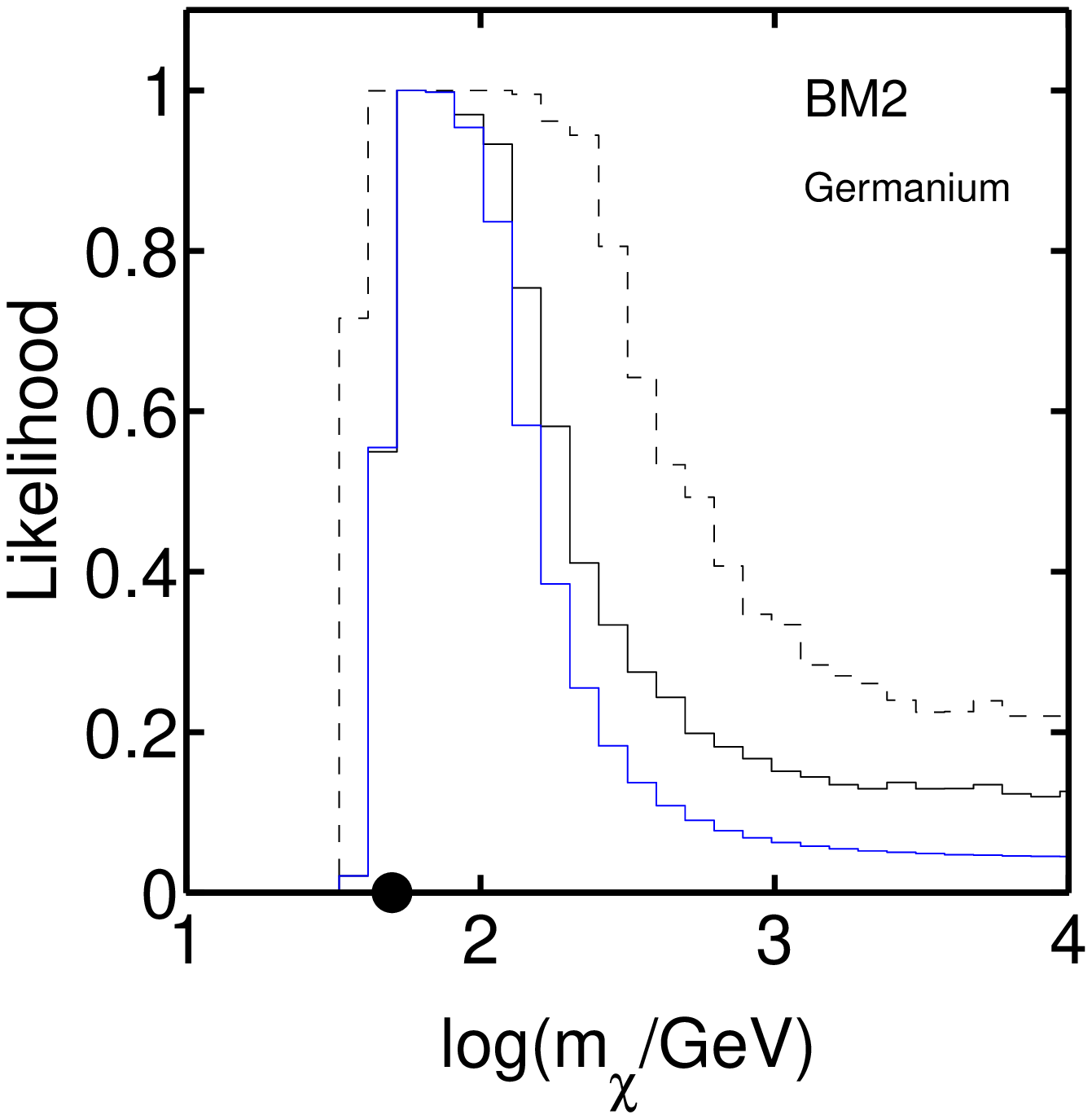}\hspace*{-0.62cm}
\includegraphics[width=0.35\textwidth]{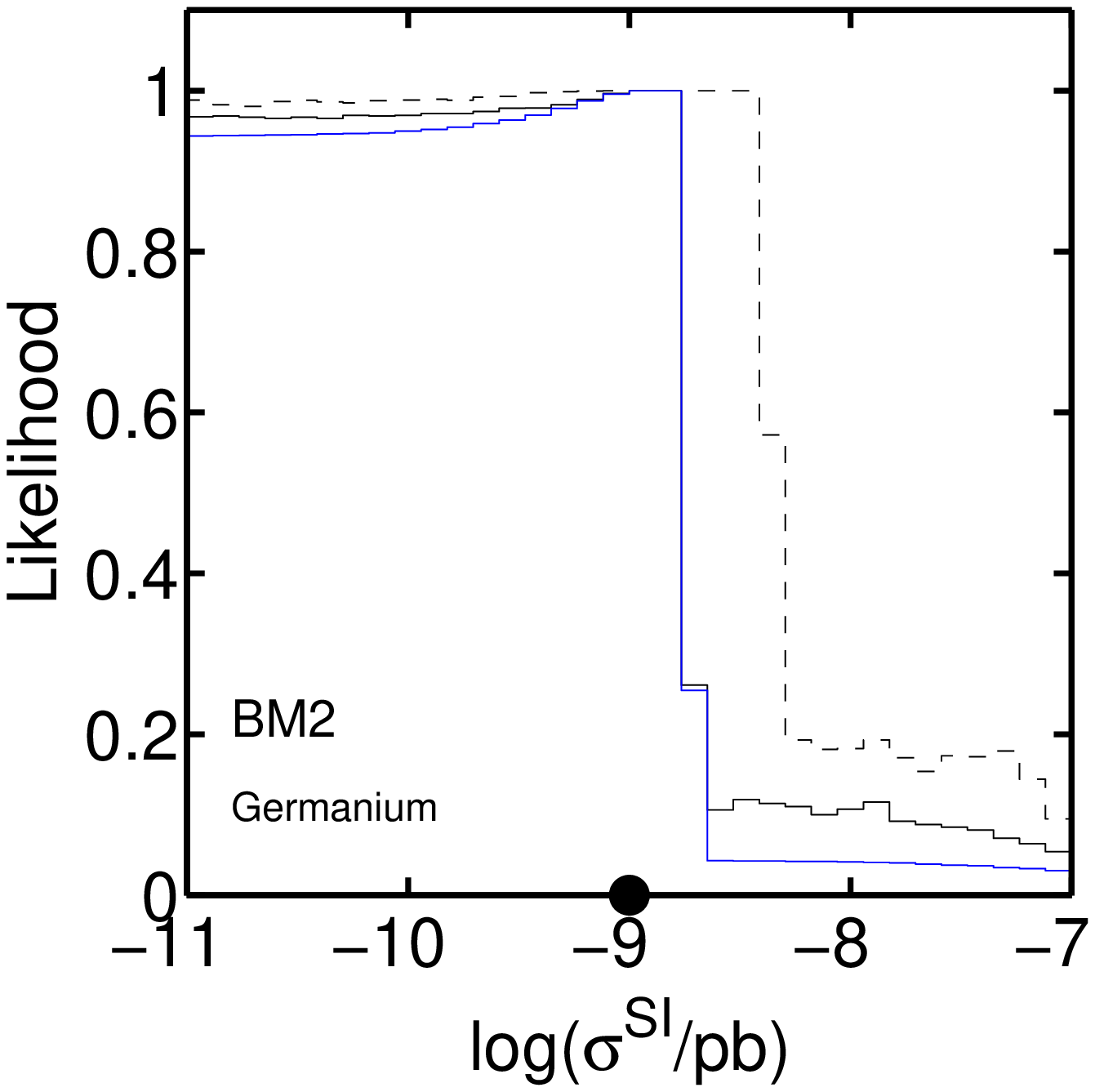}\hspace*{-0.62cm}
\includegraphics[width=0.35\textwidth]{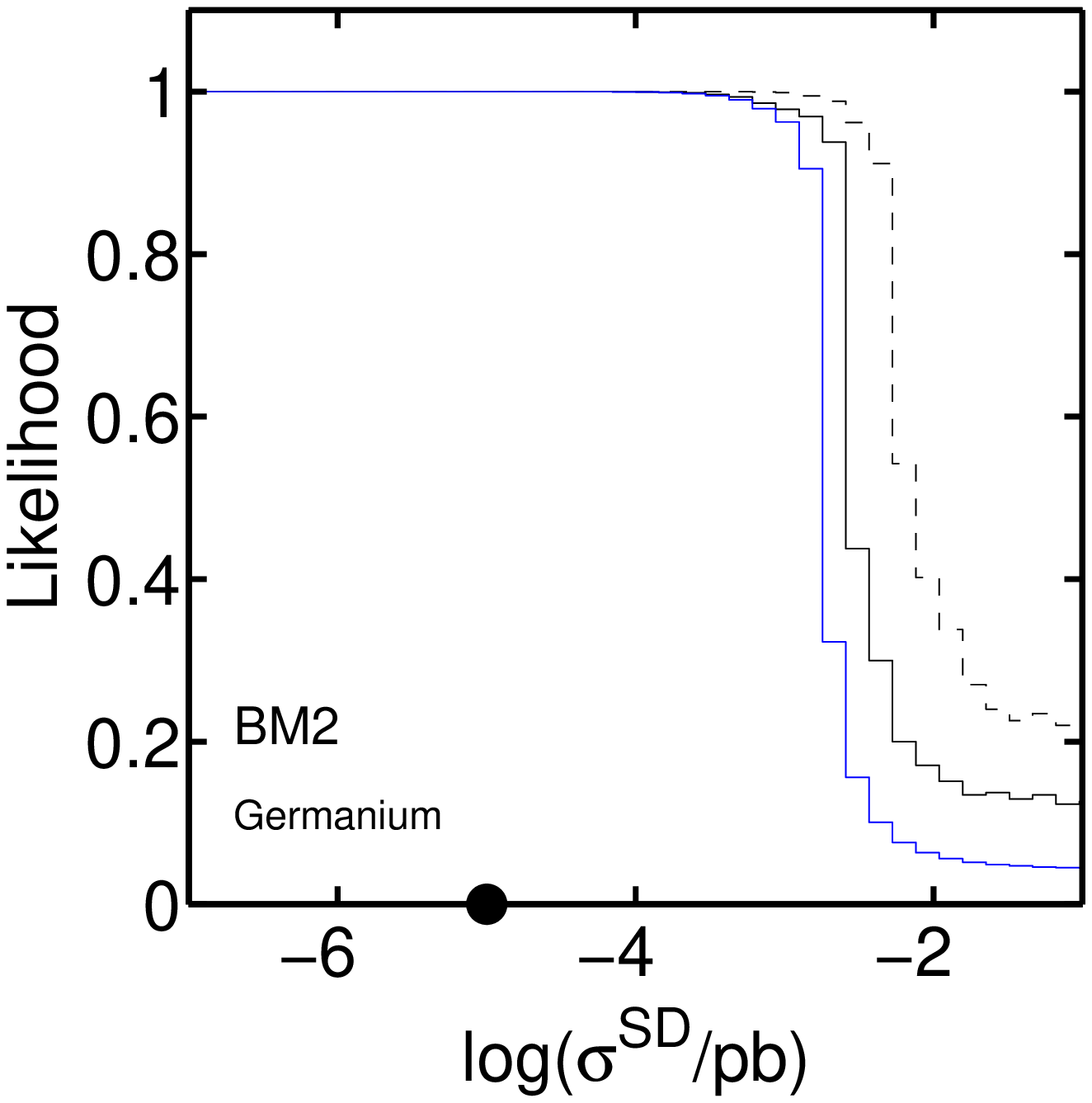}\\[-4.ex]
\includegraphics[width=0.35\textwidth]{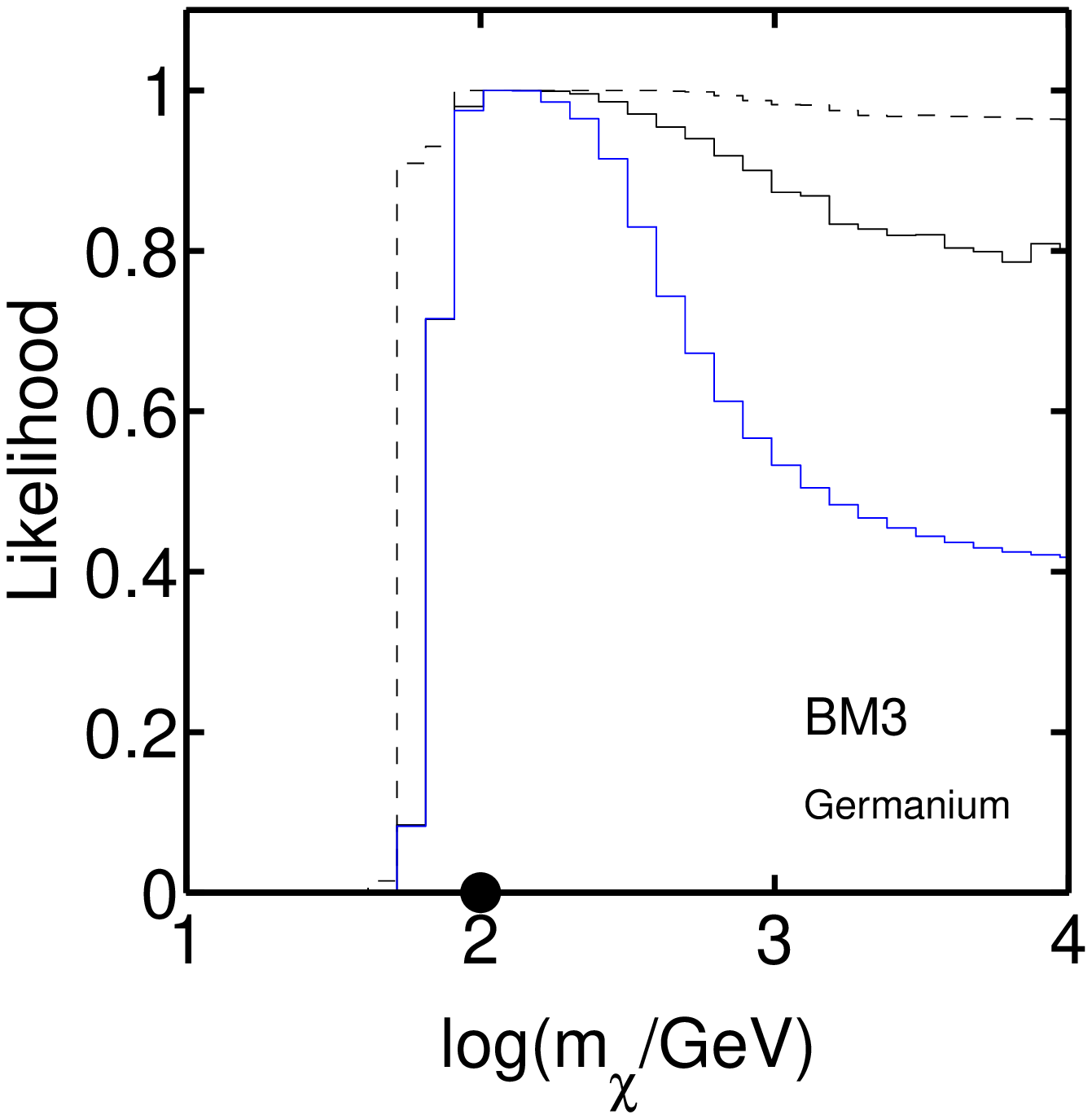}\hspace*{-0.62cm}
\includegraphics[width=0.35\textwidth]{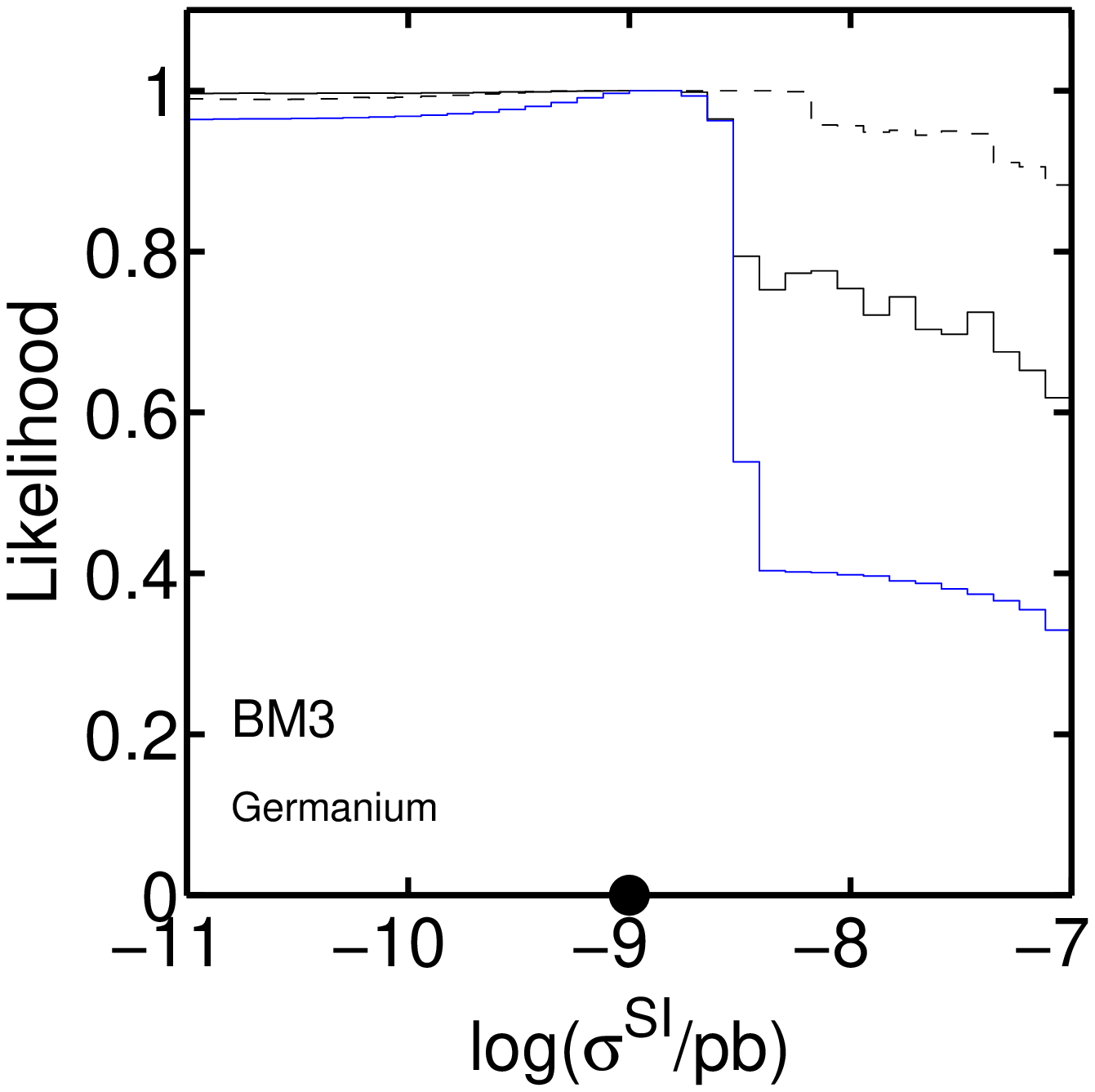}\hspace*{-0.62cm}
\includegraphics[width=0.35\textwidth]{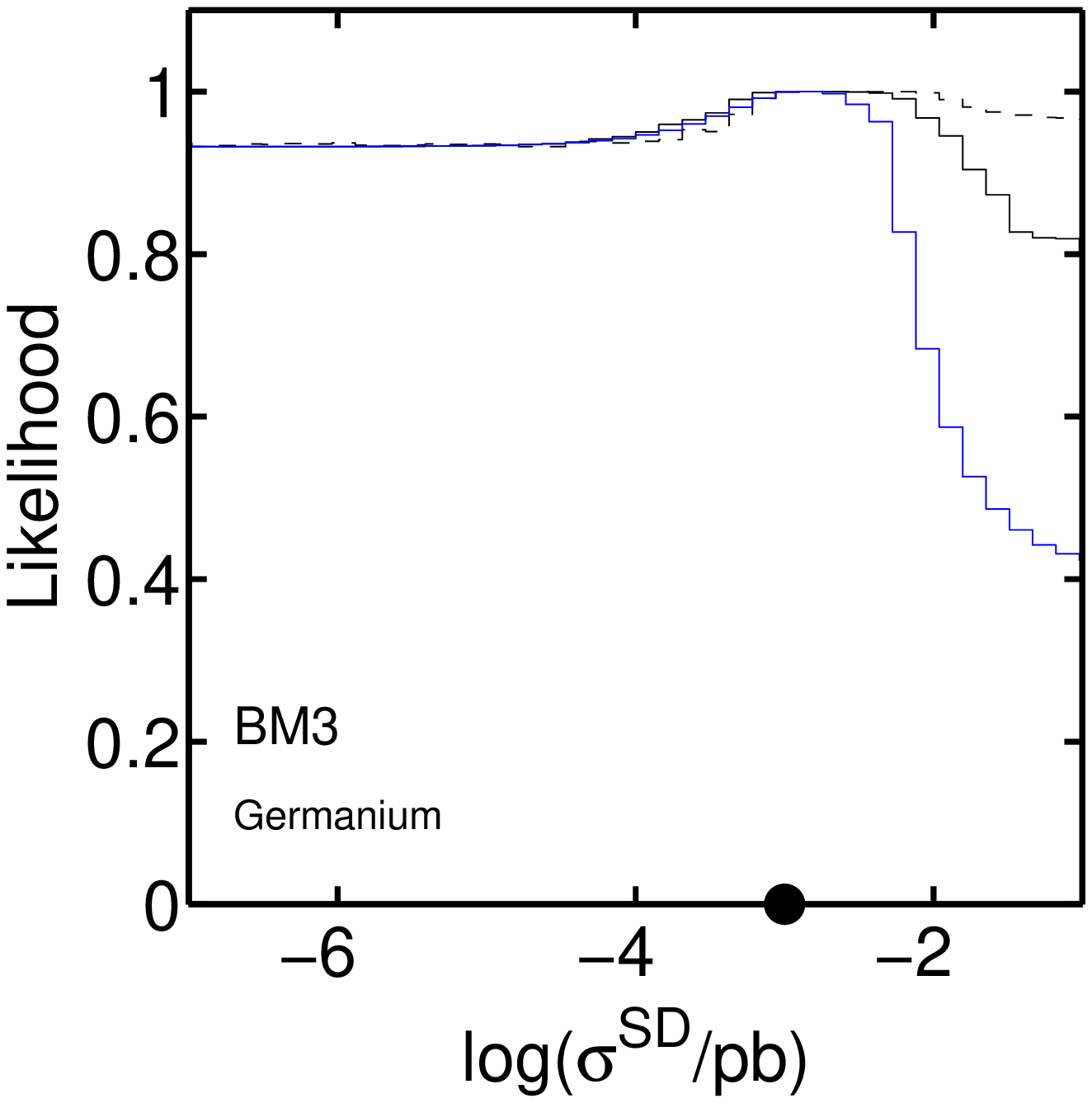}
\caption{One-dimensional profile likelihood for $\mwimp$, Ê$\sigsi$, and $\sigsd$ in BM1, BM2, and BM3 from top to bottom, respectively in the case of a germanium detector. The solid blue line corresponds to the case without uncertainties, the black solid line represents the results when nuclear uncertainties in the SDSF are included, and the dashed black line denotes the case when astrophysical uncertainties are included. The black dot represents the benchmark value of the parameters.
\label{fig:astro}}
\end{figure*}

\begin{figure*}
\includegraphics[width=0.35\textwidth]{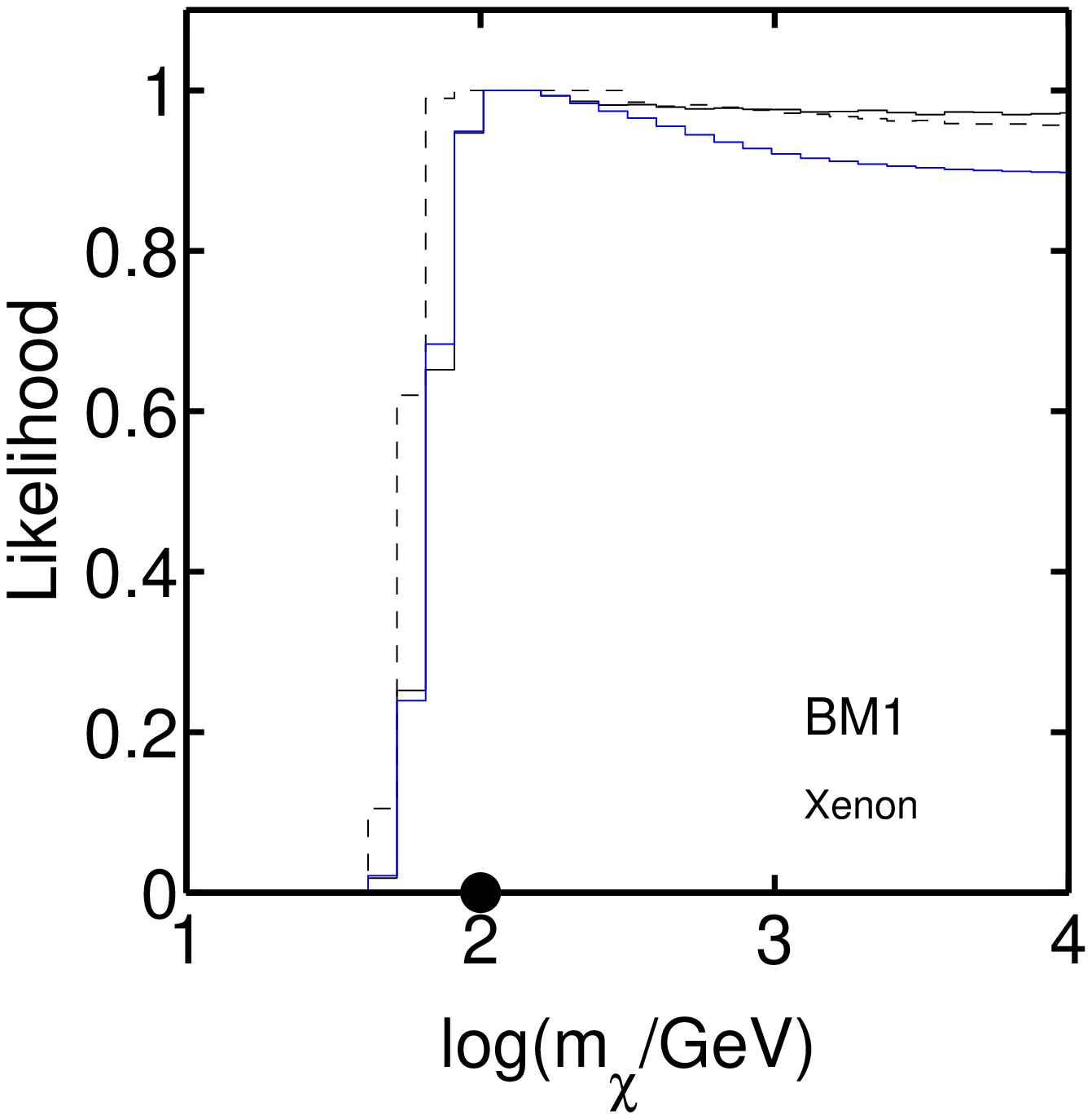}\hspace*{-0.62cm}
\includegraphics[width=0.35\textwidth]{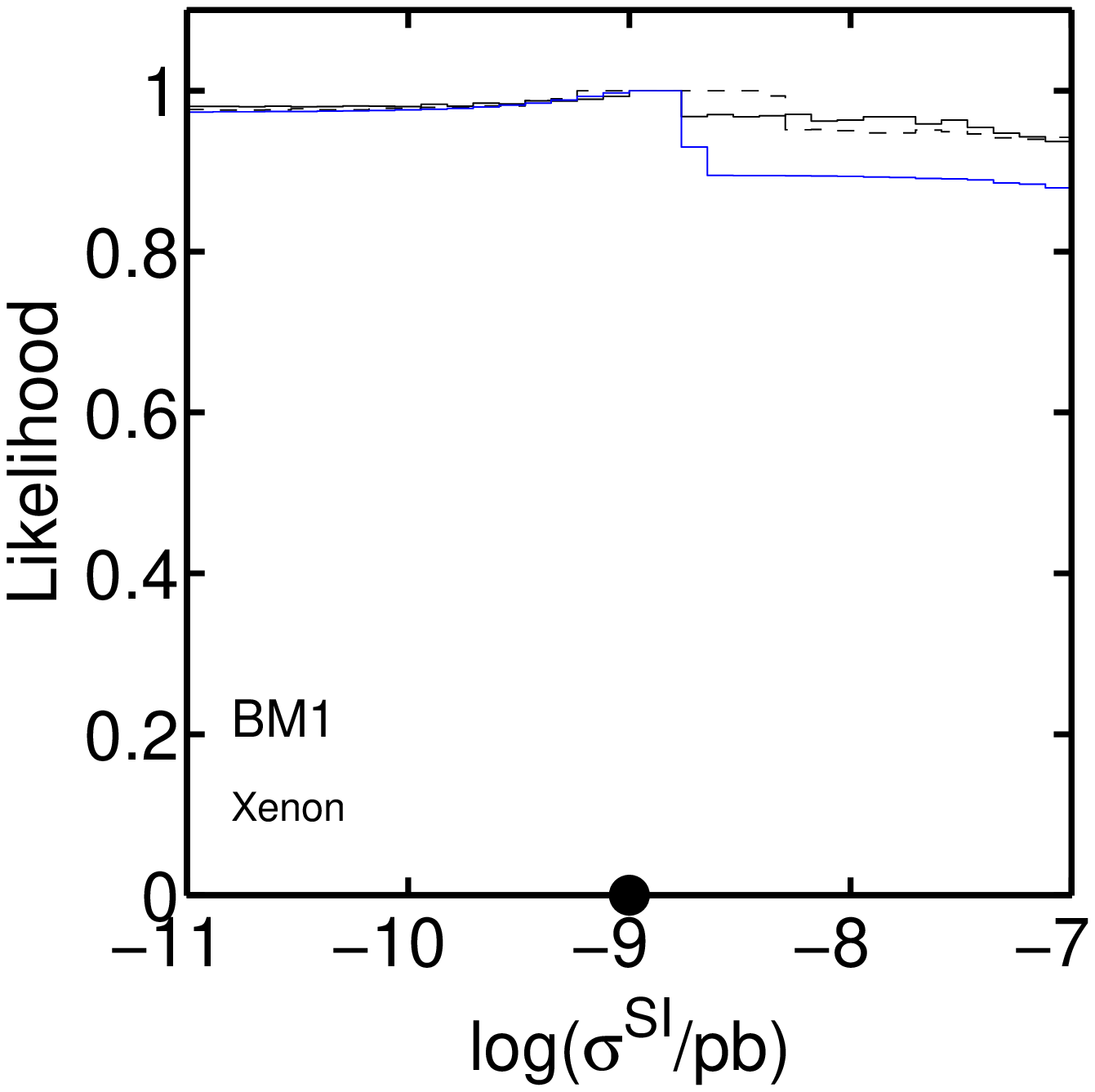}\hspace*{-0.62cm}
\includegraphics[width=0.35\textwidth]{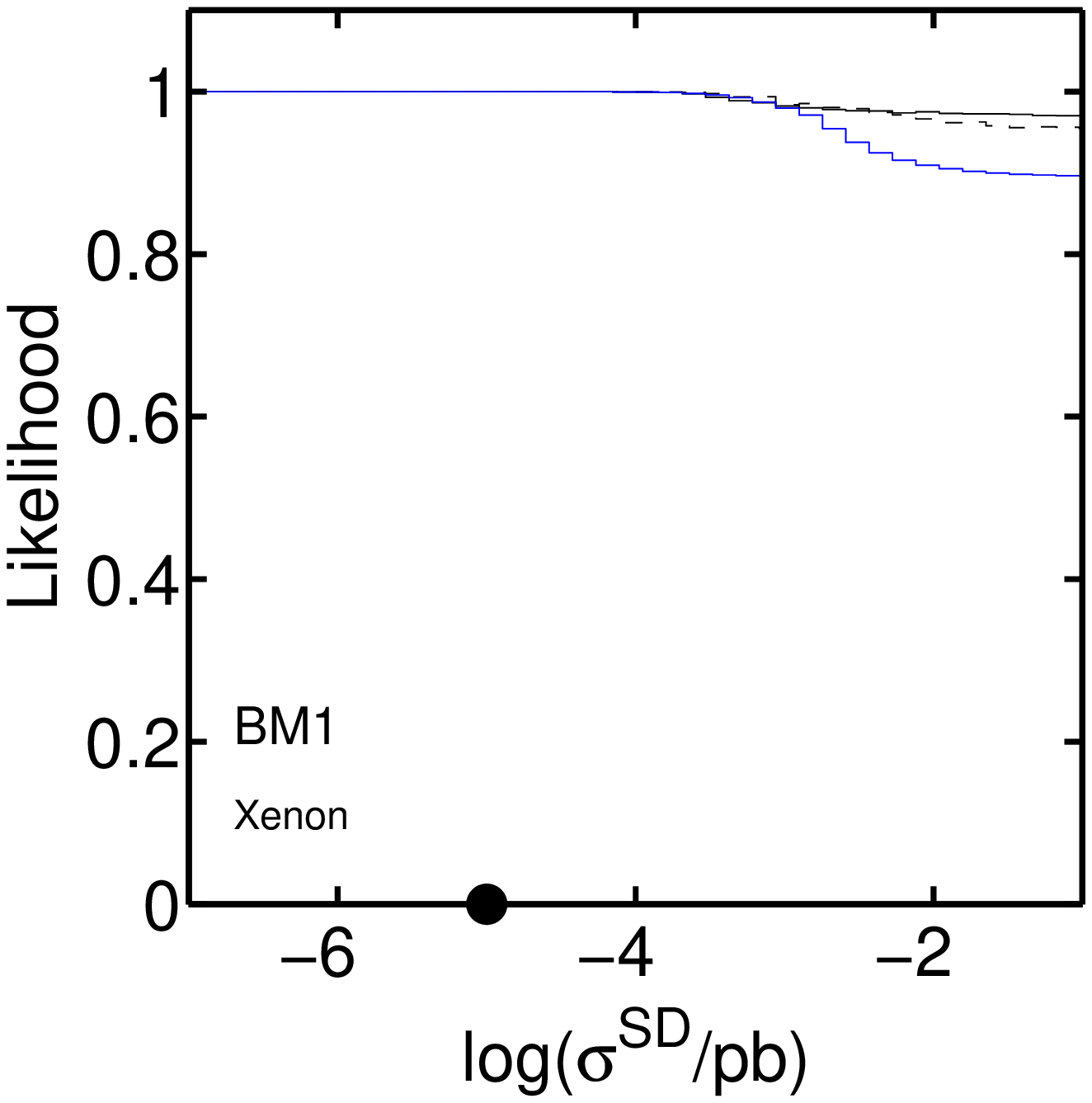}\\[-4.ex]
\includegraphics[width=0.35\textwidth]{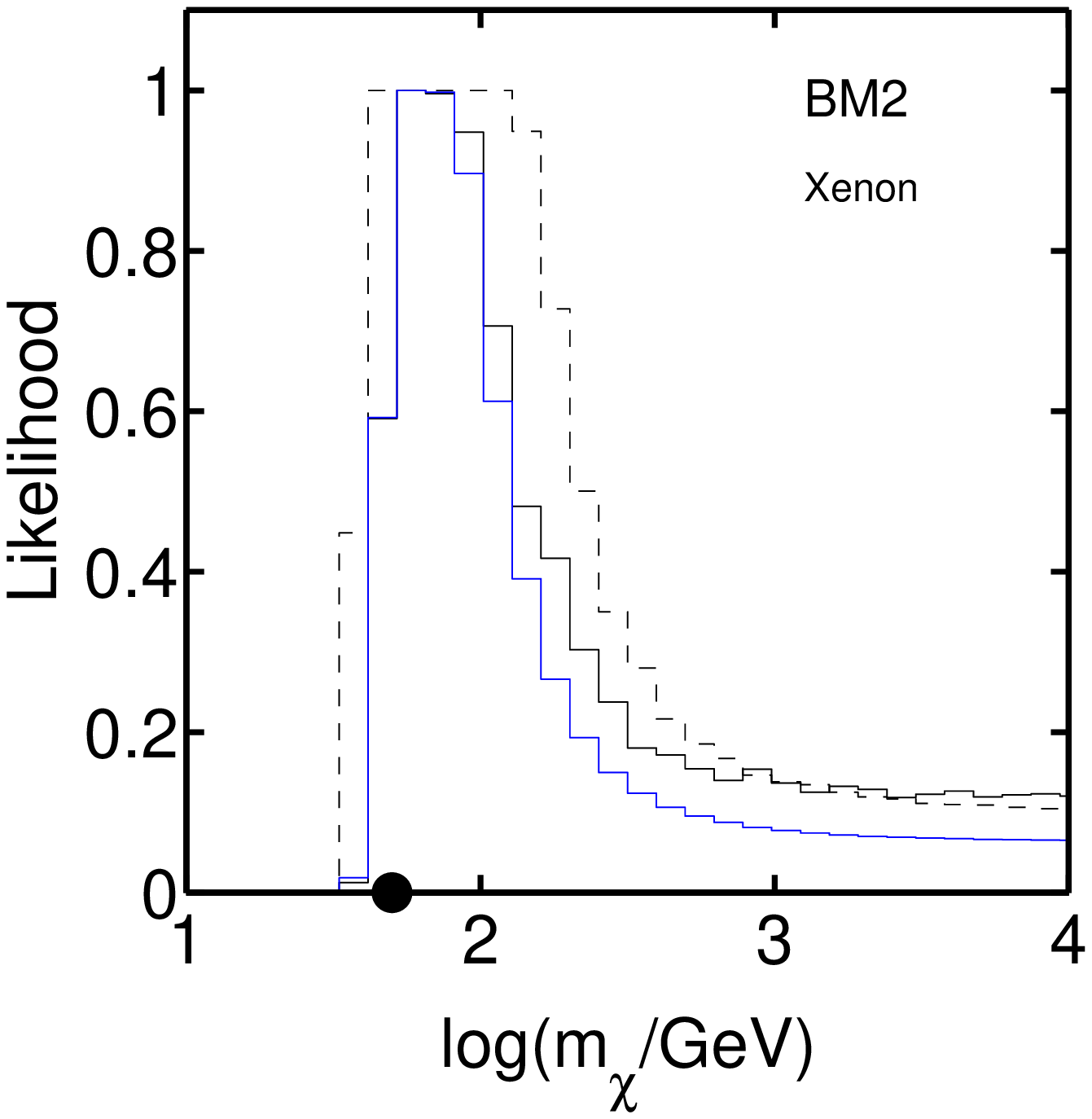}\hspace*{-0.62cm}
\includegraphics[width=0.35\textwidth]{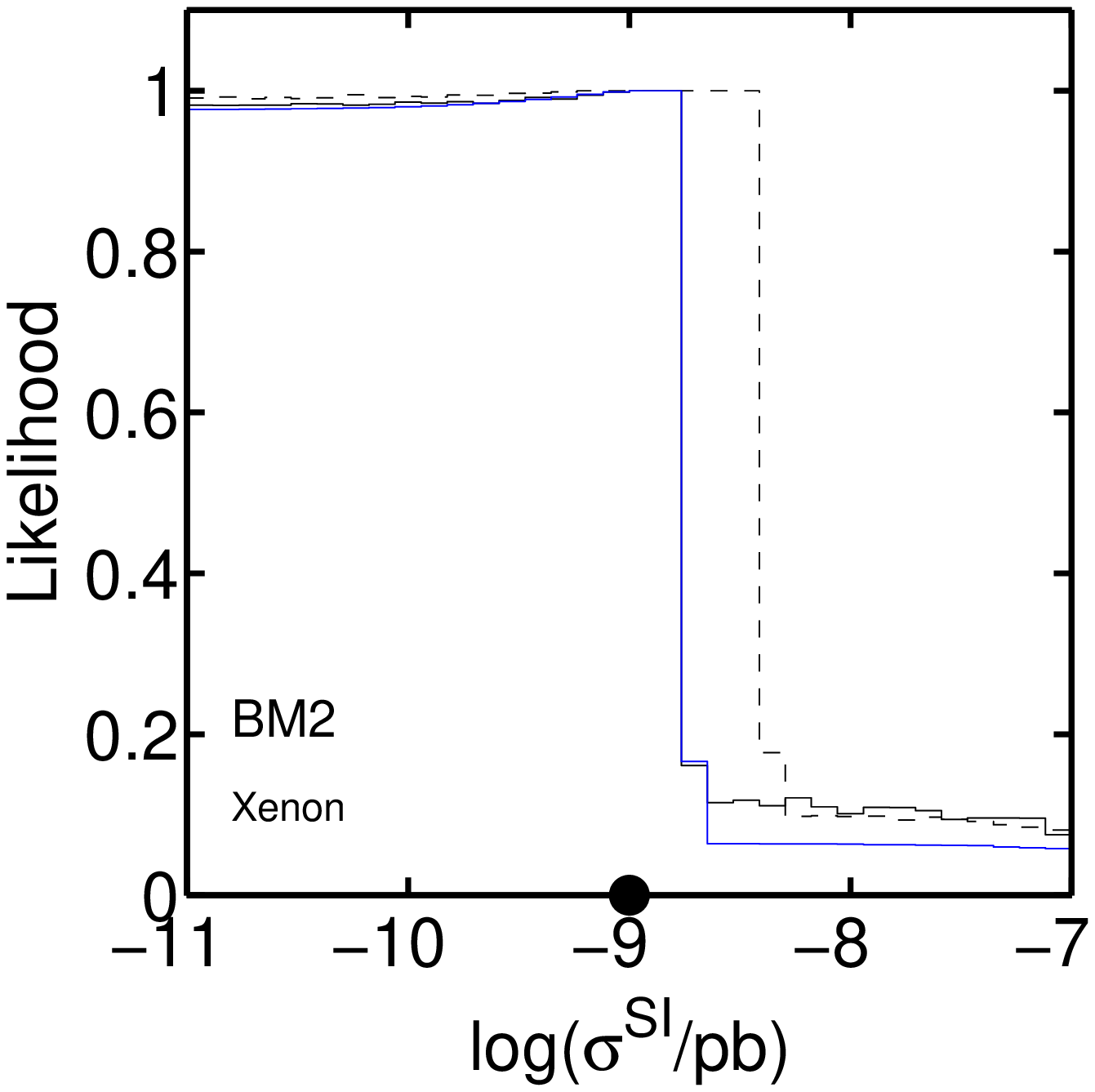}\hspace*{-0.62cm}
\includegraphics[width=0.35\textwidth]{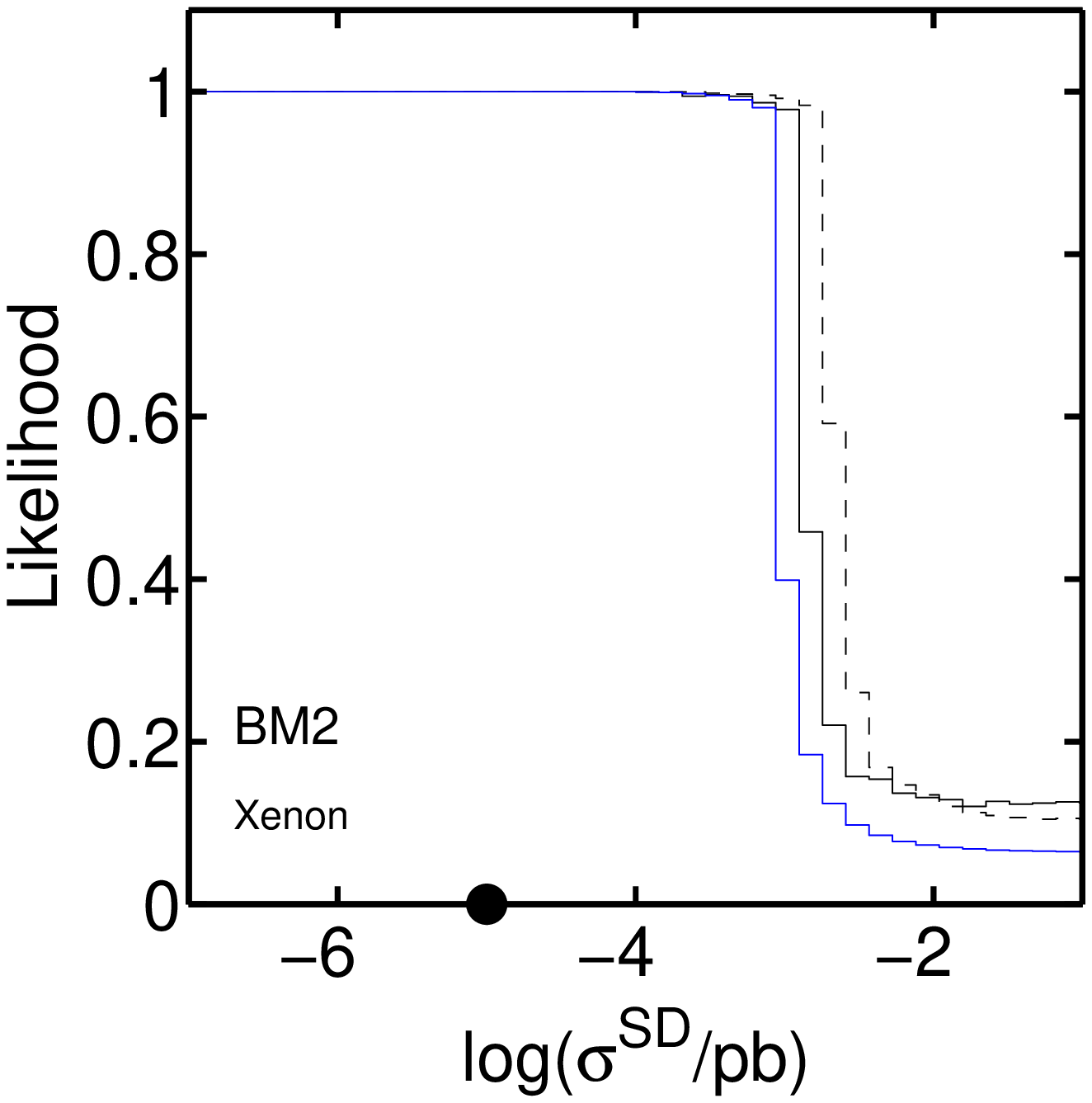}\\[-4.ex]
\includegraphics[width=0.35\textwidth]{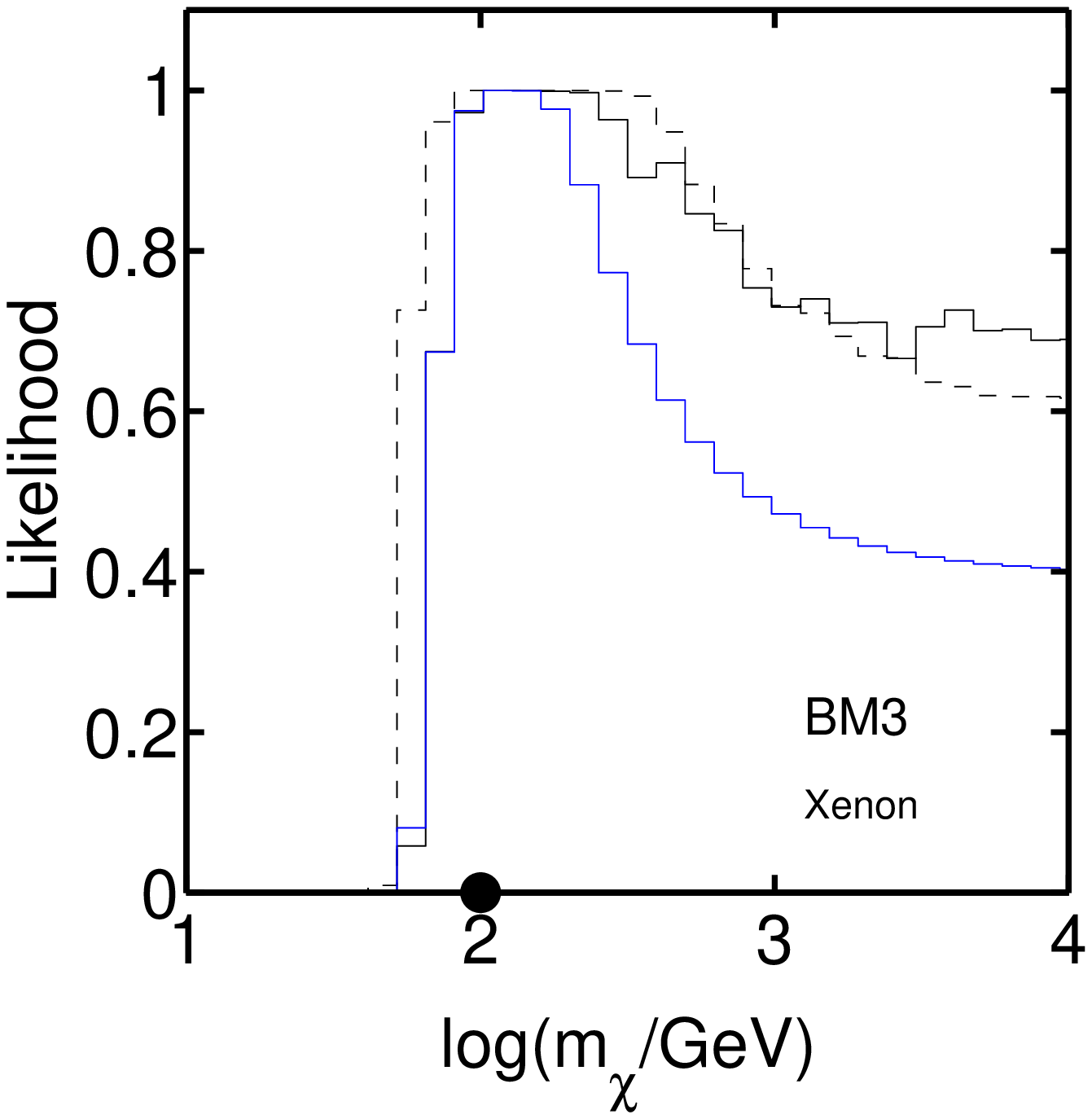}\hspace*{-0.62cm}
\includegraphics[width=0.35\textwidth]{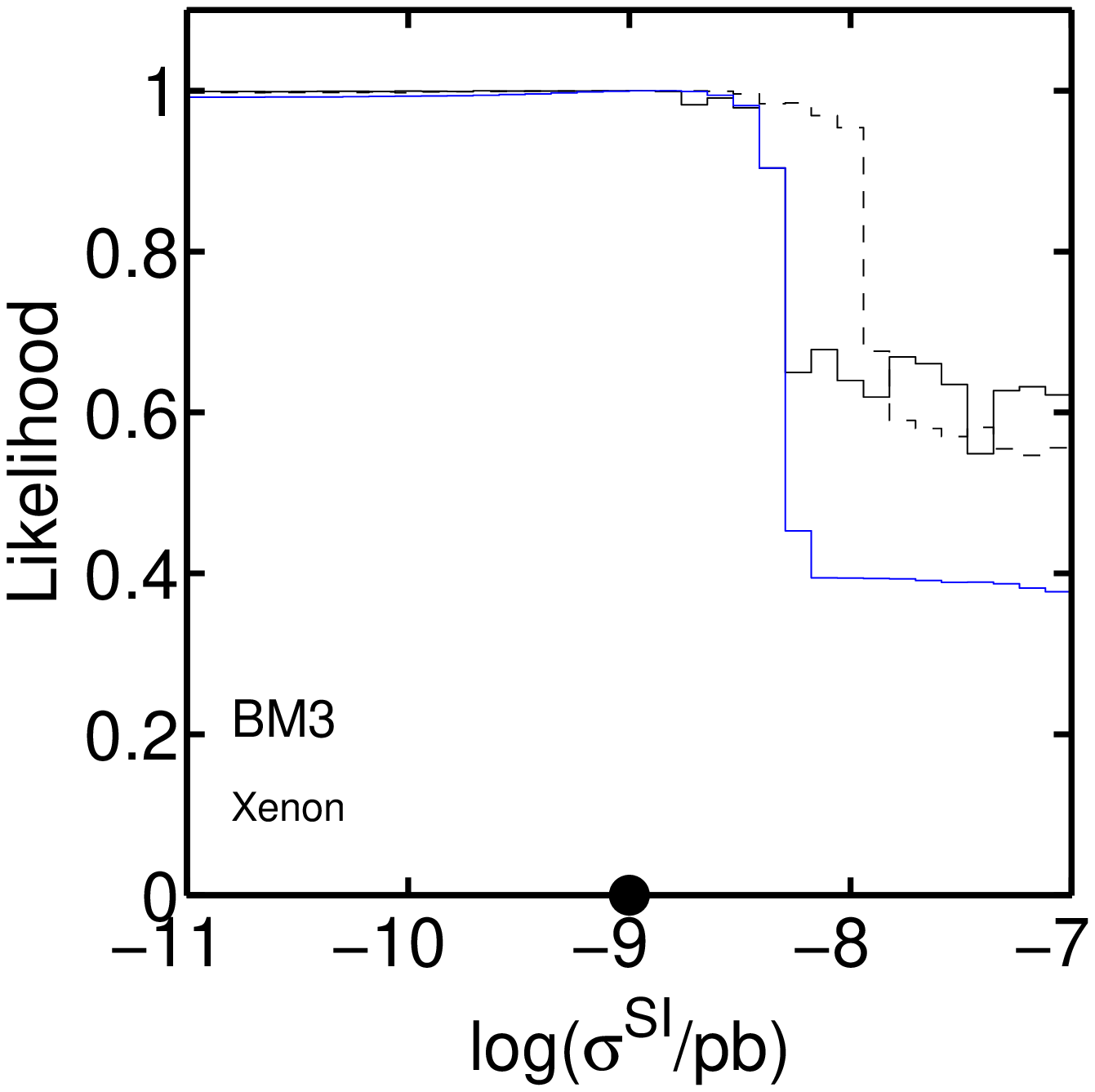}\hspace*{-0.62cm}
\includegraphics[width=0.35\textwidth]{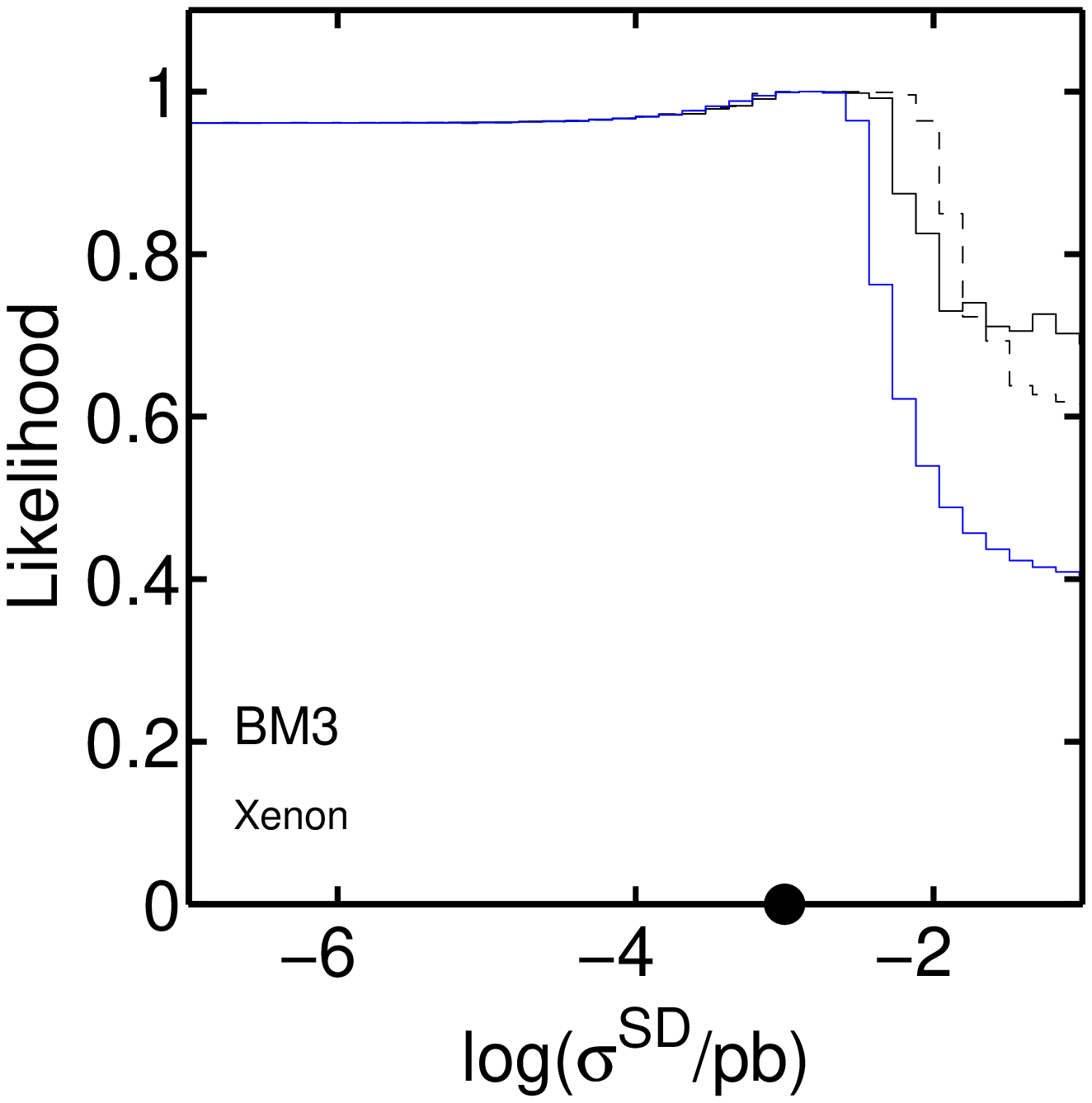}
\caption{The same as in Fig.\,\ref{fig:astro} but for the case of a xenon detector.
\label{fig:astro-xenon}}
\end{figure*}

In Figs.\,\ref{fig:astro} and \,\ref{fig:astro-xenon} we represent the one-dimensional profile likelihood for the DM parameters ($\mwimp$, $\sigsi$, $\sigsd$) for benchmarks BM1, BM2 and BM3 in the cases of a germanium and xenon detector, respectively. We display the reconstruction when no uncertainties are considered (blue line), when only nuclear uncertainties in the SDSF are included (solid black line) and when only astrophysical uncertainties are included (dashed black line). 
As noted before, the effect of nuclear uncertainties in the SDSF is more evident for BM3 than in BM1 and BM2 in both germanium and xenon, since in the latter the SD component is more important. 
The prediction for the WIMP mass is extended towards larger masses, and as we see for BM3 the effect cannot generally be neglected. Similarly, the predicted $\sigsd$ can vary significantly. In BM3 the reconstruction extends towards larger values (in BM2 and BM1 the effect is smaller).
On the other hand, astrophysical uncertainties affect both the reconstruction of the three DM parameters, $\mwimp$, $\sigsi$ and $\sigsd$ and are equally relevant, irrespectively of whether the main contribution comes from the SD or SI component. We can see how
nuclear uncertainties generally
have a smaller effect than astrophysical ones, but they can be comparable in some benchmark scenarios, especially regarding the mass reconstruction. This is the case, e.g., of BM3 in a xenon detector.

\section{Conclusions}
\label{sec:conclusions}

We have studied the effect that uncertainties in the nuclear spin-dependent structure functions have in the reconstruction of DM properties by means of direct detection experiments.

Assuming a hypothetical future observation of DM in a direct detection experiment we have systematically investigated how well its phenomenological parameters $(\mwimp,\,\sigsi,\,\sigsd)$ can be determined when uncertainties in the SD form factors of the target nuclei are taken into account. 
We focused at first on the case of a germanium target and considered two possible models describing the SDSF of its isotope $^{73}$Ge, sensitive to SD WIMP couplings. 
Using a Bayesian inference algorithm we determined for each of these models the pdf and profile likelihood of the DM parameters in a set of benchmark scenarios.
We observed that if a model is chosen to describe the SDSF of a particular nucleus, the reconstruction of the DM properties can strongly depend on the choice made (see in this sense the comparison between the predictions using the R- or D-model in Figs.\,\ref{fig:BM3_profl} and \ref{fig:BM3_pdf}). In particular, 
differences in the reconstructed values of the WIMP mass as well as the SD component of the WIMP-nucleon scattering cross-section appear.
In general these effects are more important when the SD contribution to the total detection rate is not negligible.

In the second part of the paper we have proposed a description of the SD structure functions in terms of three parameters which fit the zero-momentum value and the slope of the SDSF, and account for the presence of a high-momentum tail. 
This allows us to include uncertainties in the SDSF in the sampling of the parameter space and treat them in a consistent and systematic way.
Using this method we have computed the profile likelihood for the DM parameters for the same three benchmark points as before, in the case of a germanium-based and a xenon-based detector.

Finally, we have explicitly compared the effect of nuclear uncertainties in the SDSF with those that are associated with the parameters of the halo of dark matter. We find that uncertainties in the SDSF can even be comparable in magnitude to astrophysical ones when the SD contribution to the total detection rate is sizable.

\vspace*{2ex}
\noindent {\bf Acknowledgements}\\[0.5ex]
We have greatly benefited from discussions with A.M.~Green, B.~Kavanagh and L.~Robledo. We are grateful to A.~D\'iaz-Gil for technical support with the computing facilities at the Instituto de F\'isica Te\'orica and H.D.~Kim and the Seoul National University for allowing us to use their computational facilities in the last stages of the project.
D.G.C. is supported by the Ram\'on y Cajal program of the Spanish MICINN. 
M.F is supported by a Leverhulme Trust grant.
J.-H.H. is supported by a MultiDark Fellowship. 
M.P. is supported by a MultiDark Scholarship.
This work was supported by the Consolider-Ingenio 2010 Programme under grant MultiDark CSD2009-00064. We also thank the support of the Spanish MICINN under grant FPA2009-08958, the Community of Madrid under grant HEPHACOS S2009/ESP-1473, and the European Union under the Marie Curie-ITN program PITN-GA-2009-237920.

\end{document}